\documentclass[twocolumn]{aastex631}
\usepackage{natbib}
\usepackage{graphicx}
\usepackage{dcolumn}
\usepackage{bm}
\usepackage[nointegrals]{wasysym}
\usepackage{verbatim}
\usepackage{color}
\usepackage{amsmath}
\usepackage{tabularx}
\usepackage[utf8]{inputenc}
\usepackage{hyperref}
\usepackage{multirow}
\usepackage{gensymb}
\newcommand{\tl}[1]{{\textcolor{black}{{#1}}}}
\newcommand{\jd}[1]{{\textcolor{black}{{#1}}}}

\newcommand{\ba}{\begin{eqnarray}}
\newcommand{\ea}{\end{eqnarray}}
\newcommand{\be}{\begin{equation}}
\newcommand{\ee}{\end{equation}}
\newcommand{\LCDM}{$\Lambda$CDM}

\newcommand{\cskip}{\multicolumn{1}{c}{}}
\newcommand{\act}{\textsf{ACT}}
\newcommand{\pact}{\textsf{P-ACT}}
\newcommand{\pactlb}{\textsf{P-ACT-LB}}
\newcommand{\pactlbb}{\textsf{P-ACT-LB2}}

\newcommand{\wact}{\textsf{W-ACT}}

\newcommand{\acttt}{\textsf{ACT-TT}}
\newcommand{\actte}{\textsf{ACT-TE}}
\newcommand{\actee}{\textsf{ACT-EE}}
\newcommand{\pacttt}{\textsf{P-ACT-TT}}
\newcommand{\pactte}{\textsf{P-ACT-TE}}
\newcommand{\pactee}{\textsf{P-ACT-EE}}
\newcommand{\acttp}{\textsf{ACT+$\tau$-prior}}
\newcommand{\Planck}{{\it Planck}}
\newcommand{\WMAP}{{\it WMAP}}

\usepackage{aas_macros}

\begin{document}

\title{The Atacama Cosmology Telescope: DR6 Power Spectra, Likelihoods and \LCDM\ Parameters}
 \shorttitle{}
  \shortauthors{.}
\begin{abstract}
We present power spectra of the cosmic microwave background (CMB) anisotropy in temperature and polarization, measured from the Data Release 6 maps made from Atacama Cosmology Telescope (ACT) data. These cover 19,000 deg$^2$ of sky in bands centered at 98, 150 and 220 GHz, with white noise levels three times lower than \Planck\ in polarization. We find that the ACT angular power spectra estimated over 10,000~deg$^2$, and measured to arcminute scales in TT, TE and EE, are well fit by the sum of CMB and foregrounds, where the CMB spectra are described by the \LCDM\ model. Combining ACT with larger-scale \Planck\ data, the joint P-ACT dataset provides tight limits on the ingredients, expansion rate, and initial conditions of the universe. We find similar constraining power, and consistent results, from either the \Planck\ power spectra or from ACT combined with \WMAP\ data, as well as from either temperature or polarization in the joint P-ACT dataset. When combined with CMB lensing from ACT and \Planck, and baryon acoustic oscillation data from the Dark Energy Spectroscopic Instrument (DESI \jd{DR1}), we measure a baryon density of $\Omega_b h^2=0.0226\pm0.0001$, a cold dark matter density of $\Omega_c h^2=0.118\pm0.001$, a Hubble constant of $H_0=68.22\pm0.36$~km/s/Mpc, a spectral index of $n_s=0.974\pm0.003$, and an amplitude of density fluctuations of $\sigma_8=0.813\pm0.005$. Including the DESI DR2 data tightens the Hubble constant to $H_0=68.43\pm0.27$~km/s/Mpc; \LCDM\ parameters agree between the P-ACT and DESI DR2 data at the $1.6\sigma$ level. We find no evidence for excess lensing in the power spectrum, and no departure from spatial flatness. The contribution from Sunyaev-Zel'dovich (SZ) anisotropy is detected at high significance; we find evidence for a tilt with suppressed small-scale power compared to our baseline SZ template spectrum, consistent with hydrodynamical simulations with feedback.
\end{abstract}

\suppressAffiliations
\author[0000-0002-6849-4217]{Thibaut~Louis} \affiliation{Universit\'e Paris-Saclay, CNRS/IN2P3, IJCLab, 91405 Orsay, France}
\author[0000-0002-2613-2445]{Adrien~La~Posta} \affiliation{Department of Physics, University of Oxford, Keble Road, Oxford, UK OX1 3RH}
\author[0000-0002-2287-1603]{Zachary~Atkins} \affiliation{Joseph Henry Laboratories of Physics, Jadwin Hall, Princeton University, Princeton, NJ, USA 08544}
\author[0000-0002-9429-0015]{Hidde~T.~Jense} \affiliation{School of Physics and Astronomy, Cardiff University, The Parade, Cardiff, Wales, UK CF24 3AA}
\author[0000-0003-3230-4589]{Irene~Abril-Cabezas} \affiliation{DAMTP, Centre for Mathematical Sciences, University of Cambridge, Wilberforce Road, Cambridge CB3 OWA, UK} \affiliation{Kavli Institute for Cosmology Cambridge, Madingley Road, Cambridge CB3 0HA, UK}
\author[0000-0002-2147-2248]{Graeme~E.~Addison} \affiliation{Dept. of Physics and Astronomy, The Johns Hopkins University, 3400 N. Charles St., Baltimore, MD, USA 21218-2686}
\author[0000-0002-5127-0401]{Peter~A.~R.~Ade} \affiliation{School of Physics and Astronomy, Cardiff University, The Parade, Cardiff, Wales, UK CF24 3AA}
\author[0000-0002-1035-1854]{Simone~Aiola} \affiliation{Flatiron Institute, 162 5th Avenue, New York, NY 10010 USA} \affiliation{Joseph Henry Laboratories of Physics, Jadwin Hall, Princeton University, Princeton, NJ, USA 08544}
\author{Tommy~Alford} \affiliation{Department of Physics, University of Chicago, Chicago, IL 60637, USA}
\author[0000-0002-4598-9719]{David~Alonso} \affiliation{Department of Physics, University of Oxford, Keble Road, Oxford, UK OX1 3RH}
% ...

\date[]{\emph{Affiliations can be found at the end of the document}}



\setcounter{tocdepth}{2}
\tableofcontents
\vspace{0.2cm}
\section{Introduction}\label{sec:intro}

Measurements of the anisotropies in the cosmic microwave background (CMB) have been central to the establishment of the \LCDM\ cosmological model \citep[e.g.,][]{spergel2003,planck2018_cosmo}. The first anisotropy measurements were made of the CMB intensity; over the past twenty years the smaller polarization signal has been increasingly better characterized. Since the primary CMB signal is statistically isotropic and Gaussian distributed to within current measurement uncertainties, the angular power spectrum statistic captures the majority of the primordial information in the sky maps.

The \Planck\ satellite measured the CMB intensity and polarization anisotropies over the whole sky in nine frequency bands, with up to 80\% of the area used to estimate the power spectra \citep{planck_spectra:2019,rosenberg:2022,hillipop2024}. Analyses show that the six-parameter \LCDM\ model is a good fit to the \Planck\ data, and that model parameters estimated from the TT and TE spectra are consistent and have similar constraining power \citep{planck2018_cosmo}. The \Planck\ measurement is noise-dominated for scales $\ell>2000$ in intensity, and $\ell>800$ in polarization. During the past decade the ground-based Atacama Cosmology Telescope (ACT) and South Pole Telescope (SPT) experiments have extended the reach to smaller scales with increasingly refined measurements \citep[e.g.,][]{henning/etal:2018, choi_atacama_2020,dutcher2021,balkenhol2023,2024arXiv241106000G}.

Despite the success of the \LCDM\ model, there are some late-time astronomical data that disfavor the best-fit parameters derived from CMB data at the 2--5$\sigma$ level -- most notably the local Hubble constant measurement from Type Ia supernovae calibrated with Cepheid variable stars \citep{Riess_SHOES,Breuval:2024lsv} -- that may indicate a missing element from the cosmological model. A recent analysis using James Webb Space Telescope (JWST) data finds local Hubble constant measurements consistent with CMB and large-scale structure measurements \citep{freedman2024};  results for the Hubble constant from a wide array of methods are surveyed in \cite{verde2024}. As the sensitivity of the measurements improve, the data may also require the introduction of new features in the model, such as in the behavior of neutrinos, cold dark matter, dark energy, or primordial perturbations. There is therefore a strong motivation to continue testing this model with new data.

In this paper we present new power spectrum measurements, and \LCDM\ model parameters, from the ACT Data Release 6 (DR6) dataset. This includes data gathered from 2017 until the experiment’s completion in 2022. The DR6 maps cover 45\% of the sky; we use 25\% of the sky for this analysis after masking the Galaxy and extragalactic sources. We conservatively use only data gathered in the nighttime when the instrument beam was most stable. The ACT white noise levels are typically three times lower than \Planck's in polarization, and the maps have five times better angular resolution, extending cosmic variance limited E-mode measurements up to $\ell=1700$ over this region of the sky.

We find strong agreement of our spectra with those from \Planck\ over a common sky region, and a \LCDM\ model that fits the \Planck\ data is a good fit to the ACT data over this broader range of scales. We also find that \LCDM\ model parameters determined by a combination of \WMAP\ and ACT, which is independent of \Planck, are consistent with those determined by \Planck\ alone. This consistency between all of the datasets motivates forming an optimal combination for determining parameters. Our new nominal state-of-the-art CMB dataset uses the ACT data combined with larger-scale \Planck\ data. We estimate parameters from this dataset, and check that the \LCDM\ model is consistent from TT, TE, and EE independently; the EE precision is particularly improved compared to \Planck.  We then combine these data with later-time measurements of structure growth measured using CMB lensing from both ACT and \Planck\ \citep{carron2022,Qu_dr6_lensing,Madhavacheril_dr6_lensing}, and measurements of galaxy clustering using baryon acoustic oscillation (BAO) data from the Dark Energy Spectroscopic Instrument (DESI) \citep{DESI-BAO-III,DESI1_DR2:2025,DESI2_DR2:2025}.\footnote{The DESI DR2 data appeared as this work was submitted; our main results use the DR1 data, with some updates for the DR2 data reported in the main text and in Appendix \ref{apx:desi}.}  

This paper is part of a suite of ACT DR6 papers, with the maps presented in \citet[N25 hereafter]{dr6maps}. It builds on earlier ACT power spectrum and parameter analyses from data gathered from 2008-16 and released as DR1--DR4 \citep{das/2011, dunkley/etal:2011, sievers/etal:2013,louis_atacama_2017,choi_atacama_2020,aiola/etal:2020}. Companion papers present constraints on a broad set of cosmological models \citep[][C25 hereafter]{dr6extend}, covariance matrix estimation \citep{atkins/etal:2024}, beam measurement \citep{beams_inprep} and foreground modeling \citep{fg_inprep}. The broader set of ACT DR6 papers is summarized in N25 and includes noise simulations \citep{atkins/etal:2023}, CMB lensing maps and interpretation \citep{Qu_dr6_lensing, 2024ApJ...966..138M, Madhavacheril_dr6_lensing}, component-separated CMB and Compton y-maps \citep{act_ymap:2024}, studies of millimeter transients \citep[e.g.,][]{act_transient:2023}, and upcoming cluster and source catalogs. The DR6 data are publicly available and we accompany this paper with the power spectrum pipeline, \texttt{PSpipe}, that was used to derive our presented results.\footnote{\href{https://github.com/simonsobs/PSpipe}{\texttt{PSpipe}}. This includes code for estimating power spectra and covariance matrices, correcting for temperature-to-polarization leakage, fitting polarization angles, performing comparisons with \Planck, calibrating, and reproducing many of the figures presented in this paper.}

The paper is organized as follows. In \S\ref{sec:summary} we provide a summary of the highlights. In \S\ref{sec:ps} we describe the methods for estimating power spectra, \S\ref{sec:ps_result} shows the new ACT power spectra, and in \S\ref{sec:ACT_Planck} we compare our measurements to the \Planck\ data. Our blinding procedure is described in Appendix \ref{apx:blinding}. \S\ref{sec:like} describes the likelihood for the multi-frequency power spectra and the estimation of CMB bandpowers, and \S\ref{sec:params} tests the method for estimating parameters. In \S\ref{sec:results} we show constraints on the \LCDM\ model, test the degree of lensing and the contribution from Sunyaev-Zel'dovich (SZ) anisotropies. We conclude in \S\ref{sec:conclude}. 

\begin{figure*}[htp]
	\centering
	\hspace*{-5mm}\includegraphics[width=\textwidth]{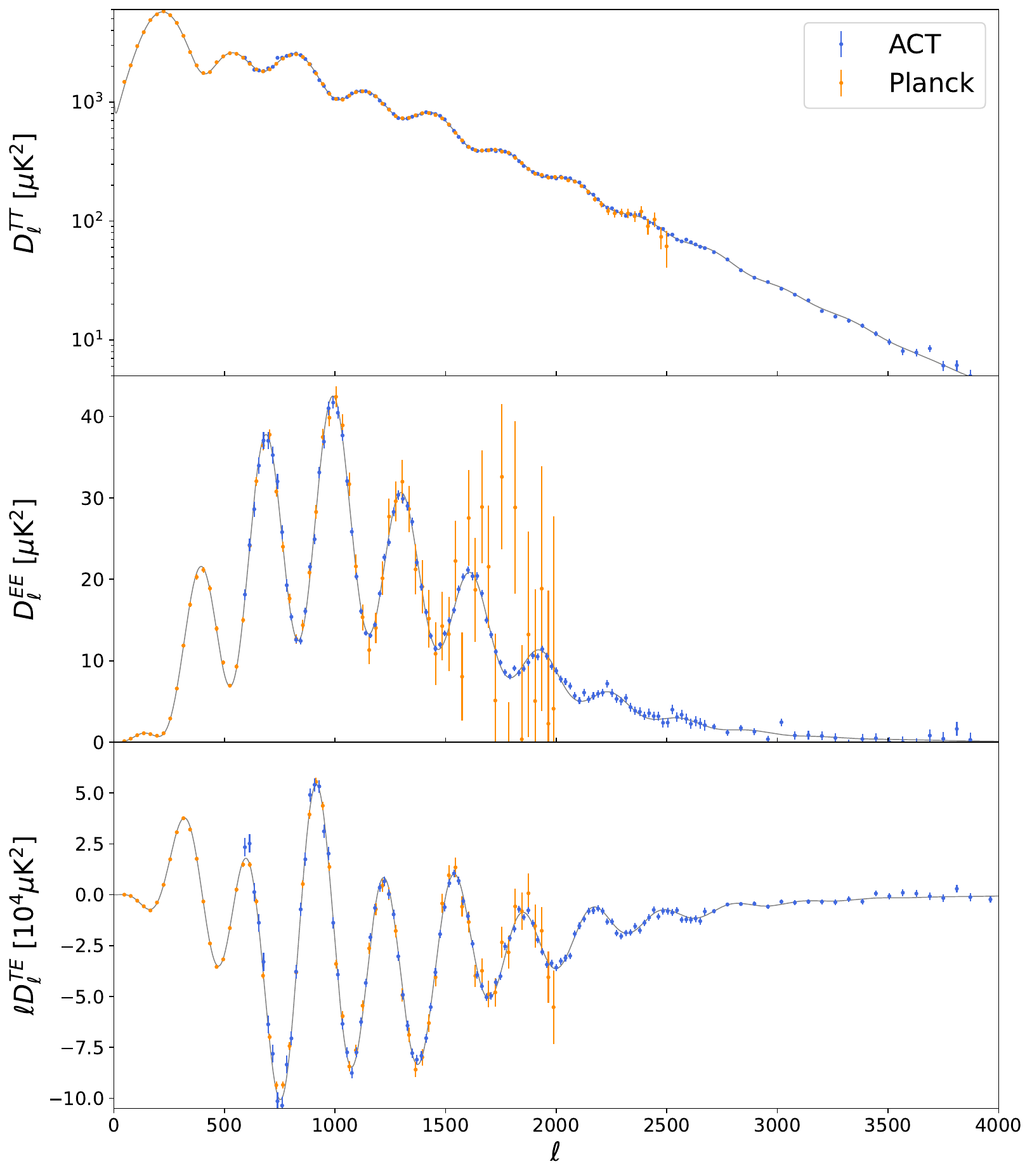}
	\caption{ACT DR6 and \Planck\ PR3 \citep{planck_spectra:2019} combined TT (top), EE (middle), and TE (bottom) power spectra. The gray lines show the joint ACT and \Planck\ (P-ACT) \LCDM\ best-fit power spectra. For plotting purposes we have subtracted the best-fit foreground power spectra. The full ACT multi-frequency spectra extend to $\ell=8500$. The \LCDM\ model provides an excellent fit to both data sets.}
	\label{fig:combined_TE_EE}
\end{figure*}

\section{Summary of key results}
\label{sec:summary}

\begin{figure*}[htp]
	\centering
	\hspace*{-5mm}\includegraphics[width=0.8\textwidth]{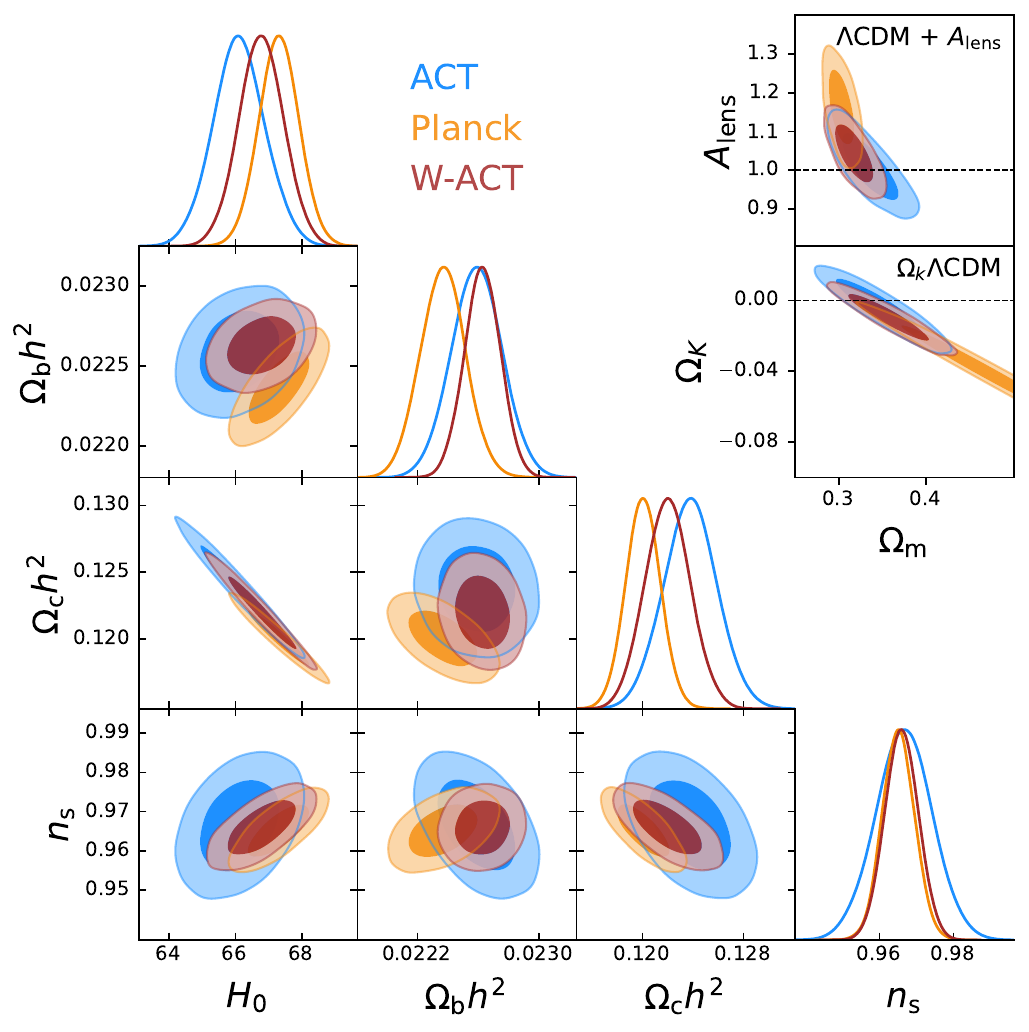}
    \includegraphics[width=0.8\textwidth]{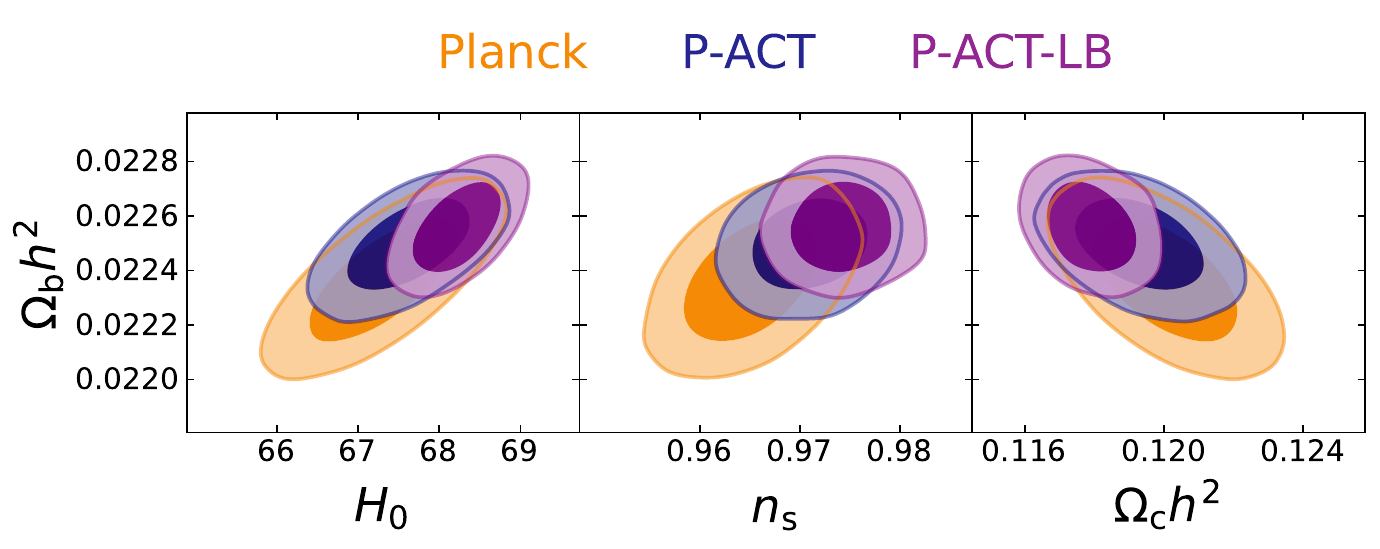}
    \caption{(Main/top) Subset of the \LCDM\ cosmological parameters estimated from ACT, and ACT combined with \WMAP\ large-scale data (W-ACT), compared to results from the \Planck\ PR3 power spectra. We show the Hubble constant, $H_0$, the baryon density, $\Omega_b h^2$, the cold dark matter density $\Omega_c h^2$, and the scalar spectral index, $n_s$. The optical depth is constrained using \Planck\ large-scale polarization data from Sroll2. (Top right) Distributions showing the correlation of the matter density, $\Omega_m$, with the lensing parameter, $A_{\rm lens}$ (defined in \S\ref{subsec:lensed_cmb}), or the curvature $\Omega_{K}$, when each are added as extensions to the \LCDM\ model; we see no departures from the expected lensing, or from spatial flatness, with ACT or W-ACT. (Bottom row) Subset of parameter distributions when ACT and \Planck\ PR3 are combined, with small-scale \Planck\ data removed (P-ACT, defined in \S\ref{subsec:actplanck}), and when including CMB lensing from ACT and \Planck, and baryon acoustic oscillation data from DESI \jd{DR1} (P-ACT-LB). A version using the DESI DR2 data is in Appendix \ref{apx:desi}.}
	\label{fig:LCDM}
\end{figure*}

In this section we highlight the main results of the paper, that are described in \S\ref{sec:ps_result}, \ref{sec:ACT_Planck} and \ref{sec:results}.

\begin{itemize}
\itemsep0em
\item We measure the TT, TE and EE angular power spectra to arcminute scales at six cross-frequencies, described in \S\ref{sec:ps_result}. Figure \ref{fig:combined_TE_EE} shows the frequency-combined CMB angular power spectra from ACT with the estimated foreground contamination removed, compared to \Planck. 
At small scales in TT the Silk-damped and lensed signal extends to scales of a few arcminutes. 

\item The EE power spectrum directly measures the velocity variations in the primordial plasma at recombination. The ACT data extend the measurement out to at least the eighth acoustic peak.\footnote{The ACT spectra have greater than unity signal-to-noise per multipole, defined as $C_\ell/\Delta C_\ell$, out to $\ell\approx2500$ in EE and $\ell\approx2700$ in TE.} We find consistent polarization spectra from \Planck\ and ACT in our overlapping sky region, described in \S \ref{sec:ACT_Planck}.

\item The TE spectrum captures the correlation between velocity and density variations in the recombination era. The ACT data have smaller errors than \Planck\ at scales $\ell>1000$ and reach arcminute scales.

\item The BB spectrum is consistent with the expected lensed \LCDM\ signal. The calibrated TB and EB spectra are best fit with a polarization rotation angle of $0.20^\circ \pm 0.08^\circ$. 
\end{itemize}

Fitting the ACT data with a CMB and foregound model, described in \S\ref{sec:like} and \ref{sec:params}, we find the following results, described in \S\ref{sec:results}:
\begin{itemize}
\itemsep0em
\item We find there is excellent agreement of the \LCDM\ model with the ACT data. We find consistent parameters estimated from ACT, either alone or combined with \WMAP\ larger-scale data, and \Planck, shown in Figure \ref{fig:LCDM}. The ACT data measure the acoustic peak scale to the same precision as \Planck.

\item We measure the baryon density with a 0.5\% uncertainty, and the acoustic peak scale to 0.02\%, when combining \Planck\ and ACT in our ``P-ACT" data combination. Parameters estimated from the polarization data are now competitive with those from the intensity anisotropy. We include CMB lensing from ACT and \Planck, and baryon acoustic oscillation data from DESI DR1 (``P-ACT-LB"). The spectral index is $n_s=0.974\pm 0.003$. 

\item The local Hubble constant is estimated to be $H_0=67.62\pm0.50$~km/s/Mpc from P-ACT, and $68.22\pm0.36$~km/s/Mpc combined with CMB lensing data and BAO from DESI \jd{DR1}\footnote{We find $H_0=68.43\pm0.27$~km/s/Mpc when substituting DESI DR1 for DR2. This combination is also reported in \cite{garcia-quintero:2025}.}.  This is in agreement with other early-universe data, and with  measurements from \cite{freedman2024}, and in strong disagreement with the measurements from \cite{Riess_SHOES,Breuval:2024lsv}.

\item We measure the anisotropy from the Sunyaev-Zel'dovich effect to have power $3.3\pm0.4~\mu$K$^2$ in the thermal signal at $\ell=3000$ at 150~GHz, and $<4~\mu$K$^2$ at the 95\% confidence level (CL) in kinematic power. We find evidence at 3$\sigma$ for a new parameter that tilts a standard template thermal SZ spectrum towards larger scales, giving a spectral shape consistent with simulations with enhanced feedback.
\end{itemize}

\section{Power spectrum methods}\label{sec:ps}

\subsection{Dataset and noise properties}\label{subsec:data_set}

As detailed in N25, the ACT Data Release 6 (DR6) comprises five years of observations collected between May 5, 2017, and July 2, 2022. These observations were conducted using the mid-frequency arrays PA5 and PA6, which operated at f150 (124–172 GHz) and f090 (77–112 GHz), and the high-frequency array PA4, which operated at f150 and f220 (182–277 GHz). Each frequency band for each array includes four independent maps (data splits), along with their associated inverse variance maps and cross-linking information. The dataset also features null test maps split by precipitable water vapor, elevation, observing time, and detector position.

This analysis uses five array-band combinations for temperature: PA4~f220, PA5~f090, PA5~f150, PA6~f090, and PA6~f150, and four array-band combinations for polarization: PA5~f090, PA5~f150, PA6~f090, and PA6~f150. We exclude PA4~f150 for both temperature and polarization analyses, as it fails multiple null tests including array comparisons with PA5~f150 and PA6~f150. PA4~f220 is not used for polarization analysis due to its limited constraining power relative to other array-bands.

\begin{table}[ht!]
\centering
\begin{center}
\begin{tabular}{ll|rr|rr}
\hline
	\textbf{array} & \textbf{band} & \multicolumn{2}{c|}{\textbf{baseline}} & \multicolumn{2}{c} {\textbf{extended}} \\
\hline
 & & \textbf{Temp} & \textbf{Pol} & \textbf{Temp} & \textbf{Pol} \\
PA4 & f220 &  $1000$ &  $...$ &   $1000$   &  $...$                \\
PA5 & f090 &  $1000$  & $1000$ & $1000$ & $500$  \\
PA5 & f150 &  $800$  & $800$ & $800$ & $500$  \\
PA6 & f090 &  $1000$  &$1000$ & $1000$ & $500$  \\
PA6 & f150 &  $600$  & $600$ & $600$ & $500$  \\
\hline
\end{tabular}
\end{center}
\caption{Minimum multipole used in this analysis, both for the baseline cut and the extended cut. The extended cut was used before unblinding the data. The maximum multipole used is $\ell_{\rm max} = 8500$.}
\label{tab:multipole_cut}
\end{table}

The multipole cuts used in this analysis are summarized in Table \ref{tab:multipole_cut}. We consider two sets of cuts: the baseline cut, employed for the main results of this paper, and an extended cut, which features a lower 
$\ell_{\rm min}$ in polarization.\footnote{The extended cut was used before unblinding the data.} These cuts were selected based on the results of various null tests, evaluated both at the spectral and parameter levels, as well as our assessment of the systematic error budget in our data. The motivations for these choices are discussed in \S\ref{subsec:covmat}, \S\ref{sec:ps_result} and \S\ref{sec:params}.

\begin{figure*}[htp]
	\centering
	\hspace*{-5mm}\includegraphics[width=\textwidth]{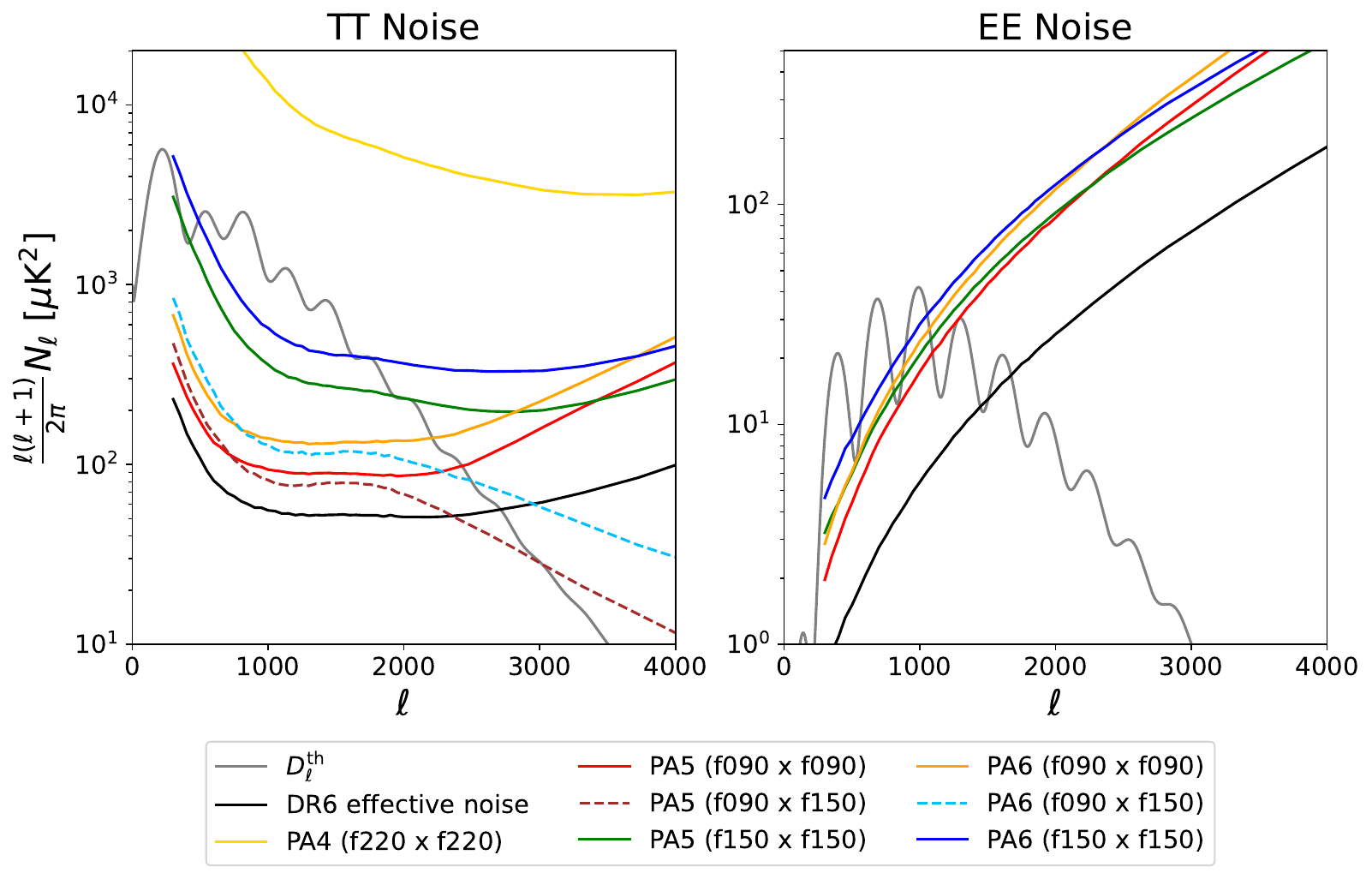}
	\caption{ACT temperature and polarization noise power spectra for each array-band used in the nominal analysis. 
 The effective DR6 noise power spectrum obtained after optimally combining the data is displayed in black. 
 At high multipoles in temperature ($\ell>4000$) the noise can be approximated as white with RMS 20 $\mu$K-arcmin (PA5 f090),
 24 $\mu$K-arcmin (PA5 f150), 23 $\mu$K-arcmin (PA6 f090),  28 $\mu$K-arcmin (PA6 f150) and  82 $\mu$K-arcmin  (PA4 f220) over $\sim 11,000 \ {\rm deg}^{2}$ of the sky. In polarization, the white noise transition starts at $\ell = 1000$. The gray lines show the best-fitting signal power spectra, estimated in \S\ref{sec:results}.\\  }
	\label{fig:noise}
\end{figure*}

The effective noise power spectra for array-bands used in this paper are shown in Figure~\ref{fig:noise}. On large scales in temperature, 
the primary source of noise is atmospheric contamination, while on small scales 
detector noise becomes the most significant contributor. When the noise power spectrum falls below the expected signal power spectrum,
the modes are said to be signal-dominated. The temperature data are signal-dominated up to multipole $\ell = 2800$, 
while the polarization data are signal-dominated up to $\ell = 1700$. At high multipoles in temperature ($\ell > 4000$), 
the noise is well approximated as white noise. In polarization, the white noise transition starts at $\ell = 1000$, 
and the root-mean-square (RMS) of the polarization noise is $\sqrt{2}$ times higher than that of the temperature noise. These noise properties are described further in N25 and \cite{atkins/etal:2023}.

\subsection{Simulation set}\label{subsec:sims}

As part of the DR6 analysis, we generated several sets of simulations based on our evolving understanding of the instrument and the sky model.
Our final simulation set consists of 1,640 simulations for each split of each array-band. Each simulation includes a Gaussian realization of the CMB and foregrounds, and a realistic noise simulation as described in \cite{atkins/etal:2023}. 

 The signal realizations are drawn from a single \LCDM\ cosmological model, and a single foreground model\footnote{An analysis of a simulation using a more complex non-Gaussian foreground model, designed to better capture realistic spatial and spectral correlations, is presented in Section~\ref{subsubsec:param_recov}.}. The values of the cosmological and foreground parameters used for the final simulation set are the best-fitting model found using a near-final iteration of our analysis. These are given in \S\ref{sec:results}. Earlier simulation sets, for example those used before unblinding, used estimates for foreground parameters based on earlier ACT data.
Since the noise model of the simulation is measured using actual ACT data, we apply the data calibration and polarization efficiencies to each noise simulation. 

As described in later sections, these simulations are used to ensure unbiased power spectra recovery, by processing them through the data pipeline. They are also used for constructing the covariance matrices used for null tests and parameter estimation. Subsets of the simulation set are used to test the effects of filtering and aberration. 

\subsection{Power spectrum estimation}\label{subsec:PS_est}
We follow a similar approach to \cite{choi_atacama_2020} for estimating the $\{T,E,B\}$ power spectra from the I, Q and U Stokes vector maps. We use the MASTER curved sky pseudo-$C_{\ell}$ method implemented 
in the public code \texttt{pspy}\footnote{ \href{https://github.com/simonsobs/pspy}{\texttt{pspy}, version 1.8.0}} to account for incomplete sky coverage and the smoothing from the instrument beam. Since the maps are produced in plate carrée (CAR)
pixelization, we use the spherical harmonic transforms (SHTs) for this pixelization, implemented as part of the \texttt{ducc}\footnote{ \href{https://gitlab.mpcdf.mpg.de/mtr/ducc/-/tree/ducc0}{\texttt{ducc}}} library. 
Unlike in \cite{choi_atacama_2020}, the sky coverage is large enough for the unbinned mode coupling matrices to be inverted exactly and applied to the
estimated pseudo-$C_{\ell}$.  Details of the masks used in this analysis are provided in Appendix \ref{subsec:analysis_mask}, and the instrument beams are described in N25 and \cite{beams_inprep}. We use 25\% of the sky after removal of regions with high Galactic emission and around extragalactic point sources with flux $>15$~mJy at f150. We rebin the power spectra using different binning schemes a posteriori to reduce correlations between multipoles and to limit the size of the data vector. The nominal binning scheme used for the cosmological parameter analysis has a minimum bin width of  $\Delta \ell_{\rm min}= 50$.

We introduce the following nomenclature:
\begin{enumerate}
\itemsep0em
\item Auto split power spectra: these power spectra can be computed either from a single split of observations or from two splits of observations taken by different array-bands at the same time. They are affected by noise bias. 
\item Cross split power spectra: these power spectra are estimated from two splits of data corresponding to different observing times.
\item Auto array-band x-spectrum: these power spectra are formed by averaging the cross-split power spectra for a single array-band, e.g., PA5~f090$\times$PA5~f090.
\item Cross array-band x-spectrum: these power spectra are formed by averaging the cross-split power spectra for two different array-bands, e.g., PA5~f090$\times$PA6~f150.
\end{enumerate}

To avoid noise bias, we only use power spectra computed from independent splits of observations in our cosmological analysis.
A single auto array-band  TT  x-spectrum is computed from the uniform average of the $n_{d}(n_{d} - 1)/2$ cross-split power spectra, where $n_{d}$ is the number of data splits for each DR6 array-band. Our nominal power spectrum dataset includes a total of fifteen auto and cross array-band x-spectra for TT,
derived from five array-bands; ten for EE, derived from four array-bands;
and sixteen for TE, 
accounting for the fact that TE and ET are not equivalent for cross array-band x-spectra.

After estimating the spectra we apply three additional corrections described in the following, to account for the transfer function from the ground pickup filter,
for residual temperature to polarization leakage,  and for the aberration arising from our own motion with respect to the last scattering surface.\footnote{While these corrections assume a fiducial \LCDM\ model, we estimate that the errors due to a mis-specified fiducial model would be significantly smaller than the experimental errors.}

\bigskip
\bigskip

\subsubsection{Ground pick up transfer function}\label{subsubsec:pick_up}

As in previous ACT analyses, we mitigate ground emission, which appears as constant declination stripes in the sky maps, by filtering out Fourier modes with $|\ell_{x}| < 90$ and $|\ell_{y}| < 50$. This filtering operation biases the power spectra. To characterize and correct for this bias, we generate 800 noise-free CMB and foreground simulations for each array-band using our best-fit signal and foreground power spectra. We then estimate all resulting power spectra, 
both before and after applying the filter. Comparing these two sets of spectra allows us to quantify the filter's effect, which we represent as a matrix $\bm{F}_{\ell}$.
This matrix is then inverted and applied to the measured power spectra, following Equation \ref{eq:cross_split} in Appendix \ref{subsec:ground_filt}. This formalism neglects coupling introduced by the filter between different multipoles. In general, the filtering operation could be described as a two-dimensional operator in multipole space, $\bm{F}_{\ell, \ell'}$. We have verified through simulations that the correlation length associated with this two-dimensional operator is significantly smaller than our bin size, making it acceptable to neglect the correlation for binned power spectra. The shape of this transfer function is shown in Appendix \ref{subsec:ground_filt}.

\subsubsection{Polarization leakage correction}\label{subsubsec:leakage}

The level of residual temperature-to-polarization leakage in the ACT DR6 data is estimated in \cite{beams_inprep} using maps of Uranus, whose emission is assumed to be polarized weakly enough to neglect \citep{2014PASP..126.1027W}.
The leakage affecting a given array $\alpha$ is represented using two functions: $\gamma^{\alpha}_{\ell, \rm TE}$, $\gamma^{\alpha}_{\ell, \rm TB}$.
The leakage affects the spherical harmonic coefficients of the polarization maps in the following way:
\ba
\tilde{E^{\alpha}}_{\ell m} &=& E_{\ell m} + \gamma^{\alpha}_{\ell, \rm TE} T_{\ell m} \nonumber \\
\tilde{B^{\alpha}}_{\ell m} &=& B_{\ell m} + \gamma^{\alpha}_{\ell, \rm TB} T_{\ell m}\,   
\ea
where $\sim$ denotes the  $\{T,E,B\}$ transforms affected by leakage. The difference between power spectra affected by leakage and the true underlying power spectra 
can be denoted by $\Delta D^{X_\alpha Y_\beta}_{\ell} =  \tilde{D}^{X\alpha Y_\beta}_{\ell}  - D^{X\alpha Y_\beta}_{\ell} $, where $X, Y \in \{T,E,B\}$. Using the ACT leakage model, this difference can be expressed as
\ba
\Delta D^{X_{\alpha} Y_{\beta}}_{\ell}  &=&   (\delta_{XE}\gamma^{\alpha}_{\ell, \rm TE}  + \delta_{XB}\gamma^{\alpha}_{\ell, \rm TB}) D^{T_{\alpha} Y_{\beta}}_{\ell}  \nonumber \\
&+&  (\delta_{YE}\gamma^{\beta}_{\ell, \rm TE}  + \delta_{YB}\gamma^{\beta}_{\ell, \rm TB})D^{X_{\alpha} T_{\beta}}_{\ell}  \nonumber \\
&+&  (\delta_{XE}\gamma^{\alpha}_{\ell, \rm TE}  + \delta_{XB}\gamma^{\alpha}_{\ell, \rm TB}) \nonumber \\
&\times &(\delta_{YE}\gamma^{\beta}_{\ell, \rm TE}  + \delta_{YB}\gamma^{\beta}_{\ell, \rm TB})D^{T_{\alpha} T_{\beta}}_{\ell},    
\ea
where $\delta$ is the Kronecker delta symbol.
We calculate these corrections using our best-fitting theory spectra\footnote{We iterated on this method while determining the best-fitting model.} and subtract them from the measured power spectra. The typical amplitude of residual temperature-to-polarization leakage is approximately 0.3$\%$, varying across array-bands. \tl{Neglecting this effect would significantly degrade null tests between different array-bands in the TE and TB spectra, and would also bias the cosmological interpretation.  It leads to a $1.7\sigma$ shift in the inferred baryon density, $\Omega_b h^2$, and a $1\sigma$ shift in the scalar spectral index, $n_s$.}

\subsubsection{Aberration correction}\label{subsubsec:ab_corr}

We account for aberration caused by our motion relative to the CMB \citep[e.g.,][]{burles/rappaport:2006}, as was also done for previous ACT analyses \citep{louis_atacama_2017,choi_atacama_2020}. We generate 300 simulations of noiseless aberrated signals and foregrounds,
and compare their resulting power spectra with those obtained without aberration. 
This allows us to derive an aberration correction for each power spectrum, which we apply to correct the data:
\ba
D^{X_{\alpha} Y_{\beta}, \rm deaberrated }_{\ell} &=& D^{X_{\alpha} Y_{\beta}, \rm aberrated }_{\ell} - \langle \Delta  D^{X_{\alpha} Y_{\beta} }_{\ell} \rangle_{\rm sims} \nonumber \\
\langle \Delta  D^{X_{\alpha} Y_{\beta} }_{\ell} \rangle_{\rm sims}  &=&  \langle D^{X_{\alpha} Y_{\beta}, \rm aberrated }_{\ell} -  D^{X_{\alpha} Y_{\beta}}_{\ell} \rangle_{\rm sims}. 
\ea
Here the average is found using the 300 simulations.
The aberration correction mostly affects the measurement of the angular scale of the acoustic peaks, shifting the inferred value of the peak position parameter, $\theta_{\rm MC}$, by $0.75''$.\footnote{Before correcting for the aberration effect, we find $100\theta_{\rm MC}=1.04093$ which decreases to $1.04056$ after applying our correction.} Given ACT's precision, not correcting for it would bias this angular scale by $1\sigma$.

\subsection{Covariance matrix} \label{subsec:covmat}

\begin{figure*}[htp]
	\centering
	\hspace*{-5mm}\includegraphics[width=0.88\textwidth]{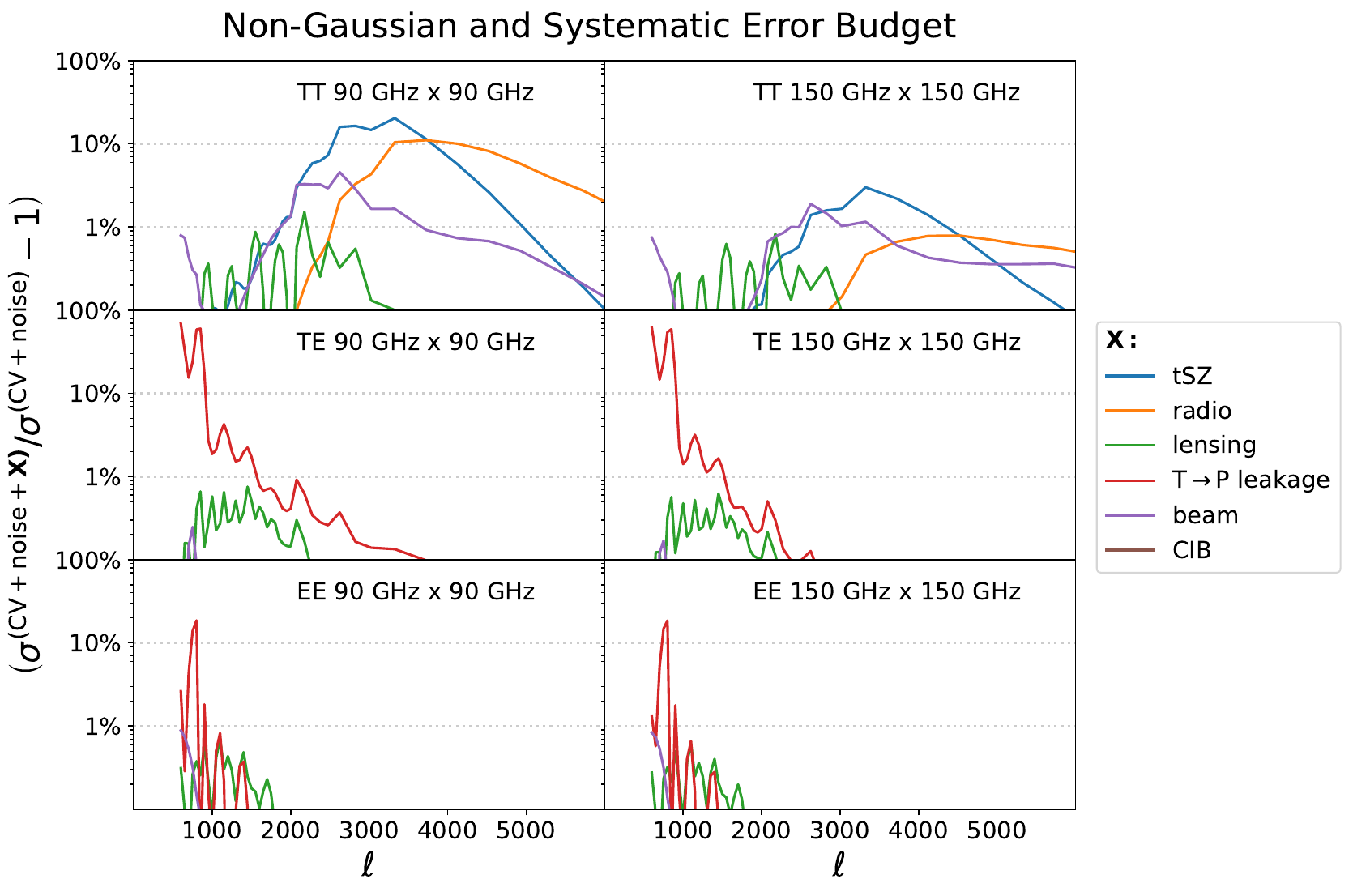}
	\caption{Relative contributions of additional error terms compared to cosmic variance and noise. The contributions to $\sigma_{\rm TT}$ are shown in the top panel for 90 GHz (left) and 150 GHz (right). For f090, uncertainties from non-Gaussian tSZ and non-Gaussian radio sources are important on small scales. The contribution from unclustered CIB non-Gaussianity is smaller than 0.1\% and not visible in the figure. The middle panel shows the contributions to $\sigma_{\rm TE}$, where uncertainties in the measurement of the leakage beam are a significant source of uncertainty on large scales. The bottom panel highlights that these uncertainties only mildly affect $\sigma_{\rm EE}$, reaching up to 15\% at $\ell = 800$. In addition to increasing errors, the additional covariance contributions also result in nonzero off-diagonal correlations.\\ }
	\label{fig:rel_contribution}
\end{figure*}

The uncertainties on the power spectra, and their associated correlations, arise from various contributions. 
Here we outline the terms contributing to the overall covariance matrix, with the first two terms discussed in detail in \cite{atkins/etal:2024}. 
 
\begin{enumerate}
\itemsep0em
\item
{\it Analytic computation of the cosmic variance and noise contribution.} The first component of the computation is the standard
    pseudo-$D_{\ell}$ analytic covariance for noise and a Gaussian-distributed signal, as described in, e.g., \cite{Couchot2017,plancklike_2015,garcia2019}. As discussed in \cite{atkins/etal:2024}, our nominal matrix assumes homogeneous noise properties for this analytic part, although we tested a second version that better accounts for inhomogeneous and correlated noise (the ``homogenous" matrix prescription is a special case of the more general ``inhomogeneous" version). While the latter analytic matrix more accurately models the distribution of simulations, \cite{atkins/etal:2024} finds that both matrices perform equally well after applying a Monte Carlo matrix correction.
   
    \item
    {\it Monte Carlo correction to the analytic computation.}  
    The nominal analytic estimate does not account for complicated noise properties, such as inhomogeneous depth and stripy correlation patterns, arising from the varying scanning strategy across the observation patch.
    Similarly, the effect of the ground pick up filter leads to an increase in error that is difficult to capture analytically. We correct the analytic estimate for both effects with a method described in \citet{atkins/etal:2024} that uses the simulations described above.\footnote{Compared to the simulations described in \citet{atkins/etal:2024}, our simulation ensemble is slightly larger --- 1640 here vs. 1600 in \citet{atkins/etal:2024} --- and includes small updates to fiducial cosmological, foreground, and systematic parameters as described in Appendix \ref{apx: changes_to_blinding_plan}.}

  \item
   {\it Beam covariance.}  
    We add a contribution to the covariance due to beam uncertainties measured from dedicated observations of Uranus, as described in \cite{beams_inprep}. At leading order this takes the form
    \ba
  && \Sigma(D^{X_{\alpha} Y_{\beta}}, D^{W_{\mu} Z_{\nu}}) _{\rm beam}= D_{\ell}^{X_{\alpha} Y_{\beta} } D_{\ell}^{W_{\mu} Z_{\nu} } \nonumber \\
   && \left[  (\delta_{\alpha \mu} + \delta_{\alpha \nu})  \left\langle  \frac{\delta B^{\alpha}_\ell \delta B^{\alpha}_\ell}{B^{\alpha}_{\ell}B^{\alpha}_{\ell}}   \right\rangle +  (\delta_{\beta \mu} + \delta_{\beta \nu})  \left\langle  \frac{\delta B^{\beta}_\ell \delta B^{\beta}_\ell}{B^{\beta}_{\ell} B^{\beta}_{\ell}}  \right \rangle  \right] \nonumber \\ 
  \ea
    where $B^{\alpha}_{\ell}$ is the beam transform of the array-band $\alpha$. Note that we assume that beam errors for different arrays are uncorrelated.\footnote{\tl{Correlation between frequency bands within a PA is expected to be small due to our selection cuts, with significantly more observations surviving at lower frequencies. Correlations between PAs are also expected to be small: PA6 observations do not overlap in time with those of PA4 and PA5, and although PA4 and PA5 do overlap, the relatively limited number of PA4 observations reduces potential atmospheric noise correlations to a negligible level. }}

    \item
    To account for the uncertainties in the measurement of the leakage beam, we generate 10,000 simulations of the leakage beam of the form:
    \ba
     \gamma^{\alpha, \rm sim }_{\ell, \rm TX} =    \gamma^{\alpha }_{\ell, \rm TX}  + \Delta \gamma^{\alpha }_{\ell, \rm TX},  
     \ea
     where  $\Delta \gamma^{\alpha }_{\ell, \rm TX}$ is a Gaussian realization drawn from a distribution with zero mean and the covariance of the leakage beam measurement.
     We then apply this simulated leakage model to a set of fiducial spectra and perform a correction assuming the mean leakage $ \gamma^{\alpha }_{\ell, \rm TX} $.
     This procedure should closely mimic the treatment applied to our actual dataset. The Monte Carlo covariance matrices $\Sigma(C^{X_{\alpha} Y_{\beta}}, C^{W_{\mu} Z_{\nu}}) _{\rm leakage}$ are then estimated from these 10,000 simulations. 
    
 \item
  {\it  Non-Gaussian lensing terms.} Following \cite{choi_atacama_2020}, we include two corrections to the covariance matrix to account for the non-Gaussian signal arising from lensing.
    The first correction accounts for the off-diagonal correlations due to lensing, which arise from a single lensing mode fluctuation simultaneously 
    affecting many $\ell$ values \citep{benoit-levy/etal:2012, peloton/etal:2017}. 
    The second correction addresses the lensing super-sample variance,  which is caused by the variation of the mean convergence over 
    the observed area \citep{manzotti/etal:2014, motloch/hu:2017}.

  \item
  {\it Non-Gaussianity in the foreground emission.} Most of the foreground components, discussed in \S\ref{sec:like}, cannot be described as Gaussian fields,
    and a correction for their connected trispectrum needs to be included in the covariance matrix. We model the non-Gaussianity of radio sources, of the cosmic infrared background and of the thermal Sunyaev-Zel'dovich (tSZ) effect.

    \textbf{Non-Gaussianity from tSZ.} Large and massive galaxy clusters at low redshift dominate the contribution to the tSZ power spectrum on large scales and produce a large multipole-to-multipole correlation. This non-Gaussian contribution to the covariance matrix of the CMB power spectrum is easily computed within the halo-model as the harmonic transform of the connected four-point correlation, i.e., the angular trispectrum. See \cite{Komatsu:2002wc} and their Eq.~(27) for the formula. We use \texttt{class\_sz} \citep{Bolliet:2017lha, Bolliet:2023eob} to compute the analytical approximation. Our analytical calculations are then benchmarked against two sets of Poissonian simulations: simulations  made with \texttt{nemo}\footnote{\url{https://github.com/simonsobs/nemo}} and \texttt{pixell},\footnote{\url{https://github.com/simonsobs/pixell}} and the simulations of \cite{Sabyr:2024bao} made with \texttt{hmpdf}\footnote{\url{https://github.com/leanderthiele/hmpdf}} \citep{Thiele:2018jdl}. 
  
    \textbf{Non-Gaussianity from discrete sources.} In our analysis, we mask point sources down to a relatively low flux threshold before taking power spectra. To include the contribution from remaining point sources, we calculate the expected trispectrum, neglecting angular clustering, for each of two components: radio galaxies and dusty, star-forming galaxies, which are most important at low and high frequencies, respectively. For the number counts of radio sources, we use the C2Ex model from \citet{tucci2011}, evaluated at 148 GHz and scaled to the other ACT frequencies as needed.  For the dusty galaxies, we use the model from  \citet{bethermin2012}, evaluated at each ACT frequency directly.  For each of these models for galaxy counts as a function of observed flux density, we  compute the trispectrum in the unclustered, or Poisson, limit, by taking the fourth moment of the distribution. 

\end{enumerate}
Each correction in steps 3 -- 6 is added to the covariance matrix resulting from step 2. The relative contributions of each correction compared to the cosmic variance and noise terms are shown in Figure \ref{fig:rel_contribution}. A major source of uncertainty arises from our measurement of beam leakage, leading to an increase in errors of over $50\%$ in the TE power spectrum for multipoles $\ell < 800$. Mitigating this large contribution has motivated our choice of multipole cut in polarization. 

For temperature, the primary source of additional uncertainty is the non-Gaussianity of the tSZ and radio sources signal; the contribution from unclustered CIB non-Gaussianity is always smaller than $0.1\%$.

The additional covariance contributions also result in nonzero off-diagonal correlations. In subsequent tests, we find that the impact of lensing and foreground non-Gaussianity on the uncertainties of our recovered cosmological parameters is at the sub-10\% level. 

\section{ACT power spectrum results}
\label{sec:ps_result}
In this section we describe the set of null tests performed on the spectra, show the multi-frequency TT, EE and TE spectra from ACT that are used for cosmological analysis, and present the BB, TB and EB spectra. 
\subsection{Null tests}
\label{subsec:nulls}
\begin{figure*}[htbp]
    \centering
        \includegraphics[width=0.4\textwidth]{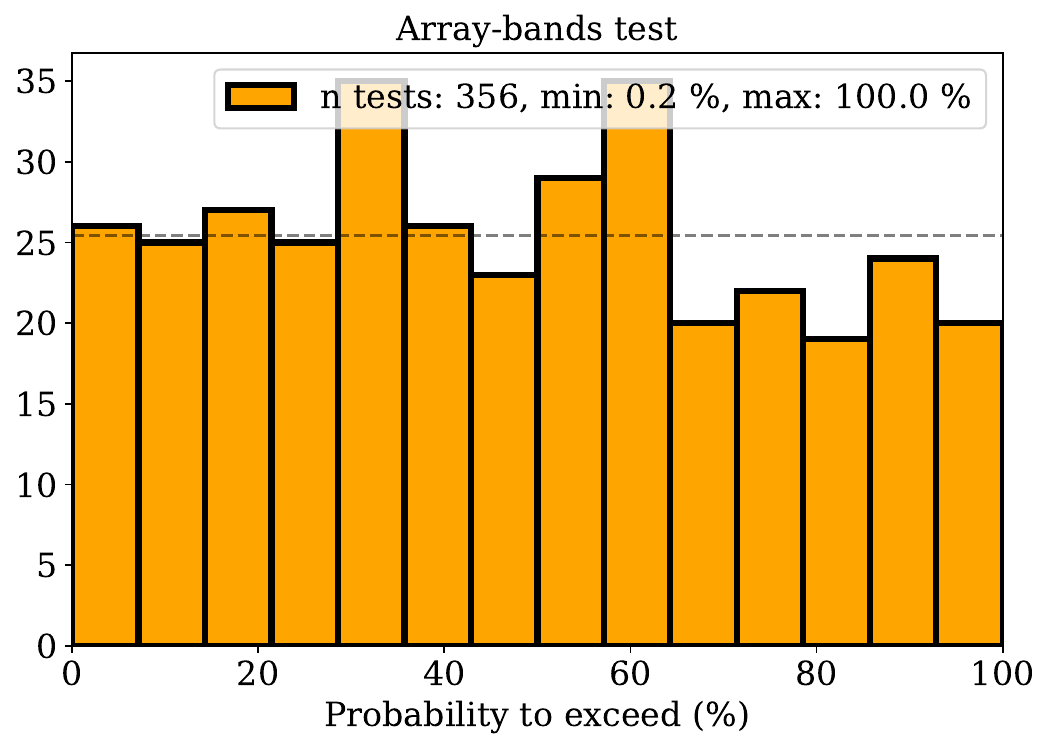}
        \includegraphics[width=0.4\textwidth]{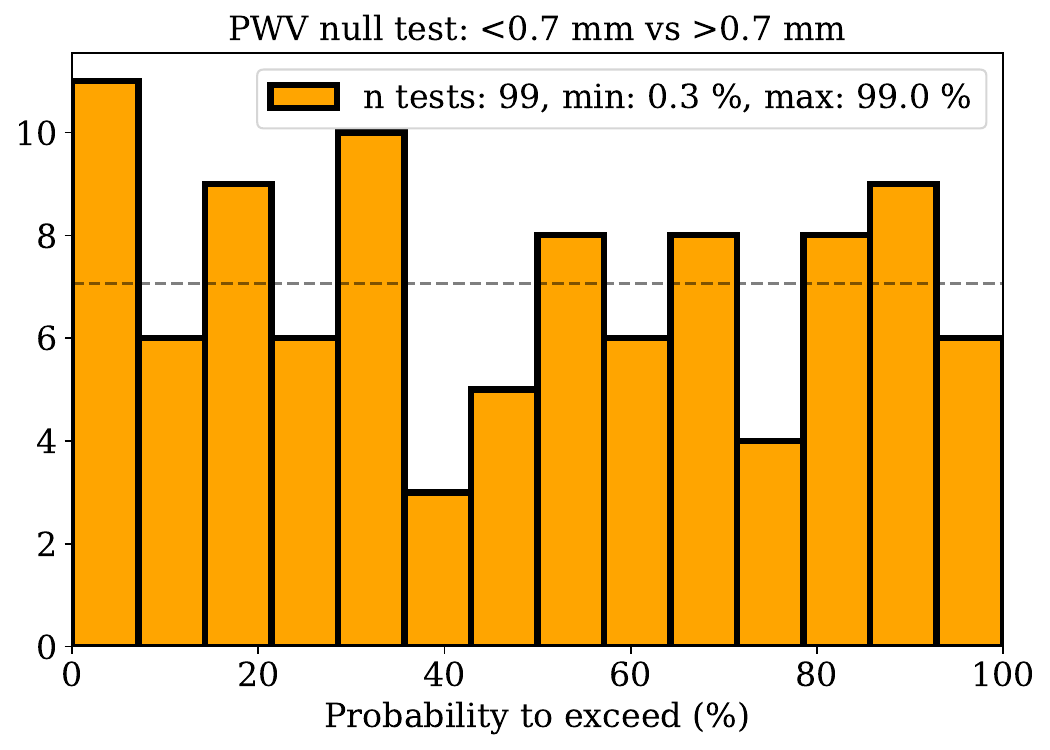}
        \includegraphics[width=0.4\textwidth]{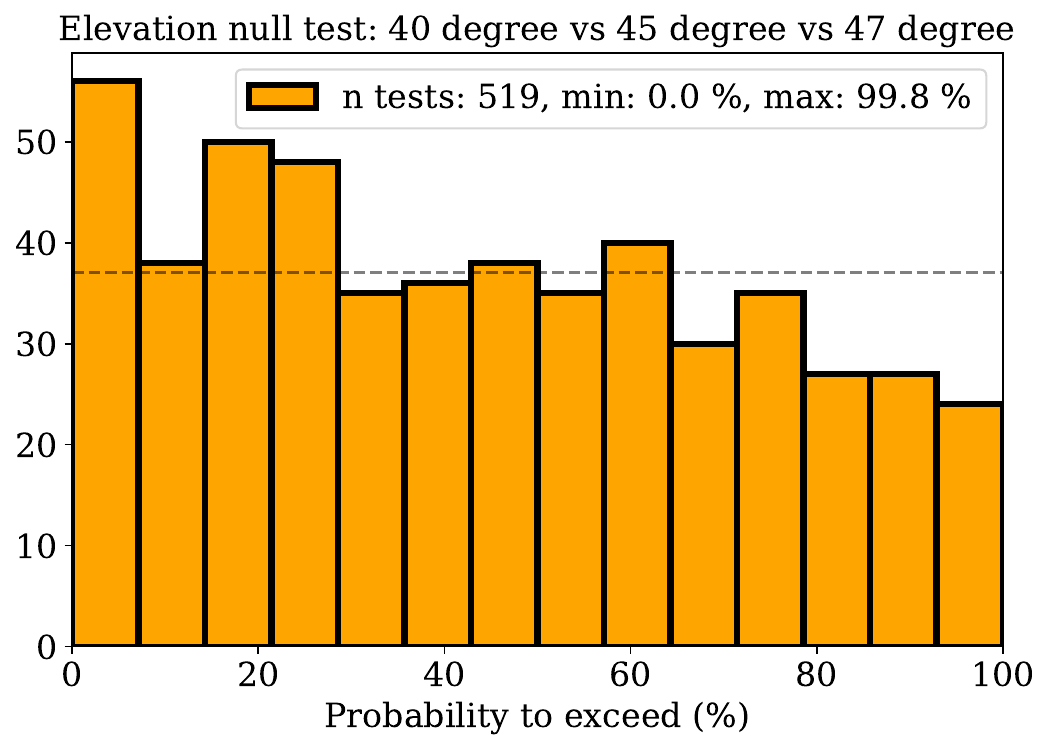}
        \includegraphics[width=0.4\textwidth]{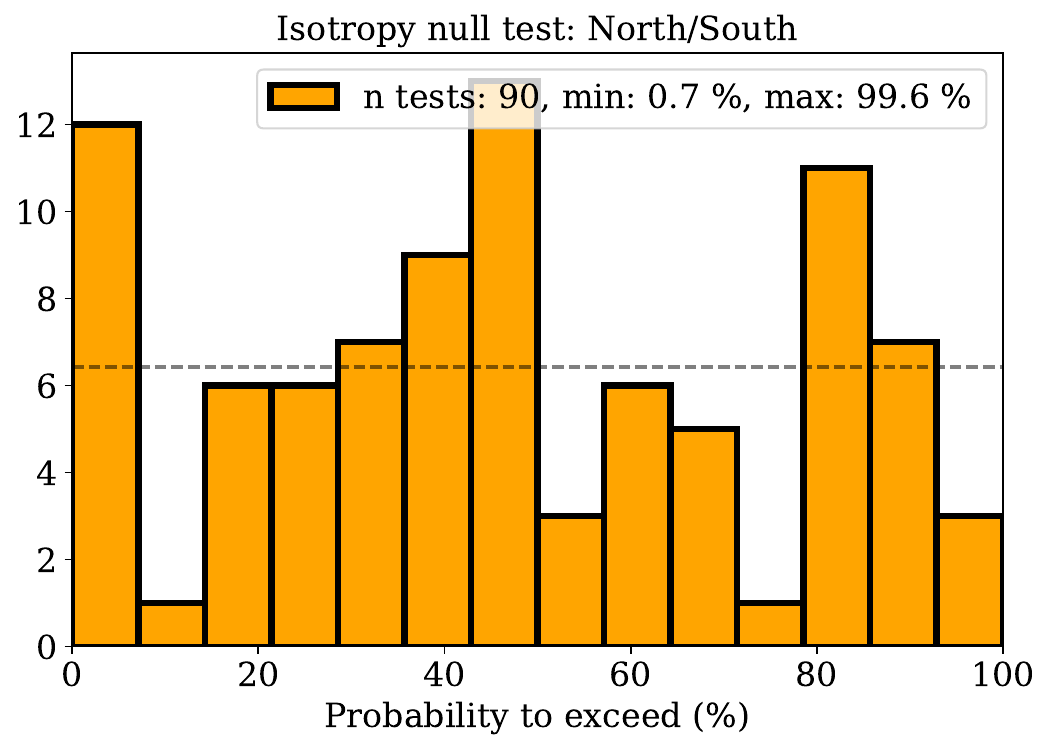}
        \includegraphics[width=0.4\textwidth]{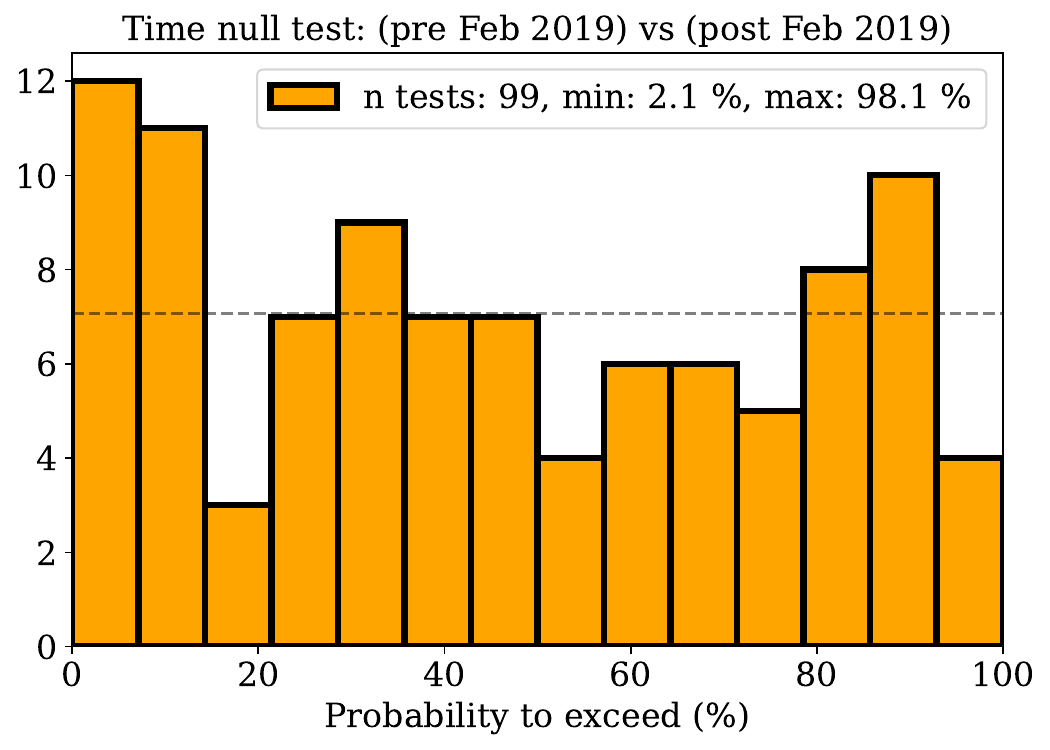}
        \includegraphics[width=0.4\textwidth]{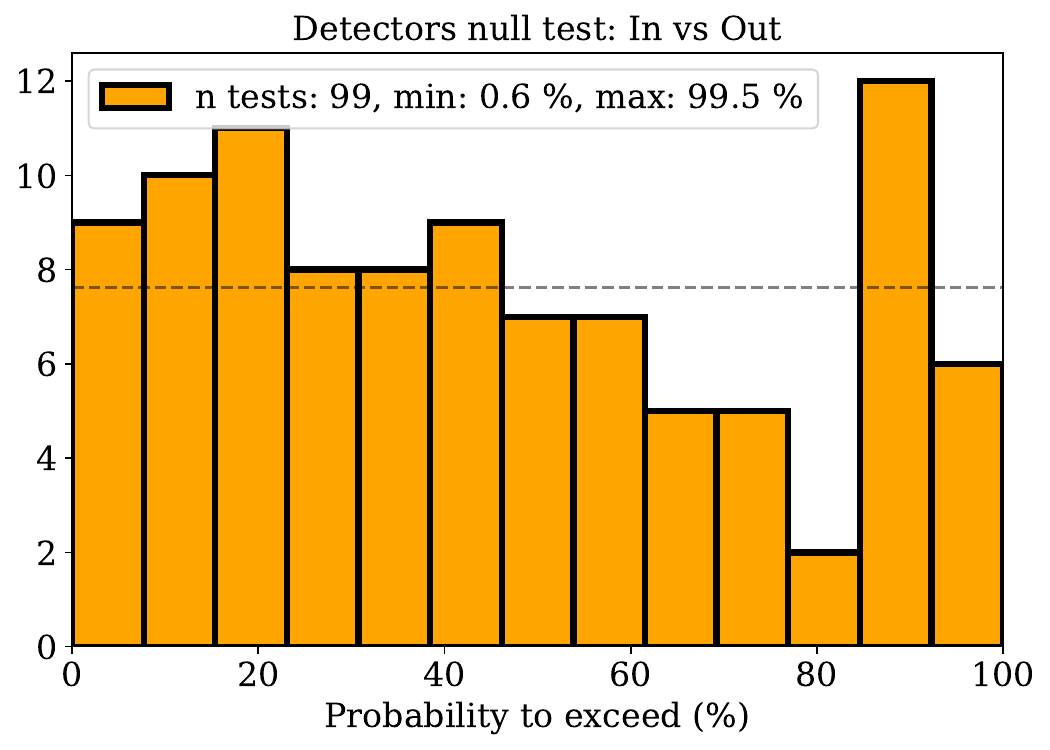}
        \includegraphics[width=0.4\textwidth]{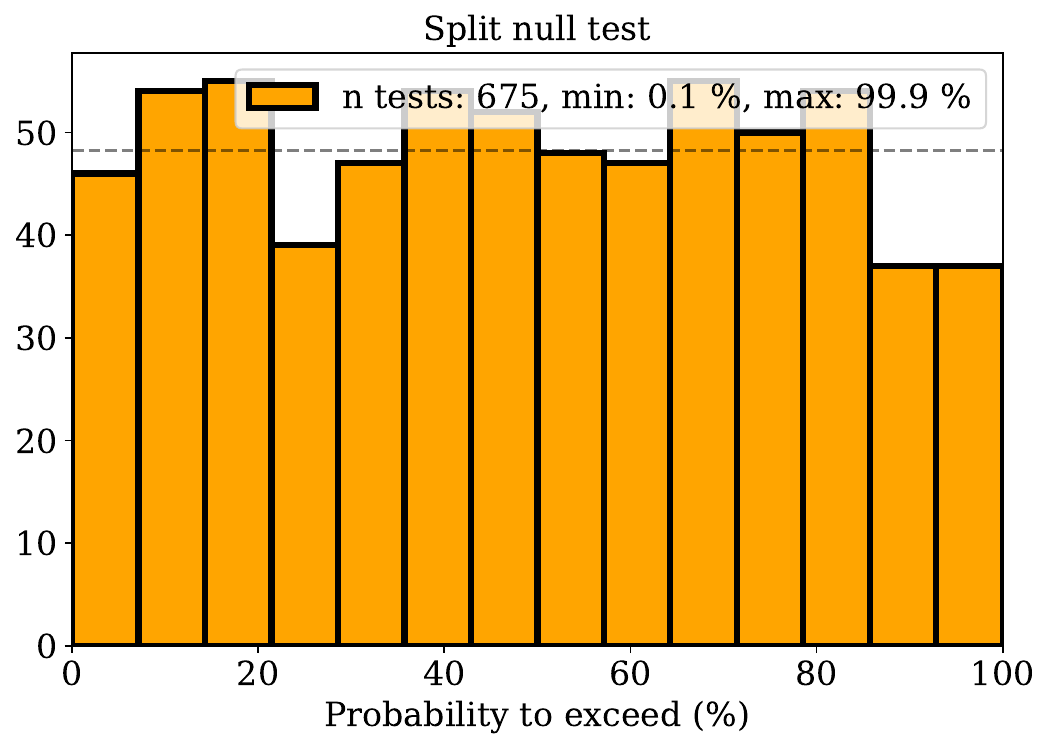}
    \caption{Histogram of the probability-to-exceed (PTE) values from all individual null tests. The dashed lines shows the expectation for a uniform distribution, given by $n_{\rm test} / n_{\rm bins}$. The lowest PTE in the elevation null test was just below 0.05\%. In total, we performed approximately 2,000 null tests on the dataset and inspected each associated residual. We do not find significant indications of failure.}
    \label{fig:mainfig}
\end{figure*}

To identify possible residual systematic effects, we assess the consistency of the power spectra of subsets of our data. As described in our blinding procedure in Appendix \ref{apx:blinding}, we performed this suite of tests before comparing our TE and EE spectra to those from \Planck, and before comparing our data to any theoretical models.\footnote{The results shown here were re-run after unblinding, with our final analysis inputs and the baseline cuts.} For each test we show the distribution of $p( \chi^{2}_{\rm null} )$ in Figure \ref{fig:mainfig}. 

\begin{enumerate}
\itemsep0em
    \item {\it Array-band null test.} In this test, we compare different auto and cross array-band x-spectra
    formed using the nominal DR6 dataset (e.g., PA5~f150 $\times$ PA5~f090, PA6~f150 $\times$ PA6~f090). 
    For temperature, we only compare spectra with similar effective
    frequency pairs to reduce the amplitude of the expected residuals, 
    thus making the test only weakly dependent on the details of the foreground model.
    In polarization, the expected residuals from comparing power spectra at different
    frequencies are anticipated to be much smaller. 
    Therefore, we compare all of them, accounting only for the expected 
    differences in Galactic dust that are described in \S\ref{subsec:fg_model}.
    In total, we form 356 tests: 14 in temperature, 45 in EE and BB, and 84 in TE, TB and EB.
    Denoting $\Delta^{XY}_{\alpha \beta \mu \nu, \ell}=  D_{\ell}^{X_{\alpha} Y_{\beta}} -  D_{\ell}^{X_{\mu} Y_{\nu}}  $
    for each test we compute 
    \ba
    \chi^{2}_{\rm null} = (\Delta^{XY}_{\alpha \beta \mu \nu, \ell} - \Delta^{XY, \rm th}_{\alpha \beta \mu \nu, \ell}) \Sigma^{-1} (\Delta^{XY}_{\alpha \beta \mu \nu, \ell} - \Delta^{XY, \rm th}_{\alpha \beta \mu \nu, \ell}) \nonumber
    \ea
     where $\Delta^{XY, \rm th}_{\alpha \beta \mu \nu, \ell}$ represent the expected residual which is computed from our foreground model and the measurement of our array-band bandpasses.
         
    \item {\it Time-split null test.} Each array-band dataset is split into 4 independent time-splits to build our cross-array-band power spectrum estimator. We assess the consistency of these 4 splits from residuals between pairs of cross-split power spectra for each array-band. Residual $XY$ power spectra for the array-band $\alpha$ are defined as $\Delta^{XY}_{\ell, ijkl, \alpha} = D_\ell^{X_{\alpha, i}Y_{\alpha, j}} - D_\ell^{X_{\alpha, k}Y_{\alpha, l}}$ where $i < j$ and $k < l$ and we use the following $\chi^2$
    \begin{equation}
    	\chi^2 = \Delta^{XY}_{\ell, ijkl, \alpha} [\Sigma^{-1}]_{\ell\ell’} \Delta^{XY}_{\ell', ijkl, \alpha},
    \end{equation}
    where we use our analytical prescription corrected from simulations to get the covariance $\Sigma$. We compute $N=15$ residuals for each pair of array-bands and spectra resulting in $N^\mathrm{tot}=675$ residuals in total. 
    
    \item {\it Isotropy null test.} The DR6 power spectrum analysis uses approximately 25$\%$ of the sky. To check for systematic effects localized in the maps,
    we split the survey into two subsets: one northern patch in Equatorial coordinates, which includes all observations above $-12$ degrees declination, 
    and one southern patch including the rest of the data. 
    We form the residuals $\Delta^{XY}_{\alpha \beta \ell}=  D_{\ell, \rm north}^{X_{\alpha} Y_{\beta}} -  D_{\ell, \rm south }^{X_{\alpha} Y_{\beta}}$ 
    and compute the null $\chi^{2}$ accounting for the estimated difference in anisotropic Galactic dust emission.
\end{enumerate}

In addition to these tests that can be performed using the nominal dataset, we also examine a set of null tests that target specific systematic effects. Performing these null tests requires the creation of new maps, as described in N25. There the data are split at the time-ordered (TOD) level to 
maximize the systematic effect in question while giving roughly equal statistical weight for each subset.

\begin{enumerate}
\itemsep0em
    \item {\it Elevation null test.} The DR6 observations are split according to the scanning elevation of the telescope. 
    This results in three datasets at mean elevations of $40\degree$, $45\degree$, and $47\degree$. 
    Each nominal cross array-band power spectrum is associated with six elevation-based cross array-band power spectra 
    (e.g., el1 $\times$ el1, el2 $\times$ el3), from which we can form 15 null tests.
    This results in 519 tests. This split is particularly sensitive to residual ground pick up contamination of the data, since the sensitivity to ground emission is expected to be dependent
    on the scan elevation.
    \item {\it PWV null test.} Here the observations are divided based on the recorded precipitable water vapor (PWV) at the time of observation, since high PWV is correlated with worse atmospheric noise.
    This results in two datasets: one with PWV less than 0.7 mm (median 0.53 mm), which represents $33\%$ of the dataset, 
    and another with PWV greater than 0.7 mm (median 1.26 mm), representing $66\%$ of the dataset.
   \item {\it Detector null test.} In this test the DR6 observations are divided according to the detector position on each wafer. ``In" corresponds to detectors near the center of the detector wafer, and ``out" to detectors farther from the center. This categorization tests for potential systematic effects localized in the focal plane. For this null test, we had to re-estimate the beam and leakage beam, as they depend on the detector location. 
   \item {\it Time null test.} The DR6 dataset spans five years of observations, from 2017 to 2022. We divide the TODs into two subsets: one covering observations before February 2019 and the other after. The rationale for this division is to account for a change in the telescope's focus that occurred in May 2018, when the secondary mirror axes were disabled. 
\end{enumerate}

\begin{figure*}[htp]
\centering
\hspace{-5mm}
\includegraphics[width=1\textwidth]{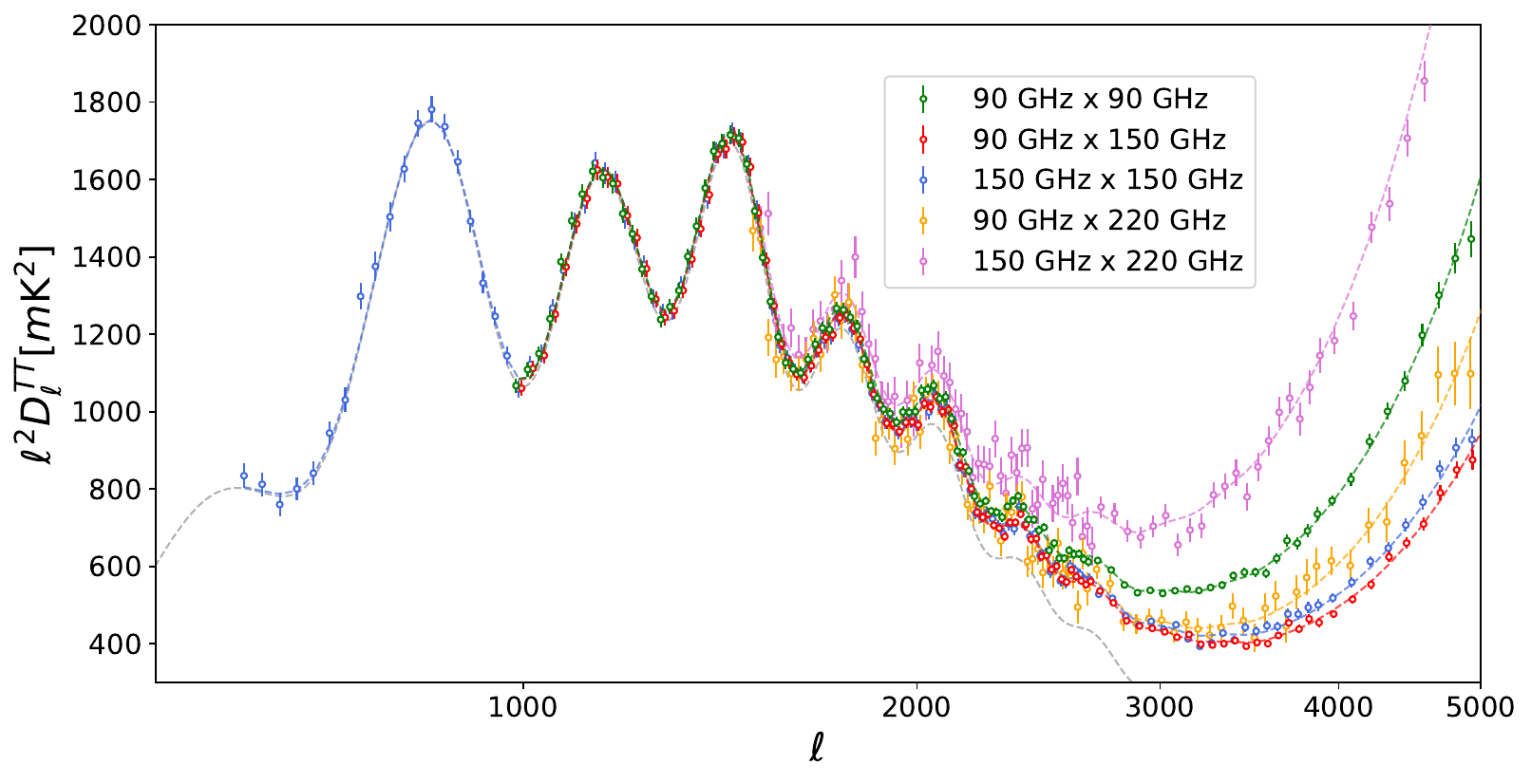}
\caption{ACT multi-frequency temperature power spectra. On small scales, the excess is due to frequency-dependent extragalactic foreground emission and secondary CMB effects. Emission from dusty galaxies dominates at higher frequencies, and from radio galaxies at lower frequencies. To improve plot clarity, data points for all spectra except 150 GHz $\times$ 150 GHz have been slightly shifted. The 220~GHz $\times$ 220~GHz power spectrum is omitted due to its large errors, and cross-spectra including 220 GHz are shown only at $\ell> 1500$. The data extend to $\ell=8500$; the full multipole range is shown in Figure~\ref{fig:TT_component}.\\}
\label{fig:freq_TT}
\end{figure*}

For each of these tests, we allowed for different overall calibration factors corresponding to the various data splits. We find that the overall calibration exhibits mild variation with elevation and PWV, remaining below $1\%$ for PA5 and PA6. However, substantial calibration variations were observed in the in/out null test, suggesting that different parts of the focal plane prefer distinct calibration factors.
This effect is particularly pronounced in the PA4 and PA5 arrays, which demonstrate variations on the order of $5\%$,
while the PA6 array-bands show variations of less than $2\%$. 

In total, we perform around 2,000 null tests on the data. Additionally, we inspect all individual residuals to identify 
any features that might not have been captured by the distribution of PTE values. This inspection informed several key analysis choices, particularly in defining the multipole cuts for the data used in the cosmological analysis.\footnote{In particular, we identified that PA4~f150 disagreed with PA5~f150 and PA6~f150 with our array-band null test and therefore decided not to use it.} Within the baseline cuts, we find no significant departures from expectations; the lowest PTE is $0.05\%$ which is within expectation given the number of tests we have conducted.\footnote{The probability of obtaining such a PTE for 1,937 independent tests is 62$\%$. However, our null tests are correlated. While a precise assessment of this correlation is beyond the scope of this paper, the PTE remains acceptable even if we conservatively assume an order of magnitude fewer independent tests. We also visually inspected this and other spectra.}
 
For the extended cuts, defined in Table \ref{tab:multipole_cut} and which we used until unblinding, we also found no significant departures from expectations in terms of PTE. However, 
after unblinding we chose to redefine our multipole cuts to those labeled ``baseline," where the minimum multiple of the polarization data is increased from $\ell=500$ to match that used for temperature ($\ell=600-1000$ depending on array-band). Described further in Appendix \ref{apx:post_unblinding_changes}, this decision was motivated by individual EE array-band null tests, and residuals to unblinded best-fit \LCDM\ models, that exhibit shaped features as a function of angular scale for the extended cuts, as well as borderline failures of parameter-level null tests between frequency bands discussed in \S\ref{subsubsec:paramnulls}.

To investigate this further, we developed a parametric model for an additional array null test, meant to capture any residual systematic differences between array-bands despite their satisfactory ensemble PTE distribution. This is detailed in Appendix \ref{apx: sys_res}. We model these systematic effects as having a non-zero slope for each array-band as a function of multipole, and 
require that the shape averages to zero over the array-bands. This decorrelates the measurement of any such systematic effects from the cosmological model. With the ``extended" multipole cuts, this exercise revealed a $\sim 3\sigma$ preference for a relative systematic difference between PA5~f090 and the other arrays in polarization, that was mitigated after applying the ``baseline" cuts instead.

\begin{figure*}[htp]
\centering
\hspace{-5mm}
\includegraphics[width=0.95\textwidth]{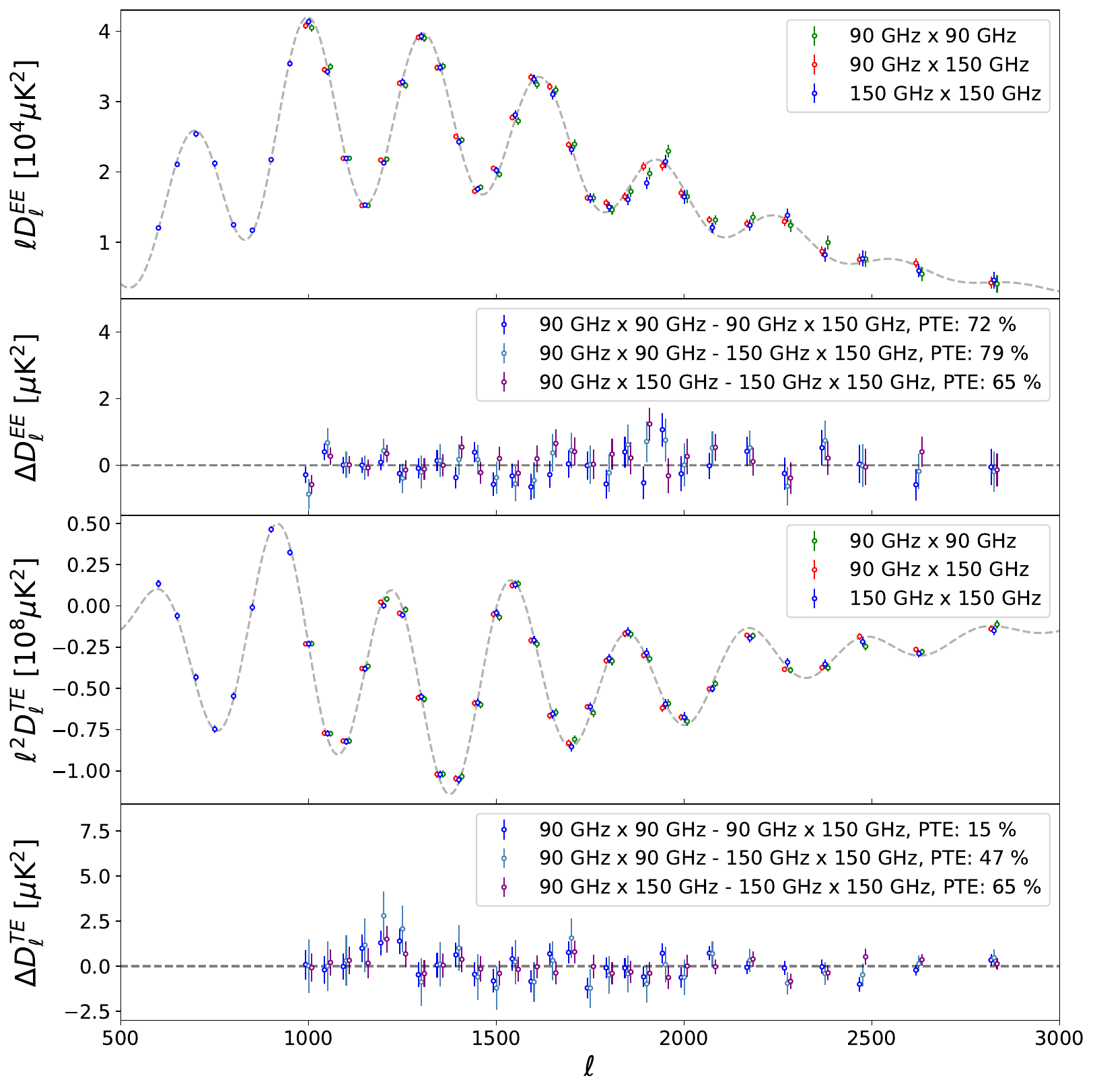}
\caption{ACT dust-subtracted multi-frequency EE and TE power spectra, and inter-frequency null tests.\\}
\label{fig:freq_TE_EE}
\end{figure*}

\subsection{Multi-frequency TT, EE \& TE spectra}

In Figure \ref{fig:freq_TT} we show the multi-frequency temperature spectra from ACT DR6, combined over arrays.\footnote{The PA4 f220$\times$PA4 f220 spectrum is not shown here given its higher noise level, but is shown individually in \S\ref{sec:results}.} The CMB signal is visible over multiple acoustic peaks, and the additional power at small scales comes from foreground and secondary emission. This is lowest for the f090$\times$f150 frequency combination, and discussed in detail in \S\ref{sec:like}. The model curve shown in Figure \ref{fig:freq_TT} is our best-fitting model estimated and discussed in \S\ref{sec:results}. 

Figure \ref{fig:freq_TE_EE} shows the per-frequency TE and EE spectra, combined over the arrays. In this case we remove an estimate of the Galactic dust, described in \S\ref{subsec:fg_model}, to compare the spectra. Residuals between these per-frequency spectra are also shown, showing no obvious frequency dependence. These are simpler to show for TE and EE than for TT, since the frequency-dependent foreground contribution is smaller. The best-fitting theoretical model is indicated, and discussed in \S\ref{sec:results}.

\begin{figure*}[htp]
	\centering
	\hspace*{-5mm}\includegraphics[width=0.7\textwidth]{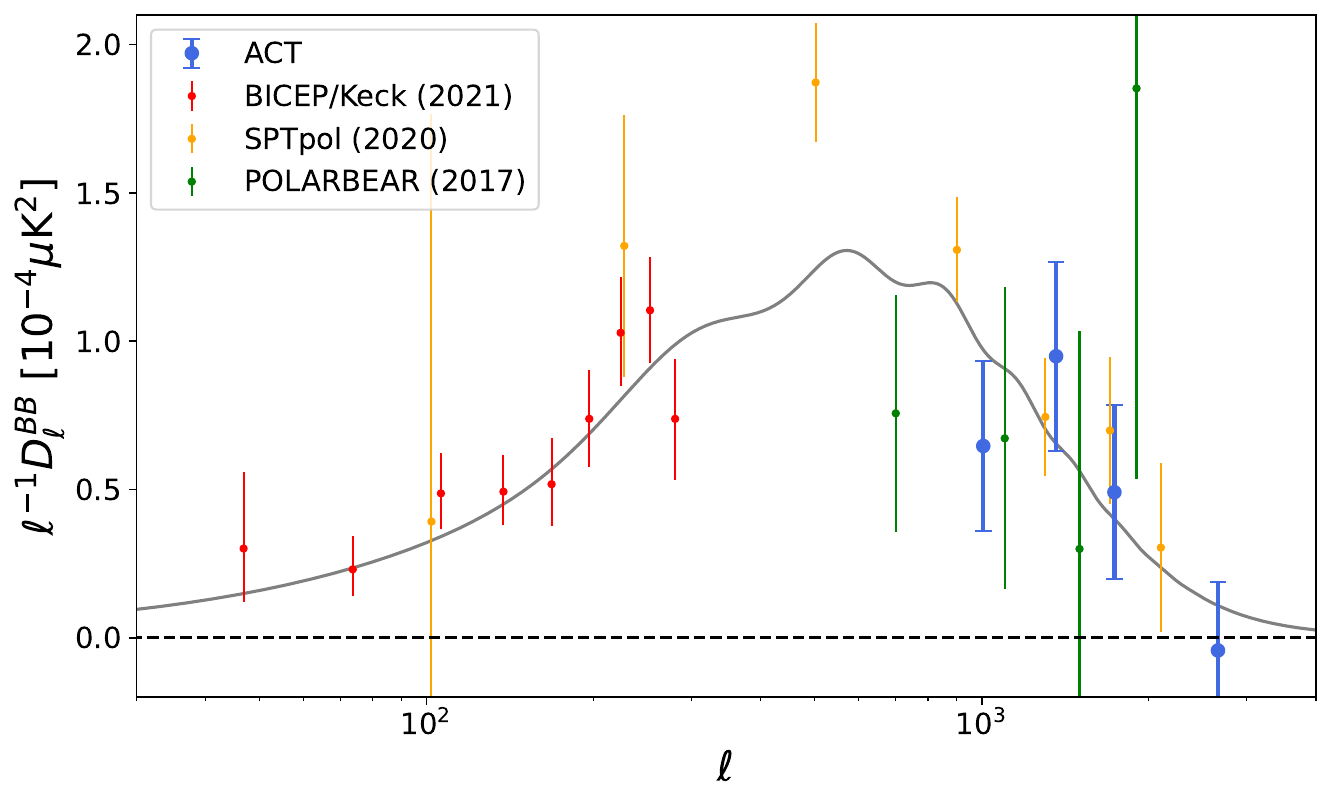}
    	\caption{Compilation of B-mode power spectra measurements from ACT, BICEP/Keck \citep{2021PhRvL.127o1301A}, SPTpol \citep{2020PhRvD.101l2003S} and POLARBEAR \citep{polarbear:2017}. The ACT amplitude is consistent with expectations from the \LCDM\ model with amplitude $A_{\rm CMB} = 0.91 \pm 0.23 $ with respect to the P-ACT best-fit cosmology. This corresponds to evidence at 4$\sigma$ for the lensed B-mode signal.\\}
	\label{fig:B_modes}
\end{figure*}

\subsection{BB power spectrum}

\begin{figure*}[htp]
	\centering

    \hspace*{-7.5mm}\includegraphics[width=0.715\textwidth]{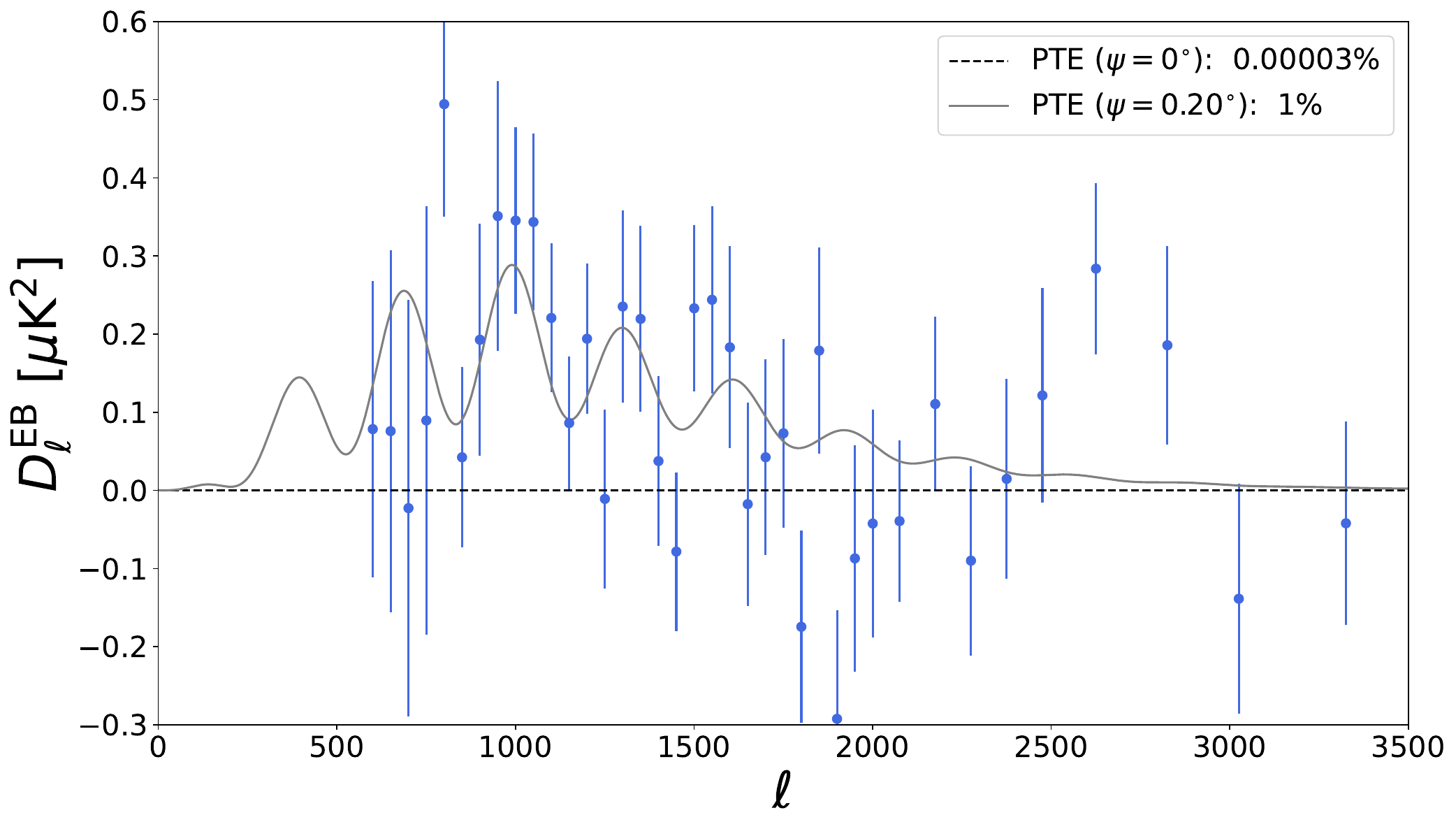}

    \vspace{-6mm}
    
	\hspace*{-5mm}\includegraphics[width=0.7\textwidth]{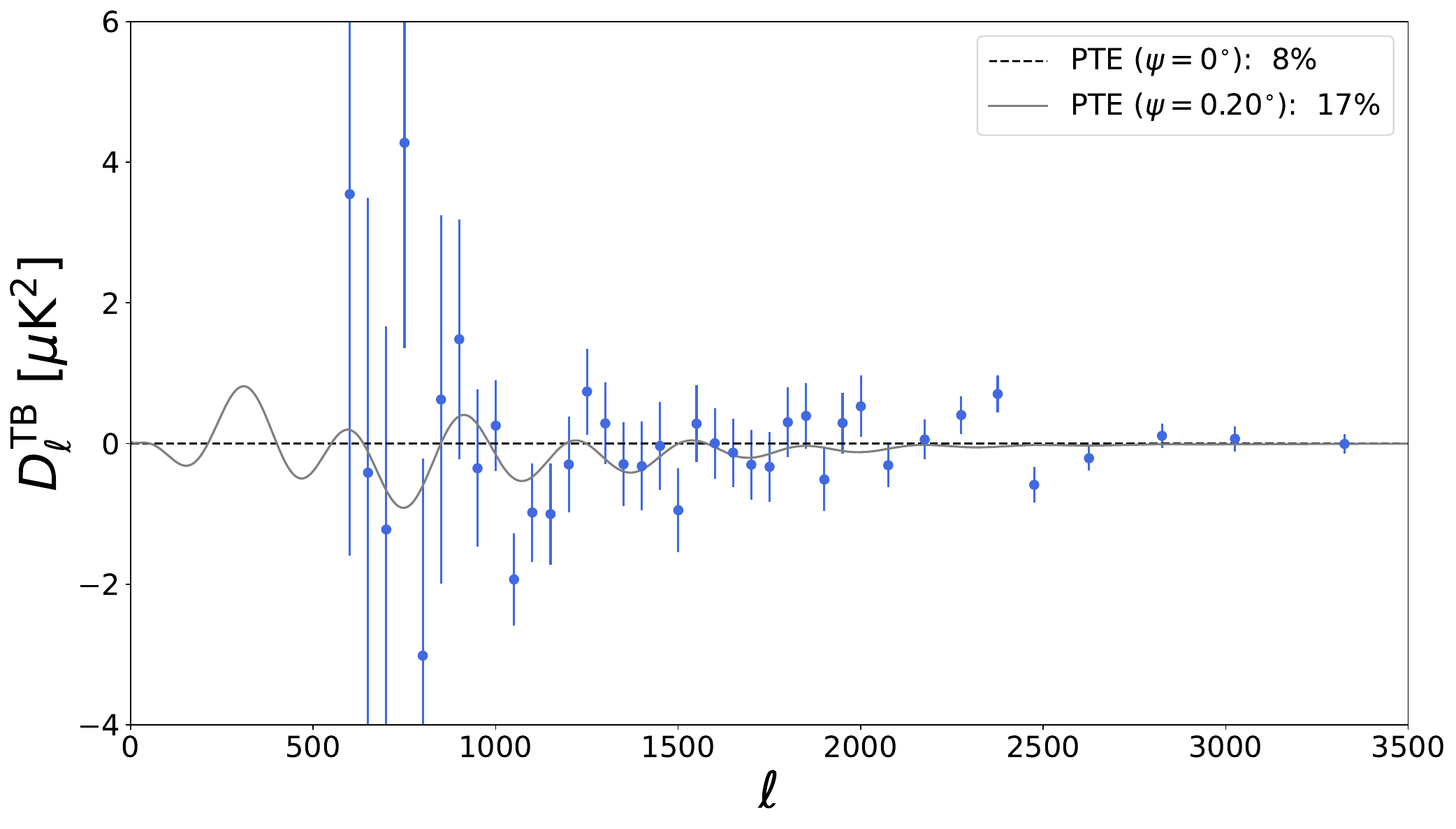}
    \hspace*{-5mm}\includegraphics[width=0.7\textwidth]{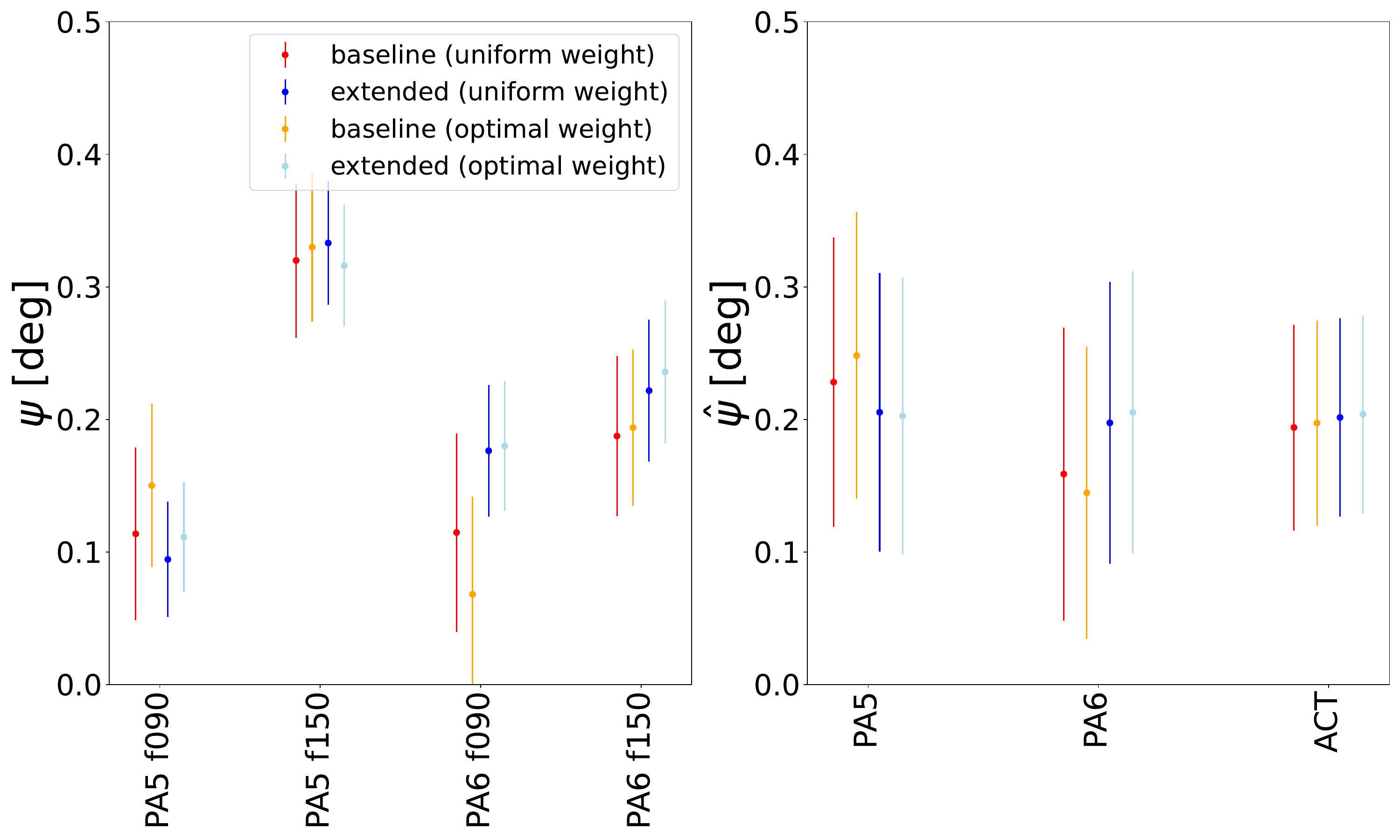}
	\caption{ACT EB (top) and TB (middle) power spectra for the baseline cut with optimal weighting. The probability-to-exceed is quoted for two different models shown: one assuming perfect calibration of the ACT polarization angles and one corresponding to an overall rotation of $0.2^\circ$ of the polarization angles. A significant preference for the latter model is observed.  The low PTE of the $0.2^\circ$ fit in EB (4\% for the extended cut with optimal weight, 1\% for the baseline cut with optimal weight and 0.1\% for uniform weight) is driven by the oscillatory feature between $\ell=1800$ and $\ell=1900$.   We also show the measurements of $\psi$ (bottom-left, statistical errors) and $\hat{\psi}$ (bottom-right, including a systematic error estimate) across different array-band combinations, comparing two data cuts and two pixel-weighting schemes. We find that the preference for non-zero rotation is stable.}
	\label{fig:angle}
\end{figure*}

The ACT scanning strategy was designed to maximize sky coverage from its observation site. This approach has led to state-of-the-art measurements of the TT, TE, and EE power spectra, but represents a suboptimal choice for detecting fainter signals such as B-modes. In Figure \ref{fig:B_modes}, we present the ACT combined BB power spectrum, with an estimate of the Galactic dust removed, alongside measurements from BICEP/Keck \citep{2021PhRvL.127o1301A}, SPTpol \citep{2020PhRvD.101l2003S}, and POLARBEAR \citep{polarbear:2017}. 
To evaluate whether our B-mode measurement aligns with the $\Lambda$CDM prediction, we model the observed B-mode power spectrum at each cross-frequency as the sum of the lensed B-mode power spectrum and a dust component:
\ba
D^{B_{\alpha} B_{\beta}, \rm model}_{\ell} = A_{\rm CMB} D^{\rm BB,  CMB}_{\ell} + D^{B_{\alpha} B_{\beta}, \rm dust}_{\ell}[a^{BB}_{\rm dust}]. \nonumber \\
\ea
Here $D^{\rm BB,  CMB}_{\ell}$ denotes the lensed power spectrum computed assuming the P-ACT cosmological parameters derived in \S\ref{sec:results} and $ D^{B_{\alpha} B_{\beta}, \rm dust}_{\ell}$ represents our dust foreground model, described in Equation \ref{eq:dust_pw}.  
We impose priors on the $a^{\rm BB}_{\rm dust}$ parameters derived in \S\ref{subsec:diff_gal} and fit for $A_{\rm CMB}$ using our 10 BB auto and cross array-band x-spectra. We find
\ba
A_{\rm CMB} &=& 0.91 \pm 0.23 \ \ \rm{(baseline  \ cut, \ PTE: 59\%)} \nonumber \\
 &=& 0.82 \pm 0.15 \ \ \rm{(extended  \ cut, \ PTE: 61\%)} 
\ea
corresponding to a $4 \sigma$ evidence for the B-mode power spectrum for the baseline cut, and a $5.5 \sigma$ detection for the extended cut.
In both cases, the amplitude of the B-mode power spectra is consistent with the $\Lambda$CDM model derived from the temperature and E-mode measurements. 

Since the B-mode signal is an order of magnitude weaker than the E-mode signal, it is significantly more vulnerable to any form of additive systematics present in the data. Establishing the consistency of our T-E-B results is therefore an important step for building confidence in our findings.

\subsection{EB \& TB power spectra}

The EB and TB power spectra serve as key observables for constraining cosmic birefringence, a phenomenon involving the rotation of linear polarization on cosmological scales due to parity-violating physics. In the \LCDM\ framework, the expected value of these power spectra is zero. However, physics beyond the Standard Model (SM) of elementary particles — such as the coupling of new pseudoscalar, axion-like fields to the electromagnetic tensor — can induce a rotation of the plane of linear polarization of photons, resulting in non-zero EB and TB power spectra \citep{carroll/etal:1990, 2022NatRP...4..452K}. Galactic emission is also expected to produce non-zero EB and TB signals \citep[e.g.,][]{huffenberger/etal:2020,clark/etal:2021}.
A challenge in interpreting these measurements is that mis-calibration of the polarization angle of the ACT detectors, characterizing the rotation of the polarization signal in the maps relative to the sky,  will also produce non-zero EB and TB, and is degenerate with an isotropic cosmic signal.

Figure \ref{fig:angle} shows the EB and TB spectra, combining the 16 EB and BE auto- and cross-array-band x-spectra that can be constructed from our four polarization array-bands: PA5~f090, PA5~f150, PA6~f090, and PA6~f150. In EB a clear non-zero excess is visible. In TB the signal is consistent with null.

To test whether these spectra are consistent with an overall rotation angle, we denote $\psi_{\alpha}$ as the mis-calibration (or birefringence) angle of the array-band $\alpha$. Assuming that EB and TB are zero at the time of decoupling and that we have negligible foreground emission, we model our observed EB and TB power spectra as
\ba
D^{E_{\alpha} B_{\alpha'}, \rm model}_{\ell} &=& D^{EE}_{\ell} \cos 2\psi_{\alpha}  \sin 2\psi_{\alpha'} - D^{BB}_{\ell} \sin 2\psi_{\alpha}  \cos 2\psi_{\alpha'} \nonumber \\
D^{B_{\alpha} E_{\alpha'}, \rm model}_{\ell} &=& D^{EE}_{\ell} \cos 2\psi_{\alpha'}  \sin 2\psi_{\alpha} - D^{BB}_{\ell} \sin 2\psi_{\alpha'}  \cos 2\psi_{\alpha} \nonumber \\
D^{T_{\alpha} B_{\alpha'}, \rm model}_{\ell} &=& D^{TE}_{\ell} \cos 2\psi_{\alpha}  \sin 2\psi_{\alpha'}  \nonumber \\
D^{B_{\alpha} T_{\alpha'}, \rm model}_{\ell} &=& D^{TE}_{\ell} \cos 2\psi_{\alpha'}  \sin 2\psi_{\alpha}.  \ea 
To estimate the $\{\psi_{\alpha} \}$, we construct a vector containing all the 16 EB and BE x-spectra from the four array-bands. We then sample the posterior distribution for the four angles simultaneously. The results are shown in the bottom-left panel of Figure \ref{fig:angle}.

To evaluate the stability of the results, we repeat this for both the baseline and extended cuts, and using two types of masks—a standard mask and an ``optimal" mask\footnote{The results obtained using this weighting scheme were not subjected to as many null tests as those presented in \S\ref{subsec:nulls}, which employed uniform weights. However, we have repeated the array-bands null test for the EB and TB power spectra and found them to be consistent.}. The optimal mask applies non-uniform pixel weighting, assigning more weight for pixels with higher signal-to-noise.   
We note that $\psi_{\rm PA5~f150}$ and $\psi_{\rm PA5~f090}$  differ by more than $3\sigma$ in the extended cuts case, while for the baseline cuts the difference is reduced to $2.4\sigma$. 

Assuming that the angle within an array should be common, we combine each pair of measurements within a given optics tube (e.g., for PA5) using the weighted average:
\ba
\psi_{\rm PA5} = \frac{ \frac{\psi_{\rm PA5~f090}}{\sigma^{2}_{\rm PA5~f090}} + \frac{\psi_{\rm PA5~f150}}{\sigma^{2}_{\rm PA5~f150}}}{\sigma^{-2}_{\rm PA5~f090}  + \sigma^{-2}_{\rm PA5~f150}}.
\ea
For the baseline cut with optimal weighting, this gives 
\ba
\psi_{\rm PA5} &=&  0.25 \pm 0.04^\circ \quad {\rm (stat-only)}\nonumber \\
\psi_{\rm PA6} &=&  0.14 \pm 0.05^\circ \quad {\rm (stat-only)},
\ea
with a combined average of  $0.20 \pm 0.03^\circ$ accounting only for statistical errors.
The theoretical prediction for the best-fitting angle of $\psi=0.2^\circ$ is shown in Figure \ref{fig:angle} together with our combined  EB and TB power spectra. Defining $\Delta \chi^{2}_{\rm EB} = \chi^{2}_{\rm EB}(\psi=0.2^\circ) - \chi^{2}_{\rm EB}(\psi=0^\circ)$, we obtain
\ba
&&\Delta \chi^{2}_{\rm EB}(\rm baseline, \ uniform \  weight) = -36 \nonumber \\
&&\Delta \chi^{2}_{\rm EB}(\rm baseline, \ optimal \ weight) = -41 \nonumber \\
&&\Delta \chi^{2}_{\rm EB}(\rm extended, \ uniform \ weight) =  -64 \nonumber \\
&&\Delta \chi^{2}_{\rm EB}(\rm extended, \ optimal  \ weight) = - 67,
\ea
for the four different cases we consider, indicating a strong preference for a non-zero polarization angle.

This analysis does not account for the potential systematic errors in the polarization angle calibration. For ACT, these are determined through a combination of metrology, modeling, and observations of planets, as described in N25. The orientation of the arrays with respect to the sky is determined using pointing information from the temperature maps. There is an uncertainty in this step of order $0.03^\circ$ per optics tube which house each of the PA4, PA5, and PA6 detector arrays. The rotation of the polarization from the sky to the detectors is then determined using an optics model of the telescope. \cite{2024ApOpt..63.5079M} conclude that the systematic uncertainty from this step is a further $\approx0.1^{\circ}$ per optics tube, estimated to within $0.04^\circ$. This error, primarily driven by uncertainties in the positioning of the lenses within each optics tube, is treated as independent for each tube. Finally the fabrication measurement is used to orient the polarization angle of the detector to the array. This is known to better than $0.01^{\circ}$.

To account for the optics-induced error of $0.1^{\circ}$ per tube we add it in quadrature to the statistical error, giving 
\ba
\hat{\psi}_{\rm PA5} &=&  0.25^{\circ} \pm 0.11^{\circ} \quad\rm{(stat+optics)} \nonumber \\
\hat{\psi}_{\rm PA6} &=&  0.14^{\circ} \pm 0.11^{\circ}\quad\rm{(stat+optics)}, 
\ea
as shown in Figure~\ref{fig:angle}. Assuming these are independent, this gives a combined
\be
\hat{\psi}_{\rm ACT}= 0.20^{\circ} \pm 0.08^{\circ}\quad\rm{(stat+optics)},
\ee
for the baseline cut with optimal weight. This is a 2.5$\sigma$ departure from zero, with some uncertainty from the optics-induced systematic error budget.\footnote{For example, a per-tube error of $0.14^\circ$ would be consistent with null at 2$\sigma$, or a lower $0.06^\circ$ error would give a more significant departure from zero.} This estimate also does not include the order $0.03^{\circ}$ pointing uncertainty.

To further evaluate the potential impact of foreground contamination, we recompute the angles in both the Northern and Southern patches, following the approach described in \S\ref{subsec:nulls} for the isotropy null test. The Northern patch has higher dust contamination than the Southern patch. We find no evidence for variation in the recovered angle between these two regions. We also test for possible variations of the angle with time, PWV, elevation, and detector position on each wafer, and find no dependence on these properties.

We do not null the polarization angles in the rest of the analysis. For E-modes, they contribute as an additional, though negligible, source of polarization efficiency, discussed in \S\ref{subsubsec:cal_and_peff}.

While the EB and TB spectra are consistent with an angle miscalibration, we do not exclude non-zero cosmic birefringence. We also note the following:
\begin{itemize}
\itemsep0em
\item Although the extended cuts are not used for any cosmological results, and while our result for $\hat{\psi}_{\rm ACT}$
does not depend on the choice of cut, we note that $\psi_{\rm PA5~f150}$ and $\psi_{\rm PA5~f090}$ are in statistical disagreement ($> 3\sigma$) 
in the extended cut case. This is unexpected given that these arrays are in the same optics tube, so at present we lack an instrument-based model to explain this discrepancy.
\item The preference for a nonzero $\hat{\psi}_{\rm ACT}$
  is primarily driven by our 150 GHz data, with only weak evidence for a non-zero angle in the 90 GHz data. This could point to frequency-dependent instrumental effects. 
\end{itemize}

Our results are consistent with previous estimates from ACT DR4 of 
$\psi = 0.07\pm0.09^\circ$ at f150 and $\psi=0.11\pm0.15^\circ$ at f090 \citep{choi_atacama_2020}, when presented using the same angle convention.\footnote{The analysis in \cite{choi_atacama_2020} used a convention such that the sign was opposite.} This was estimated from data taken with many physical replacements of the optics tubes, and so the resulting systematic error was estimated to be sub-dominant. The \Planck\ team estimate $\psi = 0.31\pm 0.05^\circ$ (stat-only) \citep{planck2016_XLIX}, and  estimate a $0.5-0.8^\circ$ systematic error from optical modeling \citep{tauber2010}. An alternative analysis from \cite{2022PhRvD.106f3503E} estimates the polarization angle by noting that any cosmic birefringence would rotate only the CMB signal, leaving the Galactic emission unrotated \citep[a method originally applied to the Planck data in][]{2020PhRvL.125v1301M}. By assuming a model for the Galactic dust EB emission, and neglecting synchrotron emission, this results in an angle estimate of $0.34 \pm 0.09^\circ$ from a joint analysis of polarization data from \Planck\ and \WMAP. A cross-correlation of ACT with \Planck\ data will be useful to explore this finding.

Given the complexity of the calibration procedures for individual experiments, a promising pathway forward, in addition to possible analyses modeling foregrounds, might involve compiling measurements from multiple telescopes, each benefiting from uncorrelated systematic errors.

\section{ACT and Planck spectra comparison}\label{sec:ACT_Planck}

In this section, we compare the ACT DR6 data with the publicly available \Planck\ data, for TT, EE and TE. We include both the \Planck\ legacy data (PR3) and the more recent NPIPE data (PR4). Our blinding strategy permitted comparisons of the temperature measurements between ACT and \Planck\ at any point during the analysis. 
However, the comparison of polarization data between ACT and \Planck\ was performed only after passing the set of internal null tests described in \S\ref{sec:ps_result}.

\begin{figure*}[t]
    \centering
    \includegraphics[width=\textwidth]{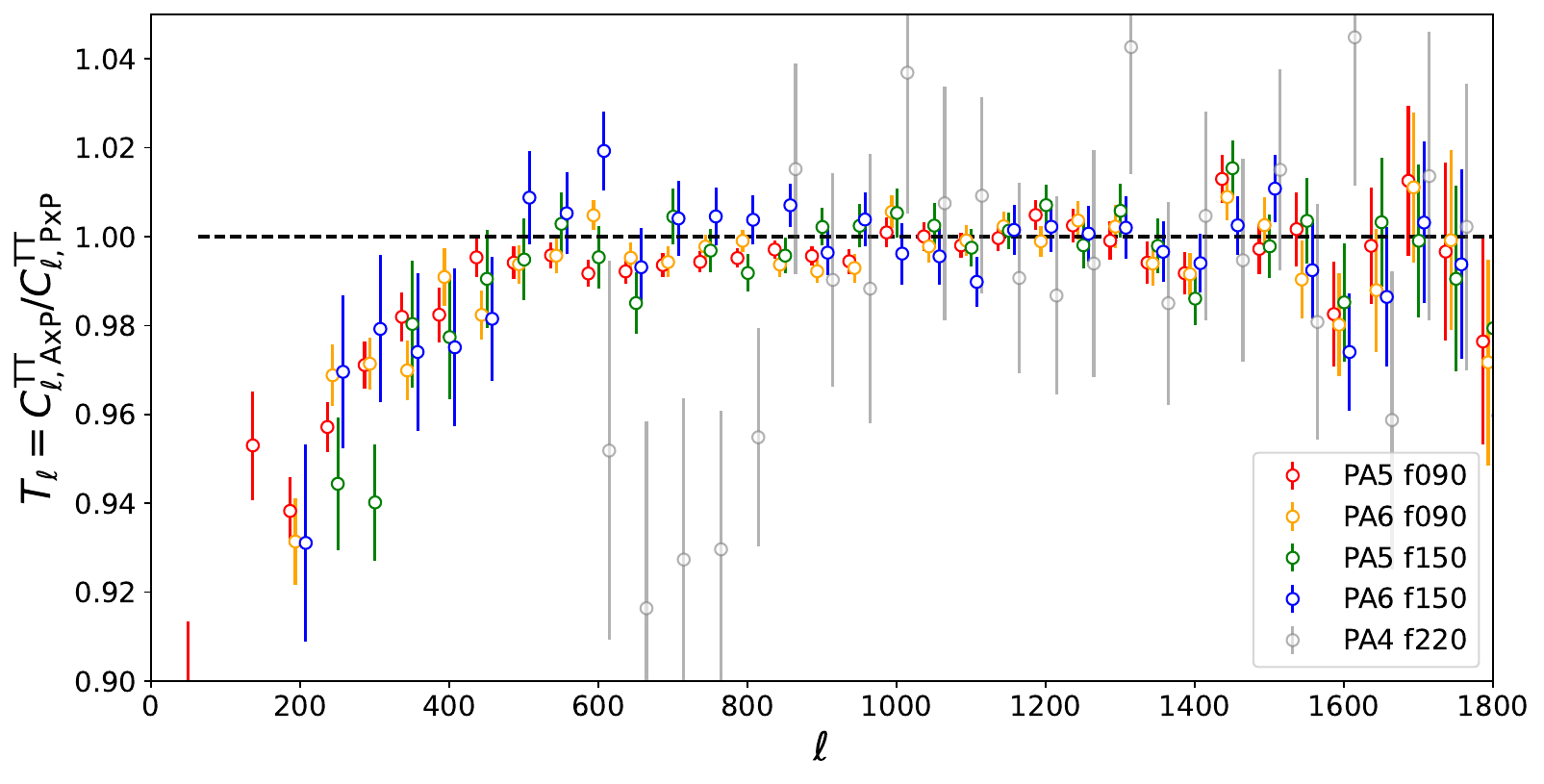}
    \caption{Estimate of temperature transfer functions of the ACT DR6 data obtained by comparing the cross-correlation between ACT and \Planck\ Legacy data with the Planck-only power spectrum. We use only the overlapping regions of the \Planck\ and ACT surveys.
    For our f090 and f150 data, we compare it with the \Planck\ 143 GHz channel data, and for our f220 data, we use the \Planck\ 217 GHz channel data. This ensures that we can measure the transfer functions up to high multipole values. These transfer functions are used to define the ACT multipole cuts: $\ell_{\rm min}=600$ for PA6~f150, $\ell_{\rm min}=800$ for PA5~f150, and $\ell_{\rm min}=1000$ for the other array-bands.\\}
    \label{fig:mm_tf}
\end{figure*}

\subsection{Planck maps: Legacy and NPIPE}

Legacy and NPIPE are alternative pipelines used to produce frequency maps from \Planck\ time-ordered data, that make different assumptions regarding detector calibration and systematic corrections. The NPIPE maps \citep{Plancknpipe} consist of two data splits per frequency, corresponding to two sets of detectors, referred to as A and B. In contrast, the Legacy pipeline \citep{planck2018_maps} splits the observation time of the satellite into two half-mission maps per frequency. The NPIPE maps benefit from an $8\%$ increase in integration time by including repointing maneuver data, resulting in lower noise levels.

\subsection{DR6 transfer function and multipole cut}

The dominant systematic in the ACT DR6 data manifests as a power loss in temperature, particularly on large angular scales.
Initially, we identified this issue by comparing the TT power spectra of different ACT DR6 array-bands. However, a more precise assessment of this power deficit can be achieved by comparing ACT data with \Planck\ data.

In Figure \ref{fig:mm_tf} we show a comparison of the cross-correlation between ACT and \Planck\ Legacy data with the \Planck-only power spectrum. To reduce sample variance and minimize the impact of potential Galactic foreground contamination, we use only the overlapping regions of the \Planck\ and ACT surveys and apply a common mask, computing the \Planck\ spectra using our analysis machinery. To account for differences in foreground emission due to differences in detector bandpasses, we subtract a foreground model from each of the ACT, \Planck, and ACT$\times$Planck spectra.

The mechanism causing this lack of power is suspected to  relate to the inter-gain calibration of the ACT bolometers, as described in N25 and \cite{naess/louis:2023} where a large-scale power deficit can be reproduced in map-making simulations and an analytic form for the transfer function can be obtained. However, we do not deconvolve this transfer function and instead use our measurement to define a multipole cut for the ACT data based on the transfer function shape. We choose $\ell_{\rm min}=600$ for PA6~f150, $\ell_{\rm min}=800$ for PA5~f150, and $\ell_{\rm min}=1000$ for the other array-bands. While it would have been more natural to use \Planck\ 100~GHz data to assess the magnitude of the transfer function affecting our 90~GHz array-bands, \Planck\ does not use its 100~GHz channel past $\ell=1000$ due to systematic contamination. Therefore, we use \Planck\ 143~GHz data as a reference for computing the transfer function of both our f150 and f090 arrays. The \Planck\ 217~GHz data is used to assess the magnitude of the PA4~f220 transfer function.

\subsection{Calibration}

N25 describes the use of Uranus observations to calibrate the ACT maps. We then refine this estimate using a calibration to \Planck, which itself calibrates to the annual dipole which causes an all-sky dipole in the CMB. We use the comparison of the ACT and \Planck\ temperature maps to infer a global calibration factor for each array-band for ACT, relative to the maps described in N25, following e.g., \cite{hajian/etal:2011,louis_atacama_2017}. We fit for an amplitude using three equivalent combinations of ACT and \Planck\ data:
\ba
^{(1)}{c }_{\rm pa X}^{\rm fY}  &=&  \left\langle \frac{C_{\ell, \rm TT}^{\rm pa X f Y \ x \ \rm pa X f Y }}{C_{\ell, \rm TT}^{\rm pa X f Y \ x \ \rm Planck }}\right\rangle \nonumber \\
^{(2)}{c }_{\rm pa X}^{\rm fY} &=& \left\langle \frac{C_{\ell, \rm TT}^{\rm pa X f Y \ x \ \rm pa X f Y }}{C_{\ell, \rm TT}^{\rm Planck \ x \ \rm Planck }}\right\rangle^{1/2}  \nonumber \\ 
^{(3)}{c }_{\rm pa X}^{\rm fY} &=&  \left\langle \frac{C_{\ell, \rm TT}^{\rm pa X f Y \ x \ \rm Planck }}{C_{\ell, \rm TT}^{\rm Planck \ x \ \rm Planck }}\right\rangle .
\ea
Here the average is performed in the range of angular multipoles $\ell \in [\ell^{\rm pa X f Y}_{\rm min}, 2000]$.
We report these calibration factors for both the Legacy and NPIPE maps in Table \ref{tab:cal} in Appendix \ref{apx:act_planck}. The calibration factors for PA5 and PA6 are stable at the $0.3\%$ level between the three methods; 
PA4 shows a larger scatter although consistent with  statistical errors. This stability of the calibration numbers across the different methods is important, as the different methods would not have to agree if the ACT or \Planck\ data were affected by additive systematic effects. We indicate  which factors are used to calibrate the maps in the table in Appendix \ref{apx:act_planck}.

The calibration of the polarization data is characterized by the product of the overall calibration and a polarization efficiency.
Since neither \Planck\ nor ACT provides a precise assessment of its polarization efficiency, we first rescale each data spectrum using a polarization efficiency parameter estimated from a fiducial $\Lambda$CDM E-mode power spectrum. This approach enables, for example, null tests that are independent of the overall amplitude of the spectra.
To eliminate this model-dependent calibration in the later stages of the analysis, we treat the polarization efficiencies as free parameters in the likelihood, sampling them jointly with each cosmological model. Instrumental uncertainties are accounted for by applying a broad $\pm10\%$ flat prior to these parameters.

\subsection{Comparison of polarization data}

Assessing the consistency between \Planck\ and ACT polarization data provides a powerful test for potential systematics affecting either experiment. 

\begin{table}[htp]
	\centering
	\hspace*{-20mm}\begin{tabular}{r|rr|rr|rr}
        \cskip  & \multicolumn{2}{c}{Planck f100} & \multicolumn{2}{c}{Planck f143}  & \multicolumn{2}{c}{Planck f217}\\
  		\hline
  		\cskip  & PR3 & NPIPE & PR3 & NPIPE  & PR3 & NPIPE\\
            \hline
            PA5~f090 & 90\% & 55\% & 84\% & 76\% & 12\% & 46\% \\
            PA5~f150 & 86\% & 49\% & 86\% & 69\% & 7\% & 59\% \\
            PA6~f090 & 86\% & 55\% & 73\% & 54\% & 15\% & 49\% \\
            PA6~f150 & 78\% & 55\% & 82\% & 78\% & 9\% & 42\% \\
            \hline
            \hline
            PA5~f090 & 43\% & 3\% & 66\% & 82\% & 36\% & 69\% \\
            PA5~f150 & 23\% & 1\% & 48\% & 65\% & 72\% & 78\% \\
            PA6~f090 & 37\% & 5\% & 65\% & 81\% & 78\% & 96\% \\
            PA6~f150 & 50\% & 9\% & 26\% & 41\% & 74\% & 78\% \\
	\end{tabular}
	\caption{PTE vaues for the comparison of the ACT DR6 and \Planck\ EE (top) and TE (bottom) power spectra computed for the same region of the sky. We use the same temperature maps when comparing the ACT and \Planck\ TE power spectra. The residuals are shown in Appendix \ref{apx:act_planck}. }
	\label{tab:p_value_AP}
\end{table}

We form residuals between the {ACT $\times$ ACT}, {ACT $\times$ Planck}, and {Planck $\times$ Planck} power spectra,\footnote{Computed on the same region of the sky.} and estimate their expected error using Monte Carlo simulations. 
For the \Planck\ legacy data, we use the 300 simulations released by the \Planck\ collaboration, while for NPIPE we use 500 available simulations.
We correct the residuals for the expected temperature-to-polarization leakage present in both the ACT and \Planck\ data, 
and account for different detector passbands by subtracting from each spectrum its estimated Galactic dust level, as described in \S\ref{subsec:diff_gal}. Overall, we find excellent agreement between ACT DR6 and the \Planck\ Legacy data, and with the NPIPE 143~GHz data. We see lower PTEs  when comparing with the \Planck\ 100~GHz NPIPE data, but no significant issues. A summary of the PTE values for each residual is given in Table \ref{tab:p_value_AP}, and all residuals are shown in Appendix \ref{apx:act_planck}.
We note that the PTE values are strongly correlated since the different residuals are formed using common data.
An assessment of this correlation is provided in Appendix \ref{apx:act_planck}.

\section{Likelihood methods}\label{sec:like}
We create two likelihoods for the ACT TT, EE and TE data, following e.g., \citet{dunkley/etal:2013,choi_atacama_2020}. The first is a likelihood for the multi-frequency spectra, given a model that includes CMB and foregrounds. The second is a CMB-only likelihood for a compressed data vector that contains an estimate of the CMB bandpowers marginalized over foreground contamination. 

\subsection{Multi-frequency likelihood}\label{sec:likelihood}

Following \citet{dunkley/etal:2013,choi_atacama_2020} we use a Gaussian likelihood to describe the multi-frequency power spectra, implemented in the \texttt{MFLike}\footnote{\href{https://github.com/simonsobs/LAT_MFLike/tree/v1.0.0}{\texttt{LAT\_MFLike}, version 1.0.0}} software developed for the Simons Observatory. The total theory model, for each of our auto and cross array-band x-spectra, is given by
\be \label{eq:dl_theory}
D_\ell^{{\rm th}, X_{\alpha} Y_{\beta}} = D_\ell^{{\rm CMB}, XY} + D_\ell^{{\rm FG},X_{\alpha} Y_{\beta}},
\ee
where $D_\ell^{{\rm CMB}, XY}$ is the lensed primary CMB (described in~\S\ref{subsec:lensed_cmb}), and $D_\ell^{{\rm FG},X_{\alpha} Y_{\beta}}$ models the frequency-dependent astrophysical foregrounds, described in the following. This is then binned, with bandpower window functions, $w$, such that $D_b^{\rm th, X_{\alpha} Y_{\beta}} = \sum_{\ell} w^{X_{\alpha} Y_{\beta}}_{b\ell}D_\ell^{X_{\alpha} Y_{\beta}}$. The model vector, $\bm{D^{\rm th}}$, ordered to match the data vector, is compared to the data, $\bm{D^{\rm data}}$, using the Gaussian likelihood
\be
-2\ln L = (\bm{D^{\rm th}}-\bm{D^{\rm data}})^T \bm{\Sigma}^{-1} (\bm{D^{\rm th}}-\bm{D^{\rm data}})
\ee
to within an additive constant. The covariance matrix $\bm{\Sigma}$ is described in \S\ref{subsec:covmat}. The data vector $\bm{D^{\rm data}}$ includes  fifteen auto and cross array-band x-spectra for TT, ten for EE,  and sixteen for TE, for a total of 1651 data points.

\subsubsection{Foreground model}{\label{subsec:fg_model}}
Our foreground model is a sum of Galactic and extragalactic foregrounds, and secondary CMB anisotropy. In this section we describe the nominal model we use, based on earlier analyses in e.g.,~\cite{dunkley/etal:2013,choi_atacama_2020}, and a set of possible extensions that we consider. 

The temperature maps receive contributions from unresolved radio point sources, cosmic infrared background (CIB) emission from dusty galaxies, thermal and kinetic Sunyaev Zel'dovich anisotropies  (tSZ \& kSZ) from electrons, and emission from Galactic dust. We neglect Galactic synchrotron, free-free and anomalous microwave emission which are expected to be negligible at our frequencies and scales. We do not include the emission from extragalactic carbon monoxide (CO) in our nominal model, which was recently identified as a possible non-negligible contaminant~\citep{Maniyar_CO, Kokron_CO} but is challenging to model. In polarization, we account for Galactic dust and radio point source emission, and neglect other possible contributions. 

In the following we describe each model term that adds power to the total TT/TE/EE spectra, written in CMB-referenced thermodynamic units for passbands that are delta functions in frequency, denoted $i,j$ hereafter; in practice they are computed by integrating across the telescope passbands.  We use the \texttt{fgspectra}\footnote{\href{https://github.com/simonsobs/fgspectra/tree/v1.3.0}{\texttt{fgspectra}, version 1.3.0}} software to produce the model spectra. Apart from the Galactic dust, and the shape of the tSZ spectrum, we use the same modeling of components as in previous ACT analyses, and they are described further in \cite{dunkley/etal:2013,choi_atacama_2020}. A summary of parameters describing the model is given in Table \ref{tab:act-foreground-parameters}, and the spectrum of each component is shown in the results \S\ref{subsec:fg}.

Unresolved radio sources are assumed to be Poisson distributed and therefore have a flat angular power spectrum. We assume that the spectral energy distribution (SED) is a power law such that
\be
D_{\ell}^{X_i Y_j},\mathrm{radio} = a_\mathrm{s}^{XY}\left[\frac{\ell(\ell + 1)}{\ell_0(\ell_0 + 1)}\right]\left[\frac{g(\nu_i)g(\nu_j)}{g^2(\nu_0)}\right] \left[ \frac{\nu_i\nu_j}{\nu_0^2} \right]^{\beta_s+2},
\ee
where $\beta_s$ is the spectral index of the radio source SED and $a_\mathrm{s}^{XY}$ is the amplitude of the radio source power spectrum at pivot scale $\ell = \ell_0$ and pivot frequency $\nu=\nu_0$. The function $g(\nu)$ converts from flux to CMB-referenced thermodynamic units. Assuming that the SED is the same in polarization as in intensity, we model the radio source emission with the same frequency dependence but separate amplitudes, $a_\mathrm{s}^{TT}$, $a_\mathrm{s}^{TE}$, $a_\mathrm{s}^{EE}$, and allow the TE amplitudes to span negative values to capture possible anti-correlations between emission in T and E. For these and other extragalactic terms we choose $\ell_0=3000$ and $\nu_0=150$~GHz, following previous analyses.

The CIB signal is sourced by contributions from clustered and Poisson-distributed dusty galaxies. 
We model the spatial statistics of the latter as for the radio sources, with a frequency dependence 
following a modified blackbody with emissivity index $\beta_p$ and temperature $T_d = 9.6\;\mathrm{K}$, with
\begin{equation}
    D_\ell^{T_i T_j}, \mathrm{CIB}{-}{p} = a_p \left[ \frac{\ell(\ell+1)}{\ell_0(\ell_0 + 1)} \right] \frac{\mu(\nu_i; \beta_p, T_d)\mu(\nu_j; \beta_p, T_d)}{\mu^2(\nu_0; \beta_p, T_d)}.
\label{eqn:cib_p}
\end{equation}
Here $\mu_d(\nu,\beta_d)\equiv \nu^{\beta_d}B_{\nu}(T_d)g(\nu)$, where $B_{\nu}(T_d)$ is the Planck function at frequency $\nu$. The clustered component of the CIB emission is given by
\begin{equation}
    D_\ell^{T_i T_j}, \mathrm{CIB}{-}{c} = a_c D_{\ell, \ell_0}^\mathrm{CIB-c} \frac{\mu(\nu_i; \beta_c, T_d)\mu(\nu_j; \beta_c, T_d)}{\mu^2(\nu_0; \beta_c, T_d)},
    \label{eqn:cib_c}
\end{equation}
where $\beta_c$ is the spectral index and $D_{\ell, \ell_0}^\mathrm{CIB-c}$ is a template normalized at $\ell = \ell_0$. We construct this template using the CIB power measured from \Planck ~\citep{Planck_CIB} up to $\ell=3000$ and continuing with a power law $\ell^{0.8}$ at smaller scales, motivated by earlier ACT and SPT measurements \citep{Das:2013zf,dunkley/etal:2013}.

The tSZ signal spectrally distorts the blackbody spectrum due to the inverse Compton scattering of CMB photons off electrons in hot gas, and we model it as 
\begin{equation}
    D_\ell^{T_i T_j},\mathrm{tSZ} = a_\mathrm{tSZ} D_{\ell, \ell_0}^{\mathrm{tSZ}}\left[\frac{\ell}{\ell_0}\right]^{\alpha_\mathrm{tSZ}} \frac{f(\nu_i)f(\nu_j)}{f^2(\nu_0)},
\end{equation}
where $D_{\ell, \ell_0}^{\mathrm{tSZ}}$ is a tSZ template normalized at $\ell=\ell_0$ and frequency $\nu=\nu_0$ from \citet{battaglia_tsz}. The parameter $\alpha_{\rm tSZ}$ is a new parameter that we introduce in the model, not previously included in the analysis of ACT data, or other \Planck\ and SPT analyses, that allows for a different scale dependence of the tSZ signal compared to the template. This is motivated by the discrepancy we observe between our nominal template shape and the predicted spectrum from the Agora \jd{or BAHAMAS simulations \citep{mccarthy/etal:2017,agora}.} The function $f(\nu)=\left(h\nu / k_B T_\mathrm{CMB}\right)\mathrm{coth}\left( h\nu / 2k_B T_\mathrm{CMB} \right) - 4$ rescales
the expected tSZ signal observed at a frequency $\nu$.

The kSZ signal is a blackbody component that adds anisotropy to the CMB signal due to Doppler scattering off moving electrons. We model it via the rescaling of a template describing the late-time kSZ from \citet{Battaglia_ksz1, Battaglia_ksz2} such that
\begin{equation}
    D_\ell^{T_i T_j},\mathrm{kSZ} = a_\mathrm{kSZ} D_{\ell, \ell_0}^\mathrm{kSZ}.
\end{equation}

A non-zero correlation is expected between the tSZ emission and the clustered component of the CIB signal \citep{addison/dunkley/spergel:2012}. We model this contribution as
\begin{align}
    D_\ell^{T_i T_j},\mathrm{tSZ}\times\mathrm{CIB} = - \xi \sqrt{a_c a_\mathrm{tSZ}}D_{\ell, \ell_0}^{\mathrm{tSZ}\times\mathrm{CIB}} \nonumber\\
    \times \left( \frac{f(\nu_i)\mu(\nu_j; \beta_c, T_d) + f(\nu_j)\mu(\nu_i; \beta_c, T_d)}{f(\nu_0)\mu(\nu_0; \beta_c, T_d)} \right),
\end{align}
where $D_{\ell, \ell_0}^{\mathrm{tSZ}\times\mathrm{CIB}}$ is a template normalized at $\ell=\ell_0$ and frequency $\nu = \nu_0$, and $\xi$ the correlation between the tSZ and CIB components. We neglect correlations between other components in the baseline model.

We model the Galactic dust emission with a power law power spectrum, with frequency dependence given by a modified blackbody spectrum,
\begin{equation}\label{eq:dust_pw}
    D_\ell^{X_i Y_j},\mathrm{dust} = a_g^{XY}\left[ \frac{\ell}{\ell_0} \right]^{\alpha_\mathrm{dust}^{XY}}\frac{\mu(\nu_i; \beta_d, T_d^\mathrm{eff})\mu(\nu_j; \beta_d, T_d^\mathrm{eff})}{\mu^2(\nu_0; \beta_d, T_d^\mathrm{eff})}.
\end{equation}
We fix the power law index to be $\alpha_g^{TT} = -0.6$ in temperature and $\alpha_g^{TE/EE}=-0.4$ for polarized emission, and set the effective dust temperature to be $T_d^\mathrm{eff}=19.6\;\mathrm{K}$ and the emissivity $\beta_d=1.5$, motivated by observations from \Planck\ \citep{planck_poldust:2018}.\footnote{The mean index measured from \Planck\ data outside the Galactic plane is $\beta_d=1.48\pm0.01$ in intensity, and $1.53\pm0.03$ in polarization \citep{planck_poldust:2018}. Some spatial variation of the power-law behavior is noted in \cite{cordovarosado2024} from a TE analysis of WISE, \Planck\ and ACT data.} We fit for different amplitudes in TT, TE and EE, for pivot scale $\ell_0=500$. 

\subsubsection{Estimates of Galactic dust emission}\label{subsec:diff_gal}

Atmospheric fluctuations limit our ability to measure the larger-scale emission from Galactic dust. To address this, we use data from \Planck\ PR3 at 353~GHz and 143~GHz to estimate the amplitude of the expected contamination. We calculate the following dust-dominated residuals in our sky region: 
\ba
\Delta D^{XY, \rm data}_\ell &=& D^{XY, \ 353 \rm GHz \times 353 \rm GHz}_{\ell, \ \rm planck} +  D^{XY, \ 143  \rm GHz \times 143  \rm GHz}_{\ell,  \ \rm planck} \nonumber \\
&-& 2  D^{XY, \ 143 \rm GHz \times 353 \rm GHz}_{\ell,  \ \rm planck} ,
\ea
a combination of spectra that does not contain CMB signal. For the EE, BB and TE power spectra we model these residuals as
\be
\Delta D^{XY, \rm model}_\ell = a^{XY}_{g}(\ell / \ell_{0})^{-0.4}\Delta^{\rm dust}_{353, 143},
\ee
where $\Delta^{\rm dust}_{353, 143}$ gives the expected frequency scaling of the residual assuming a modified blackbody integrated in the \Planck\ measured passbands, as in Equation \ref{eq:dust_pw}. 

\begin{figure}[tp]
	\centering
	\includegraphics[width=\columnwidth]{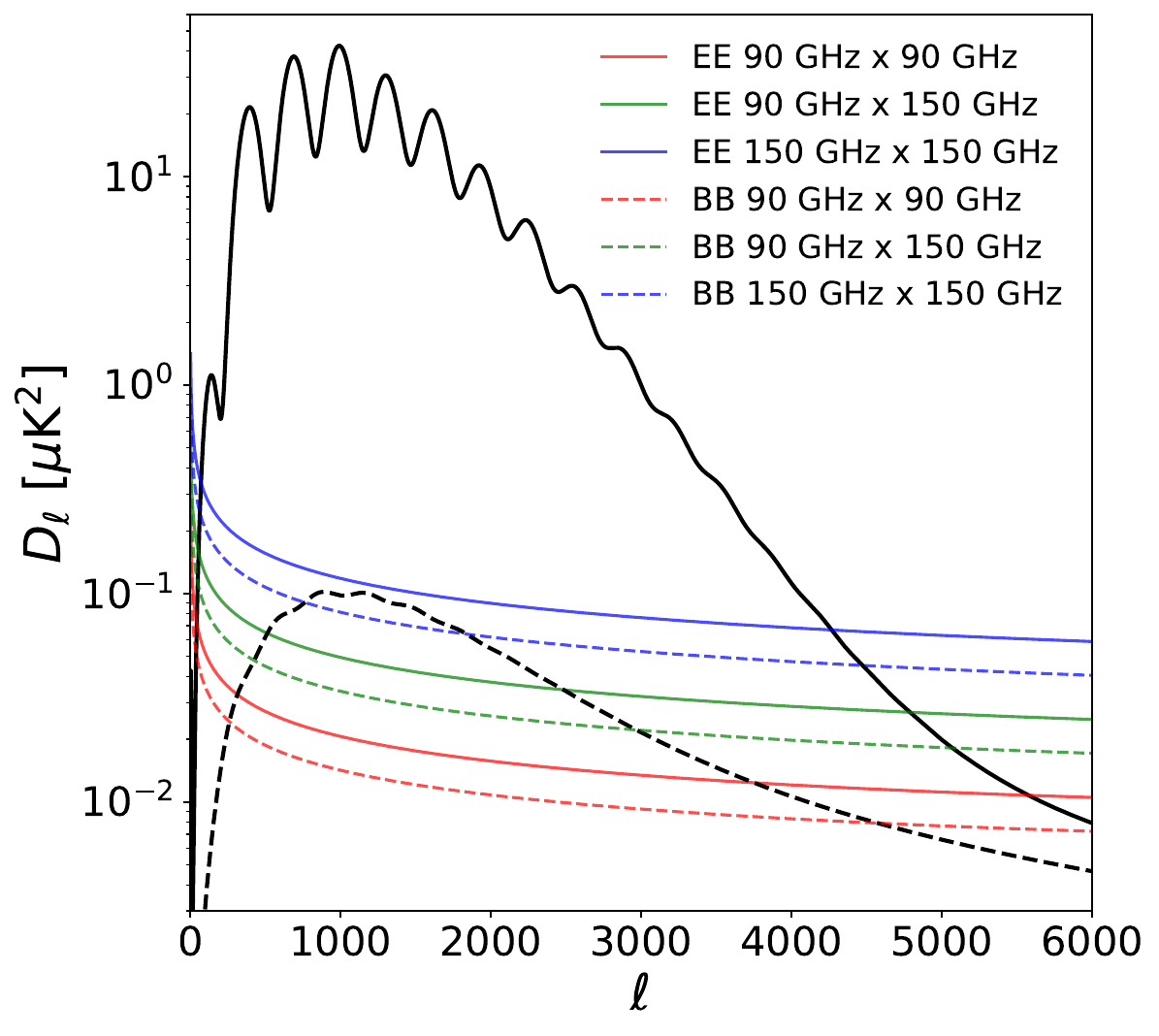}
	\caption{Model for the polarized dust emission in the region used for the ACT power spectra, estimated by fitting \Planck\  data at 353~GHz and 143~GHz, and assuming a power-law multipole dependence. The CMB EE (solid line) and BB (dashed line) power spectra for a \LCDM\ model are shown for comparison.}
	\label{fig:spec_dust}
\end{figure}

We find this power-law model provides an excellent fit to the \Planck\ data residuals in our observed sky area, with amplitudes at $\ell_0=500$ and pivot frequency $\nu_0=150$~GHz estimated as
\ba
a^{EE}_{g} &=&  0.17 \pm 0.01  \ \ ( {\rm PTE}: 38\%) \nonumber \\ 
a^{TE}_{g} &=&  0.420 \pm 0.015  \ \ ( {\rm PTE}: 69\%) \nonumber \\
a^{BB}_{g} &=&  0.11 \pm 0.01 \ \ ( {\rm PTE}: 79\%)
\ea
where PTE gives the probability to exceed. Figure~\ref{fig:spec_dust} shows this model for the polarized dust power spectra in the ACT survey area.

The temperature residual between 143~GHz and 353~GHz includes a CIB contribution; other components are expected to be negligible. To account for this we use the CIB model described in Equations \ref{eqn:cib_p} and \ref{eqn:cib_c}, such that
\ba
\Delta D^{TT}_\ell &=& a^{TT}_{g}(\ell / \ell_{0})^{-0.6} \Delta^{\rm dust}_{353, 143} +  \Delta D^{\rm CIB}_\ell (a_{c}, a_{p}). 
\ea
We fit for $a^{TT}_{g}$ while marginalizing over $a_{c}$ and $a_{p}$. This yields
\be
a^{TT}_{g} =  8.0 \pm 0.2  \ \ ( {\rm PTE}: 20\%). 
\ee
We use these measurements as Gaussian priors on the dust amplitudes, as summarized in \S\ref{sec:params}, conservatively doubling the errors for the prior widths.

\subsubsection{Additional foreground complexity}
\label{subsec:fg_ext}

We test a set of changes or extensions to the foreground model  -- motivated by comparisons to the Agora and PySM3 simulations -- that are expected to better characterize the true emission, but may add unnecessary complexity to the model given the quality of the current data. We did this after unblinding.
The set of changes we define pre-unblinding include (1) switching the tSZ, kSZ and tSZ-CIB templates from our nominal choices to those estimated from the Agora simulations (2) including both early- and late-time terms in the kSZ template, and (3) changing the Galactic dust scale-dependence slope to $\alpha=-0.7$ for temperature. 

The individual extensions we consider are to allow the radio source index, $\beta_s$, to be different for temperature and polarization (4); to marginalize over the shape of the high-ell CIB template (5); to allow the dust emissivity index $\beta_p \ne \beta_c$ for the Poisson and clustered CIB components with the standard $T=9.6$~K (6); or higher dust temperature $T=25$~K (7). We test the effect of including additional correlations between components, adding in a possible template for the CO$\times$CIB and CO$\times$CO contribution (8); and adding templates for radio$\times$SZ and radio $\times$ CIB correlations (9). Finally, we include a possible decorrelation of the radio and CIB emission with frequency (10). Additional tests we included post-unblinding were allowing the scale-dependent slope of the polarized Galactic dust to vary, and removing the priors on the dust amplitudes. 

Other than the tSZ shape, all of these model adjustments result in a shift in the estimated cosmological parameters\footnote{Tested for \LCDM\ and \LCDM$+N_{\rm eff}$ models.} by less than $0.5\sigma$, and so we do not adopt them in our baseline model. They may become important for future, deeper data sets. It was this set of tests, however, that revealed the need to include the tSZ shape parameter, $\alpha_{\rm tSZ}$, in our nominal model. The impact of including it on cosmological parameters was at the 0.5$\sigma$ level for the \LCDM$+N_{\rm eff}$ model, and a non-zero value for $\alpha_{\rm tSZ}$ was preferred at 3$\sigma$. Highlights of these tests are provided in \S\ref{sec:results}, with further details in \citet{fg_inprep}. 

\subsubsection{Instrument passbands and beam chromaticity}\label{subsec:instrument}
The DR6 array-bands are sensitive to specific frequency ranges, described by their bandpass transmission functions, $\tau^{\alpha}(\nu)$. To reliably estimate foreground parameters, we integrate each component of the foreground model across these frequencies, ensuring consistency with the instrumental response.

{In addition, we include chromatic beam window functions, $b^{\alpha}_{\ell}(\nu)$, to account for how the ACT beam varies with frequency, and the spectral distribution of foregrounds. This step is important for recovering foreground parameters, as neglecting this effect can shift parameter estimates by up to $1\sigma$.

For two array-bands $\alpha$ and $\beta$ with passbands $\tau^{\alpha}(\nu)$ and $\tau^{\beta}(\nu)$, the modeled signal of a given foreground component (``comp"), for beam-filling sources, is 
\begin{equation}
    D_\ell^{X_{\alpha}, Y_{\beta}}, \mathrm{comp} = \int d\nu_{i}  d\nu_{j} \tilde{B}^{\alpha}_{\ell}(\nu_{i}) \tilde{B}^{\beta}_{\ell}(\nu_{j}) D_\ell^{X_i Y_j}, \mathrm{comp}
\end{equation}
where $\tilde{B}^{\alpha}_{\ell}(\nu_{i})$ are normalized passbands\footnote{The normalized passband is computed as 
\be
\tilde{B}^{\alpha}_{\ell}(\nu) = \frac{b^{\alpha}_{\ell}(\nu) \mathcal{F}(\nu) \tau^{\alpha}(\nu) \nu^{-2}}{\int d\nu' b^{\alpha}_{\ell}(\nu') \mathcal{F}(\nu') \tau^{\alpha}(\nu') \nu'^{-2}}
\ee
where $\mathcal{F}(\nu) \equiv \frac{\partial \textrm{B}(T,\nu)}{\partial T}|_{T=T_{\rm CMB}} \propto \nu^2 \frac{x^2 e^x}{(e^x-1)^2}$ is the CMB-referenced thermodynamic temperature to surface brightness unit conversion factor (equal to $g^{-1}(\nu)$, where $g(\nu)$ is defined in \S\ref{subsec:fg_model}), and $b^{\alpha}_\ell$ is the beam window function that is measured from the data. More details on this derivation are in~\cite{Giardiello:2024rew}. The $\nu^{-2}$ factor is applied to the passbands since they are measured as the response to a Rayleigh-Jeans source. } and ``$ D_\ell^{X_i Y_j}, \mathrm{comp}$" are the foreground power spectra described in \S\ref{subsec:fg_model}.

As described in N25, the measurement of the ACT passbands was performed using a Fourier-transform spectrometer (FTS) based on the PIXIE design \citep{2011JCAP...07..025K} with added coupling optics that match the outgoing beam from the FTS to the input of the ACT receiver. Each detector array was measured with the FTS sampling 15--20 different pointing positions over the input window of the optics tube in order to sample across the focal plane. These data were combined and averaged, with weights given by the inverse variance of the noise. The errors on the passbands are a combination of statistical errors and optical systematic errors of the FTS and coupling optics system. The resulting uncertainties in the passbands can be well approximated as uniform shifts across the entire passband by a fixed value. For most array-bands, these uncertainties are of order 1 GHz. However, for PA4~f220, they are significantly larger, reaching up to 3.6 GHz.

Marginalizing over these bandpass uncertainties constitutes a substantial contribution to the overall error budget in the recovery of foreground parameters. To account for this effect, we introduce a bandpass shift parameter $\Delta^\alpha_\nu$ for each array-band, allowing for shifts of the form $\tau^{\alpha}(\nu) \to \tau^{\alpha}(\nu + \Delta^\alpha_\nu)$. These shift parameters are sampled jointly with the cosmological and foreground parameters, ensuring that the impact of bandpass uncertainties is fully integrated into our analysis. 

For simplicity, and while sampling the bandpass shift parameter,  the  shift is not propagated to the chromatic beam computation, as the correction to the beams due to the variation of the bandpass shift parameters are second order.  This formalism is implemented in the \texttt{MFLike}\footnote{\href{https://github.com/simonsobs/LAT_MFLike/tree/v1.0.0}{LAT\_MFLike, version 1.0.0}} software and described in \citet{Giardiello:2024rew}.

\subsubsection{Calibration and polarization efficiency}\label{subsubsec:cal_and_peff}

The calibration factors estimated in \S\ref{sec:ACT_Planck} are used to calibrate the ACT maps in intensity, such that the expected calibration factor is unity after this process. The factors that are used are highlighted in Table \ref{tab:cal}. There is an associated error for each array-band, which we conservatively double to define the prior used when sampling cosmological parameters. To model these uncertainties, we introduce five calibration parameters, denoted as $c_{\rm pa_X}^{\rm fY}$, in our data model, for $X \in \{4,5,6\}$ and $Y \in \{090,150,220\}$.

Uncertainties in the \Planck\ dipole calibration are also propagated into the analysis. To address this, we introduce a parameter, $\mathrm{cal}_{\rm dipole}$, which encapsulates the calibration uncertainties specific to \Planck\footnote{For historical reasons, this parameter is labeled as $\mathrm{cal}_{\rm ACT}$ or $\mathrm{calG}_{\rm all}$ in some of our MCMC chains.}.

In polarization, the situation is more complex. The amplitude of the spectra are influenced not only by calibration but also by the polarization efficiency of the detectors. This efficiency quantifies the detector's ability to accurately measure the polarized signal and is currently subject to significant uncertainties due to the lack of a precise, independent measurement. To account for this, we introduce a separate parameter for each array-band, $p_{\rm pa_X}^{\rm fY}$, modeling uncertainties in polarization efficiency across the four array-bands used in ACT.
The calibrated theoretical model, incorporating these factors, is computed as:

\ba \label{eq: lik_cal}
D^{\rm th, T_{\rm pa_X}^{\rm fY} T_{\rm pa_W}^{\rm fZ}}_{\ell, \rm cal} &=& D^{\rm th, T_{\rm pa_X}^{\rm fY} T_{\rm pa_W}^{\rm fZ}}_{\ell} (\mathrm{cal}^2_{\rm dipole}\, c_{\rm pa_X}^{\rm fY} c_{\rm pa_W}^{\rm fZ})^{-1} \nonumber \\
D^{\rm th, T_{\rm pa_X}^{\rm fY} E_{\rm pa_W}^{\rm fZ}}_{\ell, \rm cal} &=& D^{\rm th, T_{\rm pa_X}^{\rm fY} E_{\rm pa_W}^{\rm fZ}}_{\ell} (\mathrm{cal}^2_{\rm dipole}\, c_{\rm pa_X}^{\rm fY} c_{\rm pa_W}^{\rm fZ} p_{\rm pa_W}^{\rm fZ})^{-1} \nonumber \\
D^{\rm th, E_{\rm pa_X}^{\rm fY} E_{\rm pa_W}^{\rm fZ}}_{\ell, \rm cal} &=&D^{\rm th, E_{\rm pa_X}^{\rm fY} E_{\rm pa_W}^{\rm fZ}}_{\ell} (\mathrm{cal}^2_{\rm dipole}\, c_{\rm pa_X}^{\rm fY} c_{\rm pa_W}^{\rm fZ} p_{\rm pa_X}^{\rm fY} p_{\rm pa_W}^{\rm fZ})^{-1}. \nonumber \\ 
\ea

\subsection{CMB-only likelihood} \label{subsec:cmb-only-likelihood}
We follow the method described in \citet{dunkley/etal:2013, choi_atacama_2020}, and used in e.g., \citet{balkenhol2025}, to estimate CMB-only bandpowers from the multi-frequency dataset that have been marginalized over foreground contamination and relative calibration and passband uncertainty. We use the \texttt{MFLike} multi-frequency likelihood as described above, where instead of the 6 \LCDM\ parameters we estimate $N_{\rm data}=135$ bandpowers, for TT, TE and EE (with 45 bandpowers in each of TT, TE and EE). This method involves jointly Gibbs-sampling the bandpowers, and using the Metropolis-Hastings algorithm to sample the foreground and systematics parameters.  More details are in Appendix~\ref{apx:cmbonly}.

An estimate of the mean values of the bandpowers, and their covariance matrix, are obtained from the Gibbs samples and passed as inputs to our CMB-only likelihood, \texttt{ACT-lite}, which we approximate as Gaussian, with 
\be
-2 \ln L = (\bm{D}^{\rm th, CMB} - \bm{D}^{\rm{data}})(\bm{\Sigma}^{\rm CMB})^{-1}(\bm{D}^{\rm th, CMB} - \bm{D}^{\rm{data}})
\ee
to within an additive constant. All the frequency-dependent parameters have been marginalized over and this likelihood only includes a single overall calibration and a single polarization efficiency parameter.

\begin{figure}[htp]
	\centering
         \includegraphics[width=\columnwidth]{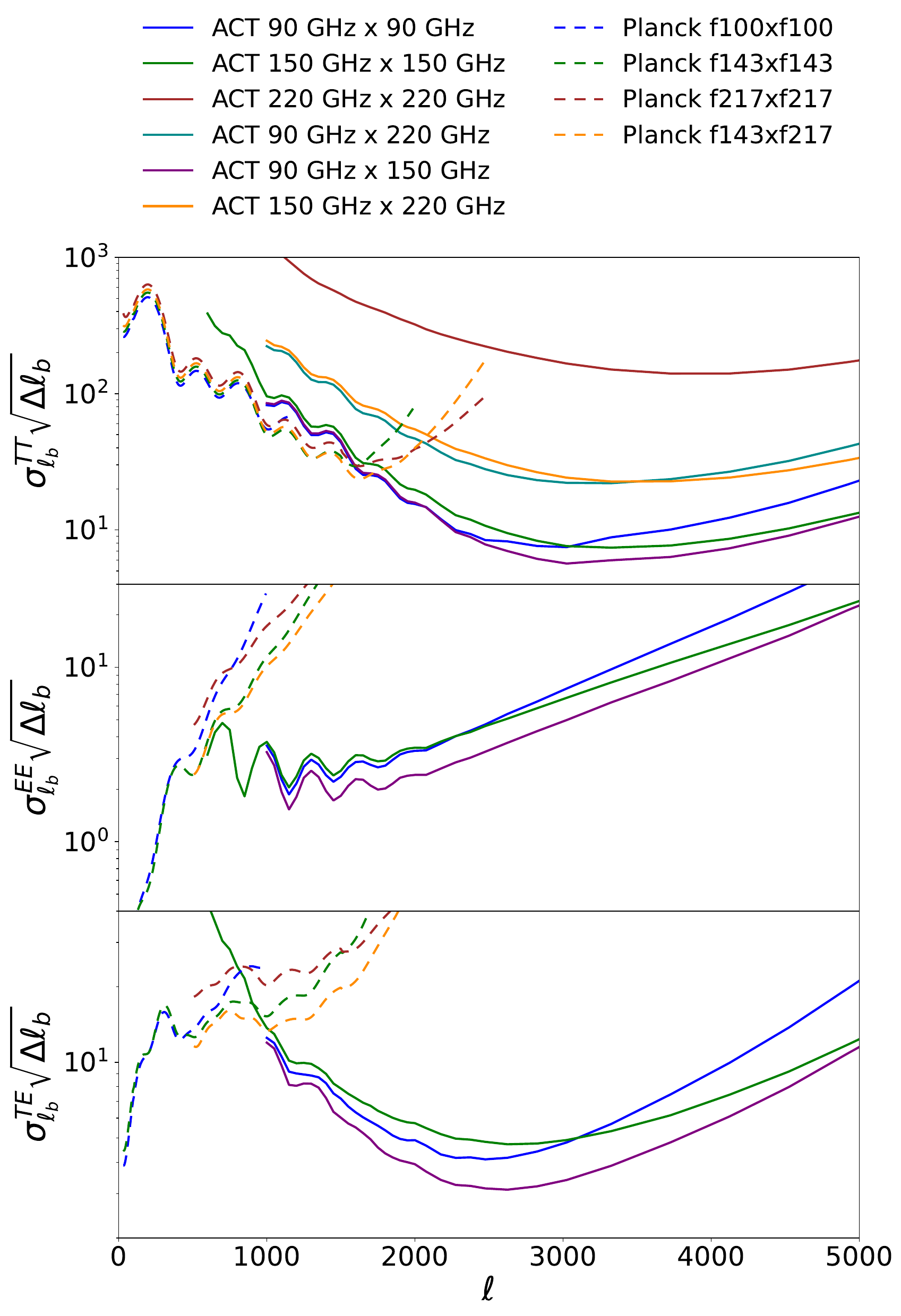}
         \includegraphics[width=\columnwidth]{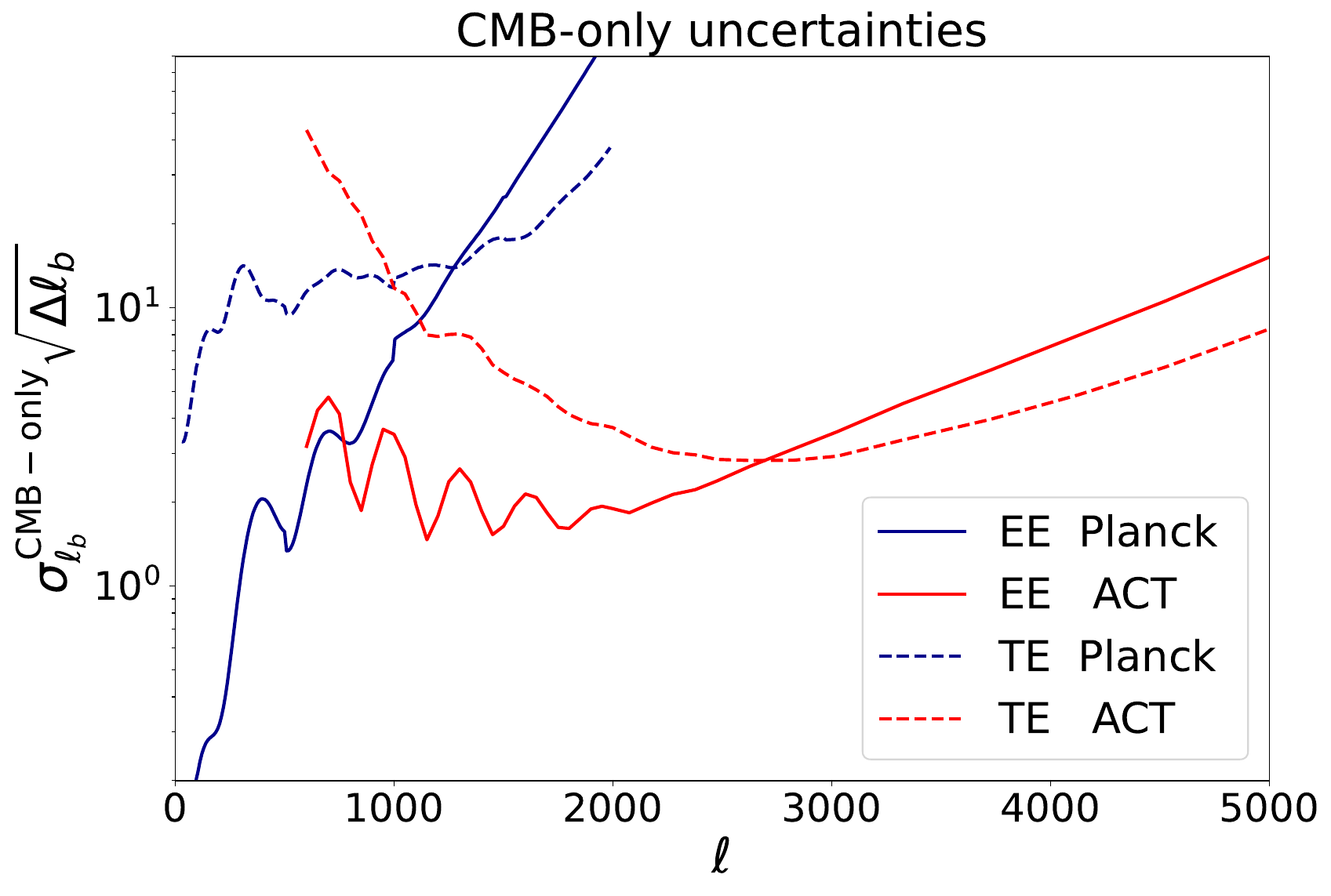}

	\caption{Comparison of \act\ and Planck uncertainties. The first three panels illustrate the error-bar comparison across all cross-frequency spectra for TT, EE, and TE. The bottom panel focuses on the comparison of CMB-only uncertainties for \act\ and \Planck\ in TE and EE. Note that TT is excluded from the CMB-only comparison as its errors are highly correlated at small scales, making it unsuitable for representation in a one-dimensional plot. The CMB-only bandpower correlations for TT, TE, and EE are shown in Appendix \ref{apx:cmbonly}.}
	\label{fig:err_compare}
\end{figure}

\subsection{Combination with Planck or WMAP}
\label{subsec:actplanck}
We combine ACT with CMB satellite data to extend the angular range to reach $\ell=2$. In analyses with ACT alone, labeled \act, we include a \Planck-derived measurement of the optical depth, using the \Planck\ \href{https://web.fe.infn.it/~pagano/low_ell_datasets/sroll2/}{\texttt{Sroll2}} likelihood~\citep{pagano/etal:2020} for low-ell polarization. In some cases, we simplify this by replacing the Sroll2 likelihood with an equivalent prior on $\tau=0.0566 \pm 0.0058$ (using the mean value recovered by Sroll2 and a symmetric errorbar); we label this combination \acttp. 

Our baseline CMB combination, labeled \pact, includes a ``\Planck$_{\rm cut}$" dataset using \Planck\ high-$\ell$ data at $\ell<1000$ in TT and $\ell<600$ in TE/EE from the \citep{planck_spectra:2019} PR3 likelihood, as well as the low-$\ell$ \Planck\ temperature likelihood and substituting in the Sroll2 likelihood for low-ell polarization. We cut the \Planck\ high-ell spectra in the multipole ranges where ACT has data, since there is substantial sky overlap between the two surveys. Figure~\ref{fig:err_compare} shows that ACT has similar or stronger constraining power to \Planck\ at the multipoles discarded. We use the \Planck\ {\texttt{plik\_lite}} likelihood which combines the 100, 143 and 217~GHz data and uses the full multipole range of the \Planck\ data to pre-marginalize over foreground contamination. We neglect correlations in TT in the $600<\ell<1000$ range where ACT and \Planck\ overlap; the \Planck\ uncertainties are smaller in this angular range. 

We do not use the NPIPE likelihoods from \citet{rosenberg:2022,hillipop2024} in this combined analysis, but note that the truncation of the \Planck\ data to $\ell<1000$ is expected to make our results less sensitive to the choice of \Planck\ likelihood. We were also motivated in this choice by the poorer agreement between the ACT and NPIPE 100~GHz ET spectra.

A more optimal approach would be to use all of the \Planck\ data (from the PR3 and/or NPIPE processing), and include the covariance between the \Planck\ and ACT multi-frequency spectra. By comparing errors estimated from a simulation when either cutting the data, or combining the full \Planck\ and ACT data and neglecting the covariance between the two data-sets, we estimated that this would improve errors over our simpler approach at the $<5$\% level on $\sigma(N_{\rm eff})$, for example. We leave this for future analyses.

To combine with \WMAP\ data from the final 9-year release~\citep{bennett/etal:2013}, we form \wact\ using the full multipole range of both datasets, as the overlap in multipoles is minimal. 
We use a Python implementation of the likelihood, \href{https://github.com/HTJense/pyWMAP}{\texttt{pyWMAP}}, discarding the low-$\ell$ \WMAP\ polarization likelihood at $\ell<23$ and substituting the \Planck\ Sroll2 likelihood to constrain the optical depth.

To ensure consistent comparisons across datasets, when we report results for the dataset referred to as ``Planck," we re-estimate the parameters using the \Planck\ high-$\ell$ PR3 Legacy likelihood in combination with Sroll2. This is an updated version of the Sroll likelihood used in \citet{planck2018_cosmo}.

\subsection{External datasets}
Constraints on some \LCDM\ parameters improve further when combining the primary CMB with other observations at intermediate and low redshifts. We obtain state-of-the-art results with \pactlb, adding CMB lensing and BAO data to \pact. For CMB lensing we use the combined ACT DR6 and \Planck\ PR4 lensing bandpowers released in a joint likelihood from ~\cite{Qu_dr6_lensing,Madhavacheril_dr6_lensing}, using data from \cite{carron2022}. For BAO we include the \jd{DR1} release of BAO in galaxy, quasar and Lyman-$\alpha$ forest tracers from the Dark Energy Spectroscopic Instrument (DESI)~\citep{DESI-BAO-III,DESI-BAO-VI}.\footnote{\jd{We also include  the updated DESI DR2 data from \cite{DESI1_DR2:2025,DESI2_DR2:2025}, which appeared as this paper was submitted.}} These datasets are described in more detail in C25.

\section{Parameter estimation, validation and null tests}\label{sec:params}

In this section we describe how we use the likelihoods to extract cosmological, foreground, calibration, polarization efficiency and passband parameters. We clarify when we add information with priors, and describe the likelihood validation and data consistency tests  performed before unblinding.  

\subsection{The lensed CMB theory}\label{subsec:lensed_cmb}
The baseline theoretical model used in this paper is \LCDM. The lensed theory spectra are computed as part of the \texttt{Cobaya}~\citep{Cobaya} software, which calls \texttt{camb}~\citep{Lewis_camb} with accuracy settings tuned to be sufficiently precise for the angular range probed by the ACT data, and with  high-precision recombination calculations and non-linear matter power spectrum modeling, as described in C25. The CMB predictions are obtained varying the six basic \LCDM\ parameters: the baryon density $\Omega_b h^2$, the cold dark matter density $\Omega_c h^2$, the acoustic scale, $\theta_{\rm MC}$, the amplitude of primordial scalar perturbations, $A_s$, the power-law spectral index $n_s$, and optical depth to reionization, $\tau$. We assume a spatially flat universe with adiabatic power-law scalar fluctuations, including three neutrinos species with a total mass sum of $0.06$~eV carried by one massive eigenstate, and a primordial Helium fraction that assumes the BBN consistency relation. Parameters including the local Hubble constant, $H_0$ in units of km/s/Mpc, the amplitude of fluctuations today, $\sigma_8$, and the matter density $\Omega_m$, are derived from these.

When parameters are used to validate the power spectrum and likelihood pipelines, we consider both the \LCDM\ model and a model that additionally varies the number of relativistic species, $N_{\rm eff}$, as a representative parameter that is more sensitive to the small-scale power spectrum than the \LCDM\ parameters.

A broad range of cosmological models are considered in C25. In this paper we test two specific departures from the \LCDM\ model using the CMB power spectrum data, quantified by the $A_{\rm lens}$ parameter and the spatial curvature $\Omega_K$. The $A_{\rm lens}$ artificially modifies the lensing potential that propagates to the CMB power spectrum, according to $C^\psi_\ell  \rightarrow A_{\rm lens} C^\psi_\ell$ \citep{Calabrese:2008}. Adjusting the spatial curvature, together with the other \LCDM\ parameters in a way to conserve the geometric degeneracy, also has the effect of modifying the degree of lensing in the spectrum compared to a flat \LCDM\ model. This test is motivated by the departure from $A_{\rm lens}=1$, or $\Omega_K=0$, that was seen at the almost 3$\sigma$ level from the \Planck\ PR3 power spectra alone \citep{planck2018_cosmo} and reduced to under $2\sigma$ with the inclusion of more \Planck\ data in \cite{rosenberg:2022,hillipop2024}.

\subsection{Parameter extraction and priors}\label{subsec:param_ext_prior}
To extract parameter constraints we run MCMC chains with \texttt{Cobaya} with theory predictions computed using $\ell_{\rm max}=9000$ and with the Gelman-Rubin convergence parameter, $R-1$, reaching values smaller than 0.01. We assume flat uninformative priors on the six cosmological parameters of the \LCDM\ model, unless replacing the Sroll2 likelihood with the simple Gaussian prior. To these we add a total of 15 foreground parameters of which 14 are freely varying (9 in TT, 2 in TE, 2 in EE, and a common radio index parameter in TT, TE and EE) and one is conditioned to be equal to another parameter (the CIB spectral indices, $\beta_p \equiv  \beta_c$). Their names in the likelihood, their definitions, and their priors are listed in Table~\ref{tab:act-foreground-parameters}. We use positive priors on foreground parameters that describe amplitudes for the measurements, but let them take negative values when analyzing simulations, to check that the input values are recovered without bias when averaged over many simulations. We impose Gaussian priors on the Galactic dust amplitudes as computed in~\S\ref{subsec:diff_gal}, and broad uniform priors otherwise. The model is then corrected with six calibration parameters, four polarization efficiency parameters and five bandpass-shift parameters, listed in Table~\ref{tab:act-systematic-parameters}. We use Gaussian priors on the calibration and bandpass parameters, and uniform bounded priors on the polarization efficiencies in the range $0.9<p<1.1$.

\begin{table*}[tbp]
	\centering
	\begin{tabular}{l|l|cc}
		\hline\hline
		\textbf{Parameter} & \textbf{Description} & \multicolumn{2}{c}{\textbf{Priors}} \\
		& & Simulations & Data \\
		\hline
		$a_{\rm tSZ}$ & Thermal SZ amplitude at $\ell=3000$ at $150$~GHz & & $\ge 0$ \\
		$\alpha_{\rm tSZ}$ & Thermal SZ template shape & & \\
		$a_{\rm kSZ}$ & Kinematic SZ amplitude at $\ell=3000$ & & $\ge 0$ \\
		$a_c$ & Clustered CIB amplitude at $\ell=3000$ at $150$~GHz & & $\ge 0$ \\
		$\beta_c$ & Clustered CIB spectral index & & \\
		$\xi$ & tSZ-CIB correlation scale at $\ell=3000$ at $150$~GHz & $-1 \le \xi \le 1$ & $0 \le \xi \le 0.2$ \\
		$a_p$ & Poisson CIB amplitude $\ell=3000$ at $150$~GHz & & $\ge 0$ \\
		$\beta_p$ & Poisson CIB spectral index & $\beta_p \equiv  \beta_c$ & $\beta_p \equiv  \beta_c$\\
		$a_s^{TT}$ & Unresolved radio sources in TT at $\ell=3000$ at $150$~GHz & & $\ge 0$ \\
		$\beta_s$ & Radio sources spectral index & $\le 0$ & $\le 0$ \\
		$a_g^{TT}$ & Galactic dust amplitude in TT at $\ell=500$ at $150$~GHz & $(8.83 \pm 0.32) \, \mu {\rm K}^2$ & $(7.95 \pm 0.32) \, \mu {\rm K}^2$ \\
		\hline
		$a_s^{TE}$ & Unresolved radio sources in TE at $\ell=3000$ at $150$~GHz & & \\
		$a_g^{TE}$ & Galactic dust amplitude in TE at $\ell=500$ at $150$~GHz & $(0.43 \pm 0.03) \, \mu {\rm K}^2$ & $(0.42 \pm 0.03) \, \mu {\rm K}^2$ \\
		\hline
		$a_s^{EE}$ & Unresolved radio sources in EE at $\ell=3000$ at $150$~GHz & & $>0$ \\
		$a_g^{EE}$ & Galactic dust amplitude in EE at $\ell=500$ at $150$~GHz & $(0.165 \pm 0.017) \, \mu {\rm K}^2$ & $(0.168 \pm 0.017) \, \mu {\rm K}^2$ \\
		\hline\hline
	\end{tabular}
	\caption{The 15 parameters of the DR6 foreground model, and their priors. See~\S\ref{subsec:fg_model} for a description of the model for these parameters. If not mentioned, we impose an uninformative, wide, uniform prior on the parameter; for most parameters a wide, non-negative prior is used. For analysis of simulations we explore the full volume space by letting amplitude parameters take negative values as well. The central values of the dust priors are different between simulations and data runs because the estimates were refined post unblinding.\\} 
	\label{tab:act-foreground-parameters}
\end{table*}

\begin{table}[tbp]
	\centering
	\begin{tabular}{l|l|c}
		\hline\hline
		\textbf{Parameter} & \textbf{Description} & \textbf{Prior} \\
		\hline
		${\mathrm{cal}_{\rm dipole}}$ & Dipole calibration & $1 \pm 0.003$ \\
		\hline
		$c_{\rm pa4}^{\rm f220}$ & \multirow{3}*{Per-frequency array} & $1 \pm 0.013$ \\
		$c_{\rm pa5}^{\rm f090}$ & & $1 \pm 0.0016$ \\
		$c_{\rm pa5}^{\rm f150}$ & \multirow{1}*{gain calibration} & $1 \pm 0.0020$ \\
		$c_{\rm pa6}^{\rm f090}$ & & $1 \pm 0.0018$ \\
		$c_{\rm pa6}^{\rm f150}$ & & $1 \pm 0.0024$ \\
		\hline
		$p_{\rm pa5}^{\rm f090}$ & \multirow{3}*{Per-frequency array} & \multirow{4}*{$0.9<p<1.1$} \\
		$p_{\rm pa5}^{\rm f150}$ &  & \\
		$p_{\rm pa6}^{\rm f090}$ & \multirow{1}*{polarization efficiency} & \\
		$p_{\rm pa6}^{\rm f150}$ & & \\
		\hline
		$\Delta_{\rm pa4}^{\rm f220}$ & \multirow{3}*{Per-frequency array} & $(0 \pm 3.6)$ GHz \\
		$\Delta_{\rm pa5}^{\rm f090}$ & & $(0 \pm 1.0)$ GHz \\
		$\Delta_{\rm pa5}^{\rm f150}$ & \multirow{1}*{bandpass shift} & $(0 \pm 1.3)$ GHz \\
		$\Delta_{\rm pa6}^{\rm f090}$ & & $(0 \pm 1.2)$ GHz \\
		$\Delta_{\rm pa6}^{\rm f150}$ & & $(0 \pm 1.1)$ GHz \\
		\hline\hline
	\end{tabular}
	\caption{The 15 nuisance parameters for our model of the DR6 instrument. See~\S\ref{sec:likelihood} and~\S\ref{subsec:instrument} for a description of the model for these parameters. We impose Gaussian priors on the gain calibrations and bandpass shifts based on our calibration of ACT with respect to the \emph{Planck} temperature maps, and measurements of the instrumental bandpass. The polarization efficiencies are free to vary within relatively uninformative flat priors.}
	\label{tab:act-systematic-parameters}
\end{table}

\subsection{Parameters from simulations and nulls}
We use the same suite of simulations as described in \S\ref{subsec:sims} to test our parameter estimation pipeline.

We process each simulation with \texttt{Cobaya} using the multi-frequency likelihood described in~\S\ref{sec:likelihood}. However, as described in~\S\ref{subsec:param_ext_prior}, in the nominal likelihood we include Gaussian priors on calibration and passband parameters, on the dust amplitudes, as well as on the optical depth to reionization. To handle these in the simulations, which artificially have no scatter in the input values, we draw a different prior mean for each simulation, from the prior distribution. This approach is equivalent to absorbing the priors into the likelihood as additional datasets, for example considering them as independent measurements of the calibrations, passbands, dust amplitudes and optical depth.

\subsubsection{Parameter recovery}\label{subsubsec:param_recov}
We first estimate the full suite of parameters on 100 simulations to check that the power spectrum and likelihood pipelines are unbiased. We do this for both the \LCDM\ model and \LCDM+$N_{\rm eff}$. This test passes, with all cosmological parameters recovered within expectation for both \LCDM\ and \LCDM$+N_{\rm eff}$ for the full ACT dataset (TT/TE/EE), and for each subset of data we explore: using single probes (TT or TE or EE), or only polarization spectra (TEEE), or single frequencies (the f090 or f150 subsets, discarding other frequencies in each case). We find foreground and other nuisance parameters to be consistent with the input values, to the level specified in our blinding procedure in Appendix \ref{apx:blinding}. 

We also estimate parameters from one non-Gaussian simulation set, to test the robustness of our parametric model. We construct sky maps from realistic astrophysical components produced using the \texttt{Agora} extragalactic and \texttt{PySM} Galactic simulations \citep{psym3,agora}. Specifically, the PySM sky model used includes the configurations $a_{1}$, $s_{5}$, and $d_{10}$ for anomalous microwave emission, synchrotron, and dust, respectively.
Despite differences between our baseline foreground model and the more complex features in these simulations --- including frequency decorrelation of components that have yet to be confirmed with real data --- we find that our pipeline remains sufficiently flexible. It does not introduce any significant bias in the cosmological parameters, with all recovered values agreeing with the input parameters to within 2.1$\sigma$, which is the largest observed deviation\footnote{The shifts with respect to the input parameters are: $\Delta \Omega_c h^{2} = -0.2 \sigma$, $\Delta \Omega_b h^{2} = 2.1 \sigma$, $\Delta \log(10^{10} A_{s}) = -1.5 \sigma$, $\Delta n_{s} =-0.5\sigma$ and $\Delta H_{0} = -0.6\sigma$.}. Since this is only one simulation, we assess this to be an acceptable agreement. These results \jd{are} presented in further detail in \cite{fg_inprep}.

\subsubsection{Parameter nulls between frequencies} \label{subsubsec:paramnulls}
We test the stability of parameters across frequencies by looking at parameter differences from the posteriors derived from the f090 or f150 subsets, for both data and simulations. To assess consistency, we first compare the data difference to the simulated suite for each cosmological parameter individually. We find that the parameter differences for the data are in agreement with the distribution of the simulations, with the largest shifts for $[\Delta {\Omega_ch^2}]^{90-150}$ and $[\Delta n_s]^{90-150}$ that are $1.4\sigma$ and $1.3\sigma$ respectively.

We also perform a more stringent test, calculating the agreement in the 5-dimensional $\Omega_b h^2 - \Omega_c h^2 - \theta - n_s - A_s$ space (or 6-dimensional for \LCDM+$N_{\rm eff}$) of the cosmological parameter differences between the solutions for f090 and f150, accounting for correlation between the two cases. We estimate the correlation between single-frequency runs via a Fisher matrix calculation on cosmological, foreground and nuisance parameters similarly to~\cite{Kable_2020}. We do this for the TE/EE and the full TT/TE/EE combination. For the baseline analysis we find good agreement, with: 
\begin{eqnarray}
&&\mathrm{PTE(\Lambda CDM, 90 \to 150, TT/TE/EE)} =50\%, \nonumber\\
&&\mathrm{PTE(\Lambda CDM, 90 \to 150, TE/EE}) = 53\%.
\end{eqnarray}

This multi-dimensional frequency consistency was one of the tests that motivated us, post-unblinding, to revise our original choices. With the pre-unblinding settings, i.e., using EE spectra with a minimum multiple of $\ell=500$, no template scaling for the tSZ, and no beam chromaticity, we find this agreement is worse, with:
\begin{eqnarray}
&&\mathrm{PTE^{blind}(\Lambda CDM, 90 \to 150, TT/TE/EE)} = 0.8\%, \nonumber\\
&&\mathrm{PTE^{blind}(\Lambda CDM, 90 \to 150, TE/EE}) = 1.7\%  
\end{eqnarray}

Pre-unblinding we used a second method to assess this consistency, described in Appendix \ref{apx: sys_res}. The passing of this test for the baseline cuts is consistent with our findings from the exploration of residual systematic effects that was described in \S\ref{subsec:nulls}.

\section{Constraints on the \LCDM\ model}\label{sec:results}
In this section we show that we can consistently fit a \LCDM\ model to CMB data from different experiments covering a range of angular scales -- ground-based ACT data,  satellite data from \Planck, or a combination of the two with ACT, \WMAP\ and \Planck\ -- and from temperature or polarization data. These are strong consistency checks that use different instruments, and with foregrounds that have distinct impacts on polarization and temperature, and on large and small angular scales.

\begin{table*}[t!]
    \centering
    \begin{tabular}{c}
        \includegraphics[width=\textwidth]{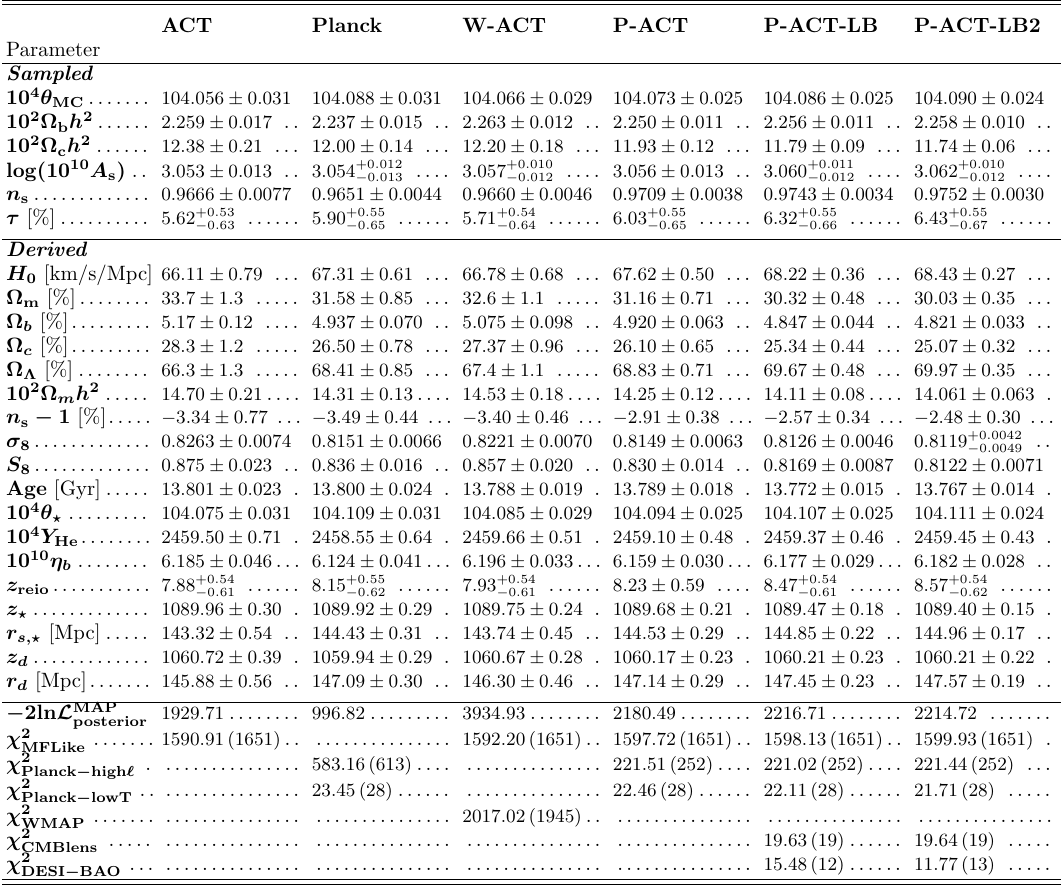}
    \end{tabular}
    \caption{\tl{Marginalized constraints on the \LCDM\ sampled and derived parameters from the ACT data (including the \Planck\ Sroll2 large-scale EE data to constrain the optical depth), and its combination with \WMAP\ (W-ACT), $\ell<1000$ \Planck\ data (P-ACT), and CMB lensing from ACT and Planck and BAO data from DESI DR1 (P-ACT-LB).  The DESI collaboration presented its second data release \citep[DR2,][]{DESI1_DR2:2025,DESI2_DR2:2025} at the same time as these results. For completeness, we also include the combination with DESI DR2, referred to as P-ACT-LB2}. Parameter definitions are given in Appendix~\ref{apx:param_tables}. The goodness of fit of the best-fitting model, with maxium posterior probability, is reported for the different datasets along with the total maximum a posteriori (MAP) value that includes contributions from the Sroll2 likelihood and informative priors. Numbers in parentheses indicate the number of data points used in the respective $\chi^{2}$  calculations. For comparison,  constraints are shown from the \Planck\ PR3 \citep{planck2018_cosmo} TT/TE/EE data that we rerun with the Sroll2 large-scale polarization data for consistency. Parameter constraints using the \Planck\ NPIPE maps in \cite{rosenberg:2022} and \cite{hillipop2024} are typically 10-20\% tighter, with comparable errors to our P-ACT combination. 
    }
    \label{tab:lcdm_params}
\end{table*}

\begin{figure}
    \centering
    \includegraphics[width=0.5\textwidth]{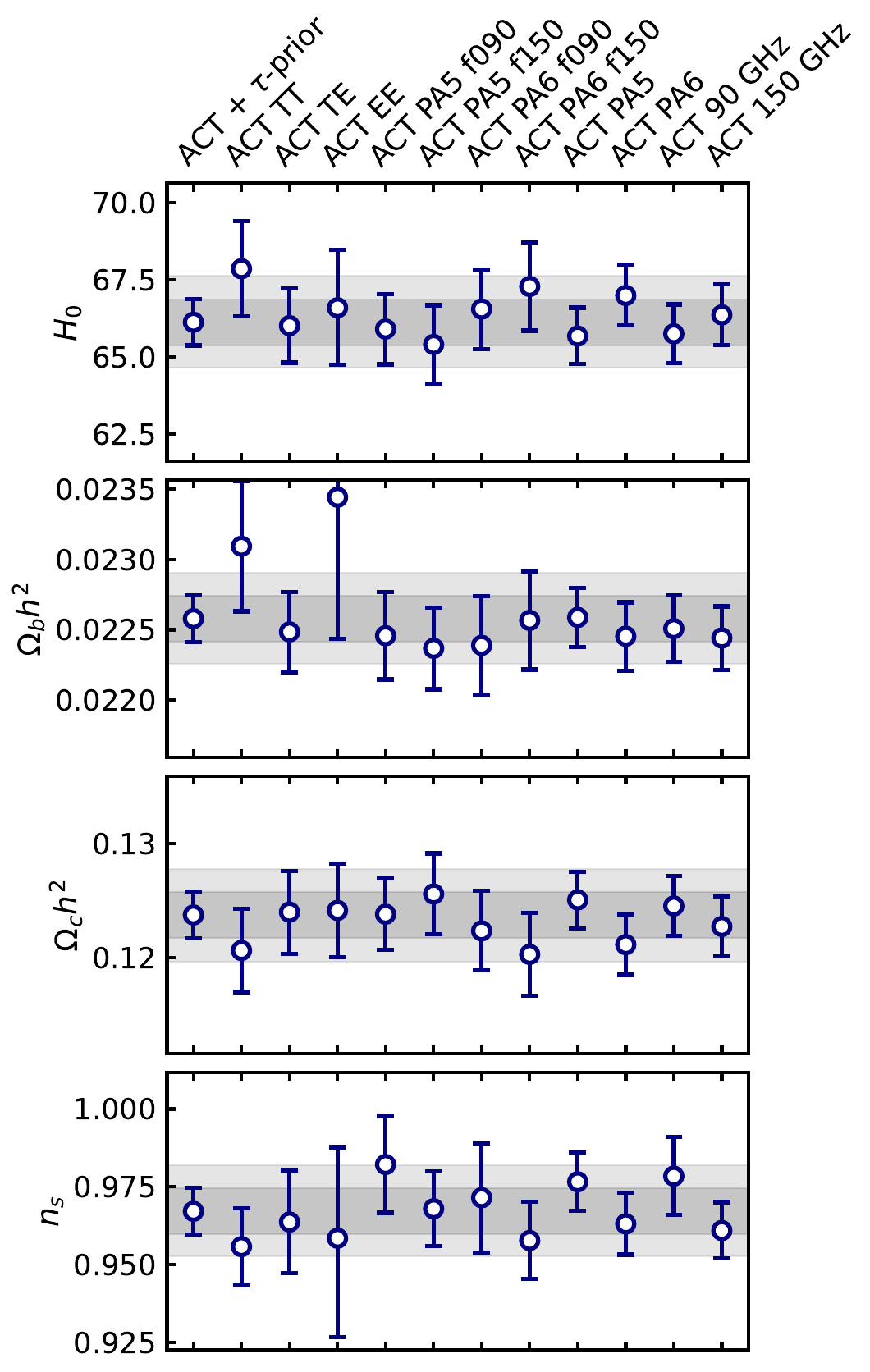}
    \caption{1D marginalized 68\% confidence levels (CL) on cosmological parameters estimated from subsets of the ACT DR6 dataset. The baryon and CDM densities are best measured by the TE spectrum, and the spectral index by the TT spectrum. The different arrays and frequencies give consistent results. All the results shown here use the same optical depth prior. The shaded band shows the 68\% and 95\% CL on the baseline ACT results.}

    \label{fig:act1d}
\end{figure}

\begin{figure*}[htb!]
    \centering
    \includegraphics[width=\textwidth]{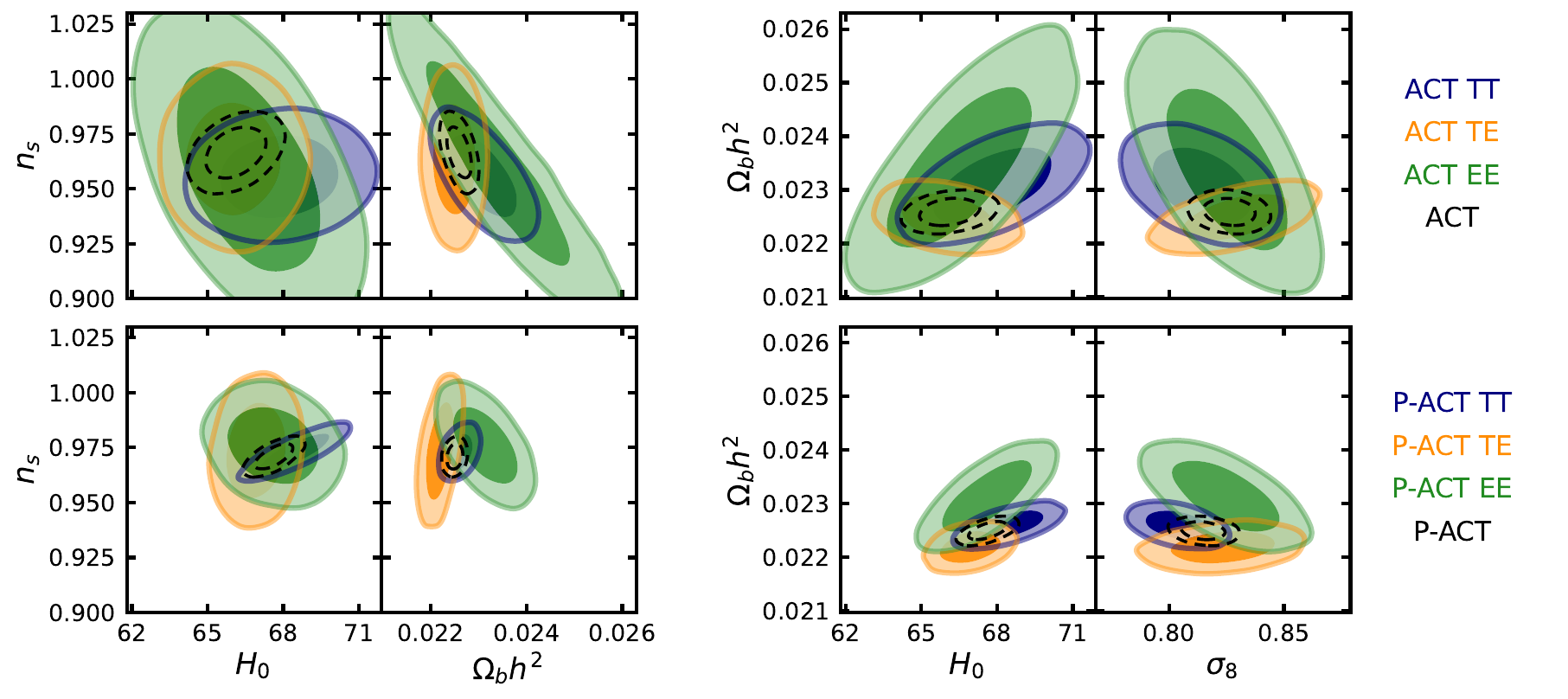}
    \caption{Cosmological parameter distributions estimated from TT, TE or EE from ACT (top) and P-ACT (bottom), including the optical depth prior. Black dashed contours correspond to the distributions estimated from TT, TE, and EE simultaneously, again for ACT (top) and P-ACT (bottom). A prior on the ACT polarization efficiencies, derived from the joint T+E fit, is imposed for the ACT (top) results. For ACT, the TE data provide the tightest constraints on the baryon density, cold dark matter density and the Hubble constant, while the TT data best measure the spectral index. The EE-only constraints are now competitive with those from TT and TE. There is less foreground contamination in the TE and EE spectra than TT; the consistent results add confidence in the model.\\}
    \label{fig:ttteee}
\end{figure*}

\subsection{Results from ACT}
We show cosmological parameter constraints from the ACT data, including the \Planck\ Sroll2 data to measure the optical depth, in Figure \ref{fig:LCDM} in the summary section. We report them in Table \ref{tab:lcdm_params}, with the full set of foreground and nuisance parameters given in Appendix \ref{apx:params}. Of the six cosmological parameters in \LCDM, the optical depth $\tau$ is measured by the \Planck\ large-scale EE data, and the overall amplitude, $A_s\exp{(-2\tau)}$, is constrained by the relative calibration to \Planck. The remaining four parameters ($\Omega_b h^2$, $\Omega_c h^2$, $\theta_{\rm MC}$ and $n_s$), and derived parameters including the Hubble constant, are measured by the ACT data. We find that the best-fitting \LCDM\ model to ACT is a good fit to the data, with $-2\ln L=$1591 for 1617 degrees of freedom (dof, for 1651 data points and 34 model parameters) for the multi-frequency likelihood, with PTE of $67\%$.\footnote{Using the CMB-only likelihood we find the same parameter constraints to within 0.1$\sigma$, as shown in Appendix \ref{apx:cmbonly}, with best-fitting $-2\ln L = 142.2$ for 128 dof (135 data points and 7 model parameters), and PTE of $18\% $.} With the additional peaks, the ACT data by themselves constrain the acoustic peak scale to the same precision as \Planck, with a 0.03\% error and a consistent measurement.

In Figure \ref{fig:act1d} we show the marginalized limits on four of the parameters in the left-most column. These can be compared to constraints from just the TT, TE and EE data, or the four different array-bands, in addition to the data split by array or by frequency. We find stability between parameters derived from each spectrum and array. The f090 and f150 data give constraints with comparable uncertainties, as do the PA5 and PA6 data.\footnote{When using TE or EE data alone, we impose priors on the polarization efficiencies matching those estimated from the joint TT/TE/EE data. The impact of polarization efficiencies is discussed in Appendix~\ref{apx:poleff}.}

Figure \ref{fig:ttteee} shows how the TT, TE and EE data from ACT measure different degeneracy directions. The E-mode polarization provides a measure of the velocity perturbations at recombination, enabling a sharper measurement of the acoustic features than for intensity \citep{galli/etal:2014}. This results in the TE data providing the tightest constraints on the baryon density, cold dark matter density and the Hubble constant. The TT data best constrain the spectral index by measuring a broader range of scales with high signal-to-noise. The EE data give consistent results with distinct parameter correlations.
For these results we impose a prior on the polarization efficiency from the combined TT/TE/EE analysis; we find that the uncertainty on $\Omega_ch^2$, and on the derived Hubble constant, approximately doubles for EE-alone when the polarization efficiency is allowed to vary in the range $[0.9,1.1]$, shown in Appendix \ref{apx:poleff}. We interpret this as being due to the radiation driving effect which amplifies the sound wave oscillations for modes that entered the horizon during radiation domination, corresponding to $\ell \gtrsim 200$. Varying $\Omega_{c}h^{2}$ alters the redshift of matter-radiation equality, changing the amount of radiation-driven amplitude boosting. For the angular scales measured by ACT, this has a similar effect to varying the overall amplitude of the EE spectrum.

\subsection{Consistency of ACT, W-ACT and Planck}
Figure \ref{fig:LCDM}, in the summary of key results, shows a comparison of \LCDM\ cosmological parameters estimated from ACT, or ACT combined with \WMAP\ larger-scale data (W-ACT), with the \Planck\ PR3 data. We find statistically consistent parameters estimated from ACT, W-ACT or \Planck, summarized in Table \ref{tab:lcdm_params}, with similar constraining power from either W-ACT or \Planck. The W-ACT dataset independently confirms that a Harrison-Zel'dovich primordial spectrum with $n_s=1$ is ruled out at more than 7$\sigma$:
\ba
n_s &=& 0.9660\pm0.0046  \ (\textsf{W-ACT})  \nonumber\\
&=& 0.9651\pm0.0044 \ (\textsf{Planck}), 
\ea
with the same value measured to within 0.2$\sigma$.
With more acoustic peaks, W-ACT also provides a slightly tighter limit on the acoustic peak scale, and a tighter limit on the baryon density, with
\ba
\Omega_b h^2 &=&  0.02263\pm0.00012  \ (\textsf{W-ACT})   \nonumber\\
&=&0.02237\pm0.00015\ (\textsf{Planck}). 
\ea
We find that the estimate for the Hubble constant is stable to the choice of dataset for the CMB power spectrum, with
\ba
H_0 &=&  66.78 \pm 0.68 {\rm ~km/s/Mpc}  \ (\textsf{W-ACT})   \nonumber\\
&=&67.31\pm0.61 {\rm ~km/s/Mpc} \ (\textsf{Planck}).
\ea 

Parameter constraints using the \Planck\ NPIPE maps are typically 10-20\% tighter than those quoted above \citep{rosenberg:2022,hillipop2024}. When comparing parameters in the four-dimensional space excluding $\tau$ and $A_s$, we find that the difference between the  parameter means estimated from ACT and \Planck\ is at the 1.6$\sigma$ level with the \Planck\ PR3 data, and at the 2.5$\sigma$ level with the \Planck\ NPIPE data. Shown in Appendix \ref{apx:params}, the $\Omega_b h^2$ - $\Omega_c h^2$ and $\Omega_b h^2$ - $H_0$ parameter combinations highlight the shift between best-fitting models most clearly.

\subsection{Joint results from ACT and Planck}
We next combine the ACT and \Planck\ PR3 data, cut at $\ell<1000$ in temperature and $\ell<600$ in polarization as described in \S\ref{subsec:actplanck}, to estimate ``P-ACT" parameters. We then add CMB lensing and BAO data to estimate ``P-ACT-LB" parameters.

\begin{figure*}[bt!]
	\centering  \includegraphics[width=1.0\textwidth]{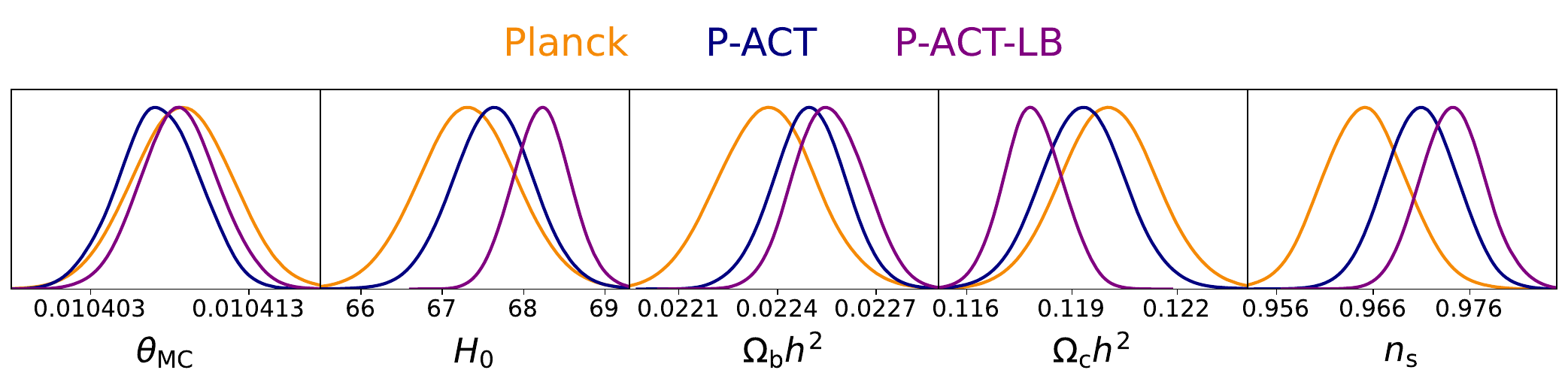}
    \caption{Cosmological parameters for the \LCDM\ model estimated from the combination of ACT and \Planck\ (P-ACT) and with the addition of CMB lensing from ACT and Planck (L) and BAO data from DESI \jd{DR1} (B). Figure \ref{fig:ns_obh2} gives an example of how the data sets combine together to reduce the uncertainty. \jd{Appendix \ref{apx:desi} shows these parameter constraints with DESI DR2.}\\}
	\label{fig:LCDM_2d}
\end{figure*}

\begin{figure}
    \includegraphics[width=\columnwidth]{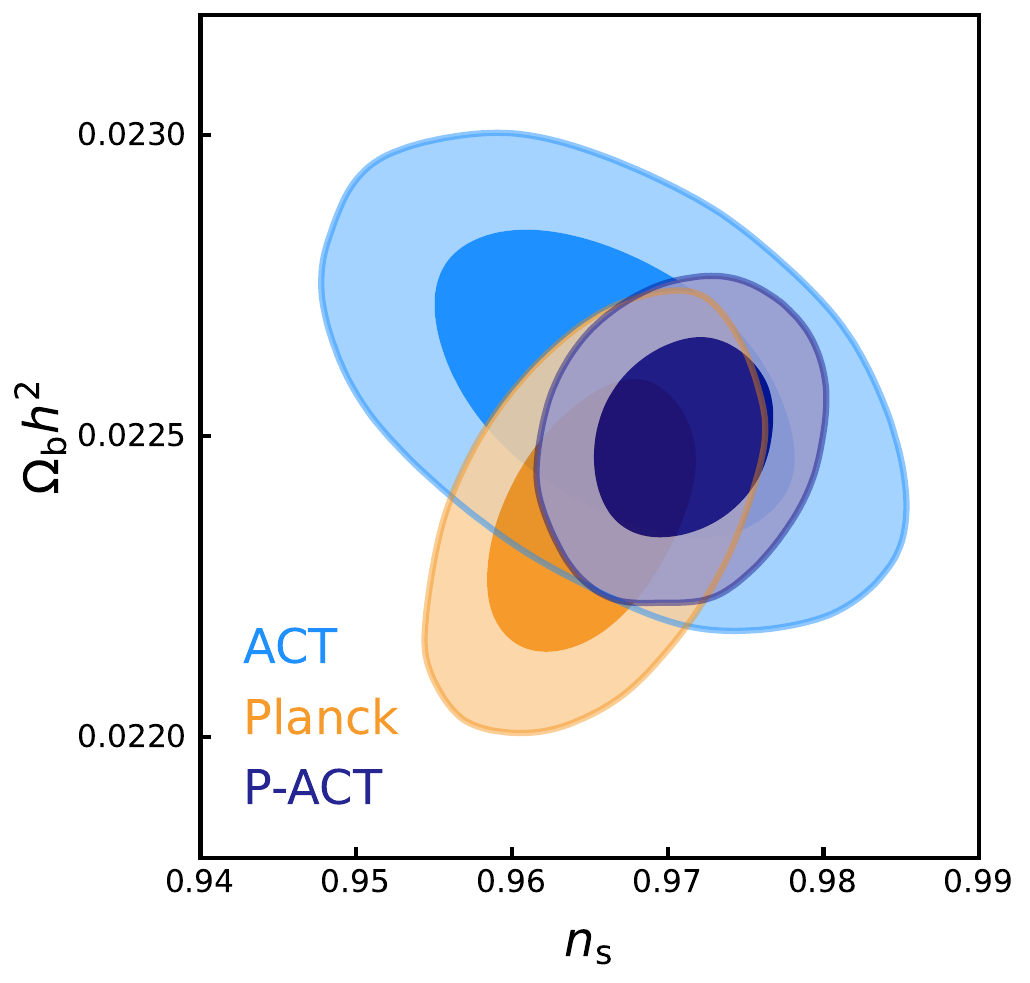}
    \caption{The marginalized posterior distribution in the $n_s-\Omega_bh^2$ plane showing how the two datasets provide complementary information. With larger scale information from \Planck\ these parameters are positively correlated; at smaller scales from ACT they both act to damp the spectrum so are anti-correlated.}
    \label{fig:ns_obh2}
\end{figure}

\begin{figure*}[htp]
\centering
    \includegraphics[width=0.95\textwidth]{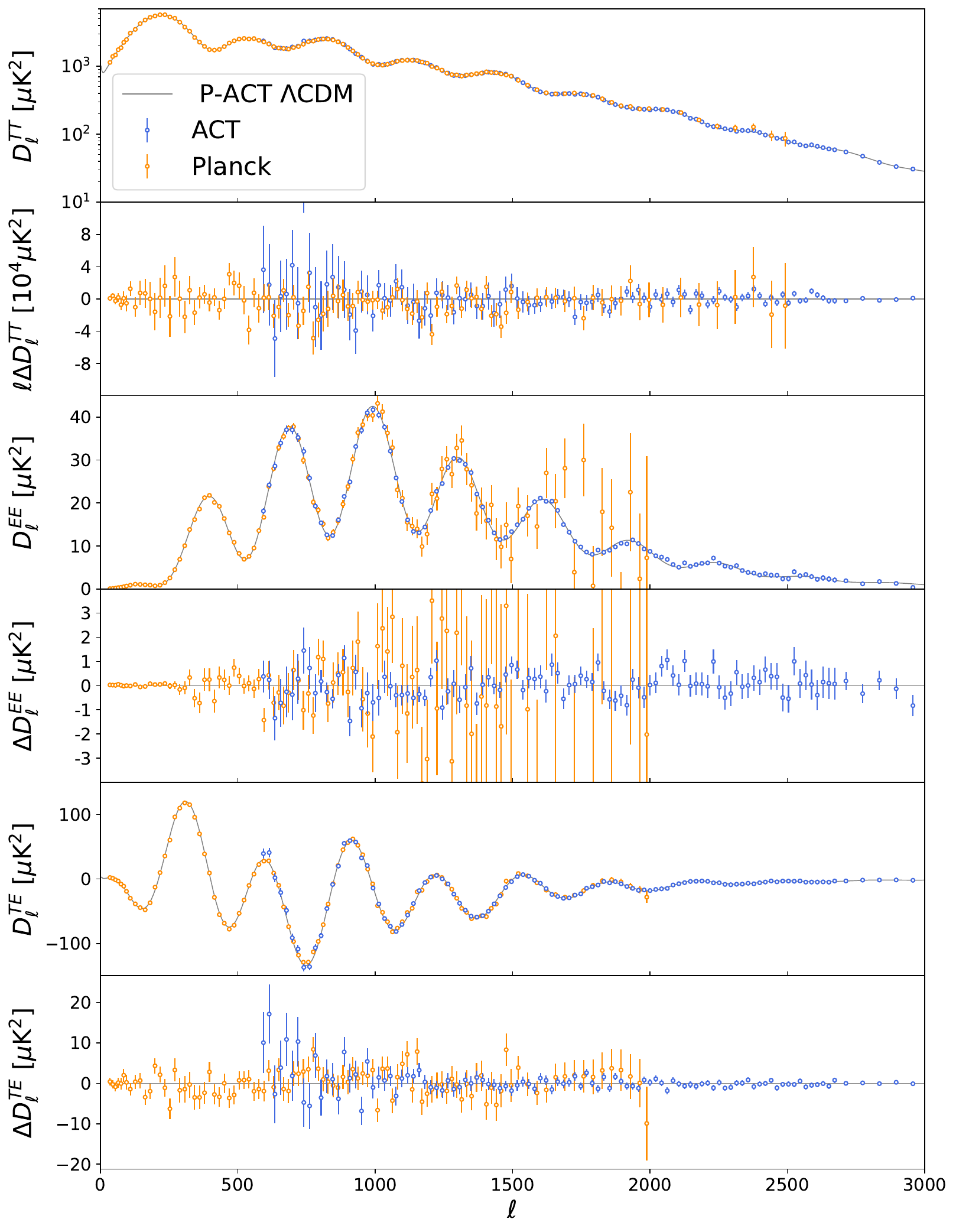}
    \caption{The ACT and \Planck\ PR3 spectra in TT, TE and EE  together with the best-fitting P-ACT \LCDM\ model, and the residuals of the spectra compared to this model. Pairs of \Planck\ data bins are combined for plotting purposes. As in Figure \ref{fig:combined_TE_EE}, the best-fit foreground power spectra are subtracted.}
    \label{fig:residuals}
\end{figure*}

The marginalized constraints for these data are given in Table \ref{tab:lcdm_params}, and in Appendix \ref{apx:param_tables}. The appendix table also includes the P-ACT-L combination that adds only CMB lensing to P-ACT. Parameter distributions are shown in Figure~\ref{fig:LCDM_2d}, and in Figure \ref{fig:LCDM} in the summary. With a broader coverage of angular scales in temperature and polarization, we find that the joint dataset excludes the part of the {\it Planck}-only parameter space from PR3 with lowest Hubble constant, baryon density and spectral index. The acoustic peak scale is reduced with P-ACT to a 0.02\% uncertainty.

 Figure \ref{fig:ns_obh2} shows an example of how the two datasets combine together. From the larger-scale information, the spectral index is positively correlated with the baryon density, as a higher baryon density will lower the ratio of the second peak to the first peak, which can compensate an increased spectral tilt. With smaller scale data, the two parameters are anti-correlated, as found in e.g, \cite{choi_atacama_2020}, with the increased small-scale power from a larger tilt being compensated by a smaller baryon density which further damps the Silk-damping tail. Combining the large and small-scale data together breaks the degeneracy and results in a 0.4\% measurement of both the baryon density and spectral index, with $\Omega_b h^2=0.02250\pm0.00011$ and $n_s=0.9709\pm 0.0038$.

 We show the TT, TE and EE spectra, together with residuals to the best-fitting P-ACT model, in Figure \ref{fig:residuals}. The goodness of fit of this model for the ACT data is 1598 for 1617 dof, with PTE of $63\%$.\footnote{For the CMB-only likelihood we find 147.7 for 128 dof, with PTE of $11\%$.} This same model is also a good fit to the \Planck\ data, as shown in the figure; the overall P-ACT data has a goodness of fit of 1842 for 1897 dof, with PTE of 81\%.\footnote{In this P-ACT PTE estimate we discard the Sroll2 data points, and the optical depth degree of freedom, to simplify the estimate.} Appendix \ref{apx:slopes} shows in more detail how the best-fitting P-ACT model differs from either the \Planck-only or ACT-only best-fit model.

The breakdown of parameters estimated from TT, TE and EE is shown in Figure \ref{fig:ttteee} and reported in Table \ref{tab:TTTEEE_PACT}. The largest difference is at the 2$\sigma$ level between the baryon density estimated from P-ACT-EE and P-ACT-TE. Parameters estimated from only the polarization data are now competitive with those from the intensity anisotropy, and with this broader angular coverage the EE-only constraints depend much less on the polarization efficiency prior.

 The overall constraints are further tightened with the inclusion of the CMB lensing data from ACT and \Planck, and BAO data from DESI DR1 (P-ACT-LB). This combination, with parameters reported in Table \ref{tab:lcdm_params} and shown in Figures \ref{fig:LCDM} and \ref{fig:LCDM_2d}, gives us state-of-the-art constraints on key quantities including the ingredients, expansion rate, age and initial conditions of the universe. We find that the joint model is a good fit to all the datasets. \jd{These broad conclusions do not change when we switch to use the BAO data from DESI DR2 (P-ACT-LB2), and we find a common \LCDM\ model reported in Table \ref{tab:lcdm_params} (also reported in \citealp{garcia-quintero:2025}); this is discussed further in Appendix \ref{apx:desi}.} 

\begin{table}[t!]
    \centering
    \begin{tabular}{lccc}
\noalign{\vskip 10pt}\hline\noalign{\vskip 1.5pt}\hline\noalign{\vskip 5pt}
\multicolumn{1}{c}{\bf } & \bf TT & \bf TE & \bf EE \\
 & \multicolumn{3}{c}{}\\
\hline
\multicolumn{1}{l}{\bf} & \multicolumn{3}{c}{}\\
{\boldmath$10^4\theta_\mathrm{MC}$} & \fontsize{8pt}{8pt}$104.089\pm 0.051$ & \fontsize{8pt}{8pt}$104.094\pm 0.038$ & \fontsize{8pt}{8pt}$104.036\pm 0.044$ \\
{\boldmath$10^2\Omega_\mathrm{b}h^2$} & \fontsize{8pt}{8pt}$2.260\pm 0.018$ & \fontsize{8pt}{8pt}$2.219\pm 0.020$ & \fontsize{8pt}{8pt}$2.314\pm 0.043$ \\
{\boldmath$10^2\Omega_\mathrm{c}h^2$} & \fontsize{8pt}{8pt}$11.73\pm 0.21$ & \fontsize{8pt}{8pt}$12.08\pm 0.20$ & \fontsize{8pt}{8pt}$12.06\pm 0.26$ \\
{\boldmath$10^2n_\mathrm{s}$} & \fontsize{8pt}{8pt}$97.30\pm 0.55$ & \fontsize{8pt}{8pt}$97.4\pm 1.4$ & \fontsize{8pt}{8pt}$97.6\pm 1.2$ \\
\hline
{\boldmath$H_0$} & \fontsize{8pt}{8pt}$68.47\pm 0.91$ & \fontsize{8pt}{8pt}$66.92\pm 0.76$ & \fontsize{8pt}{8pt}$67.6\pm 1.2$ \\
{\boldmath$10^2\Omega_\mathrm{m}$} & \fontsize{8pt}{8pt}$30.0\pm 1.2$ & \fontsize{8pt}{8pt}$32.1\pm 1.2$ & \fontsize{8pt}{8pt}$31.6\pm 1.6$ \\
{\boldmath$10^2\sigma_8$} & \fontsize{8pt}{8pt}$80.37\pm 0.92$ & \fontsize{8pt}{8pt}$82.3\pm 1.5$ & \fontsize{8pt}{8pt}$82.5\pm 1.5$ \vspace{3.5pt}\\
\hline\hline
\end{tabular}
    \caption{$68\%$ marginalized constraints on parameters estimated from TT, TE and EE power spectra from the \pact\ data combination. Here the nominal flat priors are imposed on the ACT polarization efficiencies. Figure \ref{fig:ttteee} compares a set of two-dimensional parameter constraints for these data splits.}
    \label{tab:TTTEEE_PACT}
\end{table}

\subsection{Implications for cosmological tensions}

\begin{figure}
    \includegraphics[width=\columnwidth]{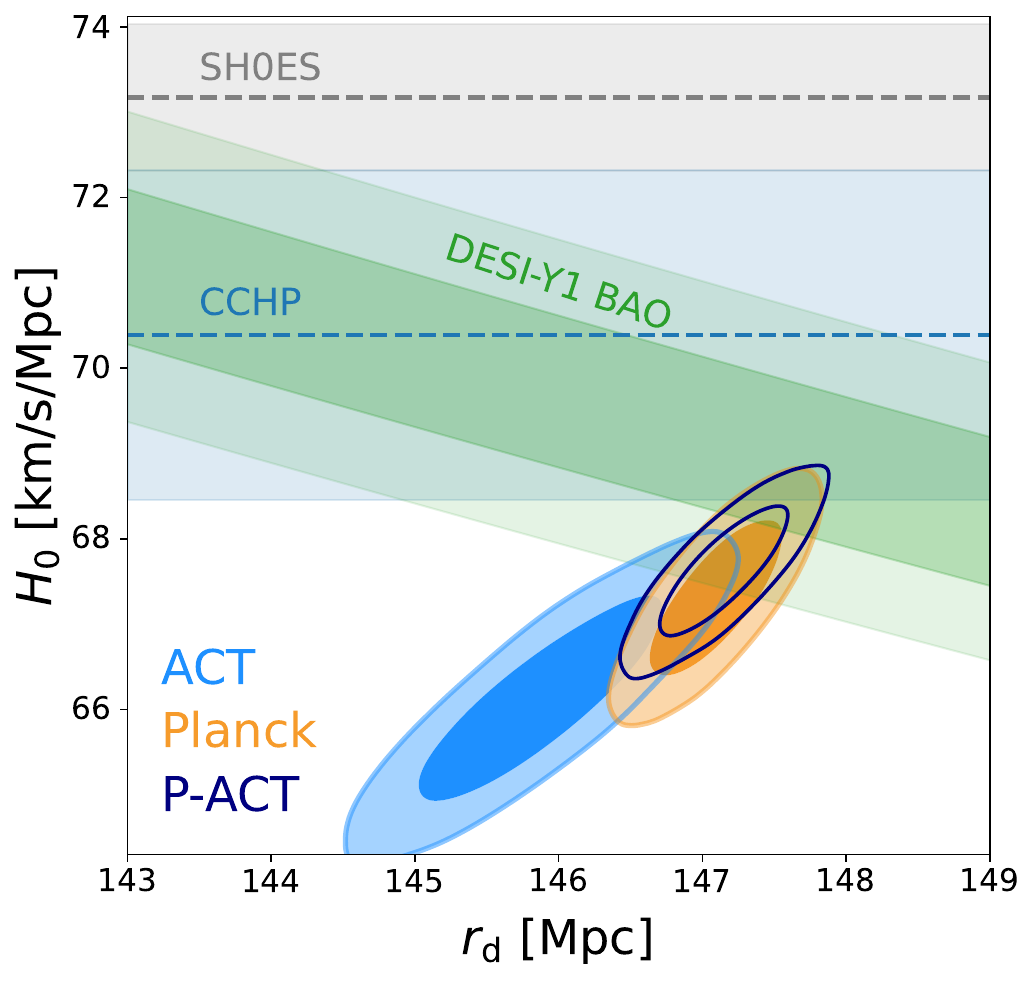}
    \includegraphics[width=\columnwidth]{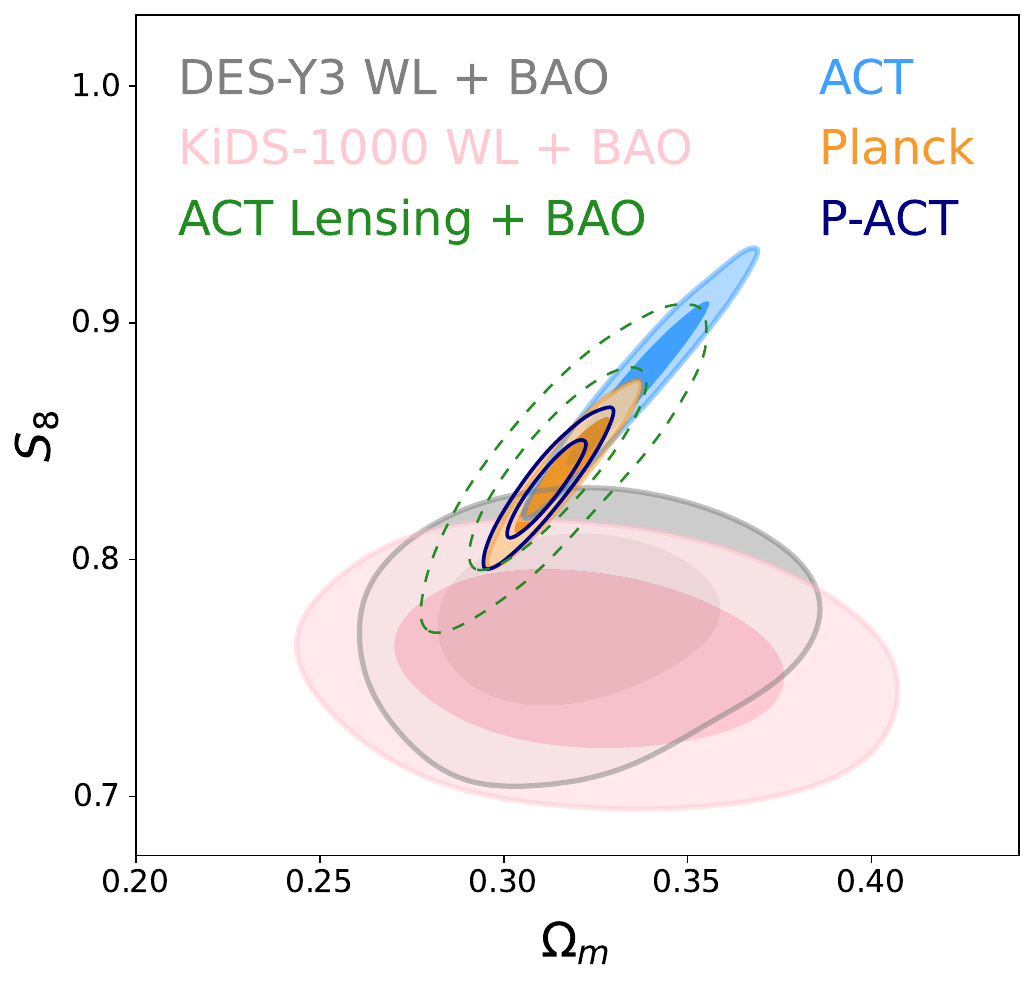}
        \caption{(Top) Constraints on $H_0$ and the sound horizon, $r_d$, from P-ACT, ACT and Planck, compared to data from DESI DR1 and direct measurements from \cite{Riess_SHOES,Breuval:2024lsv} (SH0ES) and \cite{freedman2024} (CCHP). See also \cite{verde2024} for a more complete set of $H_0$ measurements. (Bottom) Constraints on the $\Omega_m$-$S_8$ plane for the same CMB data, compared to cosmic shear results from the Dark Energy Survey (DES Y3) and the Kilo Degree Survey (KiDS-1000), and to CMB lensing results from ACT DR6, all combined with Baryon Oscillation Spectroscopic Survey (BOSS) BAO measurements (from \citep{Madhavacheril_dr6_lensing}).} 
    \label{fig:h0-s8-tensions}
\end{figure}

Our results show the continued goodness of fit of the \LCDM\ model over a broader array of multipoles and with independent datasets, and with subsets of data in temperature and polarization. Within this model, the Hubble constant is measured to be 
\be
H_0=67.62\pm0.50~{\rm km/s/Mpc} \ (\pact) 
\ee
from the CMB power spectrum, 
\be
H_0=68.22\pm0.36~{\rm km/s/Mpc} \ (\pactlb) 
\ee
when combined with CMB lensing and DESI \jd{DR1 BAO data, and 
\be
H_0=68.43\pm0.27~{\rm km/s/Mpc} \ (\pactlbb) 
\ee
combined with CMB lensing and DESI DR2 BAO data \citep[also reported in][]{garcia-quintero:2025}}. As described in \S\ref{sec:results}, the joint P-ACT dataset excludes the part of the \Planck-only parameter space, measured from the PR3 power spectra, which has the lowest
Hubble constant. As described in e.g., \cite{Knox:2019rjx,aiola/etal:2020}, the CMB data indirectly measure the Hubble constant due to its effect of scaling the distance to recombination. This distance is constrained by precisely measuring the angle of the peak positions in the CMB, which gives the ratio of the sound horizon to the distance to recombination, with the sound horizon measured primarily through the constraint on the baryon density. The relative peak heights in the CMB allow a varying Hubble constant to be distinguished from a change in the relative proportions of matter and dark energy, which also affects the distance to recombination. 

Our estimates are consistent with the result of $H_0 =70.4\pm1.9$~km/s/Mpc in \citet{freedman2024} (broken down as $H_0 = 70.39 \pm 1.22~(\rm stat) \pm 1.33~(\rm syst)\pm 0.70~(\sigma_{\rm SN})$ km/s/Mpc) that uses Tip of the Red Giant Branch (TRGB) stars as distance calibrators, but are in significant tension with the SH0ES result of $H_0=73.17\pm0.86$~km/s/Mpc \citep{Riess_SHOES,Breuval:2024lsv}.
 \cite{verde2024} reviews a broader set of recent $H_0$ measurements. The sound horizon is measured with the ACT data, within the \LCDM\ model, to 
\be
r_d =147.1 \pm 0.3~{\rm Mpc} \ (\pact).
\ee
The product of the Hubble constant and the sound horizon is consistent with the DESI DR1 BAO measurements \citep{DESI-BAO-VI}. The constraint is illustrated in Figure \ref{fig:h0-s8-tensions}. 

With the improved sensitivity of the ACT data, the inferred Hubble constant can be estimated from the TT, TE and EE data independently, as shown in Figure \ref{fig:ttteee}. We find 
\ba
 H_0  &=& 68.47\pm 0.91~{\rm km/s/Mpc} \ (\pacttt) \nonumber \\ 
 &=& 66.92\pm 0.76~{\rm km/s/Mpc} \ (\pactte) \nonumber \\ 
  &=& 67.6\pm 1.2~{\rm km/s/Mpc} \ (\pactee). 
 \ea
 The P-ACT polarization data alone (EE) now independently rule out a \LCDM\ model with a higher Hubble constant of 73~km/s/Mpc at $>4\sigma$. Such a universe can still fit the \Planck\ EE data, but is excluded by the ACT EE data. To increase the Hubble constant, the matter density is reduced to better fit the peak positions, but the model overpredicts the peak heights. This is illustrated in Appendix \ref{apx:params} and follows a similar study in \cite{aiola/etal:2020}. Within the \LCDM\ model this provides a new line of evidence for the stability of the parameters with the additional data.

 Our estimate of the Hubble constant is also consistent with results from SPT-3G data reported in \cite{dutcher2021} and \cite{balkenhol2023}. The spectra from ACT are shown together with SPT data and other recent CMB data in Appendix \ref{apx:compilation}, in an update of the summary plot presented in \cite{choi_atacama_2020}.

The other parameter that has received significant attention is the amplitude of fluctuations, quantified by $\sigma_8$ or $S_8 \equiv \sigma_8(\Omega_m/0.3)^{0.5}$, with some galaxy weak lensing analyses measuring $S_8$ lower, at typically the 2$\sigma$ level, than predicted from the CMB-derived \LCDM\ model \citep{Heymans_2021,des_y3,dalal_hsc23}. CMB lensing measurements that probe larger scales and earlier times show no evidence for a lower amplitude, for example in \cite{Madhavacheril_dr6_lensing}.

As a comparison point for the early universe prediction, we show the updated constraints on $S_8-\Omega_m$ with the P-ACT data in Figure \ref{fig:h0-s8-tensions}. The matter density $\Omega_m$ is 0.4$\sigma$ lower than for \Planck; $\sigma_8$ is dominated by the overall amplitude which ACT does not measure as well as \Planck. The two-dimensional volume shrinks to give 
\ba
\sigma_8 &=& 0.815\pm 0.006 \ (\pact) \nonumber \\
 &=& 0.813\pm0.005  \ (\pactlb)\nonumber \\
\Omega_m &=& 0.312\pm0.007\ (\pact) \nonumber \\
 &=& 0.303\pm 0.005\ (\pactlb) 
\ea
and the derived $S_8$ is $0.830\pm0.014$ for \pact. \jd{Including the DESI DR2 BAO data further tightens the matter density measurement to $\Omega_m=0.3003\pm0.0035$ (\pactlbb).}

\subsection{Lensing in the power spectrum} 

 \begin{figure}
    \centering
    \includegraphics[width=\columnwidth]{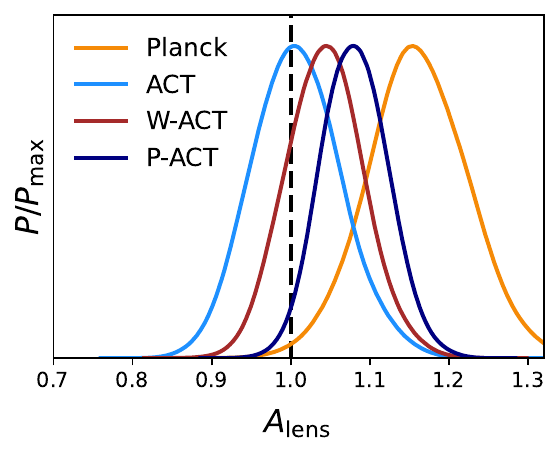}
    \includegraphics[width=\columnwidth]{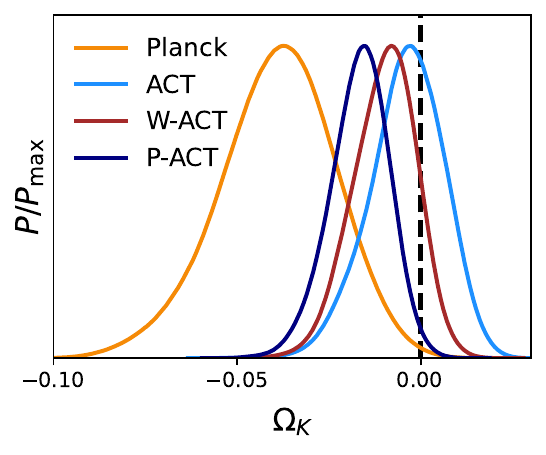}
    \caption{(Top) Constraints on a modified lensing amplitude, $A_\mathrm{lens}$ for different CMB power spectrum data; results from ACT and the combination with \WMAP\ are consistent with the $\Lambda\mathrm{CDM}$ prediction. ``Planck" here is the PR3 data; analyses with NPIPE data reduce $A_\mathrm{lens}$ to within 2$\sigma$ of unity. (Bottom) Constraints on $\Omega_K$ from the CMB power spectrum for different data combinations: the ACT data are consistent with a flat universe. }
\label{fig:lensing}
 \end{figure}

The CMB power spectrum is lensed, which has the effect of smearing the peaks and enhancing the small-scale power \citep[e.g.,][]{lewis/challinor:2006}. The PR2 and PR3 \Planck\ releases showed an excess of the lensing-like smearing effect, quantified by the $A_{\rm lens}$ parameter \citep{Calabrese:2008} described in \S\ref{sec:params}, which artificially amplifies the lensing spectrum compared to the model prediction. The PR3 analysis found $A_{\rm lens}>1$ at the almost 3$\sigma$ level \citep{planck2018_cosmo}, with $A_{\rm lens}=1.180\pm0.065$. Analyses of the NPIPE data, using larger fractions of sky, found lower departures of only 1.7$\sigma$ for the CamSpec likelihood, with $A_{\rm lens}=1.095\pm0.056$  \citep{rosenberg:2022}, or 0.75$\sigma$ using the Hillipop likelihood, with $A_{\rm lens}=1.039\pm 0.052$.~\citep{hillipop2024}}. No excess lensing was observed in the ACT DR4 data, with $A_{\rm lens}= 1.01\pm0.11$ \citep{aiola/etal:2020}.

We investigate the lensing in these new ACT spectra, varying the $A_{\rm lens}$ parameter in addition to the six \LCDM\ parameters. We find no evidence for excess lensing in the ACT power spectrum data:
 \ba
 A_{\rm lens} &=& 1.007\pm 0.057 \ \ (\act) \nonumber \\ 
 &=& 1.08^{+0.10}_{-0.12} \ \quad \quad (\acttt) \nonumber \\ 
 &=& 1.24^{+0.18}_{-0.22} \ \quad \quad (\actte) \nonumber \\ 
  &=& 0.89^{+0.10}_{-0.23} \ \quad \quad (\actee) \nonumber \\
A_{\rm lens} &=& 1.043\pm 0.049 \ \ (\wact),
 \ea 
where the final constraint includes the \WMAP\ data to better constrain other parameters from the larger scales. Figure \ref{fig:lensing} shows the  distributions for $A_{\rm lens}$ for different data combinations, and Figure \ref{fig:LCDM} includes the correlation of $A_{\rm lens}$ with the matter density. In the \Planck\ PR3 data an oscillatory residual for TT in the range $1000<\ell<2000$ was identified as driving the preference for the enhanced lensing; we do not see evidence for this in the ACT data. For $\pact$ we find $A_{\rm lens}=1.081 \pm 0.043$ which is consistent with e.g., \cite{rosenberg:2022}.
 
The high fluctuation in $A_{\rm lens}$ also manifested itself in the \Planck\ analyses as a preference for positive spatial curvature, with $\Omega_{K}<0$ ~\citep{planck2015-cosmo,planck2018_cosmo,DiValentino:2019}. This parameter is fixed to zero in the baseline \LCDM\ model. It cannot be measured using the primary CMB power spectra alone since different combination of other cosmological parameters can absorb the changes caused by non-zero flatness~\citep[see e.g.,][]{Bond_geometry, Efstathiou_degeneracy}. This geometric degeneracy is effectively broken when using the lensed CMB spectrum, or combining the CMB with lensing and/or BAO~\citep[see e.g.,][]{Efstathiou_degeneracy2,WMAP_for_curvature,sherwin/etal:2011}.

\begin{figure*}
    \centering
    \includegraphics[width=\textwidth]{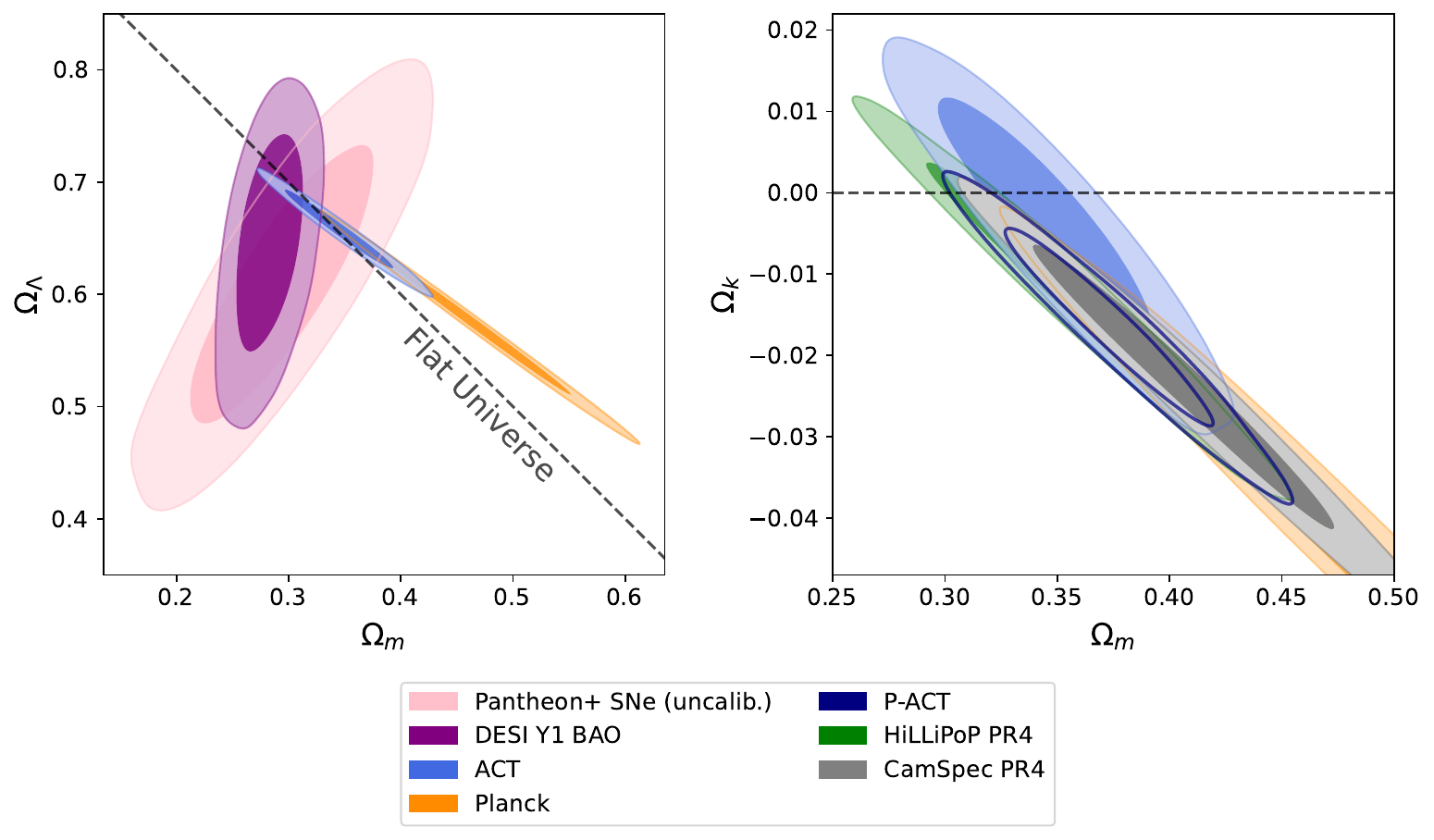}
    \caption{(Left) Measurements of background cosmological parameters, including possible non-zero spatial curvature, from DESI DR1 BAO, uncalibrated Pantheon+ supernovae (SNe), ACT, and \Planck\ PR3. Increased concordance between the BAO, SNe, and CMB results is observed with the ACT data. (Right) Results from ACT, P-ACT, and the \Planck\ NPIPE CamSpec \citep{rosenberg:2022} and HiLLiPoP likelihoods \citep{hillipop2024} are consistent with \LCDM\ (zero curvature).\\}
    \label{fig:concordance}
\end{figure*}

The \Planck\ PR3 lensed power spectrum shows a $3\sigma$ preference for non-zero curvature \citep{planck2018_cosmo}, although this is disfavored with the inclusion of large-scale structure data. As shown in Figures~\ref{fig:LCDM} and \ref{fig:lensing}, the ACT power spectra prefer a flat geometry, with the curvature parameter measured from the lensing in the power spectrum to be 
\ba
\Omega_{K}&=&-0.004\pm0.010 \ (68\%, \act) \nonumber \\
&=&-0.010\pm0.009 \ (68\%, \wact),
\ea
independently of the \Planck\ high-$\ell$ data. Further context for this result is shown in Figure \ref{fig:concordance}, where we compare lensed-CMB-only measurements with tracers of background cosmology (BAO from DESI, and uncalibrated supernovae from Pantheon+ \citep{Pantheon+, Panthoen+cosmology}). Although broad, their constraints are consistent with ACT, and all three are consistent with \LCDM\ to within $1\sigma$. Also shown is the agreement of the ACT and P-ACT measurements with the NPIPE analyses of \citet{rosenberg:2022} and \citet{hillipop2024}. The increased precision of ACT curvature measurements compared to \Planck, again due to the sensitivity of ACT data on arcminute scales, is apparent. These results from BAO, supernovae, and from the lensed CMB power spectrum adds robustness to the preferred flat geometry that is found when combining the CMB with these external datasets, shown in C25.

\subsection{SZ signal, foregrounds and calibrations} 
\label{subsec:fg}
\begin{figure*}
    \includegraphics[width=\textwidth]{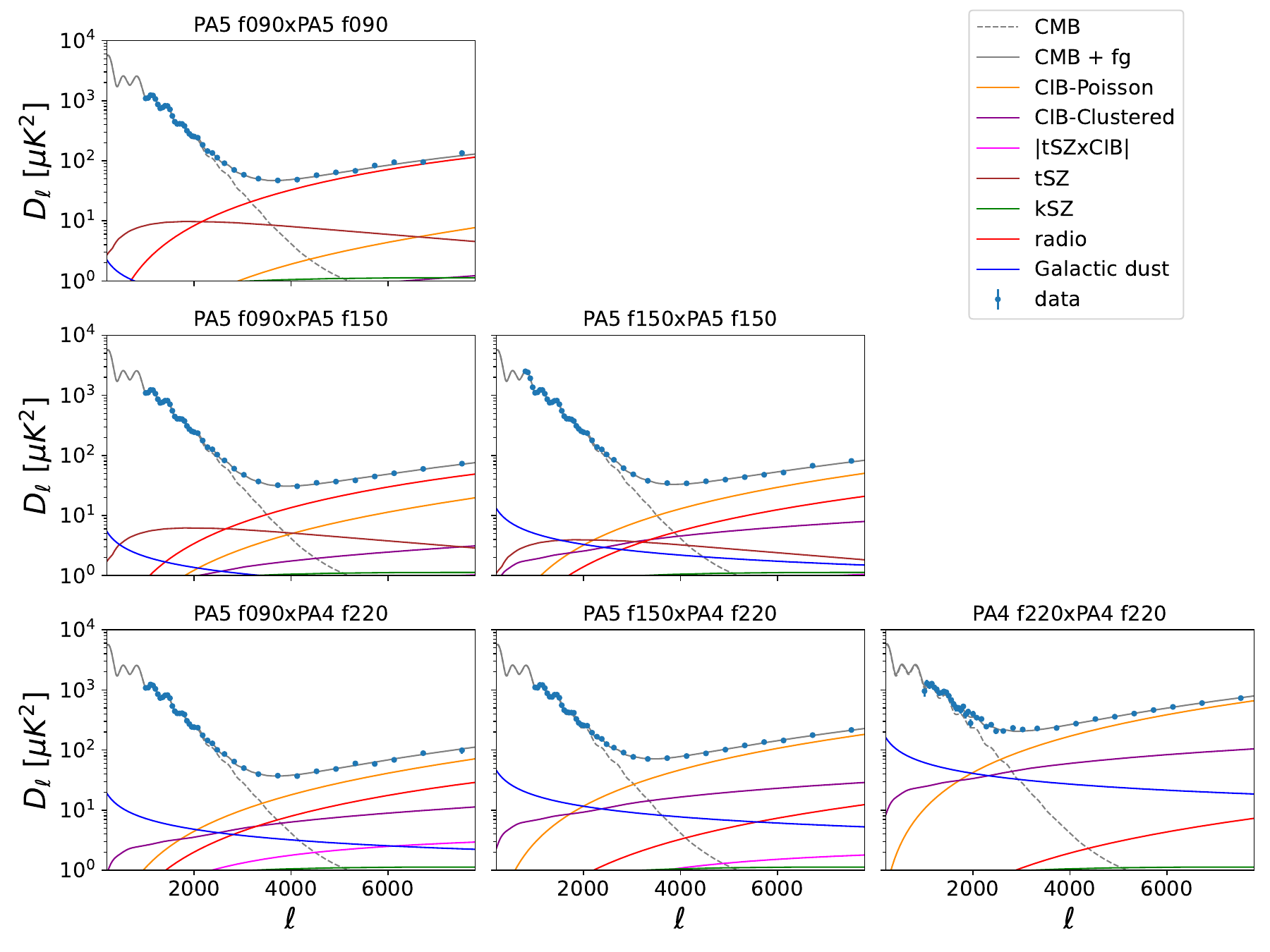}
    \caption{The foreground and secondary anisotropy components in TT for the six cross-frequencies. At f090 the dominant components are radio source emission and thermal SZ. At f220 the CIB dominates at $\ell>2000$. The cross-frequency spectra play an important role in fitting the model. We find a similar decomposition of spectra to those shown in \citet{reichardt/etal:2020,agora}  from the South Pole Telescope (SPT).}
    \label{fig:TT_component}
\end{figure*}

\begin{table}[tbp]
	\centering
	\begin{tabular}{l|ccc}
		\hline\hline
		& {\bf Nominal} & Nominal &$\beta_c \ne \beta_p$ \\
        &  {\bf $\pact$} & $\act$ & $\act$\\
		\hline
		SZ &&& \\ 
        \quad$a_{\rm tSZ}$ & $3.3\pm0.4$& $3.4\pm0.4$ & $3.0\pm0.4$\\
		\quad$\alpha_{\rm tSZ}$ & $-0.6\pm0.2$ & $-0.5\pm0.2$ & $-0.7^{+0.3}_{-0.2}$\\
		\quad$a_{\rm kSZ}$ & $2.0\pm0.9$ & $1.5^{+0.7}_{-1.1}$ & $2.4^{+0.9}_{-0.8}$\\
		CIB&&& \\ \quad$a_c$ & $3.6 \pm 0.5$ &  $3.7\pm0.5$ & $2.4^{+0.4}_{-0.8}$\\
		\quad$a_p$ & $7.7\pm0.3$&$7.7\pm0.3$ & $8.2\pm 0.4$\\
		\quad$\beta_p$ & $1.9\pm0.1$ & $1.9\pm0.1$ & $1.8\pm 0.1$\\
        \quad$\beta_c$ &  &  & $2.6^{+0.4}_{-0.3}$\\
        SZ-CIB&&& \\ \quad$\xi$ & $0.09^{+0.05}_{-0.07}$ &  $0.09^{+0.04}_{-0.08}$ & $< 0.15$\\
		Radio &&&\\ \quad$a_s^{TT}$ & $2.8\pm0.2$ & $2.9\pm0.2$ & $2.7\pm 0.2$\\
		\quad$\beta_s$ & $-2.8\pm0.1$ & $-2.8\pm0.1$ & $-2.8\pm0.1$\\
		\hline
		\quad$a_s^{TE}$  & $-0.026\pm0.012$ & $-0.025\pm0.012$ & $-0.024\pm 0.012$\\
		\quad$a_s^{EE}$ & $<0.04$ &$<0.04$ & $< 0.04$\\ 
		\hline\hline
	\end{tabular}
	\caption{Estimates from ACT and P-ACT for the nominal DR6 foreground model, not including the Galactic dust amplitudes which are prior-dominated. An example is also given of a model where the SZ parameters adjust by about 1$\sigma$, when the CIB Poisson and clustered terms are allowed to have different SEDs ($\beta_c\ne \beta_p$). Allowing for a possible correlation between radio and tSZ emission weakens the constraint on $A_{\rm tSZ}$ to $4.0^{+0.5}_{-0.6}$.}
	\label{tab:fg_params} 
\end{table}

\begin{figure*}
     \includegraphics[width=\columnwidth]{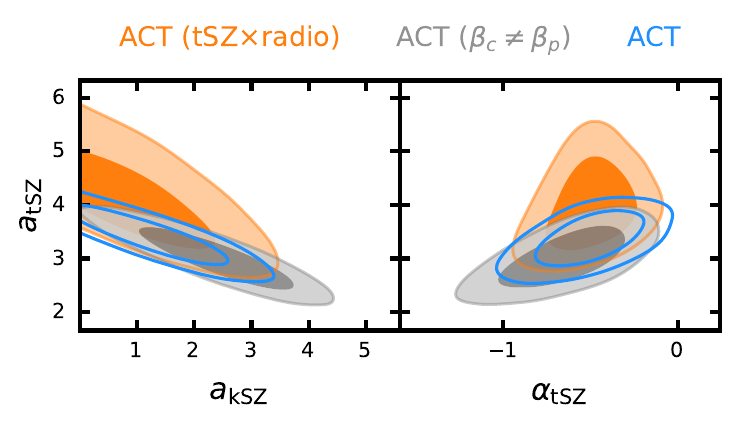}
     \includegraphics[width=\columnwidth]{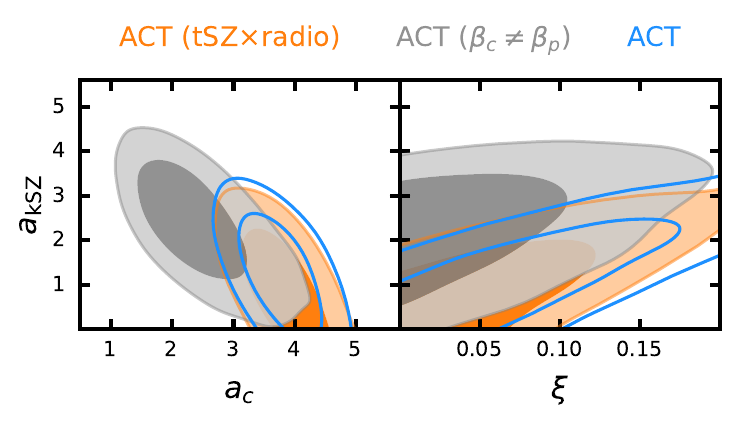}
    \caption{(Left panels) The correlations between the tSZ and kSZ amplitudes at $\ell=3000$ ($a_{\rm tsz}$ and $a_{\rm kSZ}$), and the shape parameter which scales the tSZ template by $\ell^\alpha$. (Right panels) The parameters most correlated with both the kSZ and tSZ amplitudes are the clustered CIB amplitude, $a_c$ and the degree of correlation of the tSZ and CIB, $\xi$. The equations defining these parameters are given in \S\ref{subsec:fg_model}.}
    \label{fig:sz}
\end{figure*}

\begin{figure}
    \includegraphics[width=\columnwidth]{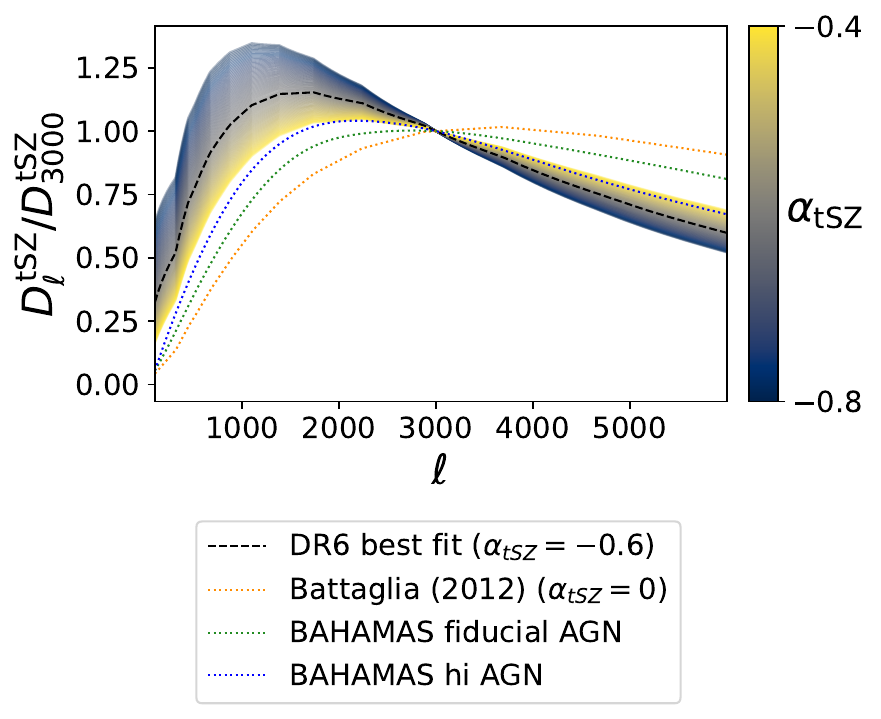}
    \caption{The shape of the estimated tSZ spectrum when its slope is allowed to vary, compared to the template from \cite{battaglia_tsz} with $\alpha_{\rm tSZ}=0$. \jd{It is a closer match to the templates from the BAHAMAS simulations \citep{mccarthy/etal:2017, 2018MNRAS.476.2999M}.}} 
    \label{fig:sz_cl}
\end{figure}

The best-fitting nominal foreground model is shown for the six cross-spectra in Figure \ref{fig:TT_component}, with a summary of the estimated parameters in Table \ref{tab:fg_params}. In temperature, we find that the data are well-fit by the sum of radio, CIB, tSZ, kSZ and Galactic dust. In performing the tests described in~\S\ref{subsec:fg_ext}, we determined that the only extension to the model that affects cosmological parameters at the 0.5$\sigma$ level is the shape of the tSZ power spectrum, which we now include in our nominal model; we find that it is correlated at the 10\% level with the baryon density and spectral index.

Of the extended models, a number of them result in shifts in foreground parameters at the $\sim1\sigma$ level, for example when allowing the clustered and Poisson part of the CIB to have different SEDs, as reported in Table \ref{tab:act-foreground-parameters}. None are significantly preferred by the data in terms of goodness of fit. The model for CO contamination is still uncertain however, and we cannot yet rule out a non-zero contribution which could impact our inference of other foregrounds. We also do not exclude other alternative foreground models.

Figure \ref{fig:sz} shows the ACT constraint on the tSZ and kSZ parameters, which are anti-correlated. The tSZ is the dominant foreground in the f090 spectrum at $\ell<2500$, but can be reduced by a corresponding increase in the blackbody kSZ component. For the nominal model we find 
\ba
a_{\rm tSZ} &=& 3.3\pm 0.4~\mu {\rm K}^2  \nonumber\\ 
a_{\rm kSZ} &<& 3.7~\mu {\rm K}^2 \ (95\%~ {\rm CL}),
\ea
for P-ACT, where these amplitudes measure $D_\ell$ at $\ell=3000$ and at f150 for the tSZ component. We find a preference for $\alpha_{\rm tSZ}$ to be non-zero at the 3$\sigma$ level, with
\be
\alpha_{\rm tSZ} = -0.6\pm0.2 \ (\pact).
\ee
with a PTE of $0.09\%$ for $\alpha_\mathrm{tSZ}=0$. Here, the negative slope steepens the SZ spectrum towards large scales, and has the effect of reducing the amplitude of tSZ at $\ell=3000$ and increasing the kSZ amplitude, compared to the best-fitting model with $\alpha_{\rm tSZ}=0$. The impact of its inclusion on other parameters is shown in Appendix \ref{apx:post_unblinding_changes}. The best-fitting tSZ spectrum is shown in Figure \ref{fig:sz_cl}, compared to the best-fitting model with $\alpha_{\rm tSZ}=0$. This steepening towards larger scales is consistent with simulations that have more feedback. Two examples are shown of the predictions from the \jd{BAHAMAS simulations with differing amount of AGN feedback, with $\Delta T_{\rm heat}^{\rm AGN} = 10^{7.8}$~K (fiducial AGN) and $10^8$~K (hi AGN)  \citep{mccarthy/etal:2017, 2018MNRAS.476.2999M}, with the $10^8$~K simulation most closely matching our best-fitting template.}\footnote{\jd{The shapes of the SZ spectra from the BAHAMAS simulations are not identical to those in the Agora simulations.}} Increasing this heating temperature enhances the gas density at large angular scales. This shape preference is consistent with findings from a recent binned reconstruction of the tSZ power spectrum in \citet{efstathiou/mccarthy:2025}, using the 90-100~GHz data from \Planck, ACT DR4 and SPT.

Within our nominal foreground model, the kSZ and tSZ amplitudes are most correlated with the tSZ-CIB amplitude, whose modeling is highly uncertain, and the clustered CIB amplitude, $a_c$, as shown in Figure \ref{fig:sz}. Since the tSZ-CIB is a negative term at f090 and f150, an increased contribution can allow more power in either tSZ or kSZ. Similarly, the SZ power can be increased by reducing the CIB contribution at f090 and f150, and adjusting the spectral index and bandcenters to conserve the CIB at f220.

In the set of extensions we consider, we find that allowing the CIB Poisson and clustered SEDs to be different has an order 1$\sigma$ effect on the SZ amplitudes but does not change the preference for negative $\alpha_{\rm SZ}$, as shown in Figure \ref{fig:sz}. However, we do not exclude other possible extensions to the foreground model that could mimic the effect of nonzero $\alpha_{\rm SZ}$. The figure also shows how allowing the radio sources and SZ to be correlated, using a scale dependent correlation template derived from Agora~\citep{agora}, results in a broadened distribution for $a_{\rm tSZ}$, although no radio-SZ correlation is detected.

\tl{Including the recently-identified CO$\times$CIB and CO$\times$CO terms at a nominal level, as estimated in \cite{Kokron_CO}, results in a non-negligible shift in the radio source amplitude, but we find this particular template amplitude is not a good fit to the overall data. Further investigations of how to model these components as a function of frequency will be valuable.}

We find that the ACT data are well fit when assuming the Galactic dust levels estimated from the \Planck\ 353 and 143~GHz data, described in~\S\ref{subsec:diff_gal}. In this model, described in Equation \ref{eq:dust_pw}, we assume fixed power-law scalings in multipole for temperature and polarization, a fixed emissivity index and dust temperature, and impose a Gaussian prior on the amplitudes at 150~GHz. We find that allowing the polarized power-law slope index to vary, and adjusting the prior on the amplitudes by refitting the \Planck\ 353-143 data for varying slope, has the effect of shifting the mean values for $n_s$-$\Omega_b h^2$ from ACT-alone by 0.2-0.3$\sigma$, and other parameters by $<0.1\sigma$. Removing the Gaussian prior on the amplitude has a similar effect. Allowing this additional freedom does not significantly improve the goodness of fit. 

We find no evidence for significant radio source emission in EE, and the data prefer a TE radio source spectrum that is negative but consistent with zero at the 2$\sigma$ level: 
\ba
a_s^{\rm TE} &=& -0.026\pm 0.012 \nonumber \\
a_s^{\rm EE} &<& 0.04 \ (95\%~ {\rm CL}).
\ea
This is the residual level after masking all sources with $>15$~mJy intensity at f150 and is consistent with ACT results in \cite{actpol_datta_2018}; without masking we find significant power from polarized sources. In the P-ACT data combination the polarized radio source emission is measured only by the ACT data at f090 and f150; it is assumed to be negligible in the \Planck\ analysis. 

Other foreground parameters are consistent with previous findings from ACT and SPT, and \jd{are} discussed further in \citet{fg_inprep}. 

We show the full correlation matrix of the cosmological, foreground, and systematic parameters in Appendix \ref{apx:params}. We find that the strongest correlation between calibration parameters and cosmological parameters for ACT alone is at the 40\% level between the overall calibration and the primordial amplitude, $A_s$. As discussed earlier, we also find an anti-correlation at the 20--30\% level between the per-array polarization efficiencies and the CDM density, $\Omega_c h^2$, discussed further in Appendix \ref{apx:poleff}, and which are reduced for P-ACT. The central frequencies of the passbands correlate strongly with foreground parameters but not with cosmological parameters. 

\section{Conclusions}
\label{sec:conclude}
This paper reports new and rigorous quantitative tests of the standard \LCDM\ cosmological model.  As in previous CMB analyses \citep{spergel2003,planck2013_cosmology,planck2018_cosmo,rosenberg:2022,hillipop2024}, the \LCDM\ model provides a remarkably good fit to the now more precise data.

With five years of data from ACT, we have extended the precision measurement of the cosmic microwave background temperature and polarization power spectrum to smaller scales than \Planck, from arcminute-resolution data covering half the sky. The spectra continue to support a \LCDM\ model, with comparable uncertainties on a consistent model seen from either ACT combined with \WMAP, or \Planck. The combination of ACT and \Planck\ data, including their CMB lensing data, further constrain the cosmological model. The maps, power spectrum pipeline and cosmological likelihoods are publicly available for further investigations.\\

Support for ACT was through the U.S.~National Science Foundation through awards AST-0408698, AST-0965625, and AST-1440226 for the ACT project, as well as awards PHY-0355328, PHY-0855887 and PHY-1214379. Funding was also provided by Princeton University, the University of Pennsylvania, and a Canada Foundation for Innovation (CFI) award to UBC. ACT operated in the Parque Astron\'omico Atacama in northern Chile under the auspices of the Agencia Nacional de Investigaci\'on y Desarrollo (ANID). The development of multichroic detectors and lenses was supported by NASA grants NNX13AE56G and NNX14AB58G. Detector research at NIST was supported by the NIST Innovations in Measurement Science program. Computing for ACT was performed using the Princeton Research Computing resources at Princeton University and the Niagara supercomputer at the SciNet HPC Consortium. SciNet is funded by the CFI under the auspices of Compute Canada, the Government of Ontario, the Ontario Research Fund–Research Excellence, and the University of Toronto. This research also used resources of the National Energy Research Scientific Computing Center (NERSC), a U.S. Department of Energy Office of Science User Facility located at Lawrence Berkeley National Laboratory, operated under Contract No. DE-AC02-05CH11231 using NERSC award HEP-ERCAPmp107 from 2021 to 2025. We thank the Republic of Chile for hosting ACT in the northern Atacama, and the local indigenous Licanantay communities whom we follow in observing and learning from the night sky.

We are grateful to the NASA LAMBDA archive for hosting our data. This work uses data from the Planck satellite, based on observations obtained with Planck (http://www.esa.int/Planck), an ESA science mission with instruments and contributions directly funded by ESA Member States, NASA, and Canada. We acknowledge work done by the Simons Observatory collaboration in developing open-source software used in this paper.

Some cosmological analyses were performed on the Hawk high-performance computing cluster at the Advanced Research Computing at Cardiff (ARCCA).  We made extensive use of computational resources at the University of Oxford Department of Physics, funded by the John Fell Oxford University Press Research Fund.
We thank Yuuki Omori for generation of and assistance with the \verb|AGORA| simulations. We thank George Efstathiou, Steven Gratton and Erik Rosenberg for useful discussions, and Alessio Spurio Mancini for help with CosmoPower.

 IAC acknowledges support from Fundaci\'on Mauricio y Carlota Botton and the Cambridge International Trust.  SA, MH and DNS acknowledge the support of the Simons Foundation. ZA and JD acknowledge support from NSF grant AST-2108126. EC, IH and HJ acknowledge support from the Horizon 2020 ERC Starting Grant (Grant agreement No 849169). OD acknowledges support from a SNSF Eccellenza Professorial Fellowship (No. 186879). JD acknowledges support from a Royal Society Wolfson Visiting Fellowship and from the Kavli Institute for Cosmology Cambridge and the Institute of Astronomy, Cambridge. RD and CS acknowledge support from the Agencia Nacional de Investigaci\'on y Desarrollo (ANID) through Basal project FB210003. RD acknowledges support from ANID-QUIMAL 240004 and FONDEF ID21I10236. SG acknowledges support from STFC and UKRI (grant numbers ST/W002892/1 and ST/X006360/1). SJG acknowledges support from NSF grant AST-2307727 and acknowledges the Texas Advanced Computing Center (TACC) at the University of Texas at Austin for providing computational resources that have contributed to the research results reported within this paper. V.G. acknowledges the support from NASA through the Astrophysics Theory Program, Award Number 21-ATP21-0135, the National Science Foundation (NSF) CAREER Grant No. PHY2239205, and from the Research Corporation for Science Advancement under the Cottrell Scholar Program.This research was carried out in part at the Jet Propulsion Laboratory, California Institute of Technology, under a contract with the National Aeronautics and Space Administration (80NM0018D0004). JCH acknowledges support from NSF grant AST-2108536, the Sloan Foundation, and the Simons Foundation, and thanks the Scientific Computing Core staff at the Flatiron Institute for computational support. MHi acknowledges support from the National Research Foundation of South Africa (grant no. 137975). ADH acknowledges support from the Sutton Family Chair in Science, Christianity and Cultures, from the Faculty of Arts and Science, University of Toronto, and from the Natural Sciences and Engineering Research Council of Canada (NSERC) [RGPIN-2023-05014, DGECR-2023- 00180]. JPH (George A. and Margaret M. Downsbrough Professor of Astrophysics) acknowledges the Downsbrough heirs and the estate of George Atha Downsbrough for their support. JK, MM  and KPS acknowledge support from NSF grants AST-2307727 and  AST-2153201.  ALP acknowledges support from a Science and Technology Facilities Council (STFC) Consolidated Grant (ST/W000903/1). ML acknowledges that IFAE is partially funded by the CERCA program of the Generalitat de Catalunya. SN acknowledges a grant from the Simons Foundation (CCA 918271, PBL). TN thanks support from JSPS KAKENHI Grant No. JP20H05859, No. JP22K03682, and World Premier International Research Center Initiative (WPI Initiative), MEXT, Japan. FN acknowledges funding from the European Union (ERC, POLOCALC, 101096035). LP acknowledges support from the Wilkinson and Misrahi Funds. KKR is supported by an Ernest Rutherford Fellowship from the UKRI Science and Technology Facilities Council (grant number ST/Z510191/1). NS acknowledges support from DOE award number DE-SC0025309. KMS acknowledges support from the NSF Graduate Research Fellowship Program under Grant No.~DGE 2036197, and acknowledges the use of computing resources from Columbia University's Shared Research Computing Facility project, which is supported by NIH Research Facility Improvement Grant 1G20RR030893-01, and associated funds from the New York State Empire State Development, Division of Science Technology and Innovation (NYSTAR) Contract C090171, both awarded April 15, 2010.

 MG and LP acknowledge the financial support from the INFN InDark initiative and from the COSMOS network through the ASI (Italian Space Agency) Grants 2016-24-H.0 and 2016-24-H.1-2018. MG is funded by the European Union (ERC, RELiCS, project number 101116027). MG acknowledges support from the PRIN (Progetti di ricerca di Rilevante Interesse Nazionale) number 2022WJ9J33.

We gratefully acknowledge the many publicly available software packages that
were essential for parts of this analysis. They include
\texttt{numpy}~\citep{harris2020array},
\texttt{scipy}~\citep{2020SciPy-NMeth},
\texttt{matplotlib}~\citep{Hunter:2007},
\texttt{healpy}~\citep{Healpix1}, \texttt{HEALPix}~\citep{gorski/etal:2005}, \texttt{pspy}~\citep{2020PhRvD.102l3538L}, 
\texttt{MFLike}, 
\texttt{mnms}~\citep{atkins/etal:2023}
\texttt{cobaya}~\citep{Cobaya}, 
\texttt{getdist}~\citep{Lewis:2019xzd}, 
\texttt{fgspectra},
\texttt{camb}~\citep{Lewis_camb},
\texttt{Astropy}~\citep{astropy:2013,
astropy:2018} and
\texttt{pixell}.

\bibliographystyle{act_titles}
\bibliography{dr6_bib, aiola_bib,choi_bib,planck_bib}

\appendix

\section{Blinding procedure}
\label{apx:blinding}

We used a two-phase blinding procedure in this analysis, to reduce the effect of confirmation bias in favor of the \LCDM\ model. When pre-defined criteria were met, we unblinded the data in these subsequent stages. The unblinding of the first phase, resulting in the comparison of the ACT polarization spectra to those from \Planck, took place on December 7 2023. The second phase unblinding, resulting in us looking at cosmological parameters, took place on April 18 2024. Here we describe the original plan as refined until October 2023, and changes to that plan before unblinding with the associated motivation. Where changes are made, we indicate them by a superscript number ($\Delta i$) in the original plan. Appendix \ref{apx: changes_to_blinding_plan} summarizes the post-unblinding changes. 

\subsection{First phase}
In the first blinded phase, we follow these rules:
\begin{itemize}
\item We allow TT, TE and EE power spectra from ACT to be plotted, but do not fit any cosmological parameters, or compare or plot them against theoretical predictions. Being able to look at the spectra allows us to identify any major problems in the data, but does not allow us to estimate precise parameters.
\item We do not look at calibrated and systematic-corrected BB spectra. We can look at the EB and TB power spectrum, with the understanding that there may be unknown systematic errors related to polarization angle.  
\item We allow ACT DR6 TT spectra to be plotted with \Planck\ TT spectra, and for correlations between ACT and \Planck\ to be computed in TT over all angular scales. This is based on the assumption that the \Planck\ TT spectrum has already been checked against \WMAP\ at $\ell<700$. 
\item We allow comparisons of all bandpowers with previous ACT DR4 data. Comparisons of BB are done by differencing DR4 and DR6 rather than looking at the spectra directly.
\end{itemize}
Before first-phase unblinding we perform a suite of internal consistency tests, and tests on a set of null maps, and calculate PTE values. We require the following criteria to be satisfied before first-phase unblinding:
\begin{itemize}
\item All baseline analysis choices made in running our pipeline, such as the range of CMB angular scales used for each array, and which arrays are included, and best-fit calibration factors, are frozen. 
\item The beam, leakage beam, and the covariance matrix, including beam uncertainties, should be sufficiently precise to pass the null tests, but can be refined before second-phase unblinding. These refinements may include a final run of the beam analysis, using the established method, and the use of an analytic covariance matrix that is tested against the simulation-based covariance. Details of our plans for these refinements are written down pre-unblinding. The polarization angle, and its systematic error, can be finalized before second-phase unblinding.
\item The distribution of PTEs for different null tests should be consistent with the distribution derived from simulations, including the number of outliers. This is to account for correlations between null tests that could make the PTE distribution non-uniform. However, this procedure may mask issues with the sims, that could also make the PTE distribution non-uniform. The distribution test should be repeated on a subset of null tests that are less correlated; the distribution should become more uniform.$^{\Delta 1}$
\item The number of null and consistency tests that fall outside the range $1\% < {\rm PTE} < 99\%$ should not be significantly inconsistent with the expectations from random fluctuations. Here PTE is measured with respect to the analytic distribution. 
\end{itemize}
The set of tests, described in \S\ref{subsec:nulls}, are (1) internal-split nulls for each array, (2) between-array nulls, (3) precipitable water vapor (PWV) null, where we expect different spectra but they should agree apart from at large scales, approximately $\ell<2000$, in TT, because we expect a stronger transfer function with increased correlated noise, (4) an elevation null, (5) an inner-outer detector null, where we expect different spectra but they should agree when the different beams are accounted for, and (6) a spatial null between different regions.$^{\Delta 2}$

In doing these tests we calculate O(2000) PTEs. 
To compute PTE we use the difference between power spectra $C^{AA}-C^{BB}$ (and this also includes the $C^{AB}-C^{BB}$ combinations, etc.).  Before first-phase unblinding the transfer functions will also be estimated from TT comparisons to \Planck\ in the typical range $200<\ell<2000$.

In this phase we also produce the same statistics for a comparison between ACT and \Planck\ TT at 100, 143 and 217 GHz, allowing for the transfer functions. If this test does not pass, this indicates that there is a systematic error in either the ACT or \Planck\ TT data at high-$\ell$. In this case we will further explore the impact of systematic uncertainties in the ACT data, but do not require this test to pass in order to proceed.

\subsection{Second phase}
Once these tests have passed for TT/TE/EE/BB, we move to the second phase. We do not require the TB/EB nulls to pass in the first phase. In the second phase we are allowed to additionally compare TE and EE to the \Planck\ 100, 143 and 217 spectra computed on the same sky region and using the same analysis pipeline, with the same requirements as above, i.e., we require that PTEs for different null tests should be consistent with the distribution derived from simulations, and the number of null tests that fall outside the range $1\%<{\rm PTE}<99\%$ should not be significantly inconsistent with the expectations from random fluctuations.
If this test does not pass, this indicates that there is a systematic error in either the ACT or \Planck\ data. We will check that the disagreement cannot be explained by including all the known beam uncertainty terms, or by accounting for passband variation within the estimated uncertainty range, or polarization efficiency. We would make a decision on how to proceed based on these findings.

If the comparison test to \Planck\ passes, before proceeding to second-phase unblinding, the following parameter-level tests must pass:
\begin{itemize}
    \item Recovery of \LCDM\ and \LCDM+$N_{\rm eff}$ parameters on our suite of idealized Gaussian sims (for ACT and ACT+Planck)$^{\Delta 3}$, to within 0.2$\sigma$. Parameters recovered from a single realistic non-Gaussian simulation should also be statistically consistent with the input parameters.$^{\Delta 4}$
     \item Test of the shift in \LCDM\ and \LCDM+$N_{\rm eff}$ parameters on subsets of the data (multipoles below or above $\ell=1000$, TT/TE/EE, combined frequency bands, array-bands), as well as for varying the k-space filter.$^{\Delta 5}$ We will only look at parameter differences for these tests.
\end{itemize}
All the products (including beam, calibration, polarization angle and covariance matrix), and theory settings, will be frozen.$^{\Delta 6}$ An estimate of the expected foreground level in BB will have been made.

\subsection{Changes to original blinding plan} \label{apx: changes_to_blinding_plan}
 
We made two updates before first phase unblinding: ($\Delta$1) The distribution of PTEs for different null tests should be consistent with a uniform distribution, rather than the distibution derived from simulations. Small (${<\sim5\%}$) adjustments to the errors would also be investigated if non-uniformity is seen. ($\Delta$2) The spatial null test between different regions is moved from first phase to second phase unblinding, since it includes an estimate for the Galactic foreground levels.

We then made a further four updates before second-phase unblinding: 
($\Delta$3) We limit the requirement to test parameter-recovery on just the ACT data, discarding the ACT+Planck test, because our default combination with \Planck\ was changed to using a cut in multipole for each dataset rather than including the ACT-Planck covariance, so we use only public \Planck\ products.

($\Delta$4) The plan required that parameters recovered from a single realistic non-Gaussian simulation should be statistically consistent with the input parameters. This was achieved post-unblinding (described in \S\ref{sec:like}), but in performing this test with simulations from Agora (extragalactic/CMB) and PySM3 (Galactic), using two different CMB realizations, we initially found 3$\sigma$ shifts in some \LCDM\ parameters compared to their input values. In diagnosing this mismatch, we determined that the foregrounds were more complicated in the simulations than we have seen with past ACT data, including the precise SED of the CIB and radio components, and may not be described at sufficiently high precision by our model; it was unclear at the time whether the differences were physically meaningful, and it was felt that actual data were needed to make further progress. Since the simulations highlighted a number of ways in which our foreground model could be modified or extended, we instead moved to use the unblinded data to test the effects of these modifications on cosmological parameters, to judge whether they matter/are needed. We stated that if any of the modifications results in cosmological parameter shifts of more than $0.5\sigma$ (tested on \LCDM\ and \LCDM+$N_{\rm eff}$), we would plan to marginalize over one or more additional foreground parameters to capture the modeling uncertainty. The tests are summarized in \S\ref{sec:like}.  With later careful reprocessing of the simulations, and the inclusion of the $\alpha_{\rm tSZ}$ parameter, consistent parameters were recovered post-unblinding, as reported in \S\ref{sec:like}.

($\Delta$5) We removed the requirement to test for consistency on TT/TE/EE or multipole ranges, as this could bias us towards \LCDM. In light of the results in Appendix \ref{apx: sys_res}, we did not perform this test between individual array-bands. We also reconstructed power spectra under more- and less-aggressive ground filters; variations in the spectra with respect to that of the the baseline filter were typically sub-percent in signal-dominated multipoles with no trend, so we also did not test parameter consistency between these spectra.   

($\Delta$6) The plan required all products to be frozen; we additionally allow for minor refinements expected to be at the $<0.2\sigma$ level in parameters. These include the final covariance matrix needing an iterative estimate of the foreground level from running parameters, and we may substitute a refined analytic covmat estimate and final theory settings, but expect those choices to have only small effects on parameters.

\section{Summary of post-unblinding changes} \label{apx:post_unblinding_changes}
Here we list all of the changes that were made after second-phase unblinding of the data.

\begin{figure}[t]
    \includegraphics[width=0.5\linewidth]{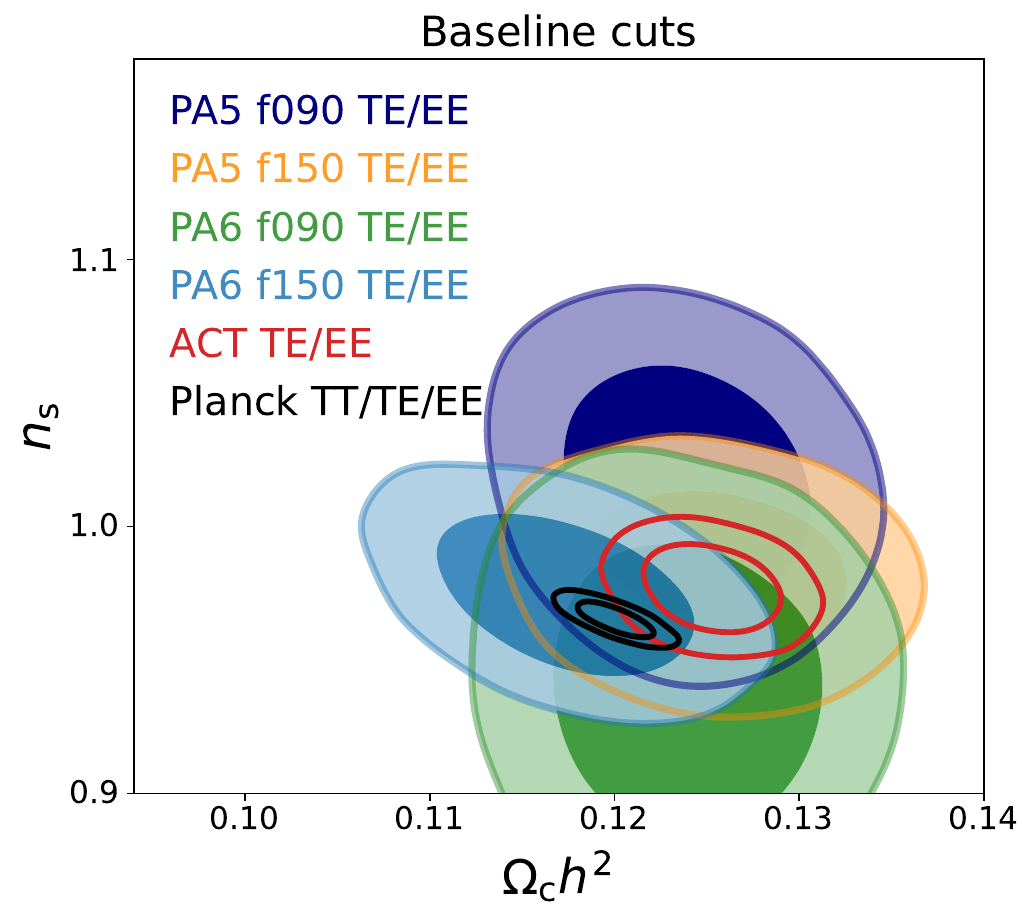}\hfill
    \includegraphics[width=0.5\linewidth]{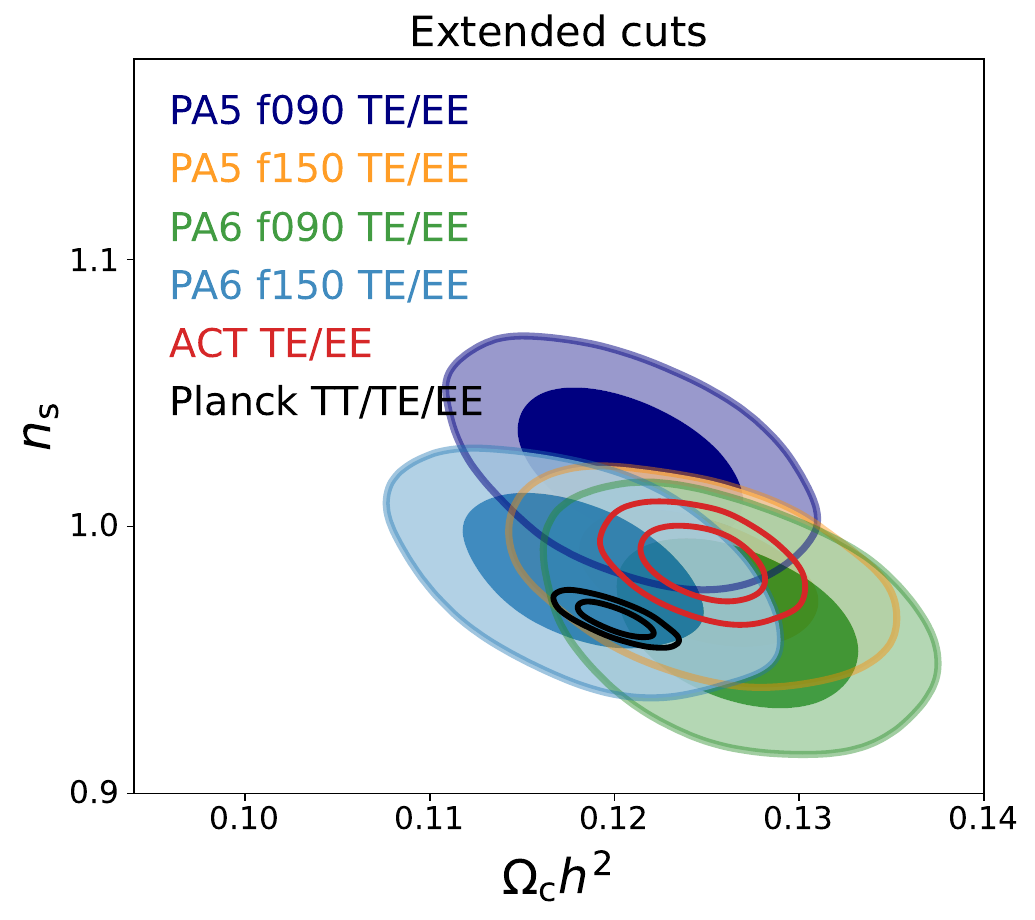}
    \caption{The constraints on the spectral index, $n_s$, and the CDM density, $\Omega_c h^2$, from the ACT TE and EE spectra for each individual array-band. Left shows the nominal multipole cuts; right shows the pre-unblinding extended cuts. The baseline cuts remove the largest scale polarization data in f090 where we see evidence for a systematic residual in PA5, shown in Figure \ref{fig: apx_spec_map_nulls}.}
    \label{fig:teee_polcut}
\end{figure}

\begin{enumerate} 
\item {\bf Polarization ell-cuts:} As described in Appendix \ref{apx: sys_res}, we adopt the same cuts in polarization as in temperature. These, and the pre-unblinding ``extended" cuts, are summarized in Table \ref{tab:multipole_cut}. The impact on the $n_s- \Omega_c h^2$ parameters derived from the TE/EE spectrum combination is shown in Figure \ref{fig:teee_polcut}. The ``extended" cuts leave in the largest $500<\ell<1000$ scales in f090 where we see evidence for a systematic residual in PA5 when comparing arrays. A steeper spectral index is preferred to fit the extended PA5~f090 data.
\item {\bf Thermal SZ template shape}:  As described in \S\ref{subsec:fg_ext}, we test a set of foreground extensions on the unblinded data to determine which, if any, are preferred by the data. The tSZ shape parameter, $\alpha_{\rm tSZ}$, is preferred at the 3$\sigma$ level so was included in the final model.
\item {\bf Beam chromaticity:} The inclusion of beam chromaticity is described in \S\ref{subsec:instrument} but was not included until after unblinding.
\item {\bf Final spectrum from 4-pass maps:} The final spectrum was computed from maps that had been run to four iterations, ``4-pass," as described in N25, to better estimate the noise model. This had a negligible effect on parameters.
\item {\bf Final covariance matrix:} The final matrix used updated foreground levels estimated from the data, the improved simulation-correction method  described in \cite{atkins/etal:2024}, and additional non-Gaussian terms. This had a negligible effect on parameters. 
\end{enumerate}

\begin{figure*}
    \includegraphics[width=\textwidth]{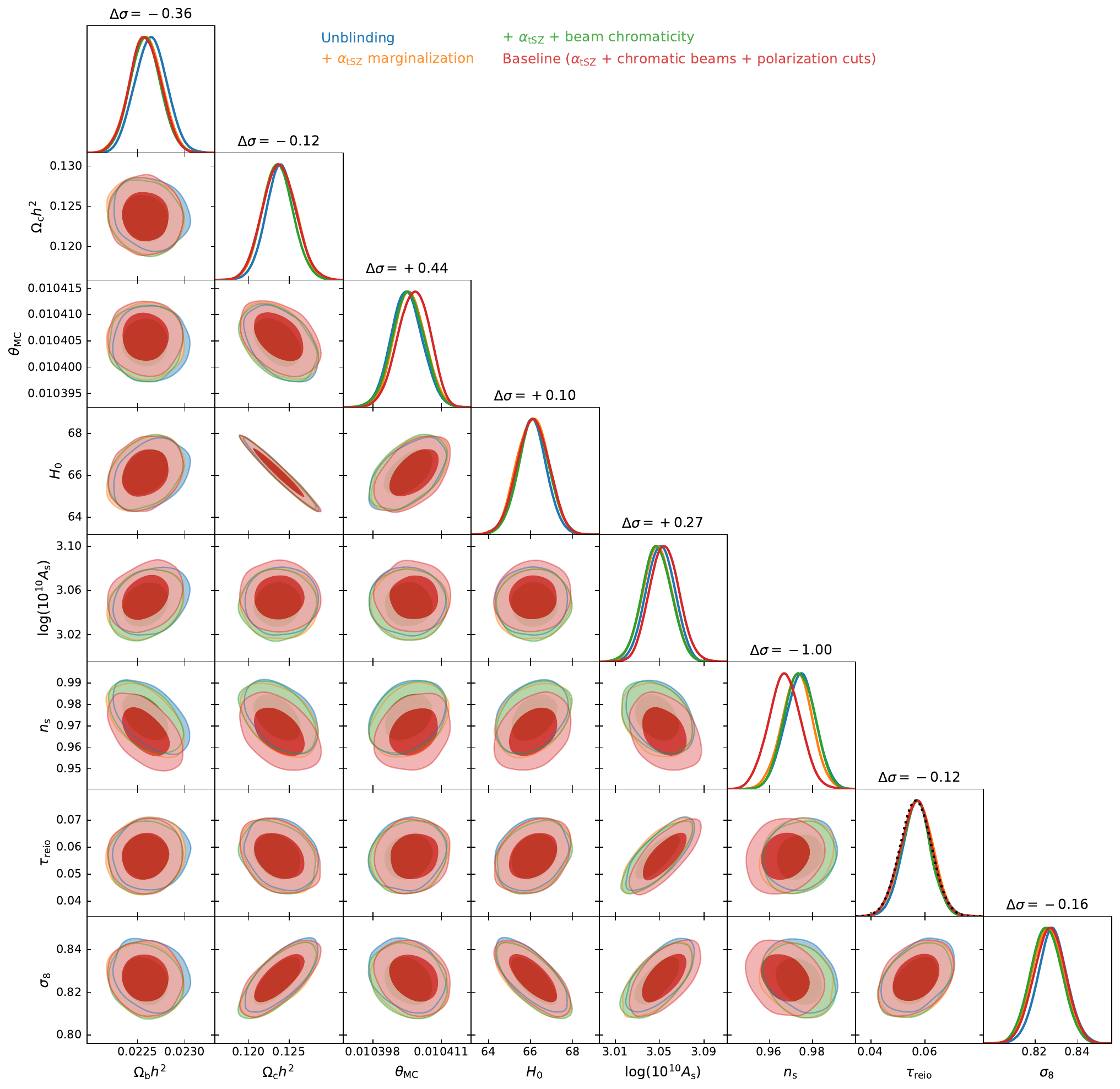}
    \caption{\LCDM\ cosmological parameters estimated at unblinding (blue), compared to our Baseline results after the set of post-unblinding changes (red). The intermediate results when just the additional tSZ shape parameter marginalization is applied to the unblinding results is also shown (orange), and then the beam chromaticity (green). The dominant effect, mainly on the spectral index $n_s$, is from the enhanced polarization multipole cuts, with a smaller effect from including the tSZ shape parameter. The labeling indicates the mean parameter shifts from the initial unblinding to the final Baseline results.}
    \label{fig:cosmo_unblinding}
\end{figure*}

\begin{figure*}
    \includegraphics[width=0.5\textwidth]{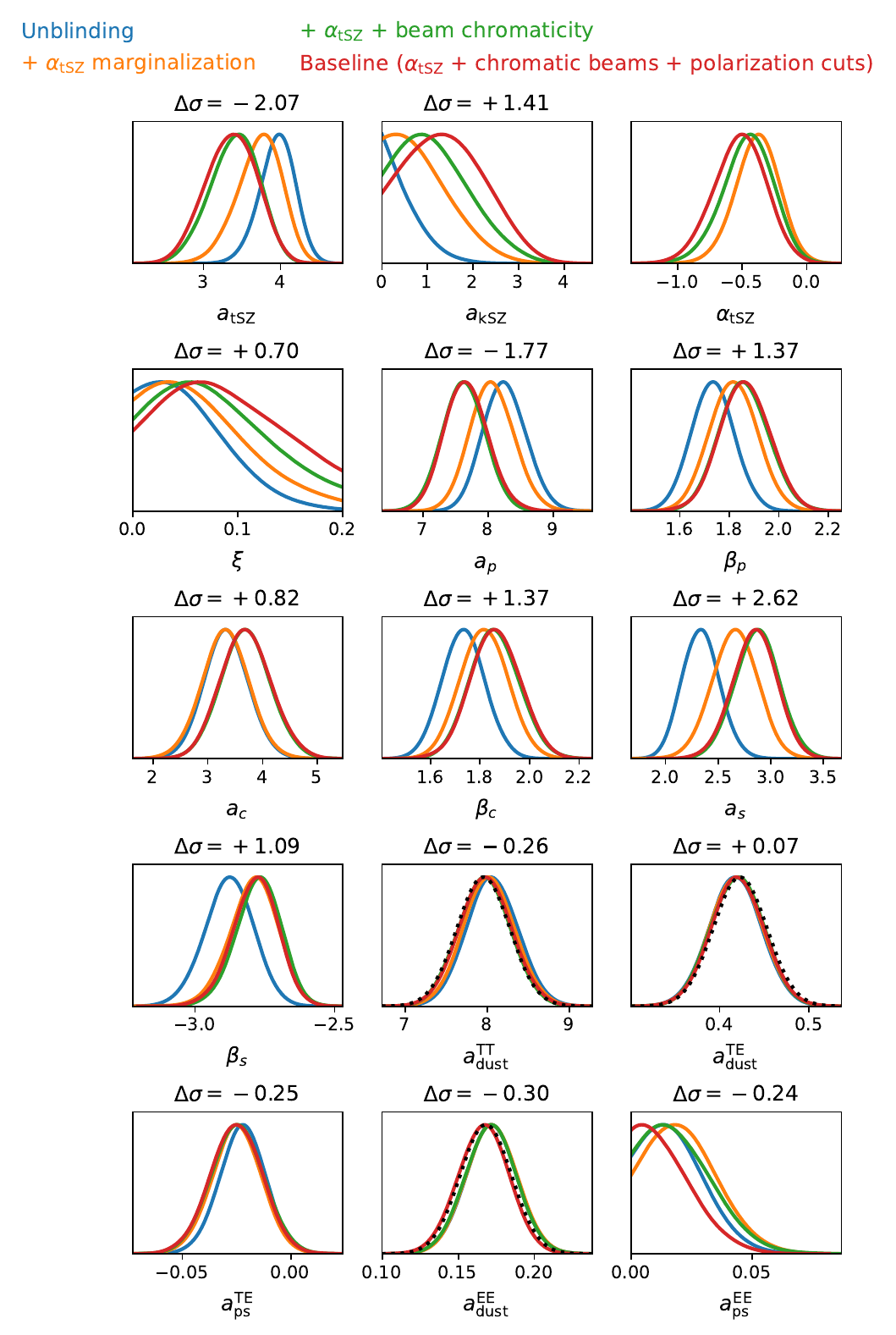}
    \includegraphics[width=0.5\textwidth]
    {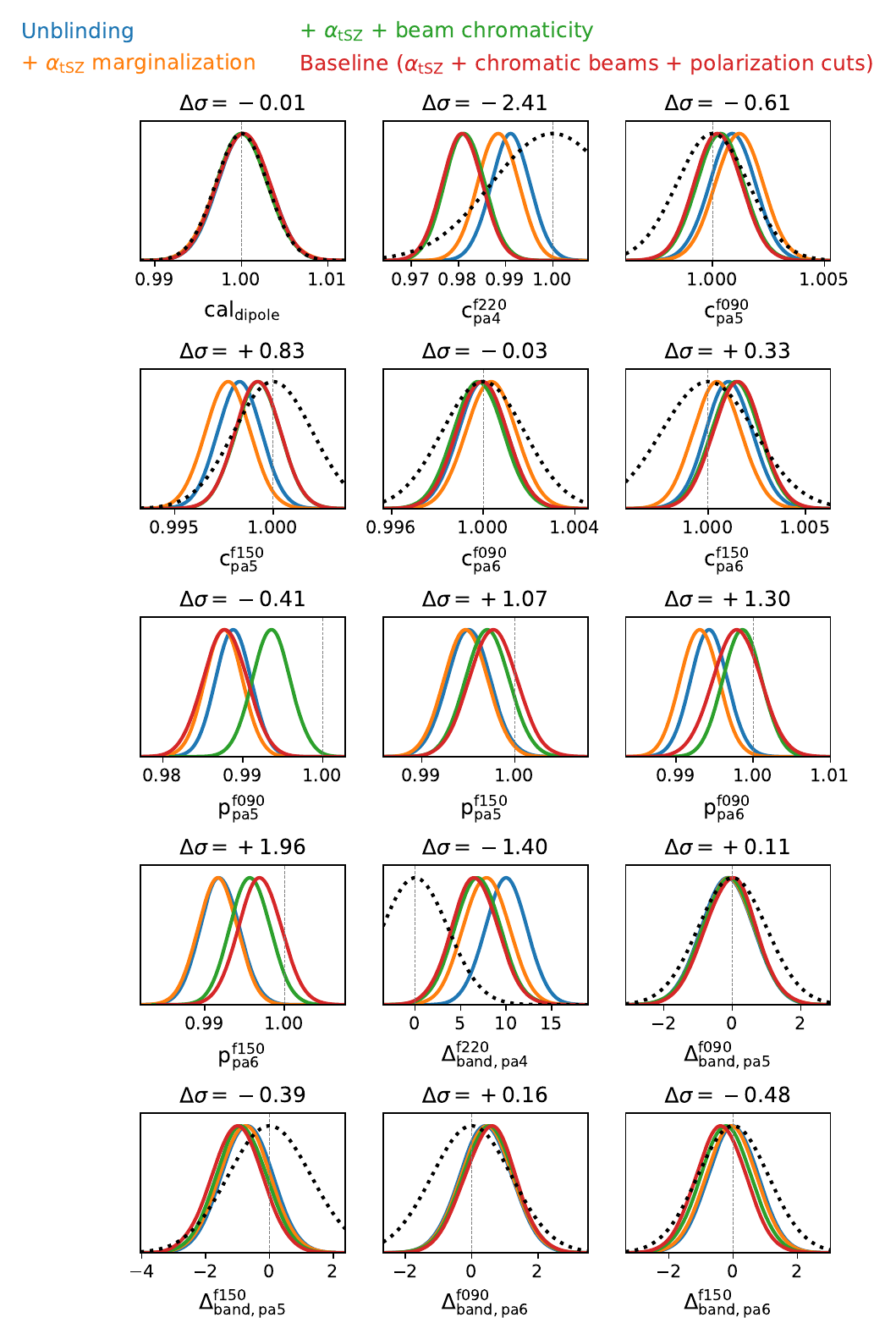}
    \caption{Foreground and systematics parameters estimated at unblinding (blue), compared to the model that includes the tSZ shape parameter (orange), then additionally the beam chromaticity (green) and our Baseline results after the set of post-unblinding changes (red). The main effects are from including beam chromaticity and including the tSZ shape parameter.For the blue and orange models the polarization efficiencies were normalized to unity using the spectra with extended cuts; for the green and red models the polarization efficiencies were re-normalized using spectra with the baseline cuts, so the efficiencies cannot be directly compared across all four cases. 
    }
\label{fig:other_params_unblinding}
\end{figure*}

The overall effect of these changes are shown in Figure~\ref{fig:cosmo_unblinding} and \ref{fig:other_params_unblinding}. The shifts in cosmological parameters come from the polarization cut and the inclusion of $\alpha_{\rm tSZ}$; the shifts in foreground parameters primarily come from the inclusion of $\alpha_{\rm tSZ}$ and the beam chromaticity. Further discussion of the foreground modeling and results \jd{are} presented in \cite{fg_inprep}.\\

\clearpage

\section{Further description of the power spectrum pipeline}
\label{apx:pipeline}

\subsection{The ACT DR6 data model}

The data model of a split, $i$, of the array-band, $\alpha$, of the ACT DR6 data as described in N25 is given by 

\ba
\begin{pmatrix}
\tilde{T}^{\alpha, i} \cr 
\tilde{Q}^{\alpha, i} \cr 
\tilde{U}^{\alpha, i}  \cr 
\end{pmatrix} (\bm{\hat{n}})
=
 W_{\rm obs}^{\alpha}(\hat{n}) \left[ \left({\bm B}^{\alpha} * \left\{
 \begin{pmatrix}
T_{\rm CMB} + T^{\alpha}_{\rm fg} \cr 
Q_{\rm CMB} + Q^{\alpha}_{\rm fg} \cr 
U_{\rm CMB} + U^{\alpha}_{\rm fg} \cr 
\end{pmatrix} +
\begin{pmatrix}
G_{T}^{\alpha, i} \cr 
G_{Q}^{\alpha, i} \cr 
G_{U}^{\alpha, i}  \cr 
\end{pmatrix}  
+
 \begin{pmatrix}
n_{\rm atm, T}^{\alpha, i} \cr 
n_{\rm atm, Q}^{\alpha, i} \cr 
n_{\rm atm, U}^{\alpha, i}  \cr 
\end{pmatrix}
\right\}
\right)  (\bm{\hat{n}}) + 
 \begin{pmatrix}
n_{\rm det, T}^{\alpha, i} \cr 
n_{\rm det, Q}^{\alpha, i} \cr 
n_{\rm det, U}^{\alpha, i}  \cr 
\end{pmatrix}  (\bm{\hat{n}})
\right],
\ea
where $*$ is the convolution operator. Here $W_{\rm obs}^{\alpha}(\hat{n})$ defines the region of observation, ${\bm B}^{\alpha}$ is a beam matrix, accounting both for the limited angular 
resolution of the telescope and the temperature to polarization leakage, $n^{\alpha, i}$ is the noise in split $i$ and array-band $\alpha$  accounting for contributions from both detector and atmospheric noise, 
for each Stokes component, ``fg" denotes the galactic and extragalactic foreground contamination, and $G_{T, Q, U}^{\alpha, i}$ represents the ground pick-up contamination of the maps. 
We use a common beam and mask for each split for a given array-band, $\alpha$. 

The data are also affected by model errors at the mapmaking stage, leading to additional biases that we refer to as transfer functions:
\ba
\begin{pmatrix}
\tilde{T}_{{\cal M}}^{\alpha, i} \cr 
\tilde{Q}_{{\cal M}}^{\alpha, i} \cr 
\tilde{U}_{{\cal M}}^{\alpha, i}  \cr 
\end{pmatrix} = 
{\cal M_{\alpha}} 
\begin{pmatrix}
\tilde{T}^{\alpha, i} \cr 
\tilde{Q}^{\alpha, i} \cr 
\tilde{U}^{\alpha, i}  \cr 
\end{pmatrix}.  
\ea
Correcting for this effect would require a precise instrument model. However, since it affects only the largest angular scales of the maps,  we mitigate it by using suitable multipole cuts for each array-band.

\subsection{Masks}{\label{subsec:analysis_mask}}

\begin{figure*}[htp]
    \centering
    \includegraphics[width=\linewidth]{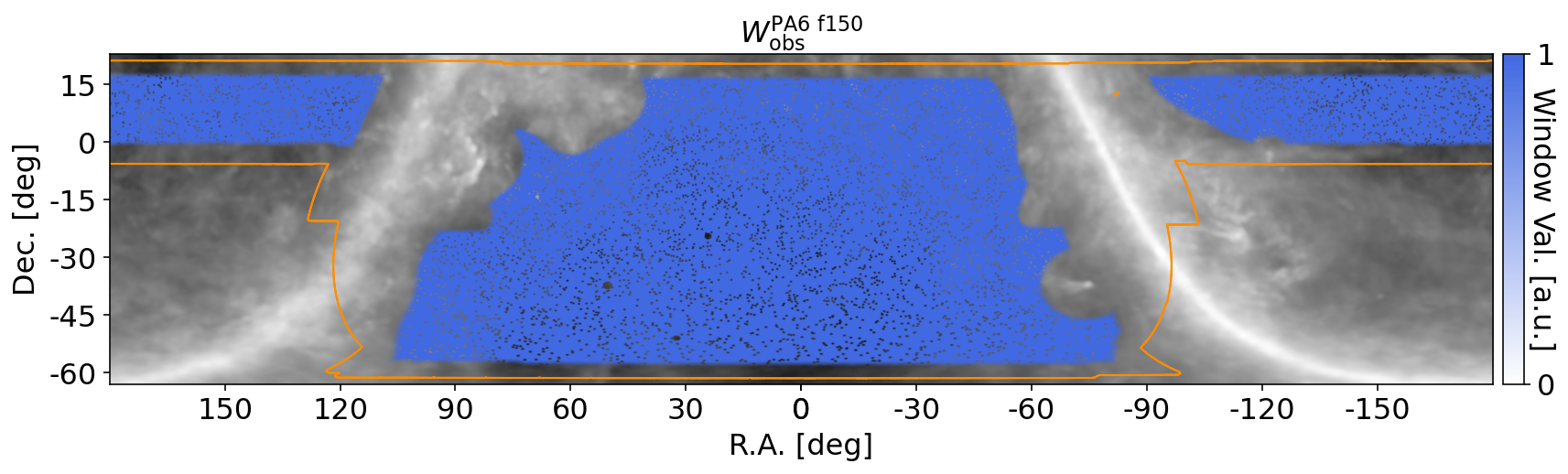}
    \caption{The observation window, $W_{\rm obs}^{\alpha}$, for PA6 f150, is shown in blue. Visible are the apodized boundaries as well as apodized point-source and extended-source holes. The maximum extent of the ACT DR6 data is outlined in orange. Data inside the orange region but not highlighted in blue --- predominantly close to the edge of the ACT footprint or the Galactic plane --- is cut from the analysis. The gray background is the logarithm of the dust intensity from \citet{planck_fg_2016}.}
    \label{fig:w_obs}
\end{figure*}

We use a set of masks in order to reduce instrumental and foreground contamination of the cosmological data:

\begin{enumerate}
    \item  $W^{\alpha}_{\rm edges}$:  A mask is created by identifying the map's edges 
and removing all pixels within a 0.5$^\circ$ angular distance from the borders. 
This step ensures that the analysis is not affected by the complex noise characteristics of these less reliably observed pixels. 
    \item  $W_{\rm Galactic}$:  
To minimize contamination from Galactic foreground emission, we apply a Galactic mask (G70) derived from \Planck's high-frequency
measurements at 353 GHz. We extend this in order to mask some extra bright dust clouds near the Galactic plane.
    \item $W_{\rm  sources}$:  We additionally construct a point source mask by identifying sources with flux brighter than 15 mJy at 150~GHz,
    leading to a total of $\sim 10^{4}$ masked sources. We do not mask Sunyaev-Zel’dovich (SZ) clusters  and instead include them in our foreground model. We additionally mask 
    extended sources that were not picked up in our original point source detection algorithm,
    such as nearby galaxies or nebulae.
\end{enumerate}

Note that while $W^{\alpha}_{\rm edges}$ depends on the array-band we consider, we use the same $W_{\rm Galactic}$ and $W_{\rm  sources}$ for all array-bands. The resulting sky area after masking is $\approx 10,000~{\rm deg}^2$.

\subsection{Ground filter}\label{subsec:ground_filt}
\begin{figure*}[htp]
    \centering
        \centering
        \includegraphics[width=0.5\textwidth]{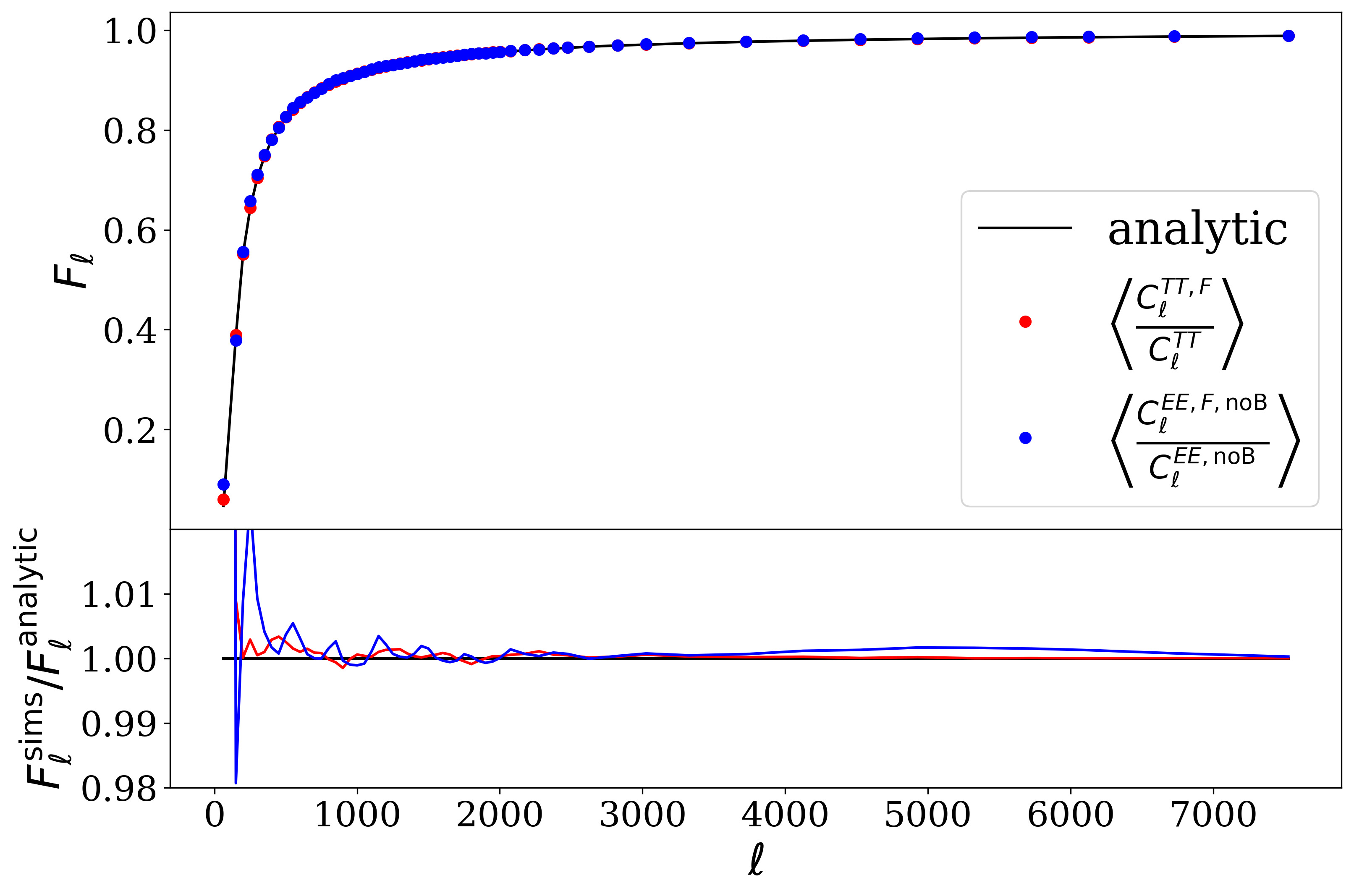}
    \hspace*{0.04\textwidth}
        \centering
        \includegraphics[width=0.35\textwidth]{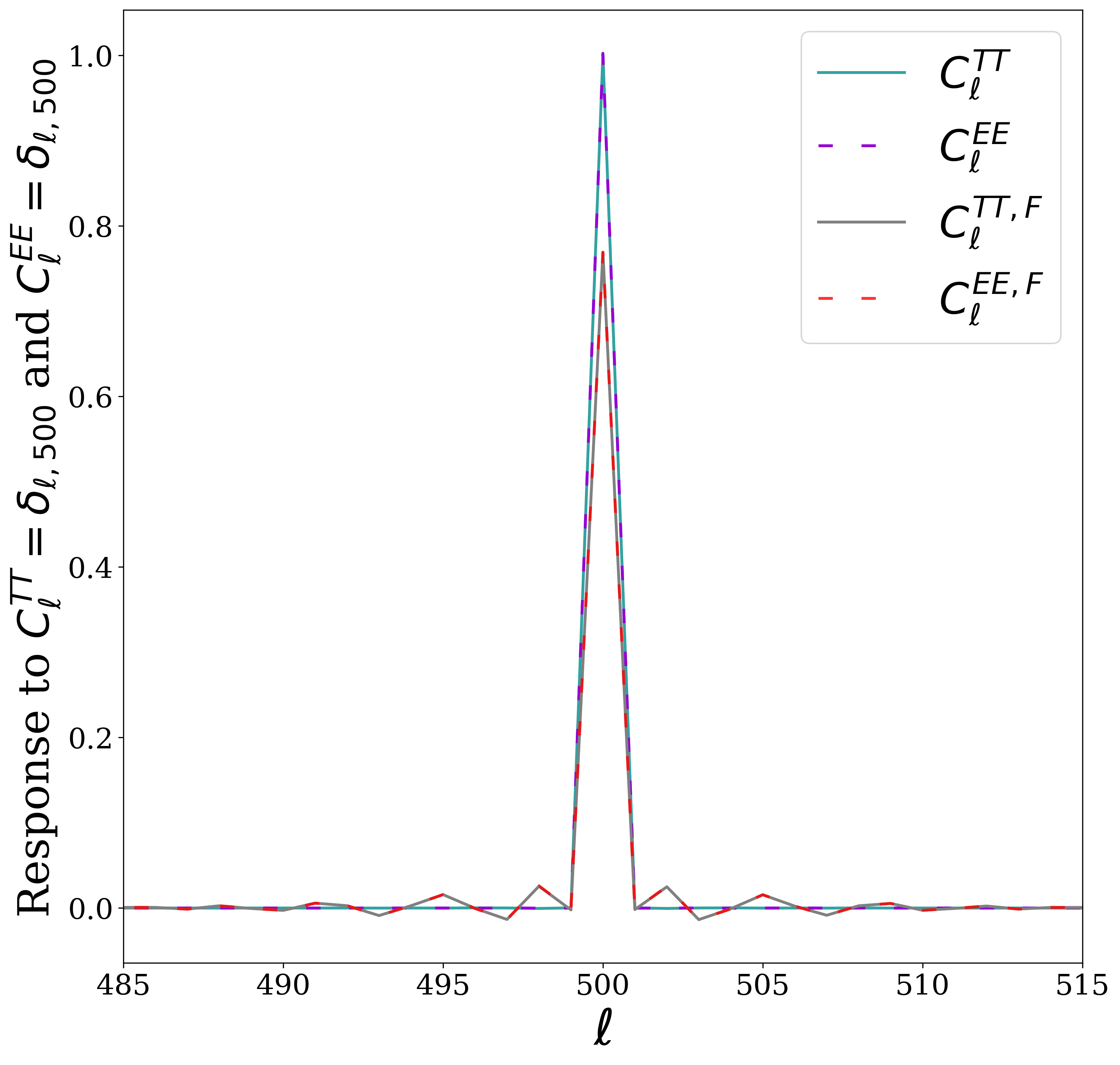}
    \caption{(Left) Elements of the filtering matrix, $\bm{F}_{\ell}$, estimated using 800 signal-only simulations of the ACT DR6 data. We show in black our analytic estimate for the filter's effect.
    (Right) Mode mixing introduced by the Fourier-space filter estimated using single-mode simulation; the effect is minimal, with correlation length smaller than our typical bin size.  }
    \label{fig:kspace_combined}
\end{figure*}

We remove the ground signal using a flat-sky Fourier-space
filter $ F_{\bm{\ell}}$, removing  modes with $|\ell_{x}| < 90$ and $|\ell_{y}| < 50$. The maps are cylindrical
(plate carrée) projections in equatorial coordinates, so $x$ and $y$ correspond to right ascension and declination. 
This filtering operation is non-local,  spreading power from each pixel across the map, so localised pixels with large values can contaminate a large fraction of the sky. 
We therefore perform two operations prior to Fourier-filtering: applying an apodized mask\footnote{With an apodization length of 2 degrees.} $W^{\alpha}_{\rm filter}  = W^{\alpha}_{\rm edges} W_{\rm Galactic}$ that removes pixels from the
edges of our region of observation and from the Galactic plane, and subtracting sources with a flux above 15 mJy at 150 GHz from the maps:
\ba
\begin{pmatrix}
\tilde{T}^{\alpha, i} \cr 
\tilde{Q}^{\alpha, i} \cr 
\tilde{U}^{\alpha, i}  \cr 
\end{pmatrix}_{\rm filtered, sources-sub.}
=
{\rm FFT}^{-1} \left(F_{\bm{\ell}} \ . \
{\rm FFT}  
\begin{pmatrix}
W_{\rm filter} (\tilde{T}^{\alpha, i} - \tilde{T}^{\alpha}_{\rm sources})\cr 
W_{\rm filter} (\tilde{Q}^{\alpha, i} - \tilde{Q}^{\alpha}_{\rm sources}) \cr 
W_{\rm filter} (\tilde{U}^{\alpha, i} - \tilde{U}^{\alpha}_{\rm sources}) \cr 
\end{pmatrix} \right).
\ea

After the filtering operation we apply an additional mask $W^{\alpha}_{\rm extended}$ that cuts an additional 2$^\circ$ near the edges of $W_{\rm filter}$ to reduce the contamination of pixels near the maps' discontinuities. The final mask for a given array-band is given by $W_{\rm obs}^{\alpha} \equiv W_{\rm  sources}W^{\alpha}_{\rm extended}$; an example for PA6 f150 is shown in Figure \ref{fig:w_obs}.\\

To correct the power spectra for the filter bias, we represent its effect in matrix form, and estimate the different elements using Monte Carlo simulations. The filter is
\ba
{\bm F}_{b}=
\begin{pmatrix} 
F^{TT,TT}_{b}& 
0 &
0 &
0 &
0 &
0 &
\cr
0& 
F^{TE,TE}_{b} &
0 &
0 &
0 &
0 &
\cr
0& 
0 &
F^{TB,TB}_{b} &
0 &
0 &
0 &
\cr
0 & 
0 &
0 &
F^{EE, EE}_{b} &
0 &
F^{BB, EE}_{b} &
\cr
0 & 
0 &
0 &
0 &
F^{EB, EB}_{b} &
0 &
\cr
0   & 
0 &
0 &
F^{EE, BB}_{b} &
0 &
F^{BB, BB}_{b} &
\end{pmatrix}, \nonumber \\
\ea
with
\ba
F^{TT,TT}_{b} &=& \left\langle \frac{C^{TT, {\cal F}}_{b}}{C^{TT}_b} \right\rangle; \
F^{EE,EE}_{b} = \left\langle \frac{C^{EE, {\cal F}, \rm no B}_{b}}{C^{EE, \rm no B}_b} \right\rangle; \
F^{BB, BB}_{b} = \left\langle \frac{C^{BB, {\cal F}, \rm no E}_{b}}{C^{BB, \rm no E}_b} \right\rangle \nonumber \\
F^{EE,BB}_{b} &=& \left\langle \frac{C^{BB, {\cal F}, \rm no B}_{b}}{C^{EE, \rm no B}_b} \right\rangle; \
F^{BB, EE}_{b} = \left\langle \frac{C^{EE, {\cal F}, \rm no E}_{b}}{C^{BB, \rm no E}_b} \right\rangle \nonumber \\
F^{TE,TE}_{b} &=& F^{TB,TB}_{b} = \sqrt{F^{TT,TT}_{b} F^{EE,EE}_{b}}; \
F^{EB,EB}_{b} =  F^{EE,EE}.  
\ea
Here, the upper script ${\cal F}$ denotes quantities estimated after the filtering operation.
The filtering operation can introduce E-to-B leakage. To characterize this,
we use simulations that natively contain no E or B modes that we denote noE and noB respectively. We also note that using $F^{TE,TE}_{b} = \sqrt{F^{TT,TT}_{b} F^{EE,EE}_{b}}$ 
for correcting  $C^{TE}_{\ell}$ leads to a small bias that we correct for using an additive term estimated from simulations. We check that applying these corrections is enough to recover unbiased power spectra in both Gaussian simulations (\S\ref{subsec:sims}) and the Agora simulations \citep{agora}. These include non-Gaussian point-sources for which we make a dedicated source catalog.

A comparison of the matrix elements estimated from simulations with an analytic estimate is shown in the left panel of Figure \ref{fig:kspace_combined}. Here, the analytic estimate of the effect of the filter is obtained by simply counting how many modes are masked by the flat-sky Fourier-space filter in each power spectrum bin: $F_{b}^{\rm analytic} = \int_{\bm{\ell} \in b} F_{\bm{\ell}} d\bm{\ell} / \int_{\bm{\ell} \in b} d\bm{\ell}$. The agreement is within approximately $1\%$, except at the largest angular scales.

In reality, the filter also leads to mode mixing: a mode corresponding to a given angular scale will transform 
into a mixture of modes after Fourier filtering. To quantify this effect, we generate simulations from power spectra that are non-zero
only for a single multipole at $\ell=500$. We then filter these simulations and
compute their resulting power spectra. As shown in the right panel of Figure \ref{fig:kspace_combined}, the mode mixing 
introduced by the filter is minimal, with a correlation length smaller than our typical bin size of $\Delta \ell = 50$.

\subsection{Spectra estimation}

Once each split has been filtered, we apply our final mask $W^{\alpha}_{\rm obs} = W^{\alpha}_{\rm extended} W_{\rm sources}$ 
consisting of the product of apodised edge mask, Galactic mask and point source mask. We then compute the corresponding 
spherical harmonic transform coefficients $\tilde{a}^{X_{\alpha}, i}_{\ell m}$ with $X \in {T,E,B}$ and form pseudo spectra
\ba
\tilde{D}^{X_{\alpha, i} Y_{\beta, j}}_{\ell}= \frac{\ell (\ell + 1)}{2\pi}  \frac{1}{2\ell +1} \sum_{m} \tilde{a}^{X_{\alpha}, i}_{\ell m}  \tilde{a}^{Y_{\beta}, j *}_{\ell m} .
\ea
We  de-bias each pseudo spectrum from the effect of the beam and masking using the standard Master
mode coupling matrix $M_{\ell \ell'}^{X_{\alpha}Y_{\beta}; W_{\mu}Z_{\beta}}$: 
\ba
\hat{D}^{X_{\alpha, i} Y_{\beta, j}}_{\ell} = \sum_{WZ} (M^{-1})_{\ell \ell'}^{X_{\alpha}Y_{\beta}; W_{\alpha}Z_{\beta}} \tilde{D}^{W_{\alpha, i} Z_{\beta, j}}_{\ell'}.
\ea
We then bin each spectrum and correct for the effect of the Fourier filter described above:
\ba
D^{X_{\alpha, i} Y_{\beta, j}}_{b} &=& \sum_{WZ} (F^{-1})^{X_{\alpha}Y_{\beta}; W_{\alpha}Z_{\beta}}_{b} \frac{1}{\ell^{\rm high}_{b}-\ell^{\rm low}_{b}} \sum^{\ell^{\rm high}_{b}}_{\ell = \ell^{\rm low}_{b} } \hat{D}^{W_{\alpha, i} Z_{\beta, j}}_{\ell} \nonumber \\
&=&  \sum_{WZ} (F^{-1})^{X_{\alpha}Y_{\beta}; W_{\alpha}Z_{\beta}}_{b}  \sum_{\ell \in b} P_{b \ell } \hat{D}^{W_{\alpha, i} Z_{\beta, j}}_{\ell}.
\label{eq:cross_split}
\ea
Finally, we form an average of these pseudo power spectra for auto and cross-array band x-spectra:
\ba
D^{X_{\alpha} Y_{\beta}}_{b, \rm cross} &=& \frac{1}{n_{\rm cross}} \sum^{n_{d}}_{i=1} \sum^{n_{d}}_{j=1} D^{X_{\alpha, i} Y_{\beta, j}}_{b} (1 - \delta_{ij}).
\ea
Here $n_{d}=4$ is the number of data splits for each DR6 array-band. $n_{\rm cross} = n_{d}(n_{d}-1)$ is the number 
of cross-split power spectra.
In the particular case of $\alpha = \beta$ and X=Y, there are only $n_{\rm cross} = \frac{n_{d}(n_{d}-1)}{2}$ independent cross-split power spectra
since, for example, $D^{X_{\alpha, 0} X_{\alpha, 1}}_{b} = D^{X_{\alpha, 1} X_{\alpha, 0}}_{b} $.

\section{Null tests}  \label{apx: sys_res}

\subsection{Details of pre-unblinding f090-f150 parameter nulls}

The results for the f090-f150 parameter consistency test are described in \S\ref{sec:params}. In practice, pre-unblinding, we used an alternative way to compute the f090-f150 consistency, by comparing the 5-dimensional data difference with the distribution of the simulations, and calculating the PTE of the data value compared to the average mean and covariance from the suite of simulations. However, to converge on these numbers would require a large suite of $\sim$1000 simulations which we found impractical for full MCMC explorations. Instead we used parameter best-fit estimates for each simulation, but found these also gave unstable results. Broadly, we found that the TE/EE and TT-only PTEs were acceptable but the total TT/TE/EE PTE fell in a range at the sub-\% to a few \% value at most, with a similar trend for \LCDM+$N_{\rm eff}$. Since single-frequency runs are susceptible to unconstrained foregrounds; we tested simplifying the foreground model, which resulted in small changes in the PTE values, raising the total TT/TE/EE agreement to percent level. We deemed this test on the threshold of failing, but not at high significance, so we proceeded to unblind. We later updated the method for testing consistency to that reported in the main text.  

\subsection{Results from residual systematics test}

This test was motivated after unblinding by the borderline f090-f150 parameter consistency described in \S\ref{sec:params} and above, in addition to a noticeable shape in some ACT array-array null spectra in EE (shown in this Appendix), and an apparent residual for the \pact\ best-fit \LCDM\ model in one of the EE spectra, when using the extended cuts in polarization. 

\begin{figure}
    \centering   \includegraphics[width=\linewidth]{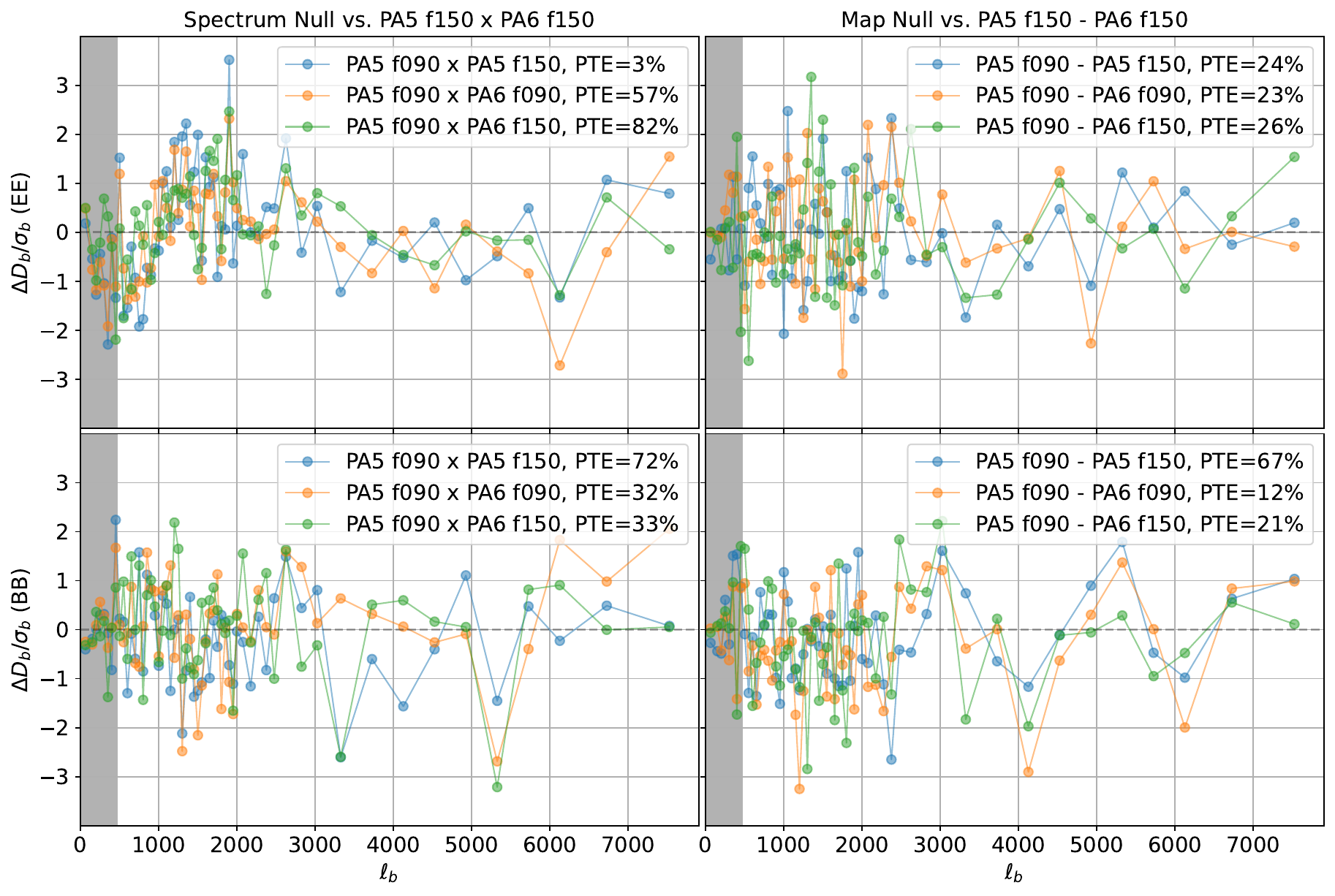}
    \caption{A selection of spectrum-level and map-level null tests in polarization. While the PTEs of these tests are consistent with statistical scatter, a coherent feature is visible for $\ell \lesssim 1500$ in the upper-left panel (spectrum-level EE tests). The feature is not visible in spectrum-level BB, nor any polarization map-level null test, suggesting a possible multiplicative, not additive, systematic. The feature is also not visible in most EE spectrum-level null tests; rather, we have highlighted a few containing only one PA5~f090 leg. The vertical gray band indicates the scales we discard in the ``extended" set of multipole cuts.}
    \label{fig: apx_spec_map_nulls}
\end{figure}

We parameterize a multiplicative correction to individual array-band spectra to quantify the significance of a possible systematic residual. We use this model of a systematic residual only to constrain relative systematics between array-bands, and thus the model can be thought of as a ``parametric" null test. The model preserves the array-averaged ACT spectrum and is nearly independent of cosmological constraints. These relative systematics are not included in our nominal data model, but instead provide a way to test whether additional data cuts are statistically motivated.

We model a general, multiplicative systematic residual for array-band $\alpha$ as the following:
\begin{equation} \label{eq: rel_syst_map}
    \begin{pmatrix}
        \tilde{T_{\ell m}^{\alpha}} \\
        \tilde{E_{\ell m}^{\alpha}}
    \end{pmatrix} = 
    \begin{pmatrix}
        1 + \delta_\ell^{T_\alpha} & 0 \\
        \gamma_\ell^{\alpha} & 1 + \delta_\ell^{E_\alpha}
    \end{pmatrix}
    \begin{pmatrix}
        T_{\ell m}^{\alpha} \\
        E_{\ell m}^{\alpha}
    \end{pmatrix},
\end{equation}
where tildes denote quantities affected by the systematic. Here the $\delta_\ell$ parameters capture beam-like errors and $\gamma_\ell$ a leakage-like error, although this is just an analogy and we do not ascribe such systematics to a mismodeled beam. We ignore the subdominant effect of ``E-to-T" leakage. This model propagates into the theory spectra as:
    \begin{equation} \label{eq: rel_syst}
        \begin{pmatrix}
            \tilde{C_\ell}^{T_\alpha,T_\beta} \\
            \tilde{C_\ell}^{T_\alpha,E_\beta} \\
            \tilde{C_\ell}^{E_\alpha,E_\beta}
        \end{pmatrix} = 
        \begin{pmatrix}
            (1 + \delta_\ell^{T_\alpha})(1 + \delta_\ell^{T_\beta}) & 0 & 0 & 0 \\
            (1 + \delta_\ell^{T_\alpha})\gamma_\ell^{\beta} & (1 + \delta_\ell^{T_\alpha})(1 + \delta_\ell^{E_\beta}) & 0 & 0 \\ 
            \gamma_\ell^{\alpha}\gamma_\ell^{\beta} & \gamma_\ell^{\alpha}(1 + \delta_\ell^{E_\beta}) & (1 + \delta_\ell^{E_\alpha})\gamma_\ell^{\beta} & (1 + \delta_\ell^{E_\alpha})(1 + \delta_\ell^{E_\beta})
        \end{pmatrix}
        \begin{pmatrix}
            C_\ell^{T_\alpha,T_\beta} \\
            C_\ell^{T_\alpha,E_\beta} \\
            C_\ell^{E_\alpha,T_\beta} \\
            C_\ell^{E_\alpha,E_\beta}
        \end{pmatrix}.
    \end{equation}

There are two important features of this parameterization of the systematic residuals. First, as written, this model can capture any multiplicative systematic. For example, the calibration and polarization efficiencies in Equation \ref{eq: lik_cal} could be written in the form of Equation \ref{eq: rel_syst}, with each $\gamma_\ell^\alpha=0$, and $\delta_\ell^{T_\alpha}$ and $\delta_\ell^{E_\alpha}$ constants over multipole. Additional systematic residuals would therefore be at least first-order in $\ell$. Second, to enforce that the array-averaged ACT spectra be unchanged by these residual systematics, we constrain that for each type of systematic and multipole, the sum over array-bands $\alpha$ be zero: for example, $\sum_\alpha\gamma_\ell^\alpha=0$. This is easily achieved so long as the parametrization of the systematics is linear: for example, $\gamma_\ell^\alpha=\sum_i A_{\ell i} \beta_i^\alpha$, where $\beta_i^\alpha$ are a small number of parameters mapped to multipoles via a matrix $A_{\ell i}$. This constraint is then satisfied if $\sum_\alpha\beta_i^\alpha=0$.  

These considerations lead us to choose the following simple parameterization for a possible systematic residual:
\begin{equation} \label{eq: rel_syst_parameterization}
    \delta_\ell^{E_\alpha} =  
    \begin{cases}
        m^{E_\alpha}\ell + b^{E_\alpha} & \ell \leq \ell_{\rm knee}^E \\
        m^{E_\alpha}\ell_{\rm knee}^E + b^{E_\alpha} & \ell > \ell_{\rm knee}^E
    \end{cases},
\end{equation}
and analogously for $\delta_\ell^{T_\alpha}$ and $\gamma_\ell^\alpha$. Here, each array-band systematic has a unique slope and intercept, for scales larger than some $\ell_{\rm knee}$, and is flat at smaller scales where the data is less constraining. Each systematic shares an $\ell_{\rm knee}$ among array-bands. We sample systematic residual parameters either conditional on, or jointly with, the rest of the model, and enforce that $\sum_\alpha m^\alpha=0$ in the sampling.\footnote{In this more general parameterization of systematic residuals, each calibration factor in Equation \ref{eq: lik_cal} is instead captured by the $b$ parameters in Equation \ref{eq: rel_syst_parameterization}. Therefore, we do \textit{not} enforce that $\sum_\alpha b^\alpha=0$; rather, we still fit for absolute $b$ parameters.} Any significantly non-zero $m^\alpha$ would indicate the presence of a relative systematic residual shape between arrays.
\begin{figure}
    \centering
    \includegraphics[width=\linewidth]{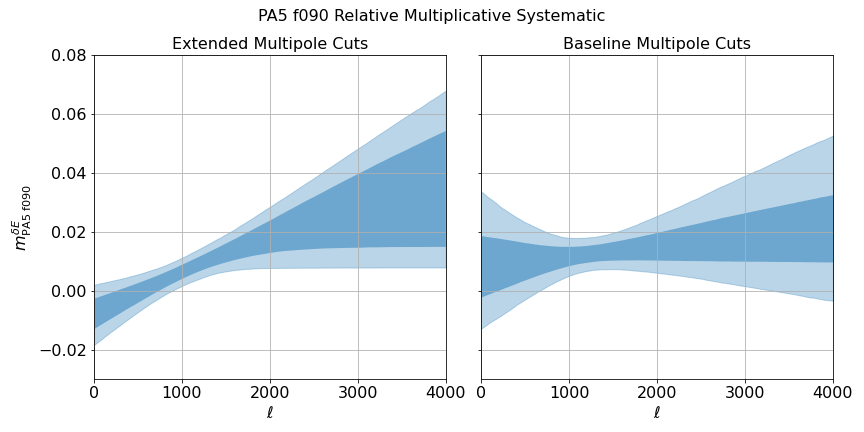}
    \includegraphics[width=0.495\linewidth]{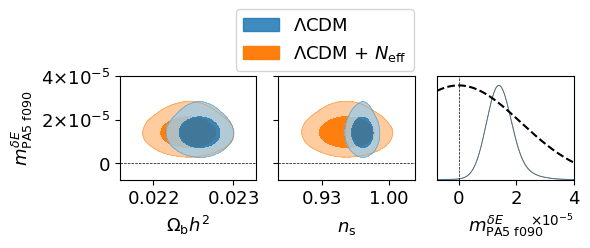}
    \includegraphics[width=0.495\linewidth]{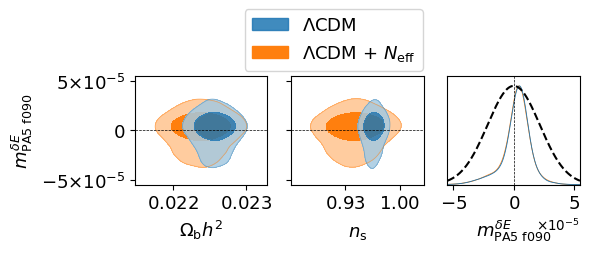}
    \caption{(Top) Posterior of systematic residuals for PA5~f090, using the ``extended" and ``baseline" cuts. Solid (faint) blue shows the $1\sigma$ ($2\sigma$) constraints. (Bottom) Posterior distribution for the PA5~f090 systematic residual, when marginalizing over standard $\Lambda$CDM and $\Lambda$CDM with a free number of relativistic species, $N_{\rm eff}$. We see that a preference for a systematic residual under the ``extended" cuts is mitigated under the ``baseline" cuts, and that these constraints on systematic effects are independent of cosmological parameters and models. The black dashed line is a wide Gaussian prior to aid chain convergence.}
    \label{fig:sys_res_results}
\end{figure}
We deduce that any measurable systematic residuals must be specific to $\delta_\ell^{E_\alpha}$.\footnote{$\delta_\ell^{T_\alpha}$ cannot be constrained independently of the signal model due to the dominant frequency-dependent foreground contributions. A set of initial constraints with this framework for several fixed cosmologies  showed $\gamma_\ell^\alpha$ to be consistent with zero for all array-bands.} Thus, in practice, we fix all $m^{T_\alpha}=0$ (we still marginalize over per-array-band calibrations) and all $\gamma$ parameters to 0, and focus on constraining $\delta_\ell^{E_\alpha}$. This adds four additional free parameters to the model, as described below.

Individual EE null spectra that exhibited coherent features are shown in the upper-left panel of Figure \ref{fig: apx_spec_map_nulls}. The feature is indicative of a multiplicative systematic: it is present in EE but not BB spectrum-level nulls (multiplicative systematics would be proportional to the underlying signal), but is not visible in either EE or BB map-level nulls. On the other hand, an additive polarization systematic would appear equally in EE and BB map-level nulls, but be suppressed in spectrum-level nulls \citep{louis_consistency}. This motivated the multiplicative residual systematic model of Equation \ref{eq: rel_syst_map}.

Aware of the post-unblinding nature of this investigation, it was desirable that quantitative statements on systematic residual significance be independent of cosmology as much as possible. As discussed, this is achieved by ensuring that for a given type of systematic, its parameterization is linear and the sum over array-bands $\alpha$ of its parameters be zero. To good approximation, this condition guarantees independence from any signal model shared by the array-bands to first-order in the systematic. This technique is therefore especially effective for polarization systematics, as foreground components that may vary over frequencies are subdominant to cosmology.

We achieve this constraint by making use of the \textit{centering matrix} $C_n\equiv I_n - (1/n)J_n$, where $I_n$ is the $n\times n$ identity and $J_n$ is the $n\times n$ matrix full of ones. $C_n$ has the property that for any $n$-dimensional vector $\vec\beta_n$, the elements of the vector $C_n\vec\beta_n$ sum to zero: $C_n$ subtracts the mean of the elements of $\vec\beta_n$ from each element of $\vec\beta_n$. Thus, sampling a vector of $N_\alpha$ ``latent" parameters $\beta^{\alpha'}$ and applying $C_{N_\alpha}$ to each sample results in the ``real" parameters $\beta^\alpha$ whose sum over alpha is zero: $\vec\beta_{N_\alpha}\equiv C_{N_\alpha}\vec\beta'_{N_\alpha}$, where $N_\alpha$ is the number of array-bands. However, this may lead to poor sampler convergence as the mean $(1/N_{\alpha})\sum_{\alpha'}\beta^{\alpha'}$ is unconstrained by the data. Instead, we take advantage of the following property of $C_n$: it is positive semi-definite, with its null space spanned by the vector of ones. Thus, it can be diagonalized by the following set of eigenvectors: the vector of ones, and $N_\alpha-1$ vectors orthogonal to the vector of ones. Removing the vector of ones from the set results in the $n\times n-1$ matrix $O_n$ whose orthonormal columns sum to zero by construction. Then, like $C_n$, $O_n$ has the property that the elements of the vector $O_n\vec\beta_{n-1}$ sum to zero for any vector $\vec\beta_{n-1}$. But, unlike for $C_n$, the elements of the latent vector $\vec\beta'_{N_\alpha-1}$ in the following definition $\vec\beta_{N_\alpha}\equiv O_{N_\alpha}\vec\beta'_{N_\alpha-1}$ are no longer degenerate, allowing the sampler to converge. In summary, we constrain the $N_\alpha$ real parameters for a given residual systematic, $\beta^\alpha$, by instead sampling the $N_\alpha-1$ latent parameters, $\beta^{\alpha'}$, and then applying $O_{N_\alpha}$.

Concretely, in terms of Equation \ref{eq: rel_syst_parameterization}, we only perform this procedure for the slope parameters $m^{E_\alpha}$. That is, we sample three new latent systematic parameters that are mapped to real systematic parameters for the four polarization maps included in DR6 via the $4\times3$ matrix $O_4$. Analogous to the null tests presented in \S\ref{subsec:nulls}, we first assume a fixed fiducial cosmological and foreground model. Across several choices of model parameters including best-fit ACT and \Planck\ \LCDM\ cosmologies, as well as setting $N_{\rm eff}=2.87$, we detect with $0.07 - 0.14\%$ PTE ($3.0-3.2\sigma$) a relative $\delta_\ell^E$ systematic between PA5 f090 and the rest of the data. This motivated a new choice of data cuts, made post-unblinding: the ``baseline" in Table \ref{tab:multipole_cut}, where polarization is cut the same as temperature. We show the measured PA5 f090 systematic residual for both data cuts in the top panel of Figure \ref{fig:sys_res_results}; in the ``extended" cuts case, a $\sim1.5\%/1,000\ell$ beam-like slope relative to the mean of the array-bands is visible. For the ``baseline" cuts, we find that all systematic residuals are consistent with zero to within $0.4\sigma$.\footnote{In this case of multiple array-bands $\alpha$ with differing multipole cuts, we ensure that the sum of systematic residuals in each $\ell$-range over only valid array-bands is zero.} The bottom panel of Figure \ref{fig:sys_res_results} shows the same results constrained jointly with cosmological and foreground parameters. In all cases, the systematic constraints are nearly independent from cosmological parameters, as intended. We note the significance of the systematic decreases to $0.24\%$ PTE ($2.8\sigma$) for \LCDM\ and $0.34\%$ PTE ($2.7\sigma$) when also marginalizing over $N_{\rm eff}$. We still assess that a conservative tightening of our multipole cuts is appropriate to mitigate this probable large-scale polarization systematic.

\section{ACT and Planck spectra comparison}
\label{apx:act_planck}
The calibration factors for both the Legacy and NPIPE maps are reported in Table \ref{tab:cal}, representing the number that the uncalibrated ACT maps should be multiplied by to match the \Planck\ amplitude. We also report the numbers chosen for the default calibration amplitude applied to the released data.\footnote{The publicly released DR6 maps, described in N25, are calibrated, i.e. they have already been multiplied by this factor.}

\begin{table*}[htb]
	\centering
	\hspace*{-20mm}\begin{tabular}{rr|rrr|rrr|r}
		\cskip & \cskip & \multicolumn{3}{c}{Legacy} & \multicolumn{3}{c}{NPIPE} & \cskip  \\
		band & arr  &  ${}^{(1)}{\rm cal }$ & ${}^{(2)}{\rm cal }$ & ${}^{(3)}{\rm cal }$  & ${}^{(1)}{\rm cal }$ & ${}^{(2)}{\rm cal }$ & ${}^{(3)}{\rm cal }$  & release  \\
		\hline
f090 & PA5 & 1.0102 $\pm$ 0.0008 & 1.0111 $\pm$ 0.0008 & 1.0118 $\pm$ 0.0009 & 1.0093 $\pm$ 0.0008 & 1.0097 $\pm$ 0.0008 & 1.0103 $\pm$ 0.0009 & 1.0111 \\
f090 & PA6 & 1.0079 $\pm$ 0.0010 & 1.0086 $\pm$ 0.0009 & 1.0088 $\pm$ 0.0010 & 1.0072 $\pm$ 0.0010 & 1.0071 $\pm$ 0.0009 & 1.0071 $\pm$ 0.0010 & 1.0086 \\
f150 & PA5 & 0.9844 $\pm$ 0.0012 & 0.9861 $\pm$ 0.0010 & 0.9877 $\pm$ 0.0011 & 0.9836 $\pm$ 0.0012 & 0.9846 $\pm$ 0.0010 & 0.9859 $\pm$ 0.0010 & 0.9861 \\
f150 & PA6 & 0.9700 $\pm$ 0.0015 & 0.9702 $\pm$ 0.0012 & 0.9696 $\pm$ 0.0012 & 0.9693 $\pm$ 0.0015 & 0.9688 $\pm$ 0.0012 & 0.9683 $\pm$ 0.0012 & 0.9702 \\
f220 & PA4 & 0.9945 $\pm$ 0.0278 & 1.0216 $\pm$ 0.0147 & 1.0478 $\pm$ 0.0065 & 0.9964 $\pm$ 0.0276 & 1.0156 $\pm$ 0.0148 & 1.0346 $\pm$ 0.0062 & 1.0435 \\

	\end{tabular}
	\caption{The calibration factors obtained by comparing ACT and \Planck\ for different array-bands and for different calibration methods.
   $^{(1)}{\rm cal }$ compare the ACT x ACT temperature power spectrum with ACT x Planck, $^{(2)}{\rm cal }$ compare ACT x ACT  and Planck x Planck and $^{(3)}{\rm cal }$ compare ACT x Planck with Planck x Planck. The calibration factors appear stable for the different calibration methods and for the NPIPE and Legacy maps. The final column represent the calibration applied to the maps we release.  }
	\label{tab:cal}
\end{table*}

\begin{figure*}
    \includegraphics[width=0.9\textwidth]{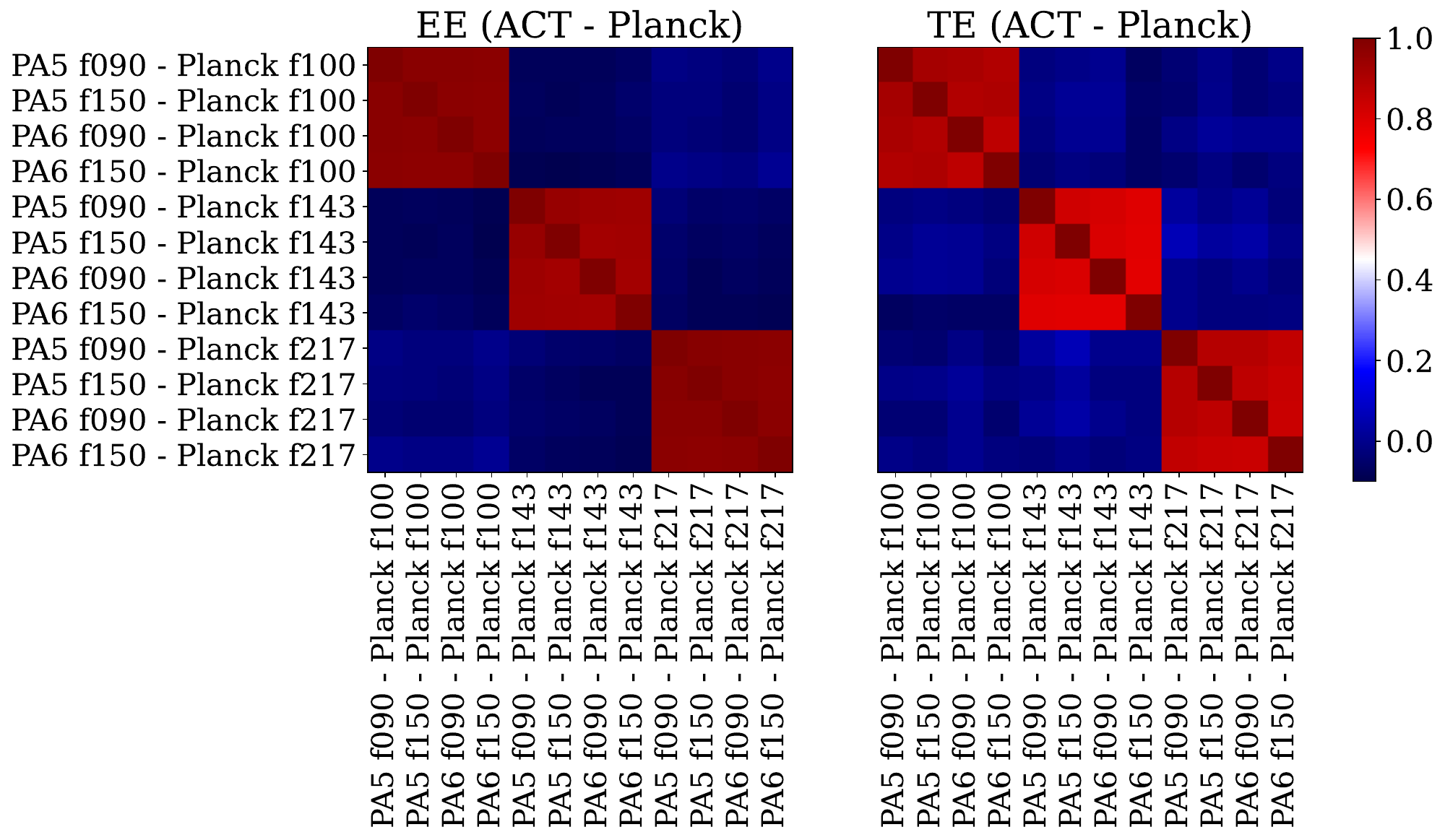}
    \caption{Correlation matrix of the p-value for the comparisons of individual spectra for EE (left) and TE (right) that we perform between ACT and \Planck. The PTE values are reported in Table \ref{tab:p_value_AP}. Each test is dominated by the \Planck\ uncertainties, leading to strong correlations for tests involving comparisons of ACT with the same \Planck\ maps.}
    \label{fig:PTE_corr}
\end{figure*}

 \begin{figure*}[htp]
	\centering
	\hspace*{-5mm}\includegraphics[width=\textwidth]{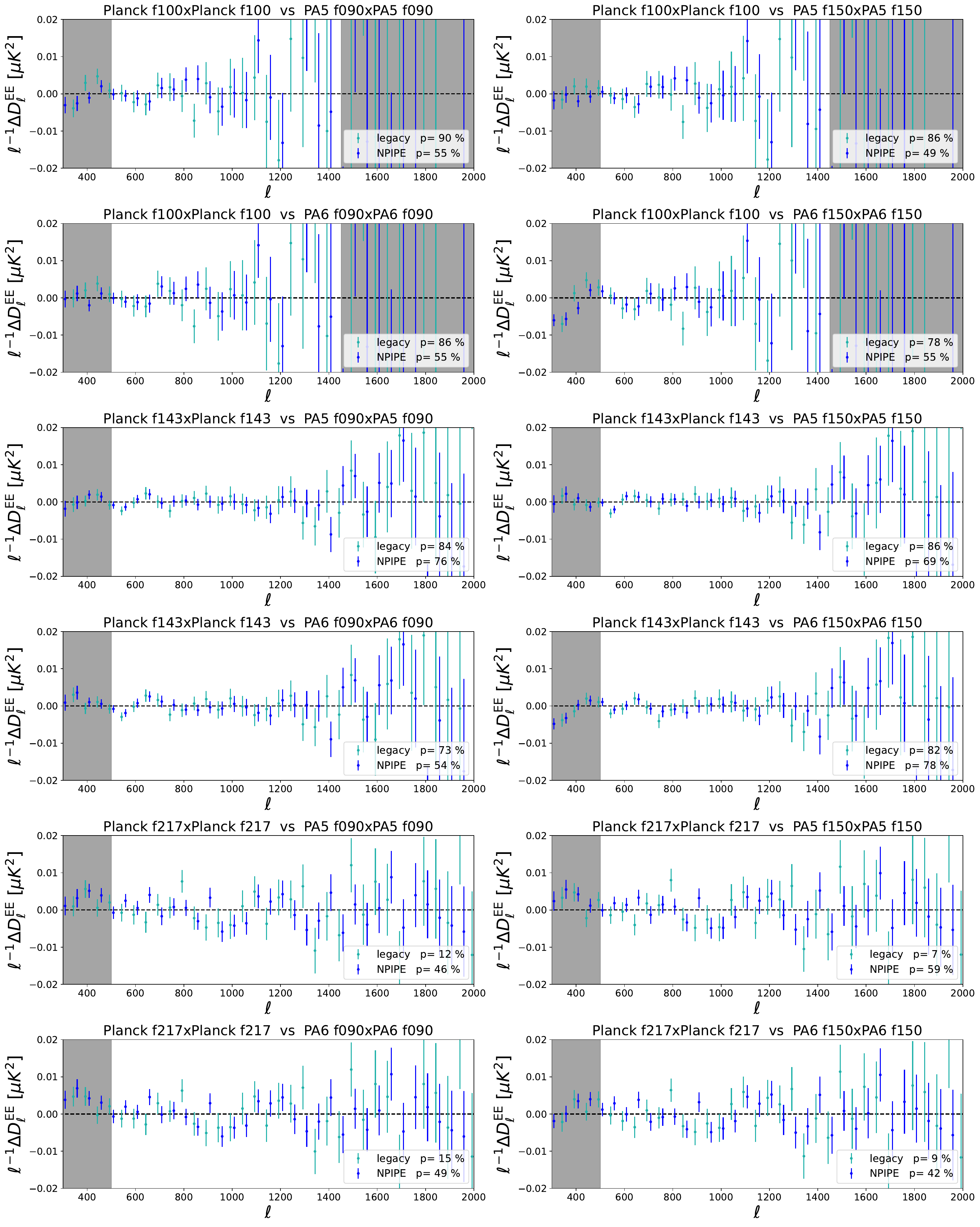}
	\caption{The per-array comparison of ACT and \Planck\ EE spectra on the common sky area, for both the PR3 ``Legacy" maps and the NPIPE maps.}
	\label{fig:EE_ACT_vs_Planck}
\end{figure*}

\begin{figure*}[htp]
	\centering
	\hspace*{-5mm}\includegraphics[width=\textwidth]{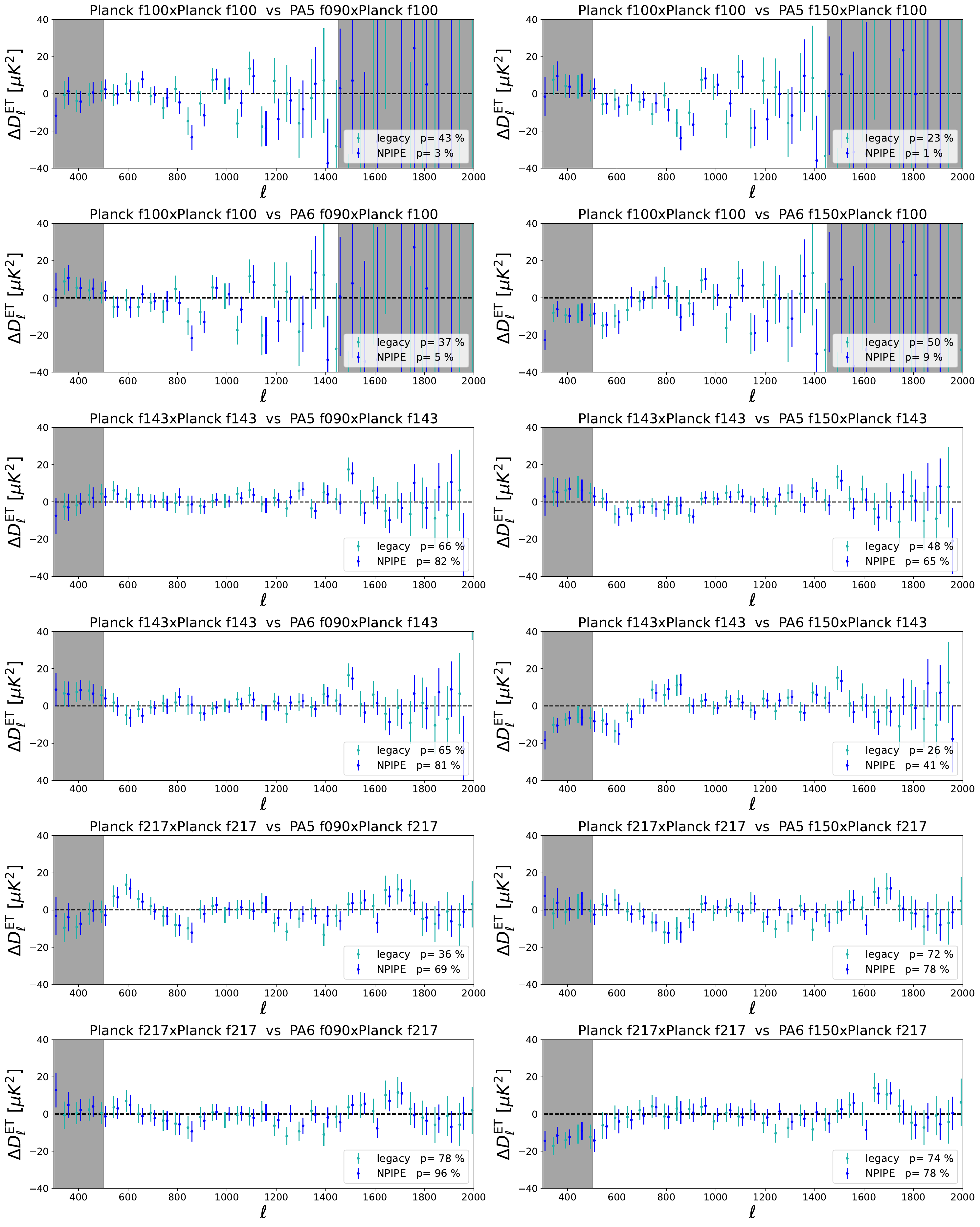}
	\caption{As in Figure \ref{fig:EE_ACT_vs_Planck} but for ET, where the T is from \Planck. There is generally good agreement, with poorer agreement at the 1-5\% PTE level between ACT and the Planck-f100 NPIPE maps.}
	\label{fig:ET_ACT_vs_Planck}
\end{figure*}

The summary statistics from comparisons of individual spectra for EE and TE for ACT and \Planck, computed on the common sky area, were reported in Table \ref{tab:p_value_AP}. The estimated dust foreground level was subtracted before comparing. Figure \ref{fig:PTE_corr}
 shows the degree of correlation between this suite of null spectra; comparisons of ACT with the same \Planck\ map are strongly correlated.
 
In Figures \ref{fig:EE_ACT_vs_Planck}
 and \ref{fig:ET_ACT_vs_Planck} we show the comparisons of individual spectra. The unshaded regions in the figure indicate the ranges used to compare the data; we used the extended range polarization cuts for this exercise, with $\ell_{\rm min}=500$. For EE we find good agreement for all the spectra, with the most sensitivity from Planck-f143. For ET, where the T is from \Planck, there is good agreement with the PR3 Legacy maps and with the f143 and f217 NPIPE maps, but poorer agreement between ACT and the Planck-f100 NPIPE maps.

\section{CMB-only bandpowers and tests}\label{apx:cmbonly}

\begin{figure}[t]
	\centering
	\includegraphics[width=\linewidth]{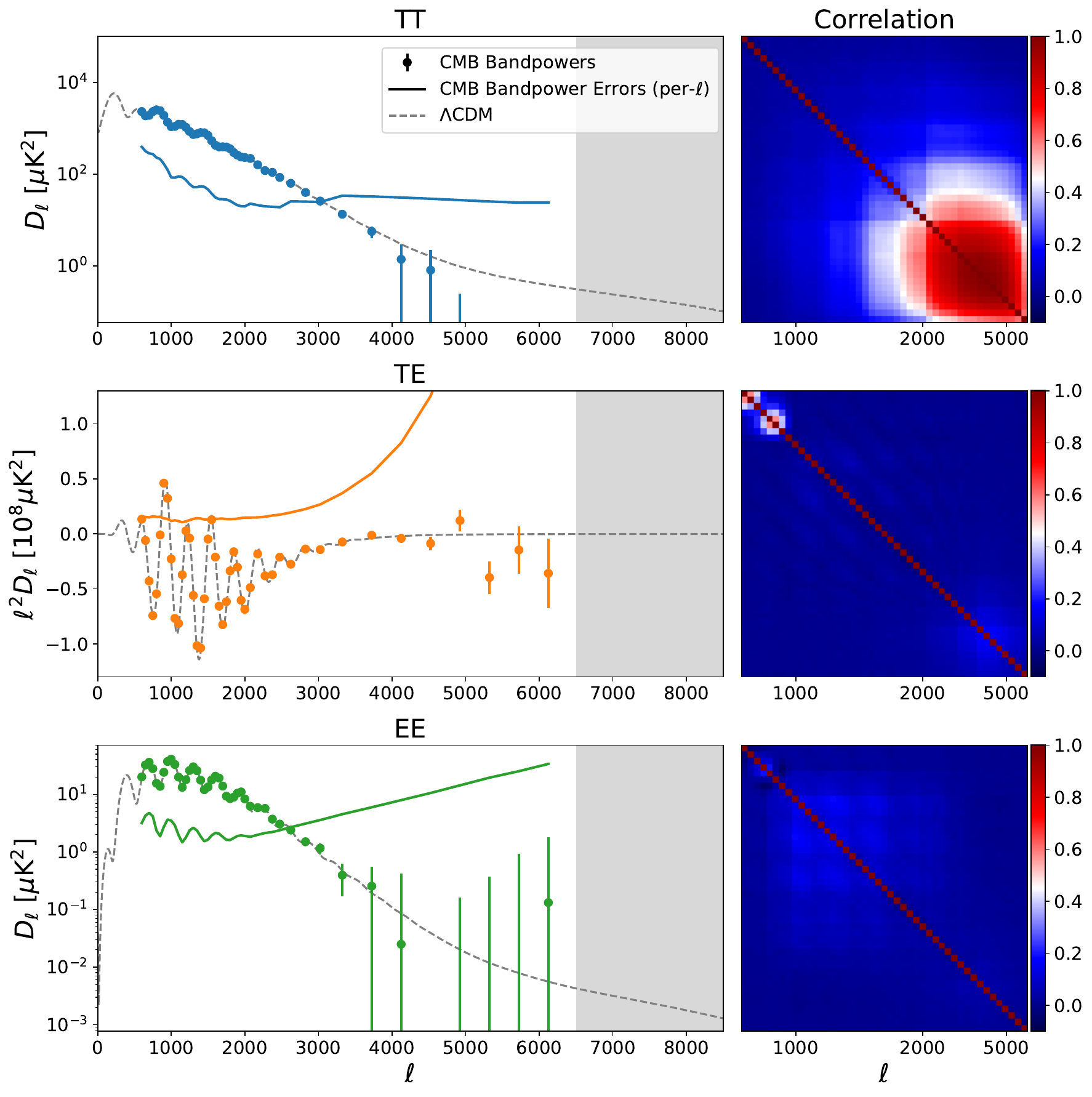}
	\caption{A comparison of the extracted CMB bandpowers (data points), the errors on the bandpowers (per multipole, solid line), and the best-fitting \LCDM\ power spectrum  (gray dashed line).  We do not extract CMB bandpowers, but do marginalize over foregrounds, in the gray region ($\ell > 6500$). Also shown is the correlation matrix for each set of bandpowers: TT is strongly correlated at high $\ell$ due to the foreground marginalization, while TE and EE are more uncorrelated.}
	\label{fig:cmb_ext}
\end{figure}

\par This Appendix provides more details about the CMB-only likelihood, \texttt{ACT-lite}, described in \S\ref{subsec:cmb-only-likelihood}, which is a compressed version of the multi-frequency likelihood, \texttt{MFLike}, described in \S\ref{sec:likelihood}. It includes a data vector for the lensed CMB bandpowers (referred to as ``CMB bandpowers") that has been pre-marginalized over the foregrounds including secondary anisotropies from kSZ and tSZ, and most systematic parameters. 

In previous ACT analyses, the CMB bandpowers were extracted to a maximum multipole of $\ell=4000$ \citep{dunkley/etal:2013,choi_atacama_2020}. In this DR6 analysis we extend the $\ell$ range of the estimated CMB to $6500$, as its signal is now non-negligible compared to the noise in the $4000<\ell<6500$ angular range. We use the full data (to $\ell=8500$) to marginalize over foreground components, including the kSZ. The foreground marginalization for the TT spectrum results in strongly-correlated bins above $\ell>2000$, which explains the negative, small-scale TT bandpower residuals in Figure \ref{fig:cmb_ext}; the TE and EE bandpowers remain minimally correlated at high $\ell$.

\begin{figure*}
    \centering
    \includegraphics[width=0.9\textwidth]{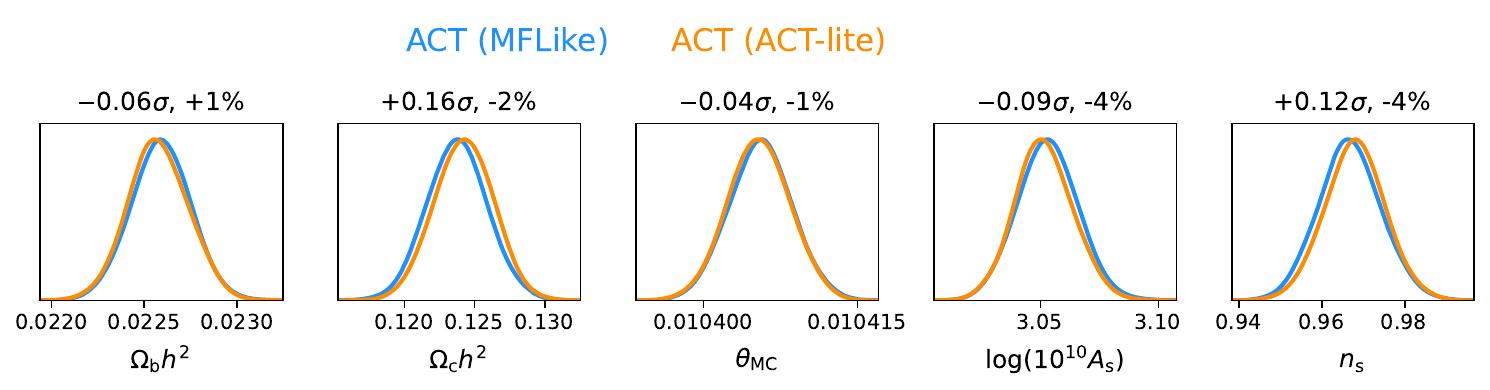}
    \includegraphics[width=0.9\textwidth]{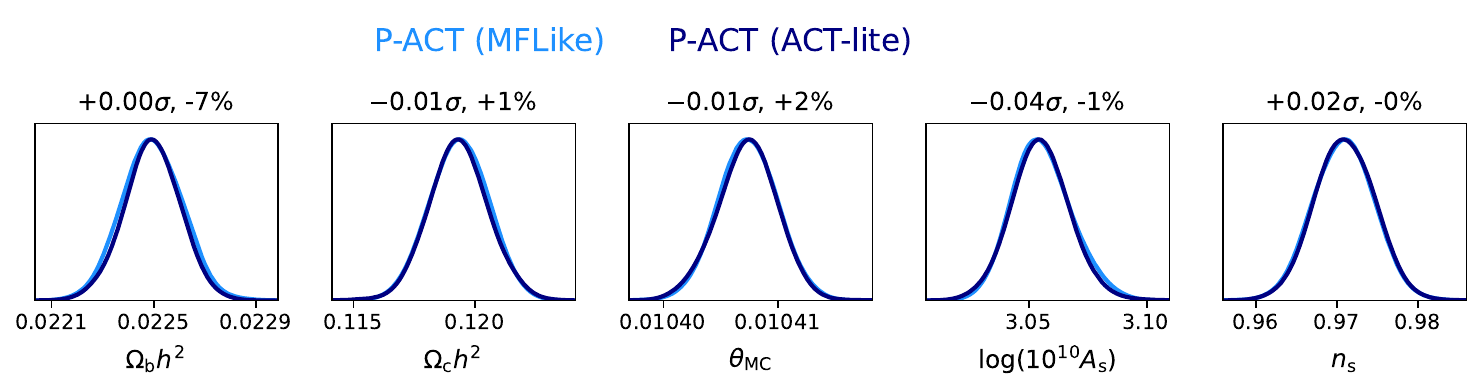}
    \includegraphics[width=0.5\textwidth]{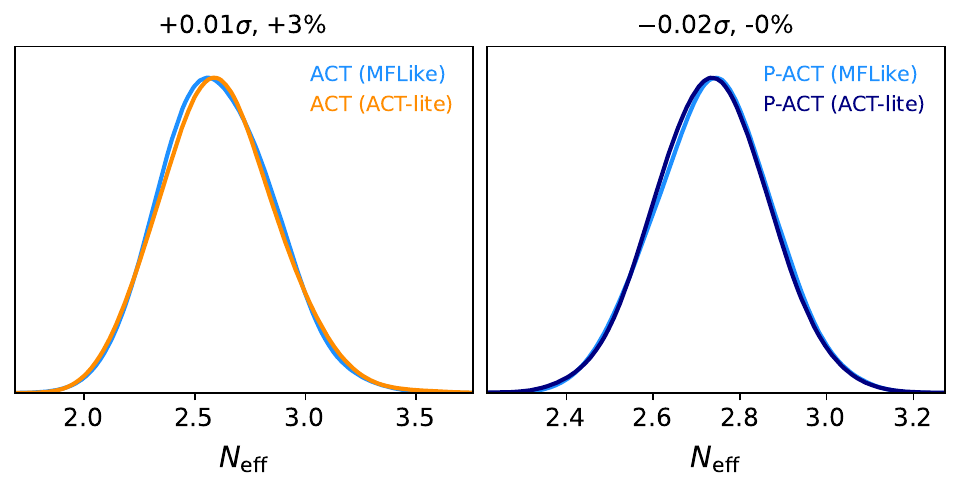}
    \caption{Comparisons of the constraints on \LCDM\ parameters estimated from the \act-alone (top row) and \pact\ (middle row) data combinations, using the full multi-frequency likelihood (\texttt{MFLike}) versus the compressed CMB-only likelihood (\texttt{ACT-lite}). Shifts between parameters are shown in terms of fraction of $\sigma$, and most are less than $0.1\sigma$. The changes in width of the distributions are noted as percentages. (Bottom) Comparison of the constraints on $N_{\rm eff}$ when adding this parameter to \LCDM\ model, for these same likelihoods. }
    \label{fig:baseline-cmbonly-comparison-lcdm}
\end{figure*}

As shown in Figure~\ref{fig:baseline-cmbonly-comparison-lcdm}, and similarly to previous applications of this method to ACT, SPT and \Planck\ likelihoods~\citep{dunkley/etal:2013,Calabrese:2017ypx,choi_atacama_2020,planck2015-cosmo}, cosmological parameters recovered from the full \texttt{MFLike} likelihood and the \texttt{ACT-lite} likelihood agree within $0.1\sigma$ for both $\Lambda$CDM and extended models, and for both ACT-alone or in combination with \emph{Planck}.  

Some of the constraints from TT, TE or EE individually will differ slightly between the two likelihoods because we perform the CMB-only marginalization only once to extract the full set of bandpowers and foreground/nuisance parameters. Since some of the foreground and calibration parameters are common between T and E, the CMB EE bandpowers are better constrained in the CMB-only likelihood compared to the full likelihood, since the T block of data is included while marginalising over foreground and calibration parameters.

\section{Further details of parameter constraints}
\label{apx:params}

\subsection{Parameter tables}
\label{apx:param_tables}

We report all of the estimated parameters, and a set of derived cosmological parameters, in Table \ref{tab:my_label}. These use the multi-frequency likelihood; cosmological parameters estimated with the CMB-only likelihood differ by less than $0.1\sigma$. The values of the parameters for the best-fitting models, found using the maximum a posteriori (MAP) point as described in C25, are reported on LAMBDA.

\begin{table}[]
    \centering
\includegraphics[width=\textwidth]{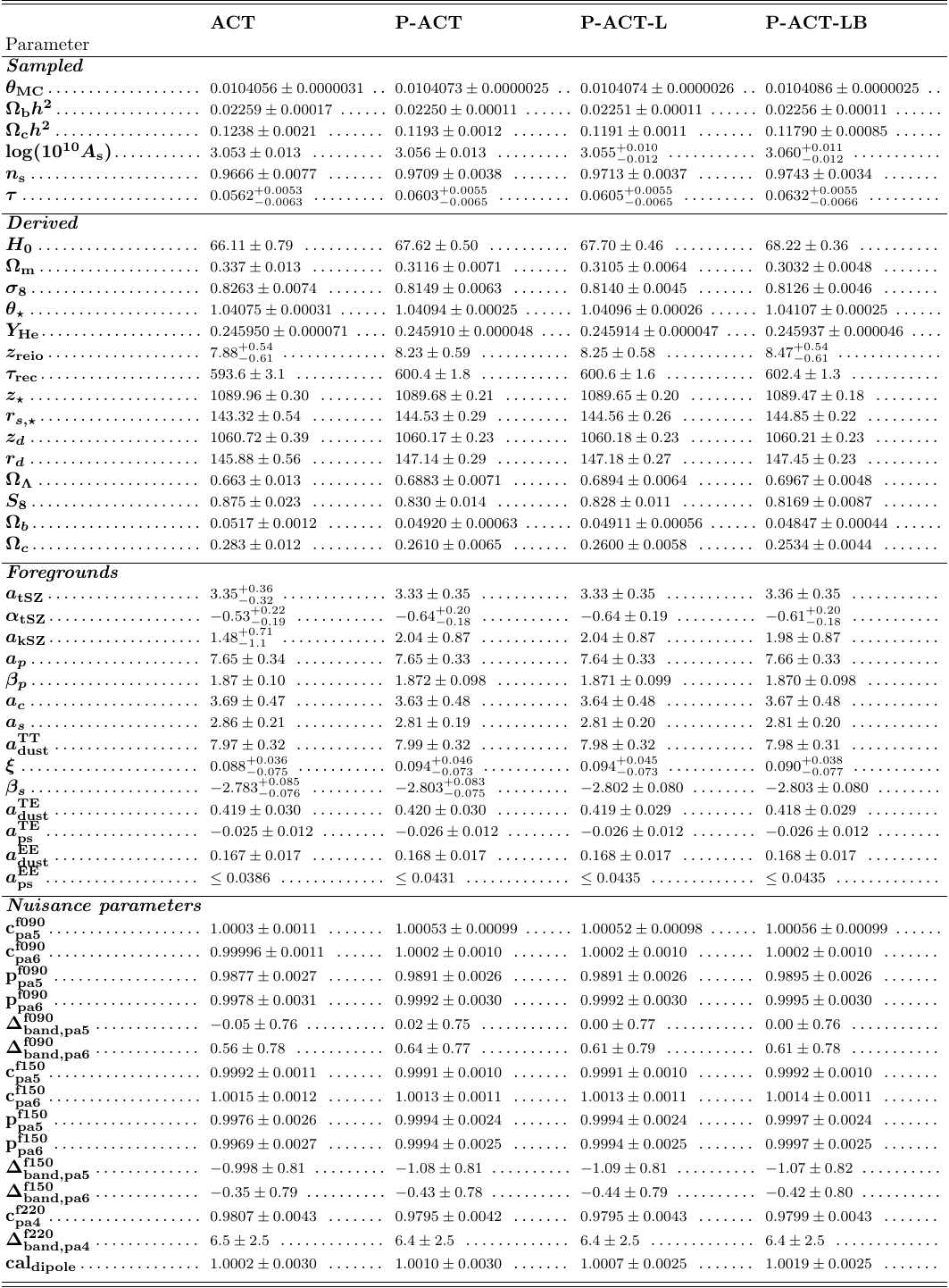}
    \caption{Estimated cosmological, foreground and systematic parameters for different data combinations.}
    \label{tab:my_label}
\end{table}

\begin{table*}[tbp]
    \centering
    	\begin{tabular}{l|l}
		\hline\hline
		\textbf{Parameter} & \textbf{Description} \\
		\hline
        \boldmath$\theta_{\rm MC}$ & \texttt{CosmoMC} approximation to the angular size of sound horizon at last scattering\\
        \boldmath$\Omega_b$ & Baryon density today ($z=0$) relative to the critical density\\
        \boldmath$\Omega_c$ & Cold dark matter density today ($z=0$) relative to the critical density\\
        \boldmath$A_s$ & Amplitude of the scalar primordial fluctuations power spectrum at $k_{\rm pivot}=0.05\;{\rm Mpc}^{-1}$\\
        \boldmath$n_s$ & Power law index of the scalar primordial fluctuations power spectrum\\
        \boldmath$\tau_{\rm reio}$ & Reionization optical depth\\
        \boldmath$H_0$ & Expansion rate today in $\rm km/s/Mpc$\\
        \boldmath$h$ & Unitless expansion rate defined as $H_0 = 100h\,{\rm km/s/Mpc}$ \\
        \boldmath$\Omega_m$ & Total matter density today ($z=0$) relative to the critical density\\
        \boldmath$\Omega_\Lambda$ & Dark energy density today ($z=0$) relative to the critical density\\
        \boldmath$\sigma_8$ & RMS matter fluctuations today in linear theory\\
        \boldmath$S_8$ & Defined as $S_8 = \sigma_8 \left(\Omega_m / 0.3\right)^{0.5}$\\
        \boldmath$\rm Age$ & Age of the Universe\\
        \boldmath$\theta_\star$ & Angular size of sound horizon at last scattering\\
        \boldmath$Y_{\rm He}$ & Fraction of Helium relative to baryonic matter\\
        \boldmath$z_{\rm reio}$ & Redshift at which the Universe is half reionized\\
        \boldmath$\tau_{\rm rec}$ & Conformal time at the end of recombination (in Mpc)\\
        \boldmath$z_\star$ & Redshift at which the optical depth equals unity\\
        \boldmath$r_{s,\star}$ & Comoving size of the sound horizon at $z=z_\star$\\
        \boldmath$z_d$ & Redshift at which baryon-drag optical depth equals unity\\
        \boldmath$r_d$ & Comoving size of sound horizon at $z=z_d$\\
        \boldmath$\eta_b$ & Baryon to photon ratio\\
		\hline\hline
\end{tabular}
	\caption{Cosmological parameter descriptions.} 
	\label{tab:cosmo_description}
\end{table*}

\begin{figure*}
    \centering
    \includegraphics[width=\textwidth]{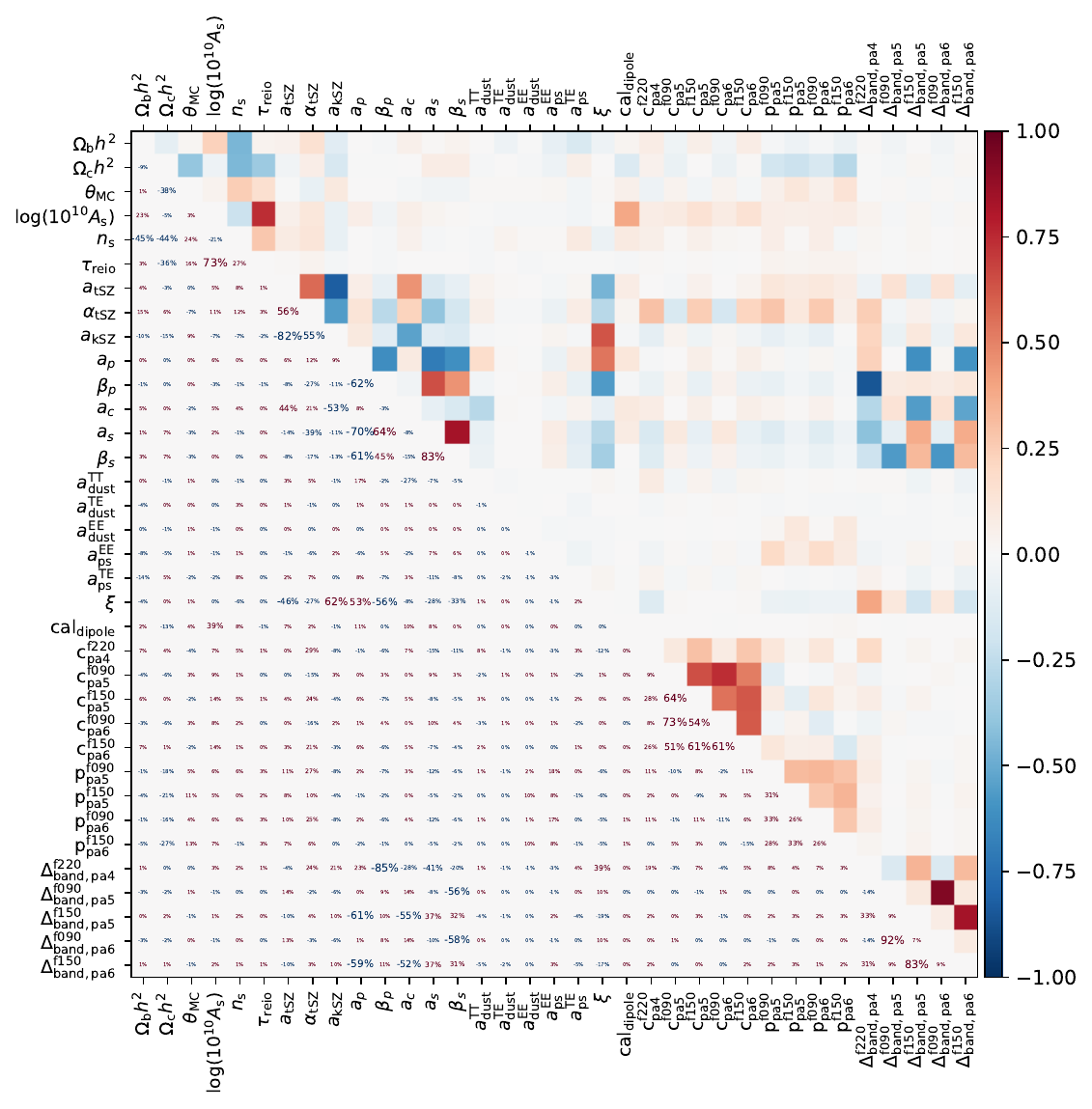}
    \caption{Parameter correlation matrix for the ACT data, combined with the \Planck\ Sroll2 low$E$ likelihood. The first block contains the six \LCDM\ cosmological parameters. This is followed by 14 foreground parameters, 10 calibration parameters, and five bandpass shift parameters. The matrix is symmetric; the lower left half shows the values of the estimated correlation coefficients, and the upper right displays them using the color scale.} 
     \label{fig:corr_matrix}
\end{figure*}

\subsection{Parameter correlations and comparison to Planck}

In Figure \ref{fig:corr_matrix} we show the full set of correlations between the cosmology, foreground and calibration parameters for the \LCDM\ model, as discussed in \S\ref{sec:results}. The lower left half of the matrix shows the values of the estimated correlation
coefficients, and the upper right displays them using the color scale. 

\begin{figure*}
\centering
\includegraphics[width=0.9\textwidth]{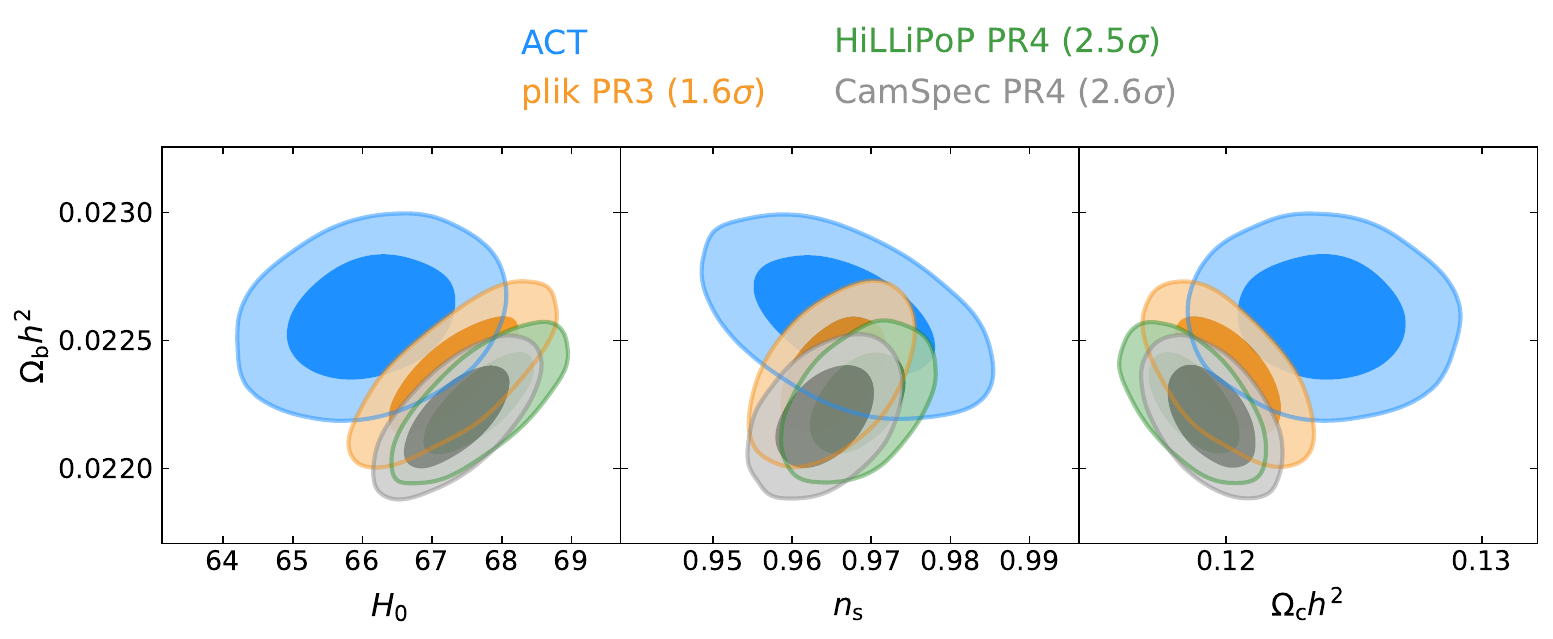}
\includegraphics[width=0.8\textwidth]{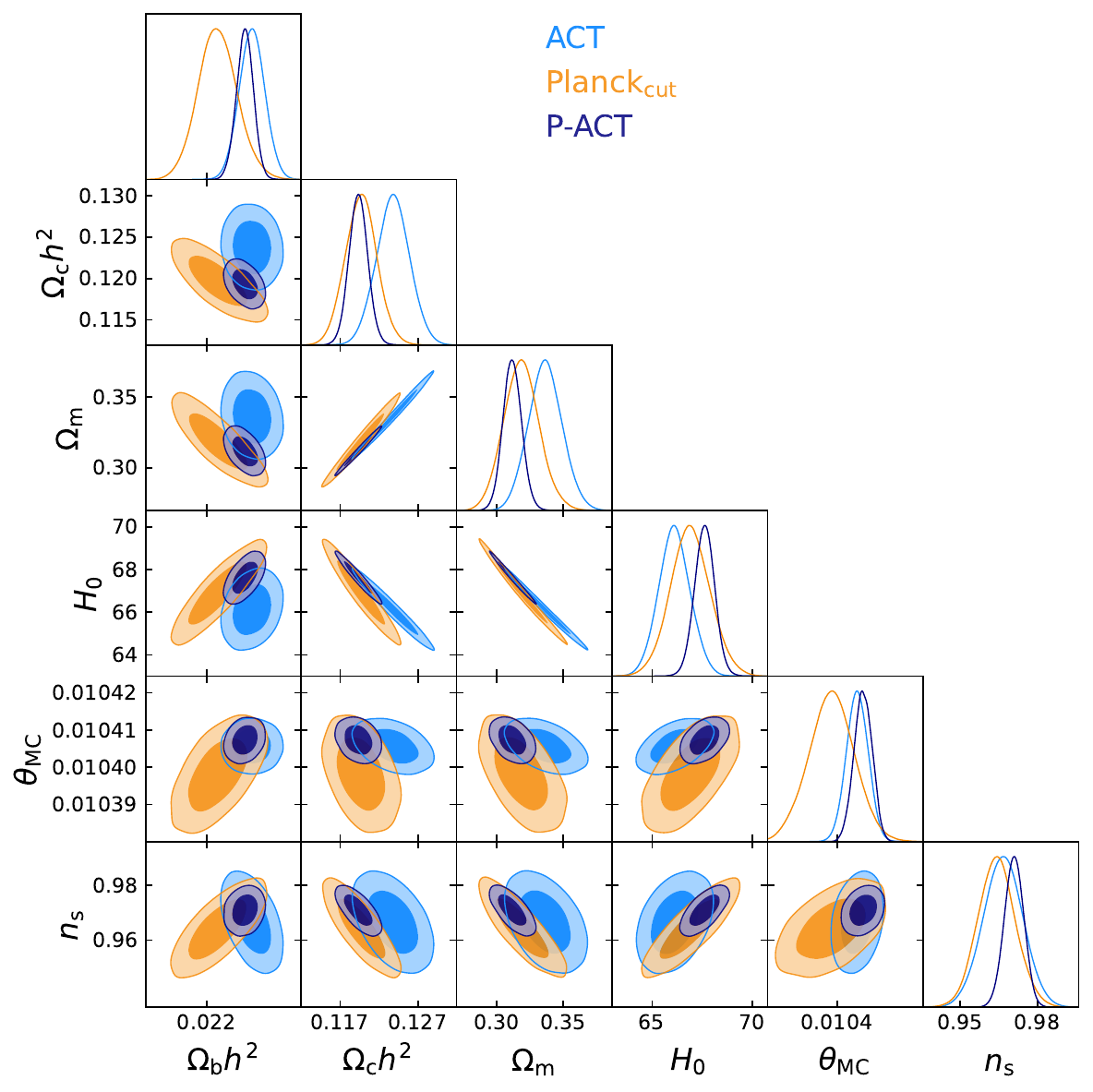}
     \caption{(Top) Subset of parameter distributions for ACT compared to \Planck, for three different \Planck\ likelihoods. In this case, to get PTE we use a 4 dimensional parameter space since $A_se^{-2\tau}$ is not constrained from ACT.  (Bottom)  Parameter distributions for ACT and the restricted-range Planck$_{\rm cut}$ (used in P-ACT), showing how the two datasets contribute to the joint constraints.}
    \label{fig:dr6_vs_plancks}
\end{figure*}

In Figure \ref{fig:dr6_vs_plancks} we show the comparison of the ACT DR6 results with those from different versions of the \Planck\ likelihoods, as discussed in \S\ref{sec:results}. The agreement between ACT and \Planck\ is closest for the Plik PR3 at $1.6\sigma$, neglecting correlations between the data and using the four-dimensional parameter distribution that discards the amplitude and optical depth; the PR4 analyses for both Camspec and Hillipop have small shifts to lower baryon and CDM densities compared to PR3, and result in an overall $2.6\sigma$ separation in the four-dimensional parameter space. 

\jd{In Figure \ref{fig:dr6_vs_plancks} we also show the comparison of parameters estimated from ACT and from the restricted-range Planck$_{\rm cut}$ PR3 data that we use in the joint P-ACT likelihood, illustrating the consistency (at the 2.1$\sigma$ level) and showing how the two datasets combine to break degeneracies.}

%

\subsection{Effect of polarization efficiency}\label{apx:poleff}

\begin{figure}
    \centering
    \includegraphics[width=\textwidth]{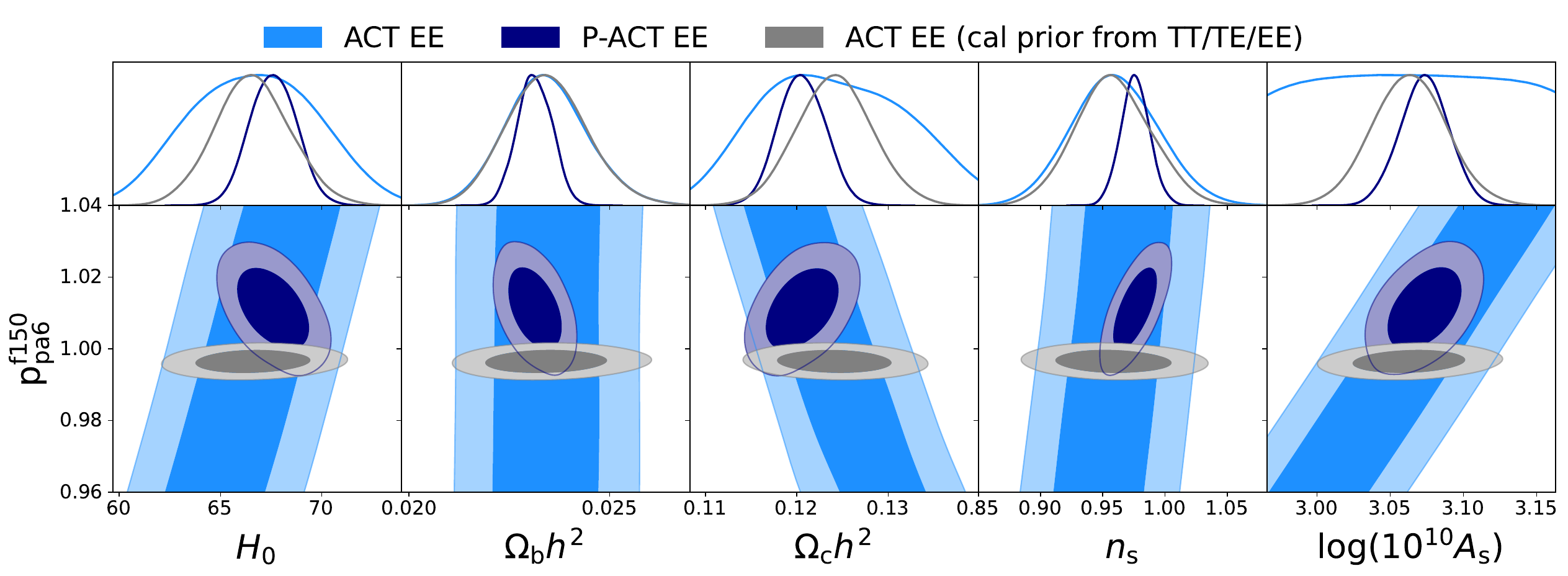}
    \caption{Marginalised posterior distributions of sampled parameters from EE, including the polarization efficiency for PA6~f150. We show constraints from \act\ only (including the Sroll2 data to measure optical depth, light blue), \pact\ (dark blue) and \act\ when using calibration and polarization efficiency priors from a full TT/TE/EE run (gray).}
    \label{fig:polar_eff_EE}
\end{figure}

While producing cosmological constraints using only the EE spectrum, we observed a strong dependency of the inferred values of $\Omega_{c}h^{2}$, and the derived $H_{0}$, on the choice of polarization efficiency calibration. The results varied depending on whether we marginalized over our broad 20\% prior or applied a Gaussian prior informed by our $\Lambda$CDM TT/TE/EE results. In the latter case, the polarization efficiency is tightly constrained because the $\Lambda$CDM model enforces a strong relationship between the amplitudes of temperature anisotropies and E-mode polarization.
The degeneracy between $\Omega_{c}h^{2}$ and the polarization efficiency parameter is illustrated in Figure \ref{fig:polar_eff_EE}. As described in \S\ref{sec:results}, this correlation arises for modes that entered the horizon during radiation domination, corresponding to scales measured by ACT. 

The figure also shows the P-ACT results for EE-only, where the degeneracy is lifted. \Planck\ does not marginalize over polarization efficiencies by default,  instead fixing them based on measurements of the temperature power spectrum and assuming the $\Lambda$CDM model (see Equation 45 of \cite{2020A&A...641A...5P}). We therefore additionally test the case where we allow the \Planck\ polarization efficiency to vary, and find that this does not broaden the P-ACT EE distributions, indicating that the degeneracy with polarization efficiency is being broken by using the broader angular range compared to ACT alone.

The default of fixing \Planck\ polarization efficiencies to \LCDM-derived values could have  implications for the recent SPT-3G results \citep{2024arXiv241106000G}. In their analysis, SPT-3G calibrated the amplitude of the E-modes by cross-correlating SPT-3G polarization maps with those from \Planck\ (see Appendix B of \cite{2024arXiv241106000G}). Calibrating E-modes to \Planck\ might introduce an unintended correlation between the cosmological results of otherwise independent experiments.

More generally, for extensions to $\Lambda$CDM, not marginalizing over instrumental uncertainties on the polarization efficiency could artificially disfavor models that predict a polarization fraction differing from that of $\Lambda$CDM. In C25 this is tested in certain cases by allowing the \Planck\ polarization efficiency to vary.

\subsection{Updated compilation of recent CMB data}
\label{apx:compilation}
\begin{figure}
\centering  \includegraphics[width=\columnwidth]{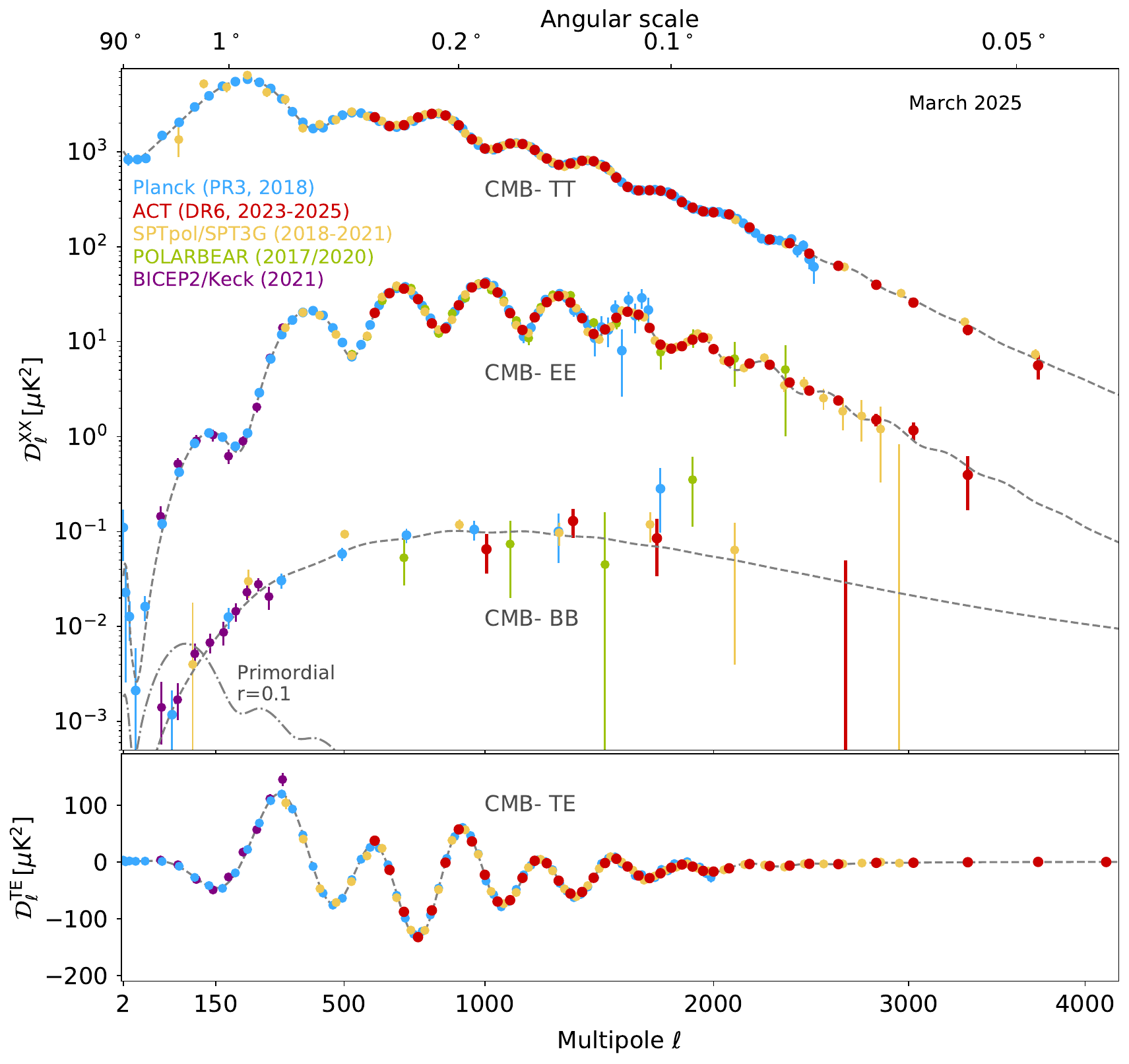}
    \caption{Updated version of Figure 25 from \cite{choi_atacama_2020}, showing the new ACT DR6 spectra compared to recent CMB power spectrum measurements.  The dashed line shows the P-ACT best-fit model; a primordial BB signal with $r = 0.1$ is indicated in dot-dashed. We show the PR3 2018 results for \Planck\ \citep{planck_spectra:2019}, which are visually similar to those in \cite{rosenberg:2022,hillipop2024} from NPIPE data. SPT data are from \cite{henning/etal:2018} for 150 GHz TT $\ell < 2000$, \cite{reichardt/etal:2020} for $\ell>2000$, \cite{dutcher2021} for TE and EE, and \cite{sayre/etal:2019} for BB. The POLARBEAR data are from \cite{polarbear:2017,adachi2020}. The BICEP2/Keck data are from \cite{2021PhRvL.127o1301A}.}
    \label{fig:compilation}
\end{figure}

In Figure~\ref{fig:compilation} we show an updated version of Figure 25 from \cite{choi_atacama_2020}, illustrating the broad consistency of ACT, SPT and Planck data for TT, TE and EE, and including recent polarization data from BICEP/Keck and POLARBEAR.

\subsection{Illustration of the EE sensitivity to the Hubble constant}
\begin{figure}
\centering
    \includegraphics[width=0.75\columnwidth]{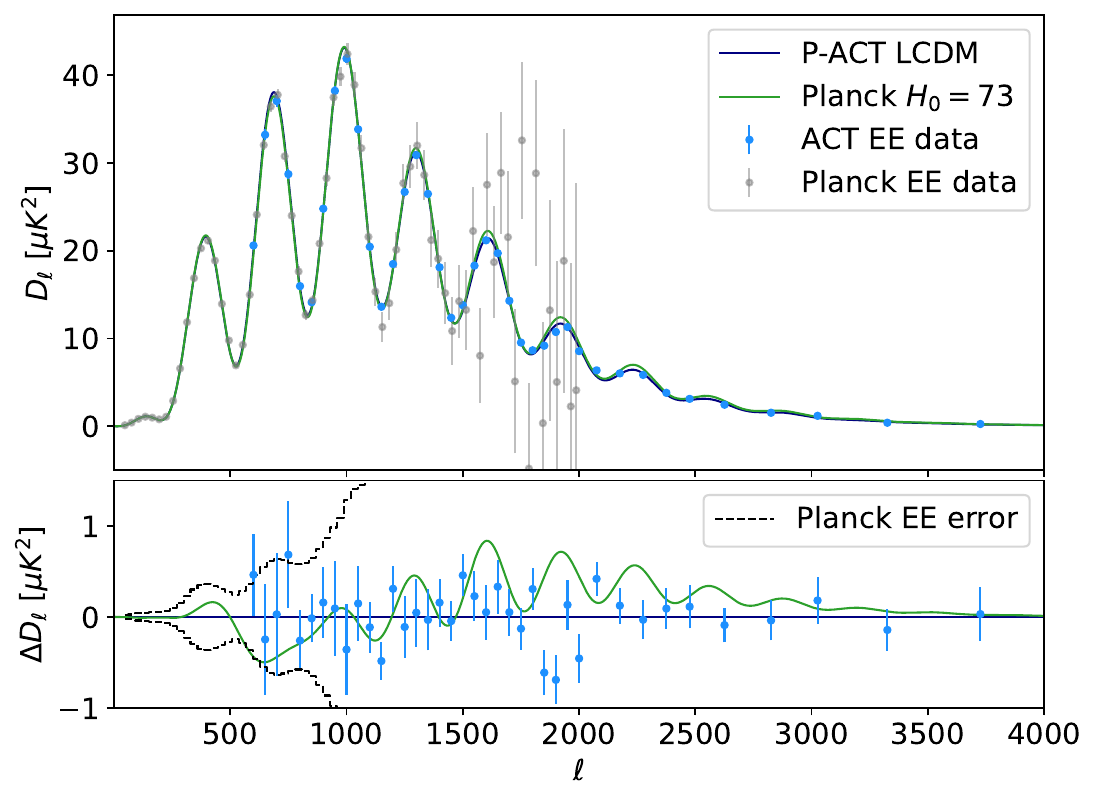}
    \caption{Illustration of how a model with $H_0=73$~km/s/Mpc fits the \Planck\ EE data but overpredicts the smaller-scale ACT EE data. The consistent estimate of the derived Hubble constant from CMB polarization data represents a new strengthening of the \LCDM\ model.}
    \label{fig:ee_H0}
\end{figure}

Following a similar approach to \cite{aiola/etal:2020}, Figure~\ref{fig:ee_H0} shows how the ACT EE data rule out a \LCDM\ model with a Hubble constant of 73~km/s/Mpc at $>4\sigma$. The model adjusts the matter density and can fit the larger-scale Planck EE data, but overpredicts the ACT EE power spectrum.

\subsection{Relative slopes in power spectra}\label{apx:slopes}

\begin{figure}
    \centering
    \includegraphics[width=0.9\textwidth]{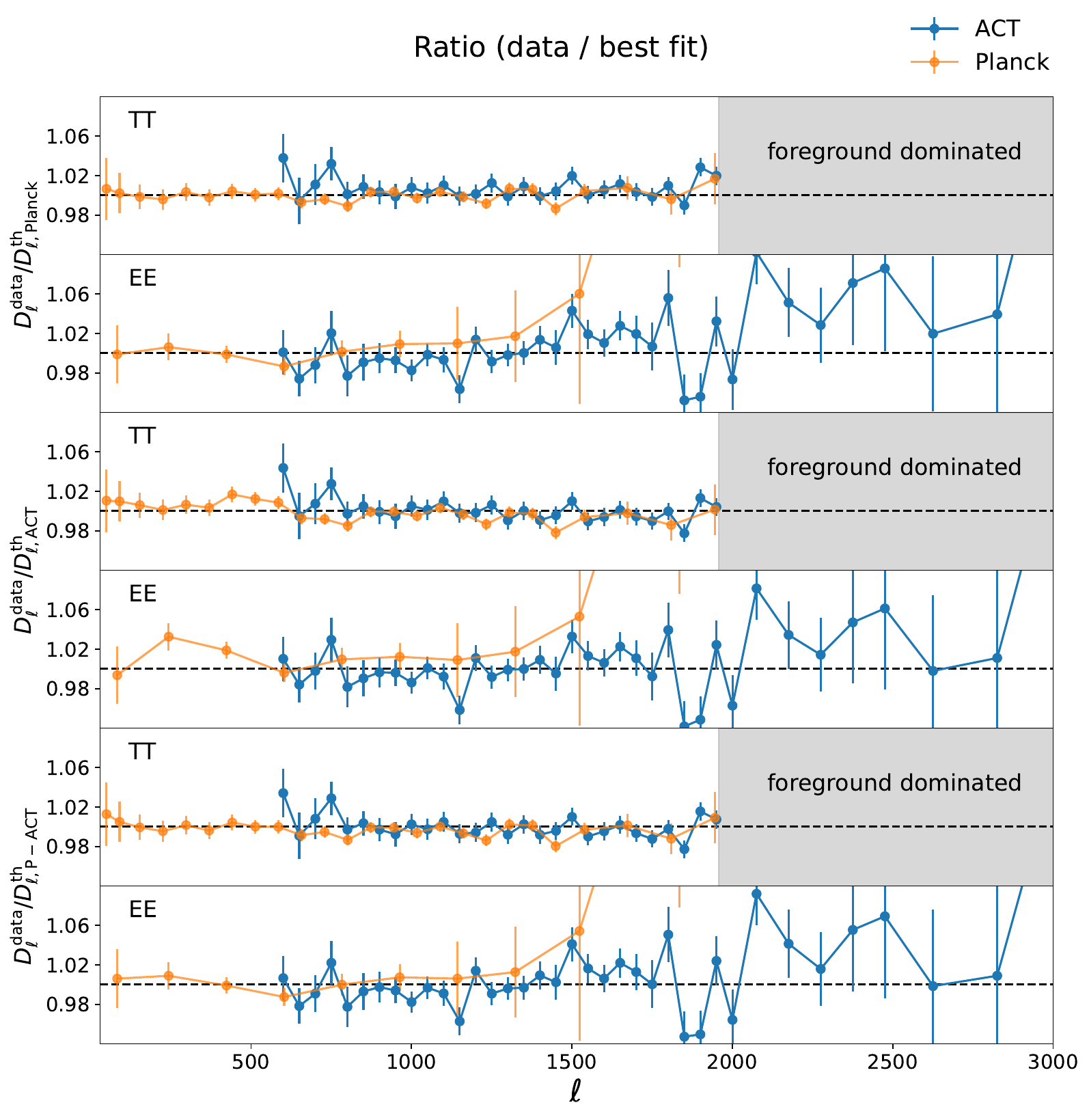}
    \caption{Ratio of the ACT and \Planck\ temperature and E-mode power spectra with respect to the best-fit models from \Planck\ (Table 2 of \cite{planck2018_cosmo}, TT,TE,EE + lowE), ACT, and the combined Planck+ACT (P-ACT) analysis. We do not plot temperature data above $\ell=2000$ due to the strong bin-to-bin correlations introduced by the foreground marginalization (see Figure \ref{fig:cmb_ext}). \Planck\ and ACT exhibit good agreement on overlapping scales. However, the \Planck\ best fit, when extrapolated to small scales, mildly underestimates the small-scale power measured by ACT, while the ACT best fit, when extrapolated to large scales, underestimates the large-scale power measured by \Planck. In contrast, the P-ACT best-fit model is a good fit to both datasets across all scales.}
    \label{fig:ratio_bf}
\end{figure}

\begin{figure}
    \centering
    \includegraphics[width=0.7\columnwidth]{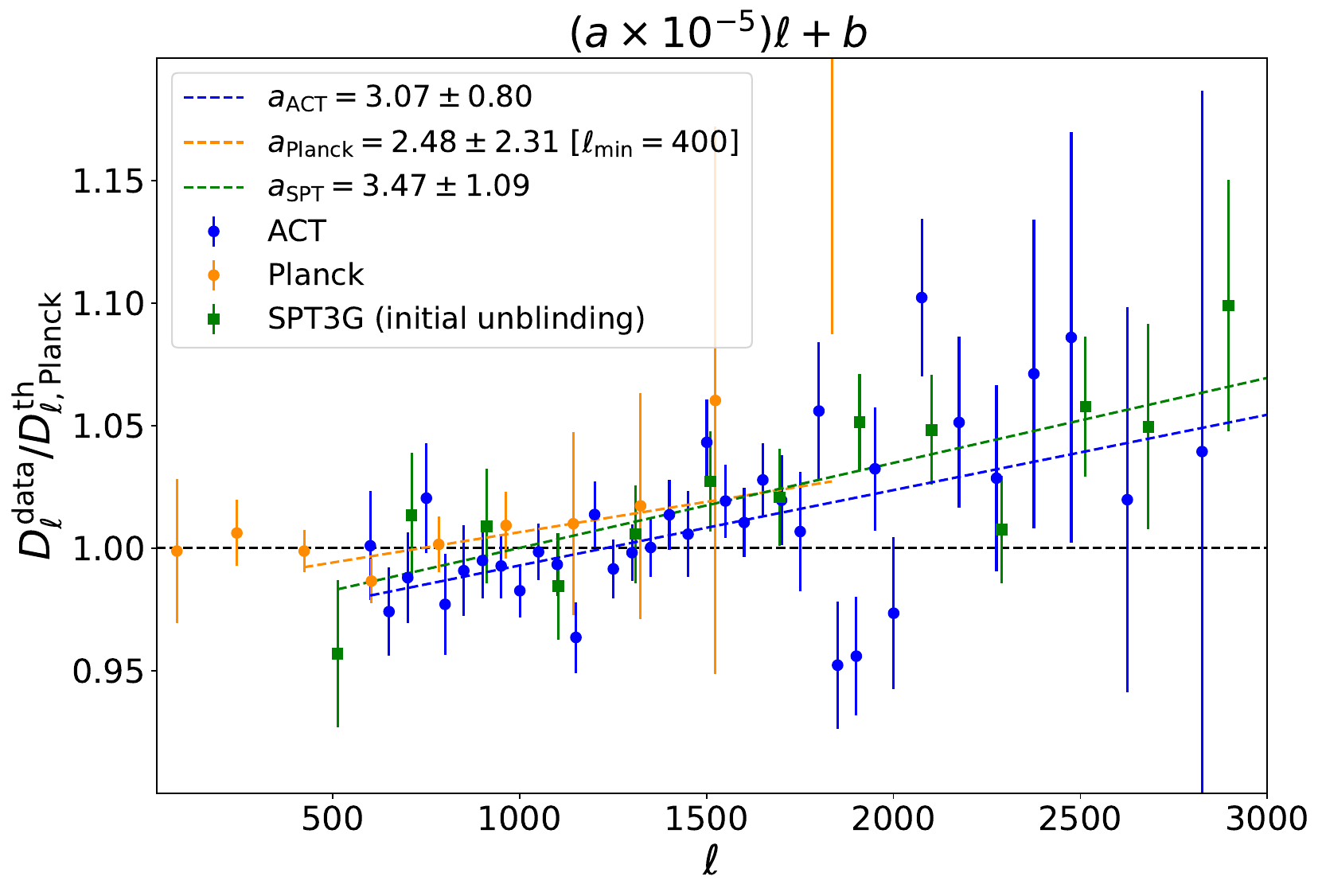}
    \caption{Zoom-in on the top EE panel of Figure \ref{fig:ratio_bf}, showing the ratio of the ACT and \Planck\ EE power spectra with respect to the \Planck\ TT-TE-EE best-fit cosmological model. We also include SPT-3G data taken from a figure in  \cite{2024arXiv241106000G}, corresponding to the pre-unblinding E-mode measurement described in that analysis. To guide the eye, we fit this ratio with a linear model, finding a consistent slope for the ACT (baseline) and SPT-3G (initial unblinding) data. We note that the \Planck\ EE data, fitted from $\ell_{\rm min} = 400 $ which roughly matches the $\ell_{\rm min}$ of ground based experiments, do not disfavor this slope. We can also quantify the degree of consistency of the ACT EE data and the \Planck\ data at the parameter level (i.e., taking into account \Planck\ uncertainties). Using the same 4-dimensional parameter space as in Figure \ref{fig:dr6_vs_plancks}, we find the difference between \LCDM\ parameters fit only to ACT EE data, compared to those from the same model fit to all \Planck\ TT/TE/EE data, is $2.3\sigma$.}
\label{fig:ratio_bf_EE_with_SPT3G}
\end{figure}

In Figure \ref{fig:ratio_bf}, we show the ratio of the ACT and \Planck\ temperature and E-mode power spectra with respect to the best-fit models from \Planck, ACT, and P-ACT. For \Planck, we use a rebinned version of the CMB-only temperature power spectrum. We find that the two measurements are in good agreement on the scales they have in common.

However, both the \Planck\ and ACT best-fit model parameters fail to accurately fit the other dataset on scales outside their respective measurement ranges. Specifically, the \Planck\ best-fit model mildly underestimates the small-scale power measured by ACT, while the ACT best-fit model underestimates the large-scale power measured by \Planck. Such a finding is not surprising, as the best-fit models derived from each experiment do not propagate the uncertainties associated with parameter determination. These limitations can lead to noticeable mismatches when extrapolating a model beyond the scales that dominate its constraints. In contrast, the P-ACT \LCDM\ best-fit model, which combines information from both datasets, effectively captures the features of both experiments and provides a good fit across all scales.

In Figure \ref{fig:ratio_bf_EE_with_SPT3G}, which is a zoom-in version of the top EE panel of Figure \ref{fig:ratio_bf}, we also include data taken from a
figure in \cite{2024arXiv241106000G}, an SPT-3G analysis. The figure shows the ratio of the SPT-3G initial unblinding E-mode spectrum measurement
described in \cite{2024arXiv241106000G}, to the \Planck\ best-fit spectrum. In that analysis the E-mode power spectrum was later corrected by introducing nuisance parameters to account for a potential mismatch between the SPT-3G temperature and polarization beams. While we are cautious about drawing strong conclusions, we note that the shape observed in the SPT-3G initial unblinding data, with respect to the \Planck\ $\Lambda$CDM cosmology, shows some similarities to the data independently observed by ACT DR6.

\section{Updates with DESI DR2 data}
\label{apx:desi}

The DESI DR2 baryon acoustic oscillation (BAO) data appeared at the same time as this work was completed. In this Appendix we report on tests of consistency of the ACT DR6 data with the DESI DR2 BAO data \citep{DESI1_DR2:2025,DESI2_DR2:2025}, and updated joint constraints. 

We estimate parameters from the P-ACT data combination, and from the DESI DR2 BAO data, to test for consistency. We find that \LCDM\ parameters agree at the 1.6$\sigma$ significance level, quantified by the Gaussian separation in 2-dimensional space of the $\Omega_m -hr_d$ parameter combination. This is in agreement with results reported in \cite{garcia-quintero:2025}. This agreement is somewhat closer than seen in \cite{DESI2_DR2:2025} when using the Planck PR4 likelihood \citep{rosenberg:2022} as the CMB dataset, and comes from the lower $\Omega_m$ estimated from P-ACT. This in turn is connected to the $n_s-\Omega_bh^2$ degeneracy-breaking that the new ACT small-scale data provide, as shown in Figure \ref{fig:dr6_vs_plancks}.

\begin{figure*}[htp]
	\centering
	\hspace*{-5mm}
\includegraphics[width=0.4\textwidth]{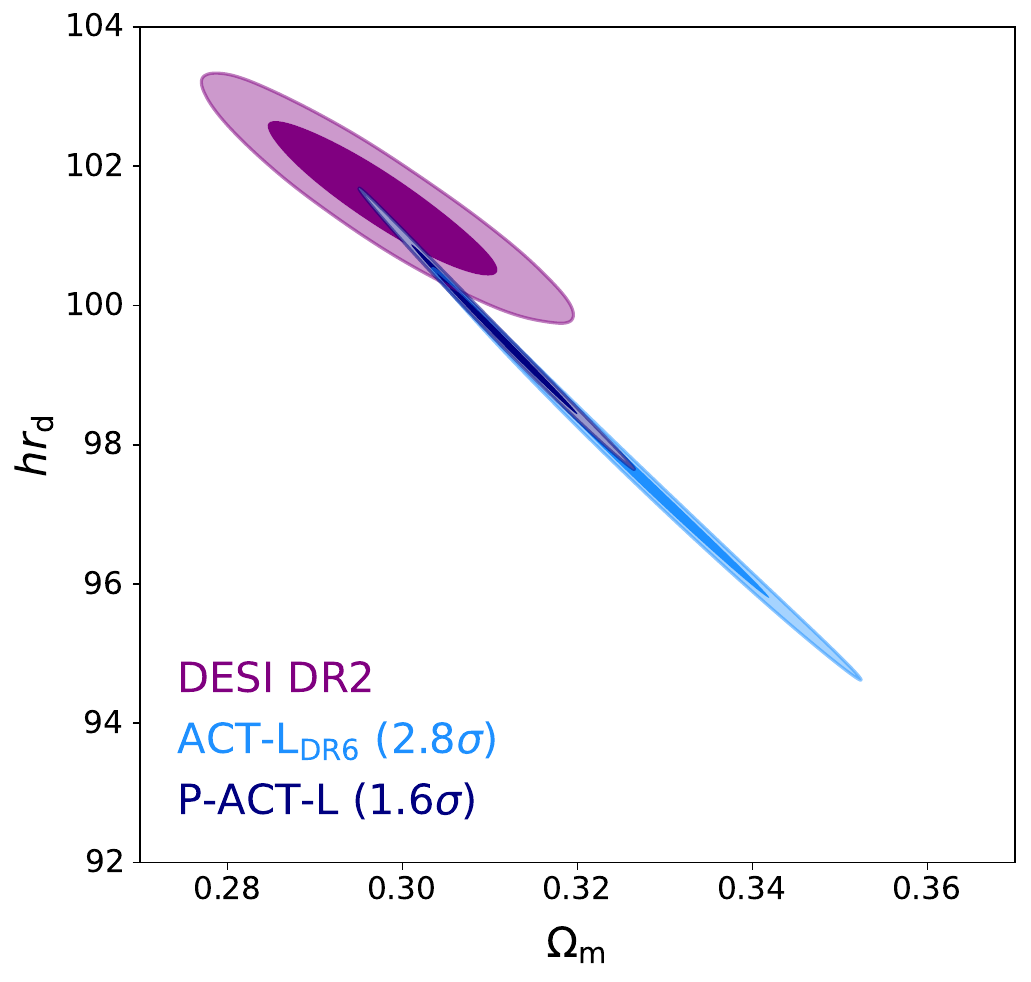}
    \caption{\tl{Comparison of parameters from different CMB data combinations and DESI DR2 in the $h r_{d}-\Omega_{m}$ plane. Here the CMB data include the temperature, polarization and lensing power spectra. P-ACT is broadly consistent with the DESI DR2 result (at the 1.6$\sigma$ level for P-ACT or P-ACT-L), while ACT-L and DESI DR2 differ at the $2.8 \sigma$ level. This difference is partly driven by the $\tau$ measurement used in our analysis. Without the \texttt{sroll2} measurement, the parameters from ACT broaden in this space. }}
	\label{fig:cmb_bao}
\end{figure*}

When using the ACT data without the Planck data, but still including the Planck \texttt{sroll2} measurement of the large-scale EE optical depth, we find that \LCDM\ parameters differ by $2.8\sigma$ to those preferred by DESI DR2, confirming results reported in \cite{garcia-quintero:2025}. This can be seen in Figure \ref{fig:cmb_bao}. Without the \texttt{sroll2} measurement, the parameters from ACT-alone would broaden in this space. Any viable model should be able to fit both the ACT data and the large-scale CMB data from WMAP or Planck. Figure \ref{fig:dr6_vs_plancks} shows that ACT and the `Planck$_{\rm cut}$' data used in P-ACT are themselves consistent, motivating their combination. 

\begin{figure*}[htp]
	\centering
	\hspace*{-5mm}
    \includegraphics[width=0.8\textwidth]{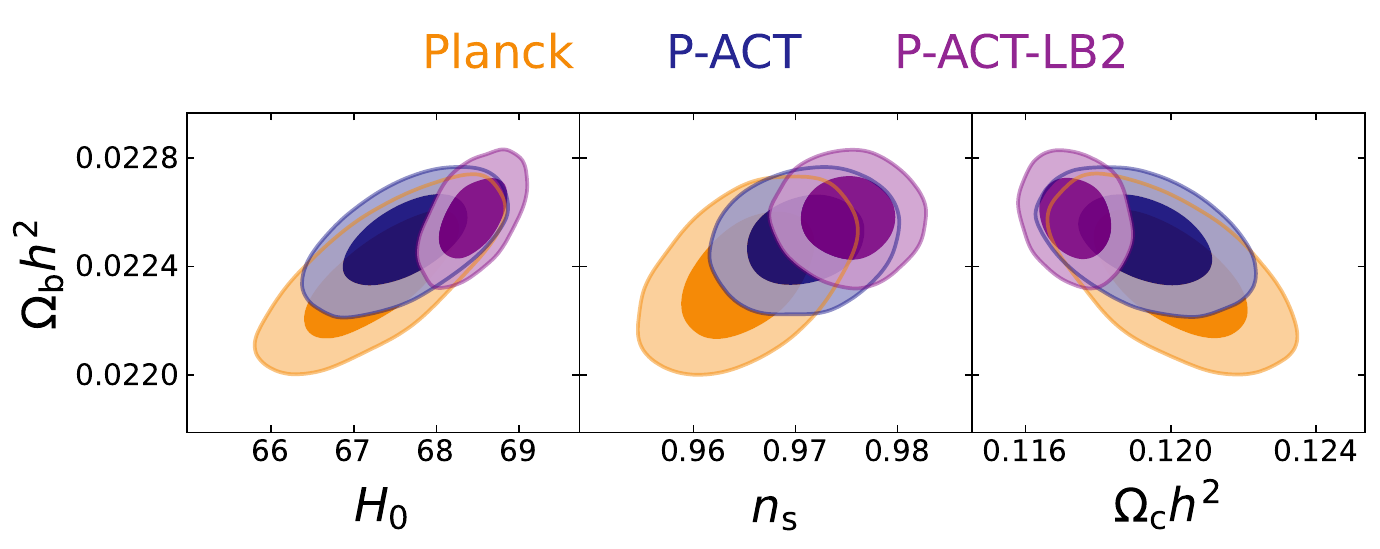}
    \centering  \includegraphics[width=1.0\textwidth]{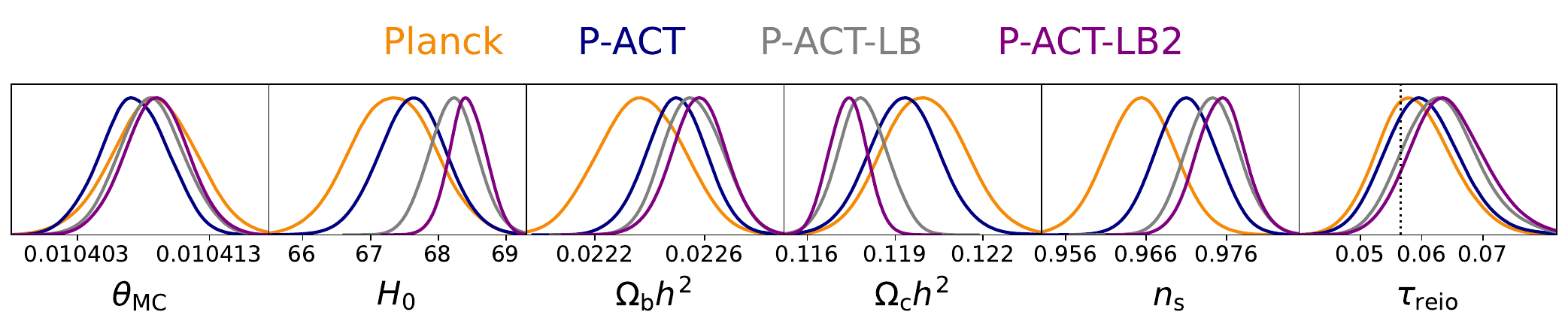}
    \caption{\jd{Update of the lower panel of Figure \ref{fig:LCDM} (top), and Figure \ref{fig:LCDM_2d} (bottom), for P-ACT combined with CMB lensing from ACT and \Planck, and baryon acoustic oscillation data from DESI DR2 (P-ACT-LB2). We additionally show the optical depth in the lower panel, to demonstrate how the combination with DESI data prefers an optical depth whose mean value is $1.3\sigma$ higher (for DR2; $1.1\sigma$ for DR1) than that estimated only from the \texttt{sroll2} EE data.}}
	\label{fig:LCDM_DR2}
\end{figure*}

When combining P-ACT and DESI DR2, we find the constraints shown in Figure \ref{fig:LCDM_DR2} and reported in Table \ref{tab:lcdm_params}. The joint model has a PTE of 0.13, indicating a good fit to the data. Following studies in \cite{sailer/etal:2025,jhaveri/etal:2025} we note that the best-fitting optical depth, $\tau$, is $1.3\sigma$ higher for \pactlbb\ (and $1.1\sigma$ using \pactlb) than the mean optical depth estimated from \texttt{Sroll2} using large-scale EE data from Planck, as shown in Figure \ref{fig:LCDM_DR2}. C25 investigates the impact on extended cosmological models of doubling the uncertainty on the large-scale EE measurement from \texttt{Sroll2}.

\color{black}{}
\collaboration{175}{}
\allauthors
\end{document}